\documentclass[preprint,publish,3p]{elsarticle}

\usepackage{graphicx}
\usepackage{subfigure}
\usepackage{float}
\usepackage{amsmath,amsthm,amssymb}
\usepackage{color}

\usepackage{natbib}

\biboptions{sort&compress}
\journal{Journal of Computational Physics}

\begin{document}

\begin{frontmatter}



\title{A coarse-grid projection method for accelerating incompressible flow computations}


\author{Omer San\corref{cor1}}
\ead{omersan@vt.edu}
\cortext[cor1]{Corresponding author.}
\author{Anne E. Staples}

\address{Department of Engineering Science and Mechanics \\ Virginia Tech, Blacksburg, VA 24061, USA}

\begin{abstract}
We present a coarse-grid projection (CGP) method for accelerating incompressible flow computations, which is applicable to methods involving Poisson equations as incompressibility constraints. The CGP methodology is a modular approach that facilitates data transfer with simple interpolations and uses black-box solvers for the Poisson and advection-diffusion equations in the flow solver. After solving the Poisson equation on a coarsened grid, an interpolation scheme is used to obtain the fine data for subsequent time stepping on the full grid. A particular version of the method is applied here to the vorticity-stream function, primitive variable, and vorticity-velocity formulations of incompressible Navier-Stokes equations. We compute several benchmark flow problems on two-dimensional Cartesian and non-Cartesian grids, as well as a three-dimensional flow problem. The method is found to accelerate these computations while retaining a level of accuracy close to that of the fine resolution field, which is significantly better than the accuracy obtained for a similar computation performed solely using a coarse grid. A linear acceleration rate is obtained for all the cases we consider due to the linear-cost elliptic Poisson solver used, with reduction factors in computational time between 2 and 42. The computational savings are larger when a suboptimal Poisson solver is used. We also find that the computational savings increase with increasing distortion ratio on non-Cartesian grids, making the CGP method a useful tool for accelerating generalized curvilinear incompressible flow solvers.
\end{abstract}


\end{frontmatter}


\section{Introduction}
\label{sec:intro}

Computational studies of incompressible flow problems are important in basic scientific research, and for a multitude of engineering applications. In the decades since the first incompressible flow computations were performed, many successful algorithms have been proposed for computing these flows \citep{kwak2010computation,hafez2003numerical,temam2001navier,ferziger1999computational,moin1998direct,quartapelle1993numerical}. Incompressible flows are fluid flows in which the flow speed is small compared to the speed of sound. Interestingly, this simple condition introduces some computational difficulties. One basic approach to solving incompressible flow problems is to retain the full compressible form of the Navier-Stokes and continuity equations (e.g., artificial compressibility methods \citep{chorin1967numerical,choi1985application,soh1988unsteady,rogers1991upwind,mateescu1994time,tang2007fractional}).
The other basic approach is to use pressure-based methods in which the Navier-Stokes equations are simplified by treating the fluid density as constant. In this approach a different difficulty arises: the continuity equation no longer involves the time derivative of the density, which was previously used to calculate the pressure field from an equation of state, so there is no longer an independent equation for the pressure \citep{kwak2005computational}. The solution, then, is to construct the pressure field to guarantee that the continuity equation is satisfied (e.g., the vorticity-stream function formulation \citep{arakawa1966computational,tezduyar1990solution}, the marker and cell method \citep{harlow1965numerical}, or projection algorithms for the primitive variable formulation of the problem \citep{chorin1968numerical,kim1985application,brown2001accurate}). This is accomplished by solving a Poisson equation to find the pressure field \citep{chorin1968numerical,kim1985application,bell1989second,choi1994effects,zang1994non,jordan1996efficient,blasco1998fractional,strikwerda1999accuracy,yanwen1999numerical,brown2001accurate,kiris2001numerical,liu2001gauge,guermond2006overview,matsumoto2006fractional}.

This latter approach has been given much attention in the literature \citep{kwak2005computational}, and is the approach we focus on here. The computational cost per time step of the pressure-based methods is that of solving a vector-valued parabolic-type advection-diffusion equation and a scalar-valued elliptic-type Poisson equation. The number of Poisson equations that must be solved at each time step varies with the method and the problem dimensions, but for all the pressure-based methods, solving the Poisson equation takes considerably more computational resources than solving the advection-diffusion time dependent equations, especially for large scale problems and high Reynolds number flows \citep{guermond2006overview}.

One straightforward way to accelerate incompressible flow simulations is to reduce the number of grid points for the most time consuming part of the problem, the elliptic solver. The coarse-grid projection (CGP) framework was first proposed by Lentine \emph{et al.} \cite{lentine2010anovel} and successfully applied to three-dimensional flow simulations for computer games using finite volume methods on unstructured grids. Here, we apply the CGP method to the primitive variable fractional step, the vorticity-stream function, and the vorticity-velocity formulations for incompressible flow problems using finite differences on structured grids to analyze the accuracy and efficiency of the CGP framework for several benchmark flows in two and three dimensions. We also solve flow problems on distorted grids, and on generalized curvilinear grids. The cost of the flow computations is reduced by coarsening the resolution of the numerical grid on which the Poisson equation is solved by factors of two in each direction according to $M = 2^{-\ell}N$, where $N$ is the fine resolution of the numerical grid on which the advection-diffusion part of the problem is solved, and $M$ is the coarse resolution for the solution of the Poisson equation. When $\ell=0$ no coarsening is applied and the CGP method reduces to the underlying incompressible flow solver method. For just one level of coarsening, in which $\ell=1$, and $M=N/2$, we have found that, remarkably, there is no significant resulting loss of accuracy in the fine resolution field data. For each subsequent level of coarsening investigated ($\ell=2$ and $\ell=3$) there is a further gain in computational speed-up, and an associated reduction in the accuracy of the fine resolution field data. However, the results obtained by the CGP method are more accurate than those of the standard coarse simulations for all the cases. In the current study, the third-order Runge-Kutta \citep{gottlieb1998total,shu2003high} and the second-order central difference schemes are used for temporal and spatial discretizations, respectively. The method is general in nature and can be applied to any Poisson equation-based incompressible Navier-Stokes solver. The results demonstrate that it is possible to obtain high-accuracy fine resolution data at the price of a mostly coarse computation. The method can easily be applied to high Reynolds number turbulence simulations in three dimensions, where we expect the saving in computational time to be significant.

One of the important aspects of the proposed method is its flexibility and its ease of use with existing incompressible flow solvers. The method is independent of the choice of Poisson solver and the choice of solver used for the advection-diffusion part of the governing equations. Linear- or quadratic-rate acceleration can be obtained, depending on the choice of Poisson solver. In this study we use two types of efficient linear-cost Poisson solvers, the V-cycle multigrid iterative solver and the fast Fourier transform (FFT) based direct solver, as our black-box Poisson solvers. The computational savings reported in this study are greater for flow solvers that use suboptimal, quadratic-cost Poisson solvers, as is demonstrated in one of the benchmark problem in this study.

This paper is organized as follows: the mathematical models including the primitive variable formulation, the vorticity-stream function formulation, and the vorticity-velocity formulation of Navier-Stokes equations are presented in Section \ref{sec:models}. In Section \ref{sec:cgp}, the CGP method is developed and the joint flow solver algorithms are presented. In Section \ref{sec:results}, the CGP flow solver algorithms are applied to several different two-dimensional benchmark flow problems: the Taylor-Green decaying vortex problem, the evaluation of a double shear layer, the merging of a pair of co-rotating vortices, two-dimensional decaying turbulence, the Taylor-Green vortex problem on a distorted grid, and laminar flow over a circular cylinder. This section also includes the application of the CGP method to three-dimensional Taylor-Green vortex flows, in which vortex stretching and tilting occurs. The results are compared to results obtained by performing the calculations using the basic flows solvers alone to test the validity of the CGP framework. Speed-ups in computational time ranging from 2 to 42 are found. Final conclusions and some comments about the effectiveness and applicability of the CGP method are presented in Section \ref{sec:concl}.

\section{Governing equations}
\label{sec:models}
\subsection{Primitive variable (PV) formulation}
\label{sec:pe}
The primitive variable formulation of the governing equations for incompressible viscous flows in dimensionless form with index notation is:
\begin{eqnarray}
\frac{\partial u_{i}}{\partial t}+\frac{\partial u_{i}u_{j}}{\partial x_{j}}&=&-\frac{\partial p}{\partial x_{i}} + \frac{1}{Re}\frac{\partial^{2} u_{i}}{\partial x_{j} \partial x_{j}}\\
\frac{\partial u_{j}}{\partial x_{j}}&=&0
\label{eq:mom}
\end{eqnarray}
where $Re$ is the Reynolds number, $u_{i}$ is the velocity component in the $i$th direction, and $p$ is the pressure. We use the fractional step procedure \citep{chorin1968numerical,kim1985application,brown2001accurate}, in which the first (predictor) step is to solve the advection-diffusion equation to obtain an estimate of the velocity field that does not satisfy the incompressibility condition. Next, the pressure (or pressure-like quantities in other formulations) is computed by solving a Poisson equation for which the estimated velocity field supplies the source term. Finally, a pressure correction is applied, and the resulting velocity field is a divergence-free solution of Navier-Stokes equations. In the predictor step, the pressure is neglected, and the intermediate velocities are computed by integrating the momentum equations as follows:
\begin{equation}
\frac{\partial \tilde{u}_{i}}{\partial t}+\frac{\partial \tilde{u}_{i}\tilde{u}_{j}}{\partial x_{j}}= \frac{1}{Re}\frac{\partial^{2} \tilde{u}_{i}}{\partial x_{j} \partial x_{j}}.
\label{eq:ad-dif}
\end{equation}
The Poisson equation for the pressure becomes:
\begin{equation}
\frac{\partial^{2} p}{\partial x_{i} \partial x_{i}} = f(\tilde{u}_{i})
\label{eq:poisson}
\end{equation}
where $f$ is a function of the intermediate velocity field. Finally, the corrected velocity field is calculated via:
\begin{equation}
u_{i} = g(\tilde{u}_{i}, p)
\label{eq:a1}
\end{equation}
where $g$ is another function of the computed pressure and the intermediate velocity field. There are several variations of the procedure given above. Here, we use a formulation based on the third-order Runge-Kutta algorithm, which we present in detail in Section \ref{sec:pe-alg}.

\subsection{Vorticity-stream function (VS) formulation}
\label{sec:vs}
An alternative to the fractional step procedure in the primitive variable formulation is the vorticity-stream function formulation, which is obtained by taking the curl of the momentum equations \citep{quartapelle1993numerical,majda2002vorticity}. Specifically, for two-dimensional incompressible flows, the dimensionless form of vorticity-stream function formulation is:
\begin{equation}
\frac{\partial \omega}{\partial t} + \frac{\partial \psi}{\partial y}\frac{\partial \omega}{\partial x} - \frac{\partial \psi}{\partial x}\frac{\partial \omega}{\partial y} = \frac{1}{Re}(\frac{\partial^2 \omega}{\partial x^2} + \frac{\partial^2 \omega}{\partial y^2}).
\label{eq:ge}
\end{equation}
This equation is used in conjunction with the kinematic relationship between the vorticity and the stream function, which has the form of a Poisson equation:
\begin{equation}
\frac{\partial^2 \psi}{\partial x^2} + \frac{\partial^2 \psi}{\partial y^2} = -\omega
\label{eq:ke}
\end{equation}
where $\omega$ is the vorticity (defined in Eq.~\ref{eq:compw} as $\omega_z$), and $\psi$ is the stream function. There are several advantages to the vorticity-stream function formulation for two-dimensional flows. Not only has the pressure been removed from the system of equations, and continuity implicity satisfied by the definition of stream function, but also it eliminates the possible projection inaccuracies of the fractional step formulations. The vorticity-stream function formulation also allows one to reduce the number of equations to be solved.

\subsection{Vorticity-velocity (VV) formulation}
\label{sec:vv}
The Navier-Stokes equations can also be written in the vorticity-velocity formulation \citep{fasel1990numerical,fasel2002numerical}. It consists of three equations, one for each of the three components of vorticity:
\begin{equation}
\frac{\partial \boldsymbol \omega}{\partial t} + \boldsymbol u \cdot \nabla \boldsymbol \omega = \boldsymbol \omega \cdot \nabla \boldsymbol u + \frac{1}{Re} \nabla^2 \boldsymbol\omega
\label{eq:vv}
\end{equation}
where the vorticity field is defined as the curl of velocity field, $\boldsymbol \omega = \nabla \times \boldsymbol u$. The components of the vorticity vector are:
\begin{equation}
\omega_x =\frac{\partial w}{\partial y} -\frac{\partial v}{\partial z}, \quad
\omega_y =\frac{\partial u}{\partial z} -\frac{\partial w}{\partial x}, \quad
\omega_z =\frac{\partial v}{\partial x} -\frac{\partial u}{\partial y}
\label{eq:compw}
\end{equation}
where $u$, $v$, and $w$ are the Cartesian velocity components. In addition, there are three Poisson equations to compute these velocity components:
\begin{eqnarray}
\frac{\partial^2 v}{\partial x^2} + \frac{\partial^2 v}{\partial y^2}  + \frac{\partial^2 v}{\partial z^2} &=&  \frac{\partial \omega_z}{\partial x} -\frac{\partial \omega_x}{\partial z}\\
\frac{\partial^2 u}{\partial x^2} + \frac{\partial^2 u}{\partial z^2} &=&  \frac{\partial \omega_y}{\partial z} -\frac{\partial^2 v}{\partial x \partial y} \\
\frac{\partial^2 w}{\partial x^2} + \frac{\partial^2 w}{\partial z^2} &=& -\frac{\partial \omega_y}{\partial x} -\frac{\partial^2 v}{\partial y \partial z}.
\label{eq:psys}
\end{eqnarray}
In the context of the current study, the fractional step primitive variable formulation, the vorticity-stream function formulation, and the vorticity-velocity formulation have in common an elliptic inversion sub-problem that needs to be solved at every time step that takes most of the computational effort within that step.

\section{Coarse-grid projection (CGP) methodology}
\label{sec:cgp}
In almost all Poisson equation-based incompressible flow solvers, solving the Poisson equation takes considerably more computational time than solving the advective-diffusive time dependent part of the problem. Within each time step, solving the advective-diffusive part of the problem is usually of $O(N_t)$ where ``$N_t$" is the number of degrees of freedom (total grid points, $N^2$ for two-dimensional problems and $N^3$ for three-dimensional problems) of the problem. In general, the alternating direction implicit (ADI), Gauss-Seidel (GS) or  successive over relaxation (SOR) types of iterative algorithms for solving the Poisson equation are of $O(N_{t}^{2})$ \citep{saad2003iterative}. The practical consequence is that it is not feasible to use these types of iterative Poisson solvers for high resolution (and therefore high Reynolds number) computations along with long time integration. In order to accelerate these solvers, very successful multigrid algorithms have been developed that reduce the computational effort to close to $O(C_{MG}N_t)$ where $C_{MG}$ is a proportionality constant \citep{saad2003iterative,wesseling1995introduction,gupta1997comparison,zhang1998fast}. And for certain ideal problems on equally-spaced grids, fast Fourier transform (FFT)-based fast Poisson solvers (FPS) can be used that are $O(C_{FFT}N_t log(N_t))$, and are presently the fastest algorithms ($C_{FFT}log(N_t)<C_{MG}$ in the relevant resolutions) for solving Poisson equations \citep{moin2001fundamentals}. Since we design our CGP framework for structured grid problems on either Cartesian or curvilinear grids, iterative elliptic solvers based on Krylov subspace methods such as the GMRES method used in Lentine \emph{et al.} \cite{lentine2010anovel} have been not included in our analysis.

\begin{table*}
\caption{Computational efficiency of Poisson solvers. CPU times (seconds) for solving one Poisson equation.}
\begin{center}
\label{tab:ps}
\begin{tabular}{lccccccc}
  \hline\noalign{\smallskip}
  Poisson Solver & $32^2$ & $64^2$ & $128^2$ & $256^2$ & $512^2$ & $1024^2$ & $2048^2$ \\
  \hline
  Gauss-Seidel & 0.093 & 1.582 & 26.15 &  & & & \\
  ADI & 0.085 & 1.339 & 22.11 &  & & & \\
  SOR & 0.009 & 0.182 & 3.131 &  & & & \\
  V-Multigrid & 0.002  & 0.01 & 0.041 & 0.175 & 0.869& 3.979& 14.71 \\
  FFT-FPS     & 0.001 & 0.003   & 0.009 & 0.037 & 0.149 & 0.659& 2.781 \\
  \hline
\end{tabular}
\end{center}
\end{table*}

\begin{figure}[h]
\centering
\includegraphics[width=0.6\textwidth]{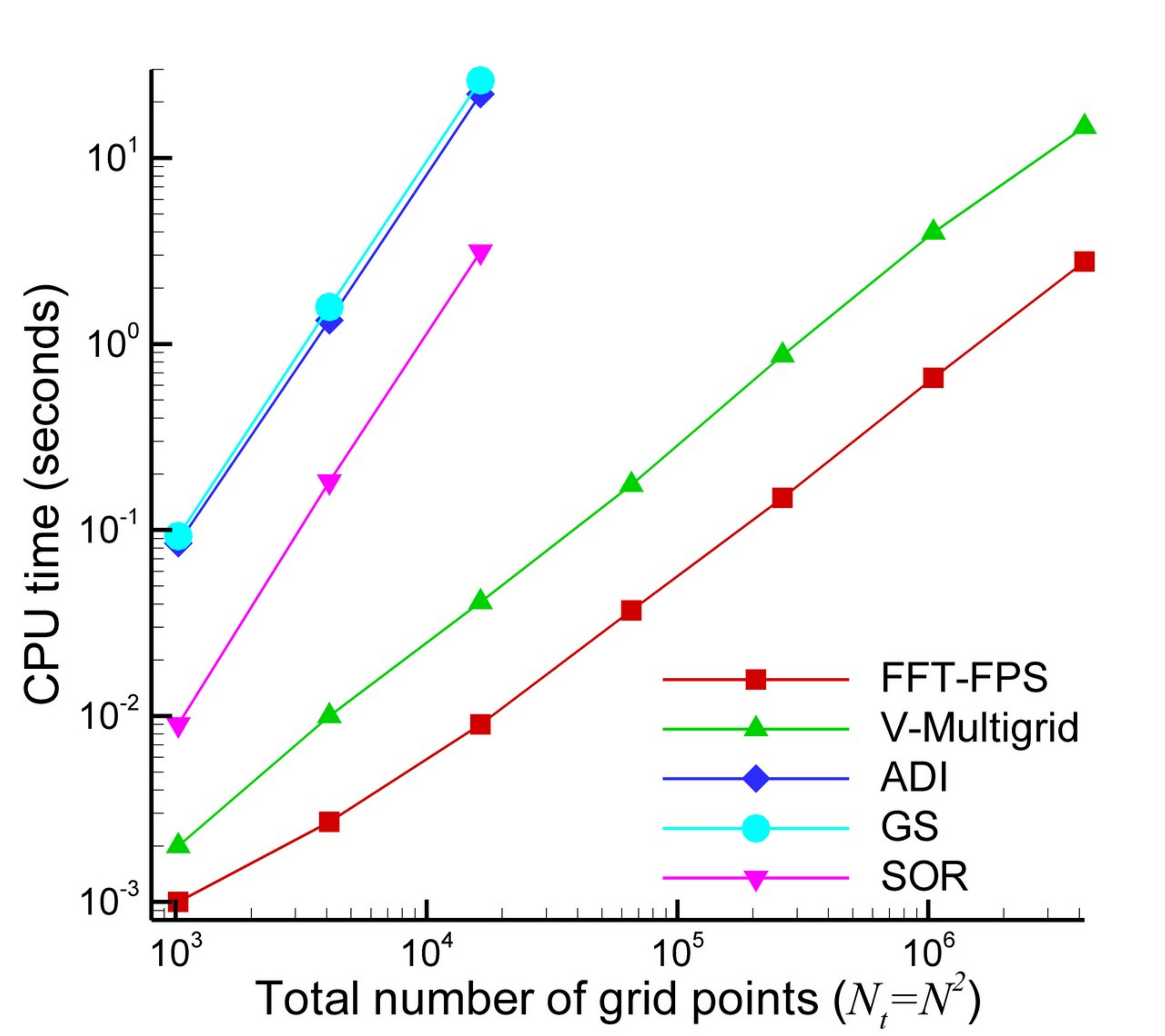}
\caption{Efficiency of Poisson solvers. ADI, GS, and SOR are classical iterative solvers that scale as $N_{t}^{2}$. FFT-FPS is the fast Fourier transform based direct solver, and V-Multigrid is the V-cycle iterative multigrid solver, which both scale as $N_t$.}
\label{fig:psolver}
\end{figure}

The computational efficiencies of different Poisson solvers are tabulated in Table~\ref{tab:ps} for a square domain with equidistant grid spacing. The computational time for solving just one Poisson equation is also illustrated in Fig.~\ref{fig:psolver}. This preliminary comparison shows that FFT-FPS is the most efficient solver for most of our test problems. Therefore, we use FFT-FPS for all cases except the distorted grid Taylor-Green vortex problem where mixed derivatives occur in the elliptic equations in the transformed generalized coordinates. The V-cycle multigrid Poisson solver is used instead for the distorted grid computations. In the multigrid solver, the Poisson equation is solved in such a way that the error in the residual is linearly smoothed. Therefore, it is an optimal linear-cost iterative solution technique for the elliptic sub-problem. Similar to the multigrid approach, in our CGP approach, we use prolongation and restriction operators between the data for the advection-diffusion part and the elliptic part of the problem. This seems to be effectively a low-pass filter on the solution to the Poisson equation. In addition to different prolongation and restriction operators, the approach presented in this study differs from Lentine \emph{et al.} \cite{lentine2010anovel}, in which the restrictions and prolongations operate on velocities, whereas in our implementation they operate on pressures.

The basic idea behind coarse-grid projection (CGP) is to use a smaller number of grid points for solving the elliptic sub-problem. The CGP approach we propose here is modular and independent of the Poisson solver that is used. Usually fast Poisson solvers are optimal for rectangular domain problems because they are fast and have minimal storage requirements. They are preferred because of their efficiency when using orthogonal coordinate systems in which there are no mixed derivatives. The general domains treated with body-fitted coordinates are out of bounds for fast Poisson solvers due to the presence of mixed derivatives in the transformed generalized curvilinear coordinates. In that case, the V-cycle multigrid solver becomes the optimal Poisson solver to use, and we use it for our benchmark problems on distorted grids. On the other hand, for the test case of flow over a circular cylinder, due to the O-grid system, we use a fast Poisson solver in which we utilize FFTs in the circumferential direction and the Thomas algorithm in radial direction, and the result is a linear-cost fast Poisson solver \cite{moin2001fundamentals}.

Fast Poisson solvers can be implemented in our computations in two different ways. One way is to perform FFTs on the equation directly which results in a spectral accuracy for the Poisson equation, and the other is to discretize the Poisson equation first and then apply FFTs, which results in the same spatial order of accuracy as the underlying finite difference scheme (second-order, in this case). According to our preliminary computations for incompressible flows, both strategies give the same results. For example, to invert a two-dimensional Poisson equation, $\nabla^2 u =f$, in a uniform rectangular domain, the second-order scheme is:
\begin{equation}
\frac{u_{i-1,j}-2u_{i,j}+u_{i+1,j}}{\Delta x^2} + \frac{u_{i,j-1}-2u_{i,j}+u_{i,j+1}}{\Delta y^2} = f_{i,j}
\label{eq:dis-poisson}
\end{equation}
where the subscripts represent the grid indices, and $\Delta x$ and $\Delta y$ are the grid spacing in $x$ and $y$ directions, respectively. The three-step procedure for the FPS is \cite{press1992numerical}:
\begin{enumerate}
\item[(i)]
Apply an inverse FFT to find the Fourier coefficients $\hat{f}_{i,j}$ from the grid values $f_{i,j}$
\item[(ii)]
Solve for the Fourier coefficients $\hat{u}_{i,j}$ according to:
\begin{equation}
\hat{u}_{i,j} = \frac{\hat{f}_{i,j} }{\frac{2}{\Delta x^2}\left[\mbox{cos}(\frac{2\pi i}{N_x}) + \lambda^2\mbox{cos}(\frac{2\pi j}{N_y}) -(1+\lambda^2) \right]}
\label{eq:fft}
\end{equation}
where $\lambda = \Delta x/\Delta y$, and $N_x$ and $N_y$ are the number of grid points in the $x$ and $y$ directions, respectively
\item[(iii)]
Apply a forward FFT to find the grid values $u_{i,j}$ from the Fourier coefficients $\hat{u}_{i,j}$
\end{enumerate}

Even though we use the fastest Poisson solver available for our computations, the most time consuming part of the computations is still the Poisson equation constraint to the velocity field (e.g., see Table~\ref{tab:cpu}). With this in mind, we propose a new multiresolution approach that reduces the number of degrees of freedom for the Poisson solver part of the problem to accelerate the computation. The procedure is as follows: first, we solve the advection-diffusion part of the problem using a fine resolution, $N$ (the resolution in one direction). Next, we restrict these data to a coarser grid with resolution $M = 2^{-\ell}N$, where $\ell$ is an integer that determines the level of coarsening ($\ell=0$ corresponds to no grid coarsening). After solving the Poisson equation on this coarser grid, we then perform a prolongation of the coarse data to the fine resolution grid for subsequent time stepping.
In the primitive variable formulation of fractional step method, the procedure is:
\begin{enumerate}
\item[(i)]
Compute intermediate velocities on fine grid using fractional step method
\item[(ii)]
Map intermediate velocity field data from fine grid to coarse grid to obtain source term for Poisson equation
\item[(iii)]
Solve pressure Poisson equation and correct velocity field so that continuity equation is satisfied on coarse grid
\item[(iv)]
Remap pressure field data from coarse grid to fine grid
\item[(v)]
Update velocities on fine grid, continue to (i) for next time step
\end{enumerate}
The procedures for the vorticity-stream function formulation and the vorticity-velocity formulation are similar to the primitive variable formulation, except that the underlying elliptic and advection-diffusion parts of the formulations are different as outlined in the preceding sections.

\subsection{Mapping operators}
\label{sec:operators}
The only modifications to the standard flow solver computational procedures are the mapping procedures from fine to coarse data and vice versa. In the current study, the full weighting averaging operation is used for data coarsening (restriction), which is given for a two-dimensional equally spaced data array as \citep{moin2001fundamentals}:
\begin{eqnarray}
\bar{\phi}_{i,j} &=&\frac{1}{16}[4\phi_{2i,2j} +2(\phi_{2i,2j-1}+\phi_{2i,2j+1}+\phi_{2i-1,2j}+\phi_{2i+1,2j}) \nonumber\\ &+&\phi_{2i+1,2j+1}+\phi_{2i+1,2j-1}+\phi_{2i-1,2j+1}+\phi_{2i-1,2j-1} ]
\label{eq:res}
\end{eqnarray}
where $\bar{\phi}_{i,j}$ and $\phi_{i,j}$ are the coarse and fine data arrays, and $i,j$ are the coarse grid indices. The extension to three dimensions is relatively straightforward. The full weighting operator for mapping from the fine scale to the coarse scale for three-dimensional data arrays is constructed by using weighting factors of eight, four, two, and one, depending of the location of the fine data \citep{zhang1998fast}. For almost all multiscale computations, the mapping from the coarse data to the fine data is a critical issue \citep{weinan2007heterogeneous,kevrekidis2009equation}. The bilinear interpolation procedure that we use is given for two-dimensional equally spaced grid as \citep{moin2001fundamentals}:
\begin{eqnarray}
\phi_{2i,2j}   &=& \bar{\phi}_{i,j} \nonumber \\
\phi_{2i+1,2j} &=& \frac{1}{2}(\bar{\phi}_{i,j}+\bar{\phi}_{i+1,j}) \nonumber \\
\phi_{2i,2j+1} &=& \frac{1}{2}(\bar{\phi}_{i,j}+\bar{\phi}_{i,j+1}) \nonumber \\
\phi_{2i+1,2j+1} &=& \frac{1}{4}(\bar{\phi}_{i,j}+\bar{\phi}_{i+1,j}+\bar{\phi}_{i,j+1}+\bar{\phi}_{i+1,j+1}).
\label{eq:pro}
\end{eqnarray}

The half mapping procedures given by Eq.~\ref{eq:res} for restriction and Eq.~\ref{eq:pro} for prolongation can be performed multiple times to obtain different levels of coarsening. The mapping procedure does not take significant computational time compared to the Poisson solver. The mapping procedures used in this study could potentially be improved by introducing higher-order spline formulas. Since we are using a second-order spatial discretization scheme, however, bilinear interpolation is suitable for this study. For three-dimensional grids, depending on the location, values of fine grid points are obtained by injection wherein the values of common points of the fine and coarse grid are directly transferred, and the values of the nearest two, four, or eight points on the coarse grid. The detailed formulas can be found in \citep{wesseling1995introduction}.

\subsection{CGP algorithm for the primitive variable formulation}
\label{sec:pe-alg}
The coarse-grid projection (CGP) method is independent from the time integration method used. It can be implemented using any time stepping algorithm, for example, the backward difference method, or a method from the Adams-Bashforth or Adams-Moulton families \citep{schafer2006computational}. Here, the CGP method is used in conjunction with the third-order Runge-Kutta method for the fractional step primitive variable formulation of the problem. The joint CGPRK3-PV algorithm is presented below. Starting with velocity field data, $u^{n}_{i}$, the velocity field is computed for the next time step according to the following sequence of computations:
\begin{eqnarray}
\tilde{u}^{(1)}_{i} &=& u^{n}_{i}+ \Delta t H^{n}_{i} \\
\tilde{u}^{(1)}_{i} &\Rightarrow&  \bar{\tilde{u}}^{(1)}_{i}\\
\frac{\partial^{2} \bar{p}}{\partial x_{i} \partial x_{i}} &=& \frac{1}{\Delta t} \frac{\partial \bar{\tilde{u}}^{(1)}_{i}}{\partial x_{i}} \\
p &\Leftarrow& \bar{p}\\
u^{(1)}_{i} &=& \tilde{u}^{(1)}_{i}- \Delta t \frac{\partial p}{\partial x_{i}} \\
\tilde{u}^{(2)}_{i} &=& \frac{3}{4}u^{n}_{i}+ \frac{1}{4}u^{(1)}_{i} + \frac{1}{4} \Delta t H^{(1)}_{i} \\
\tilde{u}^{(2)}_{i} &\Rightarrow& \bar{\tilde{u}}^{(2)}_{i}\\
\frac{\partial^{2} \bar{p}}{\partial x_{i} \partial x_{i}} &=& \frac{1}{(\frac{1}{4}\Delta t)} \frac{\partial \bar{\tilde{u}}^{(2)}_{i}}{\partial x_{i}} \\
p &\Leftarrow& \bar{p}\\
u^{(2)}_{i} &=& \tilde{u}^{(2)}_{i}- \frac{1}{4}\Delta t \frac{\partial p}{\partial x_{i}} \\
\tilde{u}^{(3)}_{i} &=& \frac{1}{3}u^{n}_{i}+ \frac{2}{3}u^{(2)}_{i} + \frac{2}{3} \Delta t H^{(2)}_{i} \\
\tilde{u}^{(3)}_{i} &\Rightarrow& \bar{\tilde{u}}^{(3)}_{i}\\
\frac{\partial^{2} \bar{p}}{\partial x_{i} \partial x_{i}} &=& \frac{1}{(\frac{2}{3}\Delta t)} \frac{\partial \bar{\tilde{u}}^{(3)}_{i}}{\partial x_{i}} \\
p &\Leftarrow& \bar{p}\\
u^{n+1}_{i} &=& \tilde{u}^{(3)}_{i}- \frac{2}{3}\Delta t \frac{\partial p}{\partial x_{i}}
\label{eq:fsCGP}
\end{eqnarray}
where $H_{i}$ is the combination of the convection and viscous diffusion terms:
\begin{equation}
H_{i} = -\frac{\partial u_{i}u_{j}}{\partial x_{j}}+\frac{1}{Re}\frac{\partial^{2} u_{i}}{\partial x_{j} \partial x_{j}}.
\label{eq:H}
\end{equation}
The CGP procedure for coarsening the grid on which the pressure Poisson equation is solved reduces the accuracy of the pressure field, as is to be expected. Any accuracy loss in the pressure field, however, does not propagate forward in time, and so a higher accuracy pressure field can be obtained at any time step by simply solving the Poisson equation on a fine grid according to:
\begin{equation}
\frac{\partial^2 p}{\partial x_{j} \partial x_{j}} = - \frac{\partial^2 u_{i}u_{j}}{\partial x_{i} \partial x_{j}}
\label{eq:fPo}
\end{equation}
which is obtained directly by taking the divergence of the momentum equation.

\subsection{CGP algorithm for the vorticity-stream function formulation}
\label{sec:vs-alg}
Similar to the algorithm for the fractional step primitive variable formulation presented in the previous section, we present here the CGP method used in conjunction with the third-order Runge-Kutta method for the vorticity-stream function formulation of the governing equations of incompressible fluid flows. Starting with the value of the vorticity, $\omega^{n}$, at the current time step, the joint CGPRK3-VS algorithm for computing the vorticity at the next time step, $\omega^{n+1}$, consists of the following steps:
\begin{eqnarray}
\omega^{n} &\Rightarrow& \bar{\omega}^{n} \\
\frac{\partial^2 \bar{\psi}^{n}}{\partial x^2} + \frac{\partial^2 \bar{\psi}^{n}}{\partial y^2} &=& -\bar{\omega}^{n} \\
\psi^{n} &\Leftarrow& \bar{\psi}^{n} \\
\omega^{(1)} &=& \omega^{n} + \Delta t G^{n} \\
\omega^{(1)} &\Rightarrow& \bar{\omega}^{(1)} \\
\frac{\partial^2 \bar{\psi}^{(1)}}{\partial x^2} + \frac{\partial^2 \bar{\psi}^{(1)}}{\partial y^2} &=& -\bar{\omega}^{(1)} \\
\psi^{(1)} &\Leftarrow& \bar{\psi}^{(1)} \\
\omega^{(2)} &=& \frac{3}{4}  \omega^{n} + \frac{1}{4} \omega^{(1)} + \frac{1}{4}\Delta t G^{(1)} \\
\omega^{(2)} &\Rightarrow& \bar{\omega}^{(2)} \\
\frac{\partial^2 \bar{\psi}^{(2)}}{\partial x^2} + \frac{\partial^2 \bar{\psi}^{(2)}}{\partial y^2} &=& -\bar{\omega}^{(2)} \\
\psi^{(2)} &\Leftarrow& \bar{\psi}^{(2)} \\
\omega^{n+1} &=& \frac{1}{3}  \omega^{n} + \frac{2}{3} \omega^{(2)} + \frac{2}{3}\Delta t G^{(2)}
\label{eq:TVDRK}
\end{eqnarray}
where
\begin{equation}
G = - \frac{\partial \psi}{\partial y}\frac{\partial \omega}{\partial x} + \frac{\partial \psi}{\partial x}\frac{\partial \omega}{\partial y} + \frac{1}{Re}(\frac{\partial^2 \omega}{\partial x^2} + \frac{\partial^2 \omega}{\partial y^2}).
\label{eq:jac}
\end{equation}

We favor the vorticity-stream function formulation for two-dimensional problems, especially for those having periodic boundary conditions. This particular formulation is an excellent model for testing the proposed CGP framework because of the absence of projection inaccuracies and errors due to boundary conditions.
\section{Results}
\label{sec:results}
To investigate the performance of the CGP method seven different flow problems are computed. First, four flow problems in periodic domains are computed: the Taylor-Green decaying vortex problem, the evaluation of a double shear layer, the merging of a pair of co-rotating vortices, and two-dimensional decaying turbulence. Problems with periodic boundary conditions are chosen both to eliminate any possible error arising from boundary condition implementations, and to be able to use fast Poisson solvers. Next, the Taylor-Green vortex problem is solved on a distorted grid, and the benchmark test problem of flow over a circular cylinder is studied to demonstrate the efficiency of CGP for non-Cartesian grid applications. Finally, the behavior of the method is tested for a three-dimensional problem, the Taylor-Green vortex problem, demonstrating the applicability of the method in three-dimensions, in the presence of vortex stretching and tilting.
\subsection{Taylor-Green vortex}
\label{sec:taylorgv}
In this section, the CGPRK3 algorithms in the vorticity-stream function formulation and primitive variable formulation are validated by solving the Taylor-Green decaying vortex problem in a two-dimensional square domain. The problem describes the two-dimensional, unsteady flow of a decaying vortex (or set of vortex arrays), and is an exact analytic solution of the unsteady, incompressible Navier-Stokes equations in Cartesian coordinates. The analytic solution in $[0, 2\pi]\times[0, 2\pi]$ domain with periodic boundary conditions is given for the velocity field by:
\begin{eqnarray}
u^{e}(x,y,t)&=&-\mbox{cos}(qx)\mbox{sin}(qy)\mbox{exp}(-2q^2t/Re)\\
v^{e}(x,y,t)&=& \mbox{sin}(qx)\mbox{cos}(qy)\mbox{exp}(-2q^2t/Re)
\label{eq:exv}
\end{eqnarray}
and for vorticity field it is given by:
\begin{equation}
\omega^{e}(x,y,t)=2q \mbox{cos}(qx)\mbox{cos}(qy)\mbox{exp}(-2q^2t/Re)
\label{eq:exw}
\end{equation}
where $q$ is an integer. We performed numerical simulations for $Re=10$ with both the CGPRK3-PV and CGPRK3-VS algorithms introduced earlier with $\Delta t=2.5\times10^{-4}$ and the Taylor array number, $q=1$. For the primitive variable formulation, the difference between the exact and computed solutions is:
\begin{equation}
e_{i,j}=|u^{e}_{i,j}-u_{i,j}|
\label{eq:erru}
\end{equation}
where $u$ is the $x$-component of the velocity. Similarly, for the vorticity-stream function formulation, the error in vorticity field is defined as:
\begin{equation}
e_{i,j}=|\omega^{e}_{i,j}-\omega_{i,j}|.
\label{eq:errw}
\end{equation}
In order to quantify the accuracy of the CGP method, we computed two different norms, the root mean squared $L_{2}$ norm:
\begin{equation}
L_{2}=\sqrt{\frac{1}{N_xN_y}\sum_{i=1}^{N_x} \sum_{j=1}^{N_y} e_{i,j}^{2}}
\label{eq:err}
\end{equation}
and the $L_{\infty}$ norm:
\begin{equation}
L_{\infty}=\mbox{Max}(e_{i,j}).
\label{eq:err}
\end{equation}
\begin{table*}
\caption{The computed error norms for the vorticity-stream function formulation algorithm (RK3/CGPRK3-VS) for $Re=10$, $\Delta t=2.5\times10^{-4}$ at time $t=1$. The first resolution number represents the number of grid points used for solving the advection-diffusion equation (vorticity-transport equation), and the second resolution number is the number of grid points used for solving the Poisson equation (kinematic relationship between vorticity and stream function). The speed-up ratio is defined as the ratio of the CPU times for the computations performed without using the CGP procedure (using the standard method) and with the CGP procedure. Results obtained using the SOR scheme for the Poisson equation are also included for comparison.}
\begin{center}
\label{tab:norm1}
\begin{tabular}{llcccc}
\hline\noalign{\smallskip}
Method & Resolution & $||\omega||_{\infty}$ & $||\omega||_{L_2}$ & CPU (s) & Speed-up \\
\hline
\emph{FFT-FPS }\\
RK3 ($\ell=0$)      & $512^2$ : $512^2$ & 4.1099E-6 & 2.0589E-6 & 2500.03 & 1.00 \\
CGPRK3 ($\ell=1$)   & $512^2$ : $256^2$ & 4.1099E-6 & 2.0590E-6 & 976.94  & 2.56 \\
CGPRK3 ($\ell=2$)   & $512^2$ : $128^2$ & 1.1188E-5 & 3.3678E-6 & 510.24 & 4.90 \\
CGPRK3 ($\ell=3$)   & $512^2$ : $64^2$  & 6.6467E-5 & 2.2245E-5 & 406.31 & 6.15 \\
RK3 ($\ell=0$)      & $256^2$ : $256^2$ & 1.6439E-5 & 8.2518E-6 & 522.67  & 1.00 \\
CGPRK3 ($\ell=1$)   & $256^2$ : $128^2$ & 1.6439E-5 & 8.2525E-6 & 218.85  & 2.39 \\
CGPRK3 ($\ell=2$)   & $256^2$ : $64^2$  & 7.6177E-5 & 2.2801E-5 & 132.42 & 3.95 \\
CGPRK3 ($\ell=3$)   & $256^2$ : $32^2$  & 5.1640E-4 & 1.7568E-4 & 114.09 & 4.58 \\
RK3 ($\ell=0$)      & $128^2$ : $128^2$ & 6.5755E-5 & 3.3132E-5 & 112.29 & 1.00 \\
CGPRK3 ($\ell=1$)   & $128^2$ : $64^2$  & 6.5755E-5 & 3.3176E-5 & 47.15  & 2.29 \\
RK3 ($\ell=0$)      & $64^2$  : $64^2$  & 2.6297E-4 & 1.3351E-4 & 27.10  & 1.00 \\
CGPRK3 ($\ell=1$)   & $64^2$  : $32^2$  & 2.6335E-4 & 1.3624E-4 & 12.16  & 2.23 \\
RK3 ($\ell=0$)      & $32^2$  : $32^2$  & 1.0511E-3 & 5.4149E-4 & 6.19  & 1.00 \\
\hline
\emph{SOR}\\
RK3 ($\ell=0$)      & $128^2$ : $128^2$ & 6.5756E-5 & 3.3133E-5 & 90334.46  & 1.00 \\
CGPRK3 ($\ell=1$)   & $128^2$ : $64^2$  & 6.5757E-5 & 3.3176E-5 & 7131.36  & 12.67 \\
RK3 ($\ell=0$)      & $64^2$ : $64^2$   & 2.6298E-4 & 1.3351E-4 & 7113.46  & 1.00 \\
CGPRK3 ($\ell=1$)   & $64^2$ : $32^2$   & 2.6346E-4 & 1.3626E-4 & 442.71  & 16.07 \\
RK3 ($\ell=0$)      & $32^2$ : $32^2$   & 1.0511E-3 & 5.4150E-4 & 431.58  & 1.00 \\
\hline
\end{tabular}
\end{center}
\end{table*}

\begin{table*}
\caption{The computed error norms for the primitive variable fractional step algorithm (RK3/CGPRK3-PV) for $Re=10$, $\Delta t=2.5\times10^{-4}$ at time $t=1$. The first resolution number is the number of grid points used for solving the advection-diffusion equations (auxiliary momentum equations in both directions), and the second resolution number is the number of grid points used for solving the pressure Poisson equation. The speed-up ratio is defined as the ratio of CPU times for the computations performed without using the CGP procedure (using standard methods) and with the CGP procedure.}
\begin{center}
\label{tab:norm2}
\begin{tabular}{llccccc}
\hline\noalign{\smallskip}
Method  & Resolution & $||u||_{\infty}$ & $||u||_{L_2}$ & CPU (s) & Speed-up\\
\hline
RK3 ($\ell=0$)     & $512^2$ : $512^2$ & 2.0550E-6 & 1.0247E-6 & 2915.10 & 1.00 \\
CGPRK3 ($\ell=1$)  & $512^2$ : $256^2$ & 2.0504E-6 & 1.0364E-6 & 1291.46 & 2.26 \\
CGPRK3 ($\ell=2$)  & $512^2$ : $128^2$ & 8.9268E-6 & 3.2635E-6 & 846.53 & 3.44 \\
CGPRK3 ($\ell=3$)  & $512^2$ : $64^2$  & 7.1230E-5 & 2.6292E-5 & 708.61 & 4.12 \\
RK3 ($\ell=0$)     & $256^2$ : $256^2$ & 8.2200E-6 & 4.0987E-6 & 661.86  & 1.00 \\
CGPRK3 ($\ell=1$)  & $256^2$ : $128^2$ & 8.6218E-6 & 4.0625E-6 & 314.74  & 2.10 \\
CGPRK3 ($\ell=2$)  & $256^2$ : $64^2$  & 6.7819E-5 & 2.5271E-5 & 227.65 & 2.91 \\
CGPRK3 ($\ell=3$)  & $256^2$ : $32^2$  & 6.4426E-4 & 2.1464E-4 & 196.49 & 3.68 \\
RK3 ($\ell=0$)     & $128^2$ : $128^2$ & 3.2878E-5 & 1.6393E-5 & 141.63  & 1.00 \\
CGPRK3 ($\ell=1$)  & $128^2$ : $64^2$  & 3.5194E-5 & 1.6566E-5 & 69.78  & 2.03 \\
RK3 ($\ell=0$)     & $64^2$  : $64^2$  & 1.3148E-4 & 6.5554E-5 & 33.35  & 1.00 \\
CGPRK3 ($\ell=1$)  & $64^2$  : $32^2$  & 1.9054E-4 & 8.5045E-5 & 16.72  & 2.00 \\
RK3 ($\ell=0$)     & $32^2$  : $32^2$  & 5.2546E-4 & 2.6192E-4 & 7.74   & 1.00 \\
\hline
\end{tabular}
\end{center}
\end{table*}

The computed $L_{2}$ and $L_{\infty}$ norms at time $t=1$ are tabulated in Table~\ref{tab:norm1} and Table~\ref{tab:norm2} for the vorticity-stream function and the primitive variable formulation solutions, respectively. Results obtained using the coarse-grid projection algorithm (CGPRK3) and the standard flow solver (RK3) without the CGP procedure are compared for different levels of coarsening. In these tables, the first resolution listed is for the fine grid on which the advection-diffusion part of the problem is solved, and the second value is the resolution of the coarse grid on which the Poisson equation is solved to constrain the fine scale field data: the vorticity field in the vorticity-stream function formulation, or the velocity field in primitive variable formulation. When these resolutions are the same, the standard method is recovered.

Three different levels of coarsening: half-coarsening ($\ell=1$, $M=N/2$), $1/4$-coarsening ($\ell=2$, $M=N/4$) and $1/8$-coarsening ($\ell=3$, $M=N/8$), were performed to investigate the behavior of the CGP method. The optimal results from the point of view of accuracy were obtained using the half-coarsening CGP method. For this level of coarsening, there is a significant reduction in computational time (by a factor of about 2.2) and no corresponding reduction in the accuracy of the flow field variables. Although it is fairly small, there is actually an increase in error from the $64^2$:$64^2$ resolution to the $64^2$:$32^2$ resolution case. For higher resolution, the pressure field is well resolved because the exact solution consists of two low frequencies. This contributes to the same level accuracy for one level of coarsening for higher resolution computations. For further levels of coarsening, there are increased savings in computational time, but also associated reductions in the accuracy of the flow field variables. The computational speed-up rate obtained using the CGP methodology was found to increase with the increasing resolution. These results are also tabulated in Table~\ref{tab:norm1} and Table~\ref{tab:norm2}. Since we are using one of the fastest available Poisson solver routines, it is reasonable to expect that the speed-up will become much more pronounced when different Poisson solvers are implemented, as will be required for general flow problems that cannot be solved on a regular grid. Table~\ref{tab:norm1} also includes the results obtained using a sub-optimal SOR-type iterative Poisson solver, showing that the computational speed-up in the CGP framework  depends highly on the choice of Poisson solver. We can also see that the accuracy of the results are independent of the choice of Poisson solver. For the rest of our analysis, we will use an optimal FFT-FPS except for in the distorted grid problem, where there are mixed derivatives in the corresponding Poisson equation in curvilinear coordinates. In that case, we use a V-cycle multigrid solver.

\begin{figure*}
\centering
\mbox{
\subfigure{\includegraphics[width=0.5\textwidth]{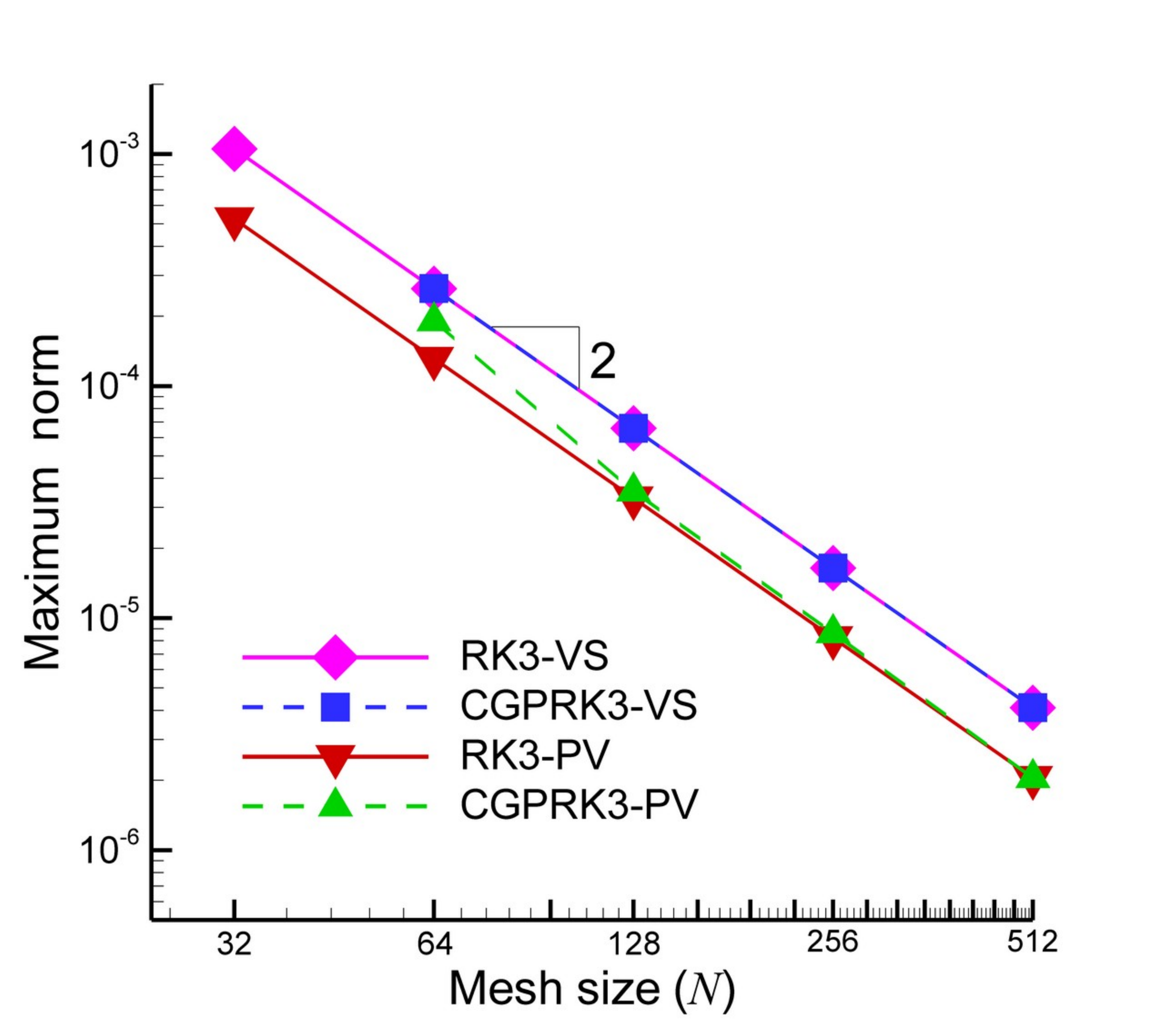}}
\subfigure{\includegraphics[width=0.5\textwidth]{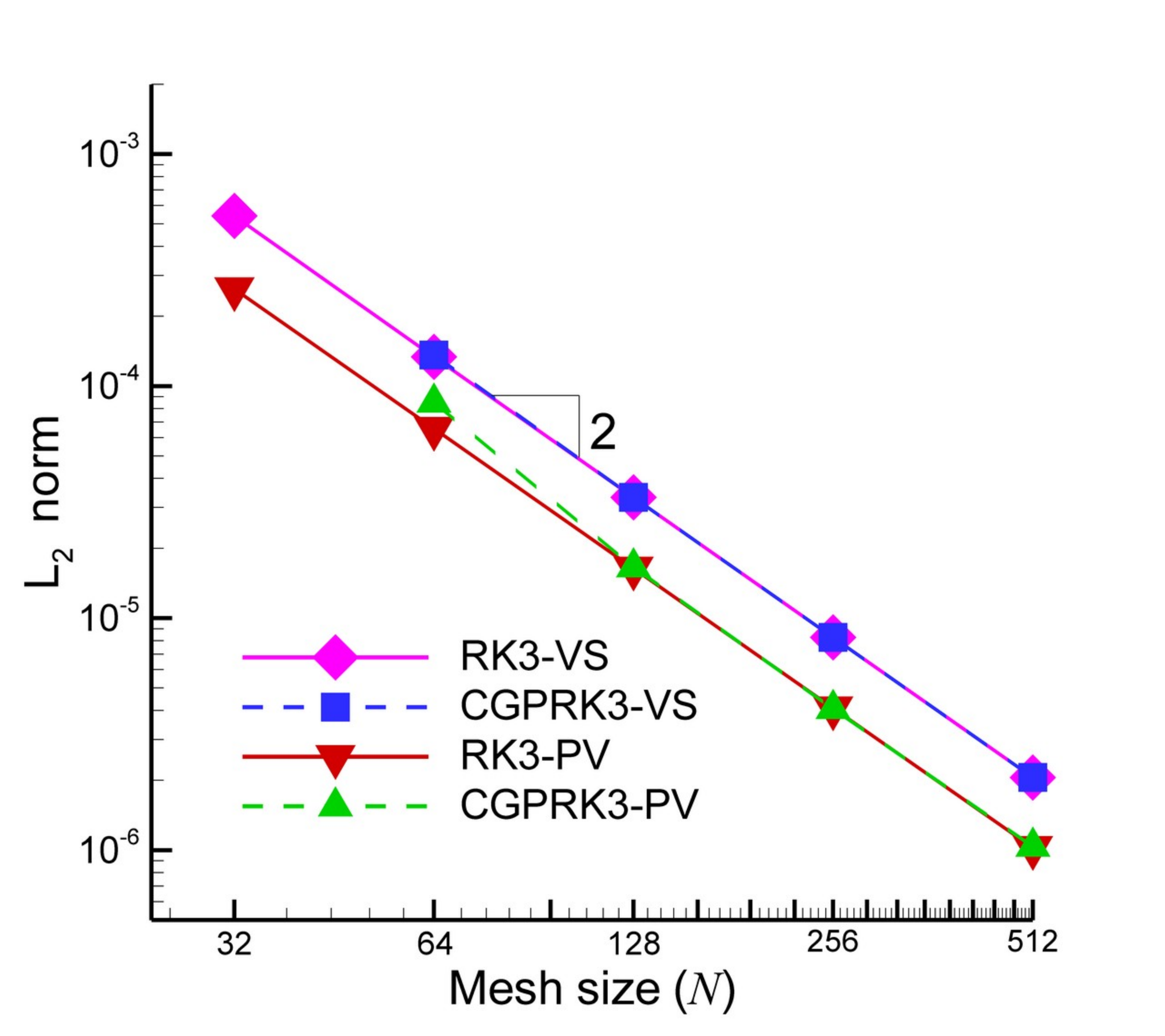}}
}
\caption{Convergence in the exact errors for the Taylor-Green decaying vortex problem using both standard and CGP methods: (left) $L_{\infty}$ norms, and (right) $L_{2}$ norms.}
\label{fig:norms}
\end{figure*}

It can be seen from the results presented in Table~\ref{tab:norm1} and Table~\ref{tab:norm2} that second-order accurate results are obtained in practice for both of the CGP formulations. For example, from Table~\ref{tab:norm1}, the slope of the maximum infinity norm and the averaged $L_2$ norm are:
\begin{equation}
n_{\infty}=\frac{ln\bigg(\frac{||\omega||^{512^2}_{\infty}}{||\omega||^{256^2}_{\infty}}\bigg)}{ln(\frac{h}{2h})} = 1.9999, \quad n_{L_2}=\frac{ln\bigg(\frac{||\omega||^{512^2}_{L_2}}{||\omega||^{256^2}_{L_2}}\bigg)}{ln(\frac{h}{2h})} = 2.0028.
\label{eq:err}
\end{equation}
The spatial convergence properties of the standard RK3 methods and the CGPRK3 methods for both the vorticity-stream function and primitive variable formulations are also illustrated in Fig.~\ref{fig:norms} by showing that a single level of coarsening strategy produces a similar convergence rate. A linear reduction rate is obtained with the slope of all curves being 2, verifying the second-order spatial scheme. It can be seen that there is a slight increase in the error norms for the low resolution computations in the fractional step primitive variable formulation. This can be attributed to a small projection error in the fractional step procedure, showing that the CGP is also somewhat solver dependent. We demonstrate, however, that there is a significant benefit to using the CGP method in both formulations.

The Taylor-Green vortex problem is one of the simplest incompressible flow test cases, having a smooth decaying field. We solve it to be able to perform an exact error analysis by comparing our results to the analytical solution of the problem. In the following sections, the CGP method is applied to more challenging flow problems.

\subsection{Double shear layer}
\label{sec:doublesl}
The double shear layer problem was introduced by Bell, Colella and Glaz \cite{bell1989second} to test the performance of a projection method, and studied later by Minion and Brown \citep{minion1997performance} using various schemes to determine the effects of the grid resolution on solutions of the unsteady, incompressible Navier-Stokes equations. It is a benchmark problem for testing the accuracy and resolution of a time dependent numerical method. The double shear layer problem with periodic boundary conditions in a square domain $[0,2\pi]\times[0,2\pi]$ is subjected to the following initial conditions \cite{yanwen1999numerical}:
\begin{eqnarray}
u(x,y,0) &=&
\bigg\{ \begin{array}{c} \mbox{tanh}[\sigma(y-\frac{\pi}{2})] \\ \mbox{tanh}[\sigma(\frac{3\pi}{2}-y)] \end{array} \begin{array}{c} \quad  \mbox{if} \quad y \leq \pi  \\ \quad \mbox{if} \quad y > \pi \end{array} \\
v(x,y,0) &=& \delta \ \mbox{sin}(x)
\label{eq:dluv}
\end{eqnarray}
where the constants $\sigma$ and $\delta$ determine the thickness of the shear layer and the amplitude of the initial perturbation, respectively. The initial vorticity field to be evolved using the vorticity-stream function formulation algorithm can easily be obtained analytically from the expressions above:
\begin{equation}
\omega(x,y,0) =
\bigg\{ \begin{array}{c}
\delta \ \mbox{cos}(x) -\sigma \ \mbox{cosh}^{-2}[\sigma(y-\frac{\pi}{2})] \\
\delta \ \mbox{cos}(x) +\sigma \ \mbox{cosh}^{-2}[\sigma(\frac{3\pi}{2}-y)] \end{array} \begin{array}{c} \quad \mbox{if} \quad y \leq \pi  \\ \quad  \mbox{if} \quad y > \pi \end{array} \\
\label{eq:dlw}
\end{equation}

\begin{figure*}
\centering
\mbox{
\subfigure{\includegraphics[width=0.33\textwidth]{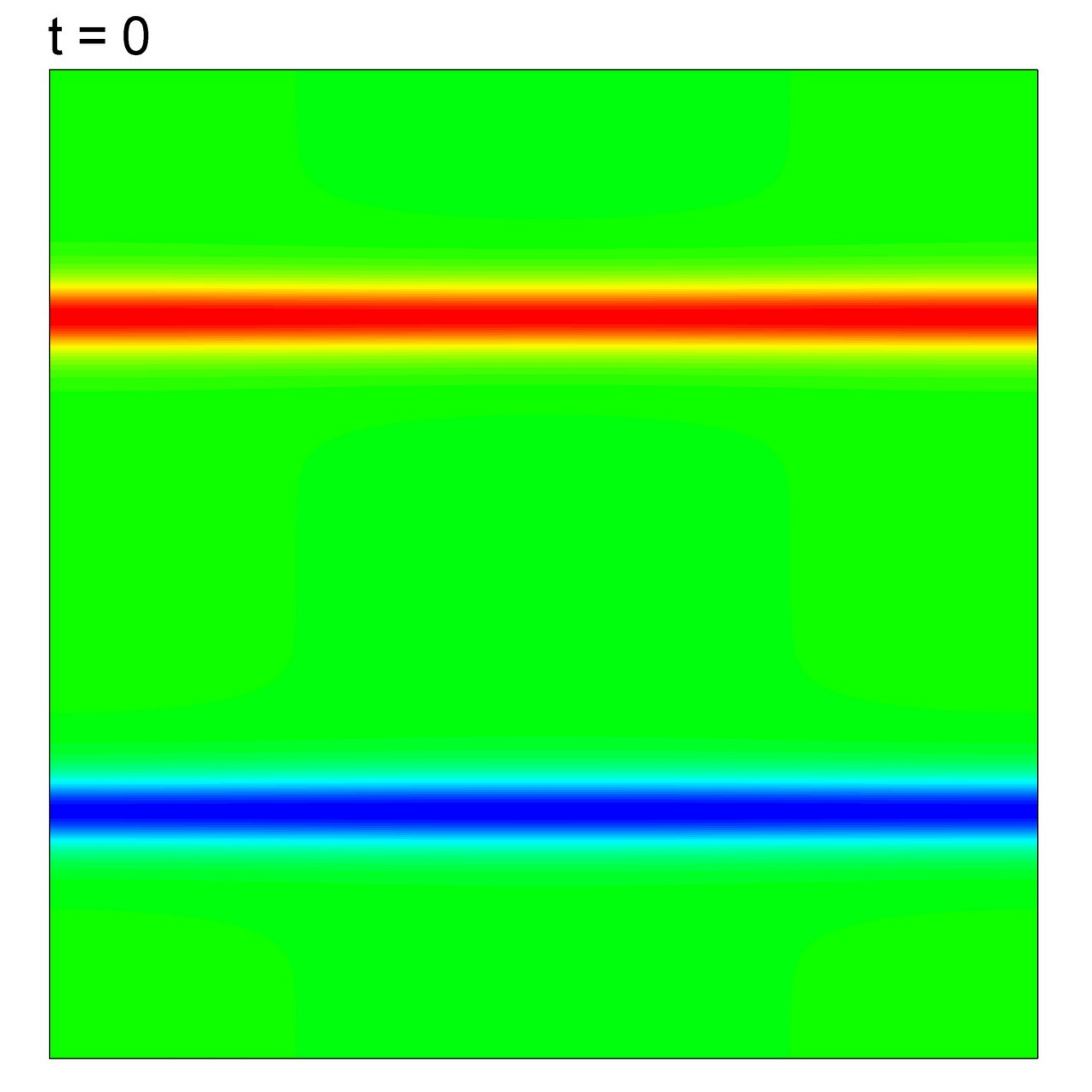}}
\subfigure{\includegraphics[width=0.33\textwidth]{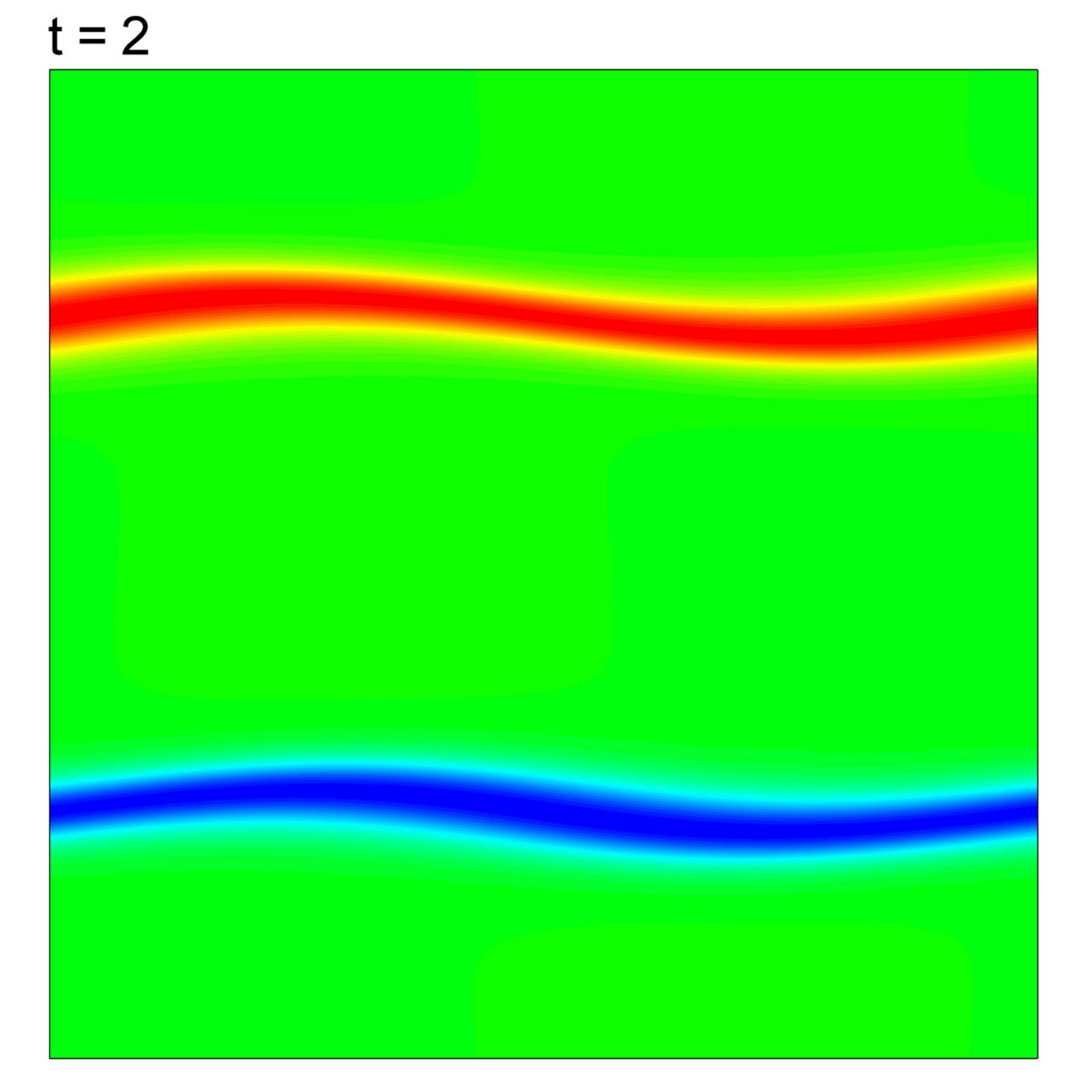}}
\subfigure{\includegraphics[width=0.33\textwidth]{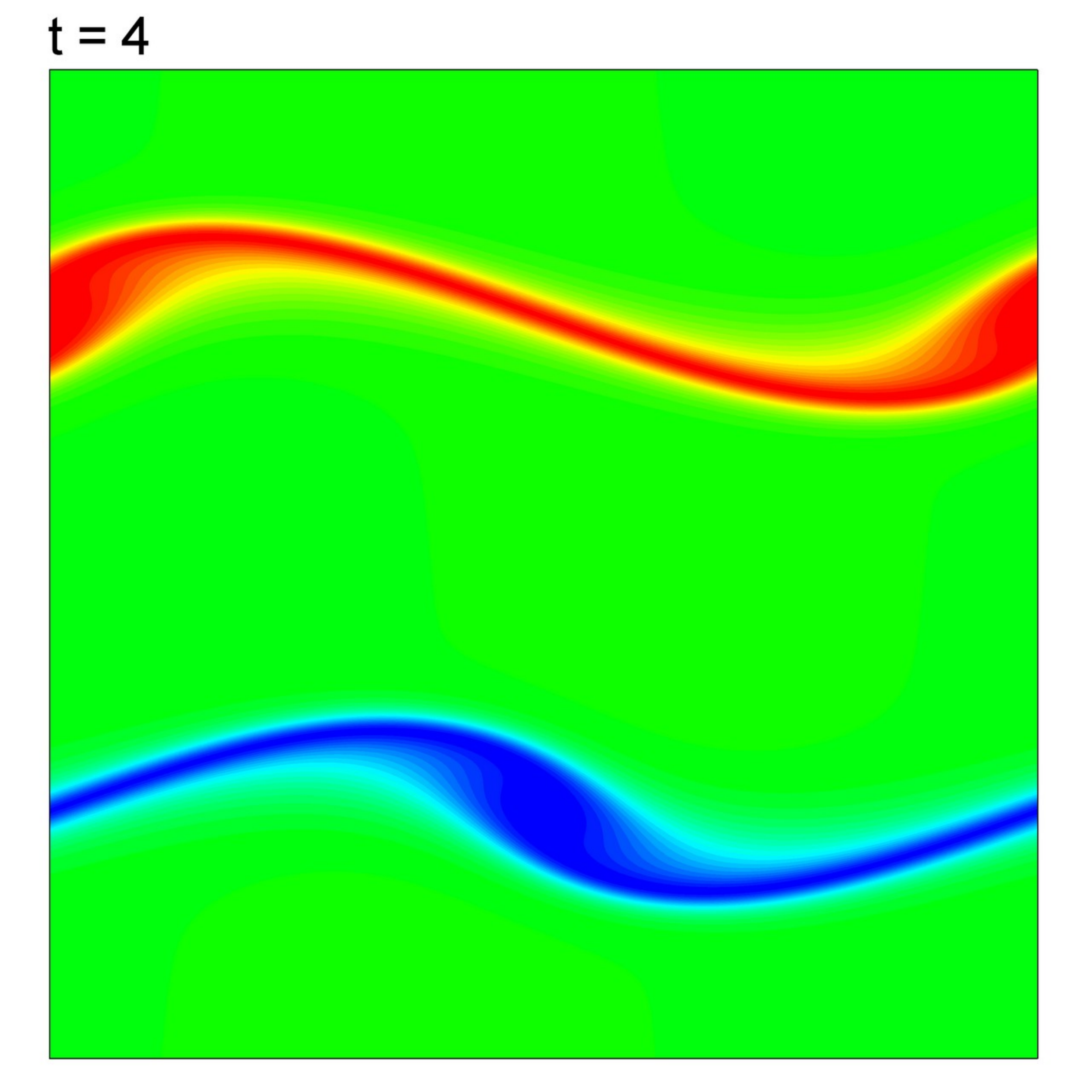}}
}
\mbox{
\subfigure{\includegraphics[width=0.33\textwidth]{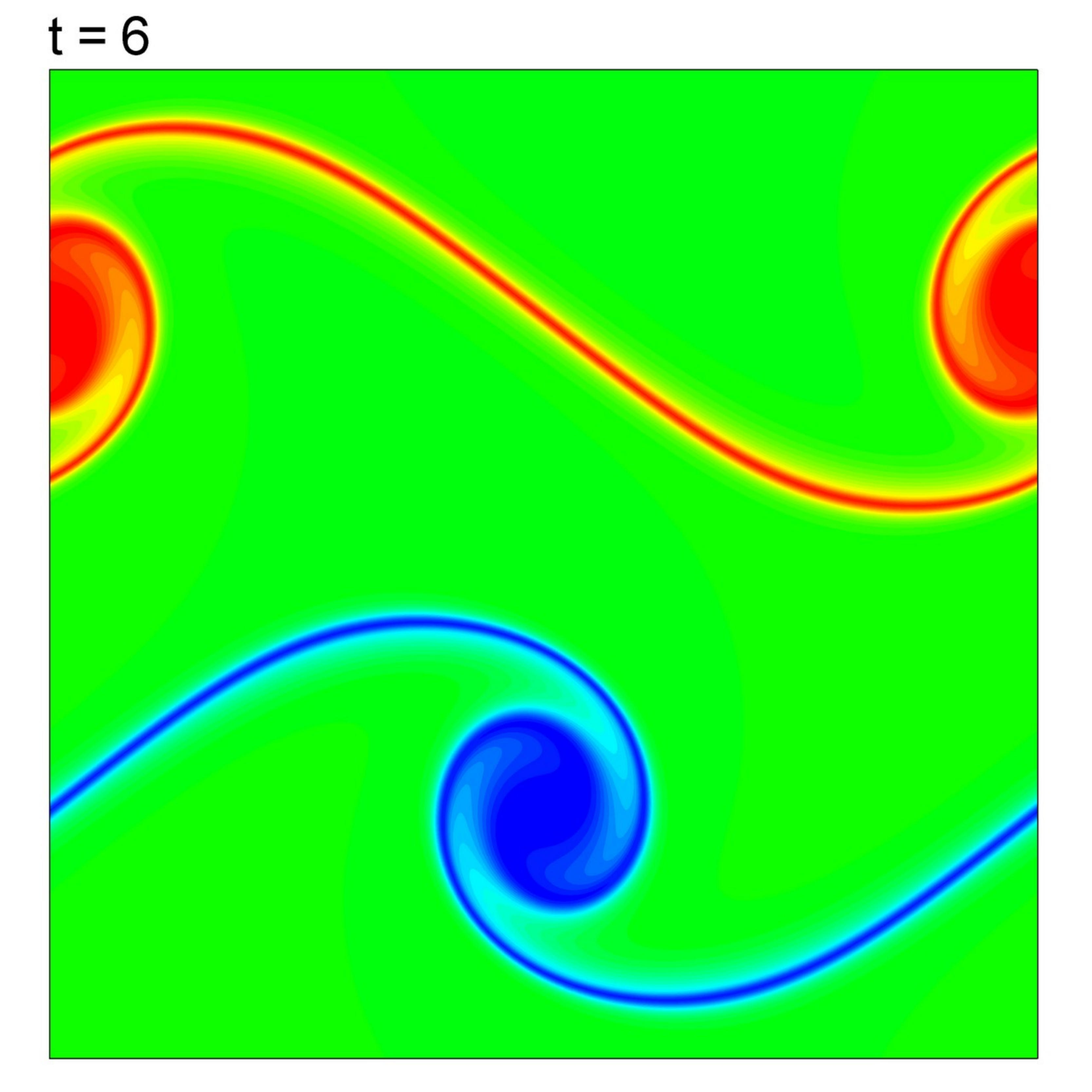}}
\subfigure{\includegraphics[width=0.33\textwidth]{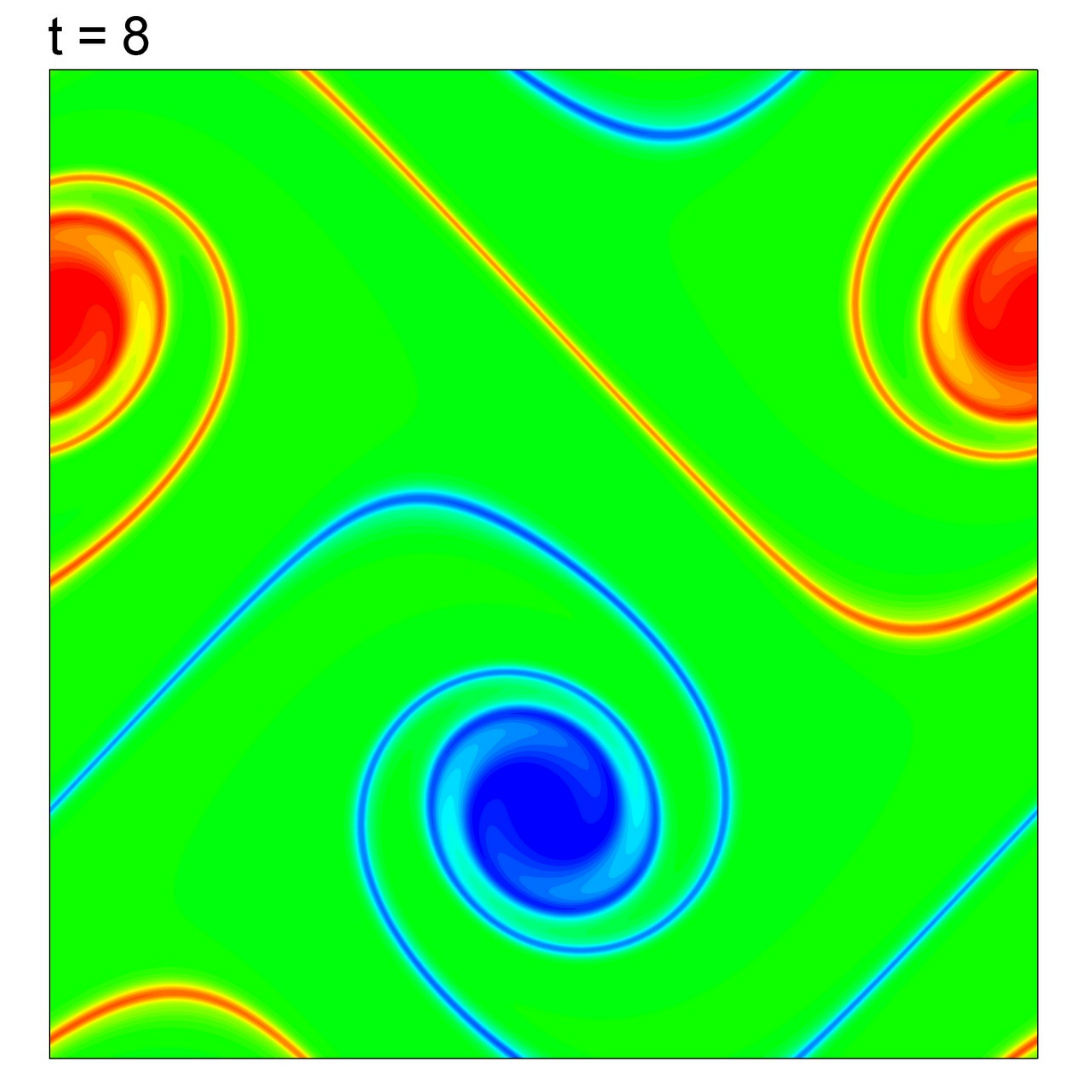}}
\subfigure{\includegraphics[width=0.33\textwidth]{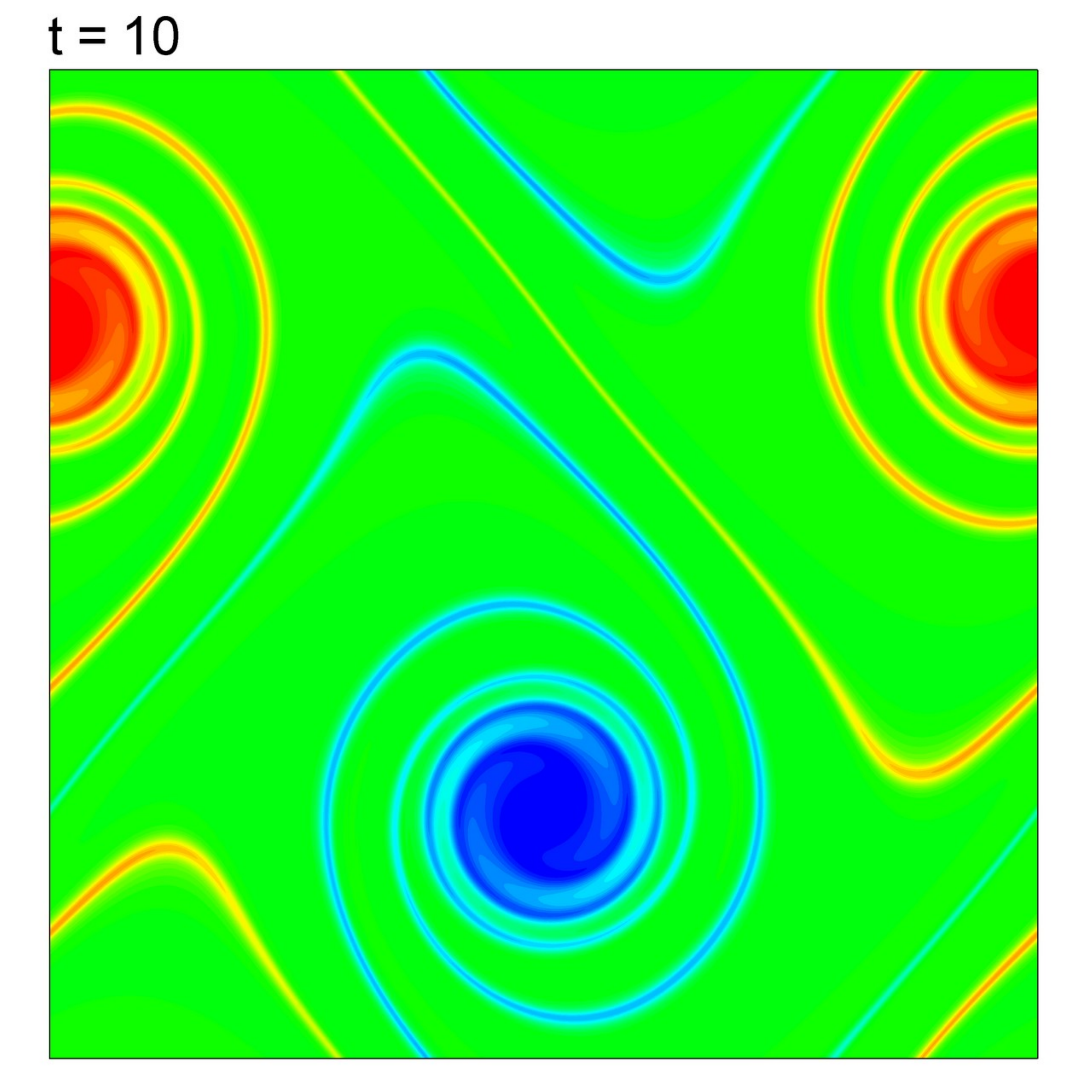}}
}
\caption{The vorticity field at different times obtained by the CGPRK3 method on $1024^2:512^2$ resolution grids.}
\label{fig:dsl-time}
\end{figure*}

In our computations, the thickness parameter is $\sigma = 15/\pi$, the perturbation amplitude is $\delta=0.05$, the Reynolds number is $Re=10^4$, and the time step is $\Delta t = 10^{-3}$. The perturbed shear layer rolls up into a single vortex and the shear layers become thinner and thinner as the flow evolves in time as shown in Fig.~\ref{fig:dsl-time}.

\begin{table*}
\caption{Total CPU times for the double shear layer problem and their component percentages (i.e., percent CPU times for the advection-diffusion part, the Poisson part, and the mappings between these two parts) for the CGP and standard algorithms for various resolutions.}
\begin{center}
\label{tab:cpu}
\begin{tabular}{llccccc}
\hline\noalign{\smallskip}
Method & Resolutions  & \% Adv-Dif & \% Poisson & \% Map & CPU (hr)  \\
\hline
RK3 ($\ell=0$)    &$1024^2$ : $1024^2$  &  6.8  & 93.2 & -    & 10.24   \\
CGPRK3 ($\ell=1$) &$1024^2$ : $512^2$   &  29.9 & 64.2 & 5.9  & 2.40   \\
CGPRK3 ($\ell=2$) &$1024^2$ : $256^2$   &  26.0 & 59.5 & 14.6 & 1.26   \\
CGPRK3 ($\ell=3$) &$1024^2$ : $128^2$   &  74.1 & 6.8  & 19.1 & 1.02   \\
RK3 ($\ell=0$)    &$512^2$ : $512^2$    &  10.1 & 89.9 & -    & 1.74    \\
CGPRK3 ($\ell=1$) &$512^2$ : $256^2$    &  33.8 & 59.6 & 6.6  & 0.53    \\
CGPRK3 ($\ell=2$) &$512^2$ : $128^2$    &  61.9 & 22.8 & 15.3 & 0.29    \\
RK3 ($\ell=0$)    &$256^2$ : $256^2$    &  12.5 & 87.5 & -    & 0.36    \\
\hline
\end{tabular}
\end{center}
\end{table*}

\begin{figure*}
\centering
\mbox{
\subfigure{\includegraphics[width=0.33\textwidth]{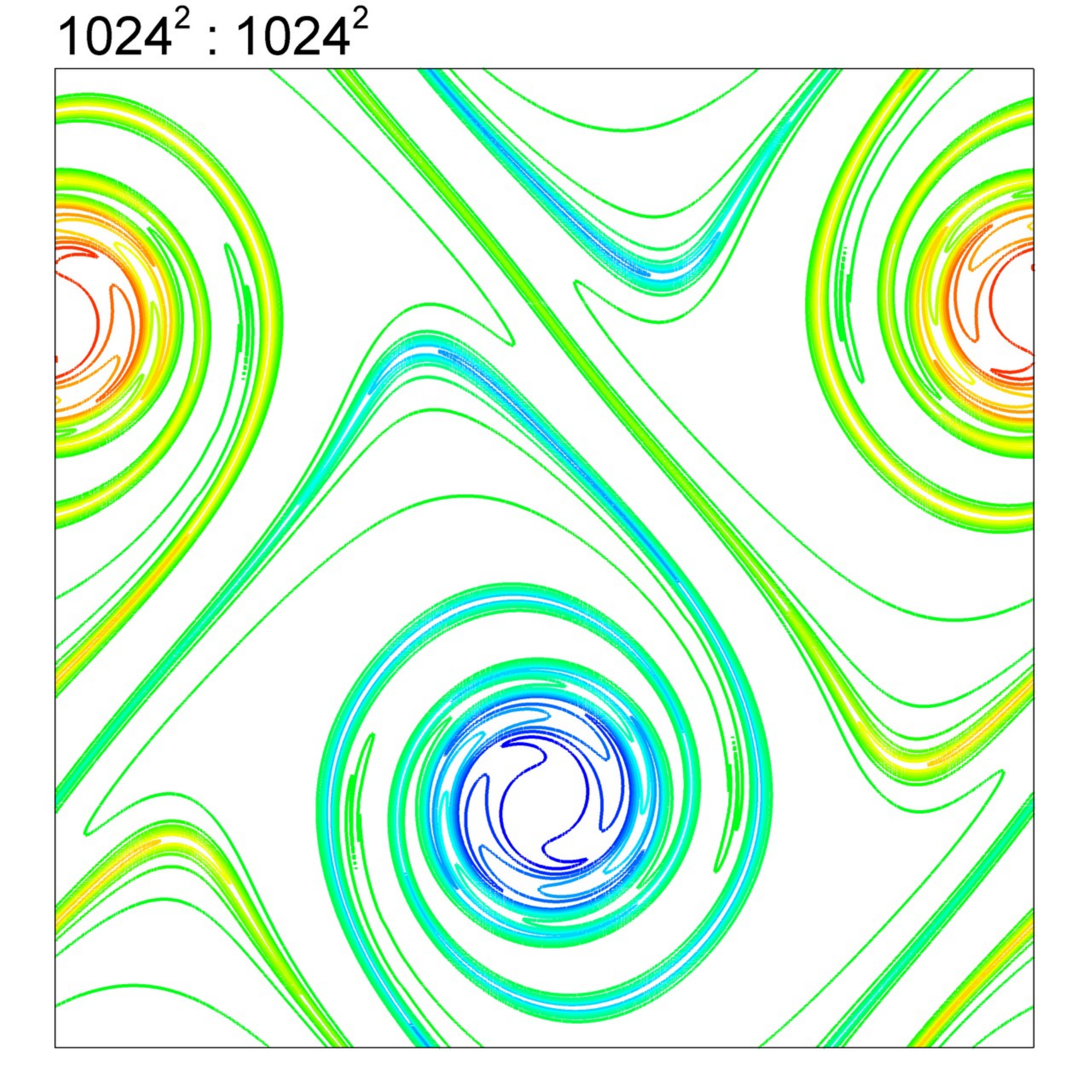}}
\subfigure{\includegraphics[width=0.33\textwidth]{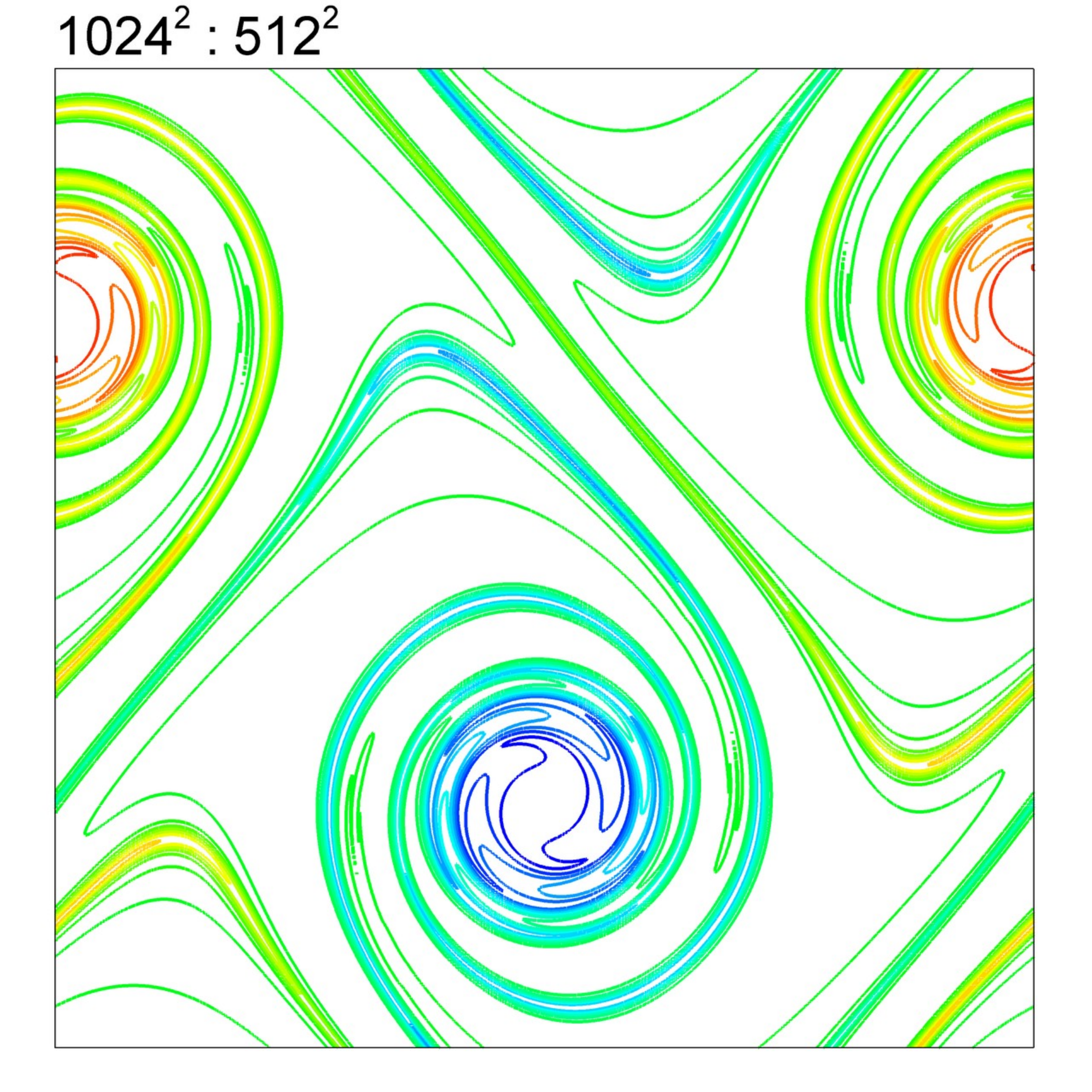}}
\subfigure{\includegraphics[width=0.33\textwidth]{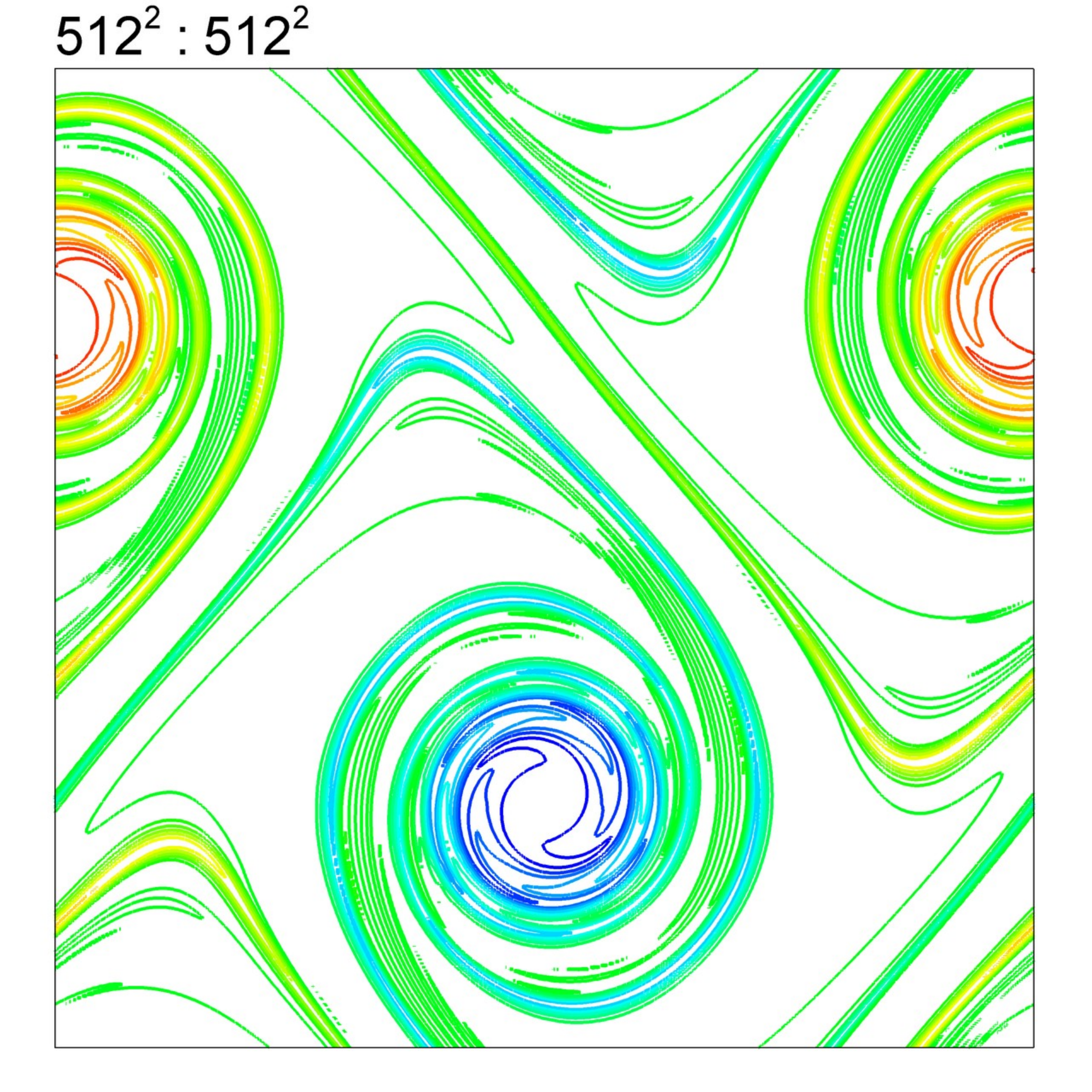}}
}
\mbox{
\subfigure{\includegraphics[width=0.33\textwidth]{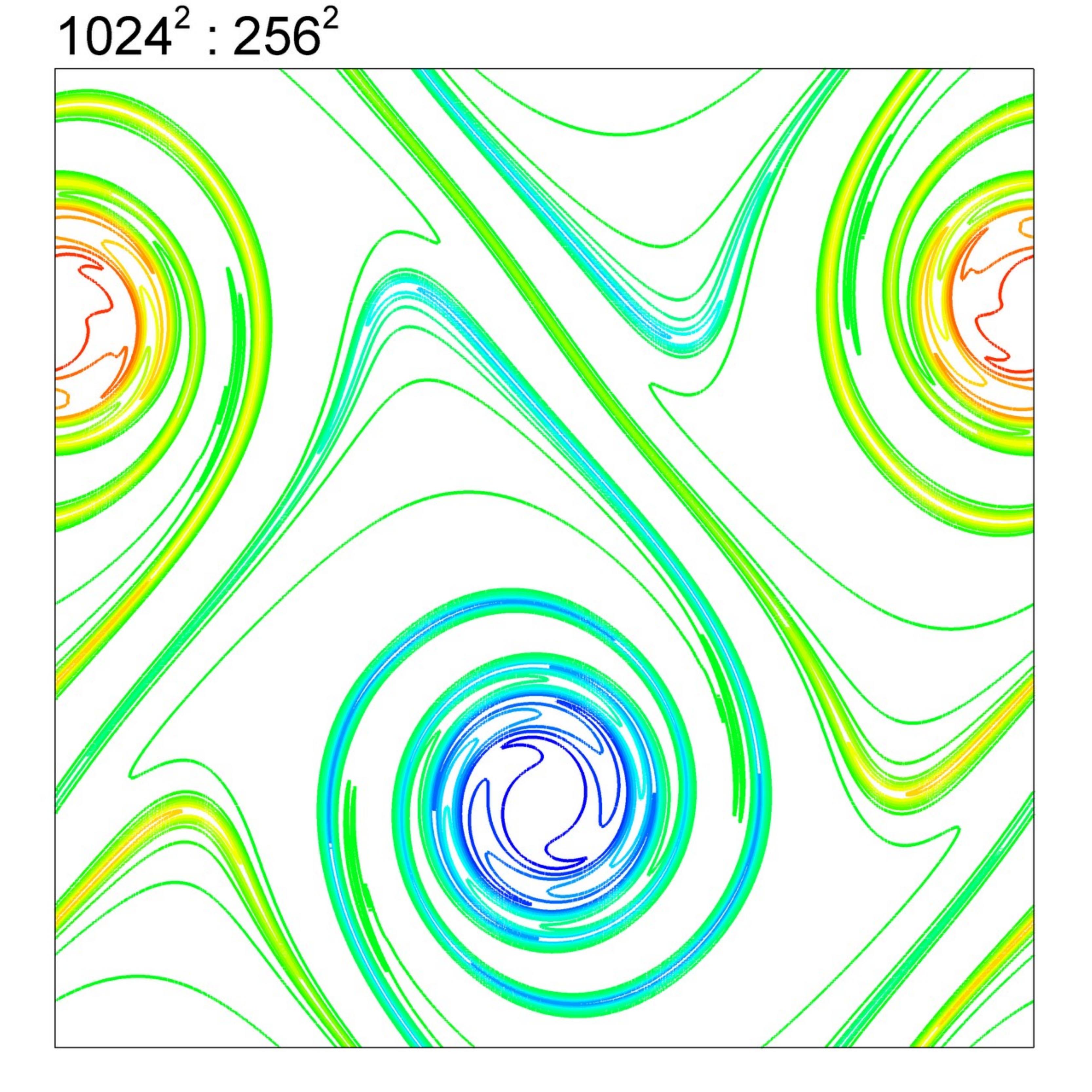}}
\subfigure{\includegraphics[width=0.33\textwidth]{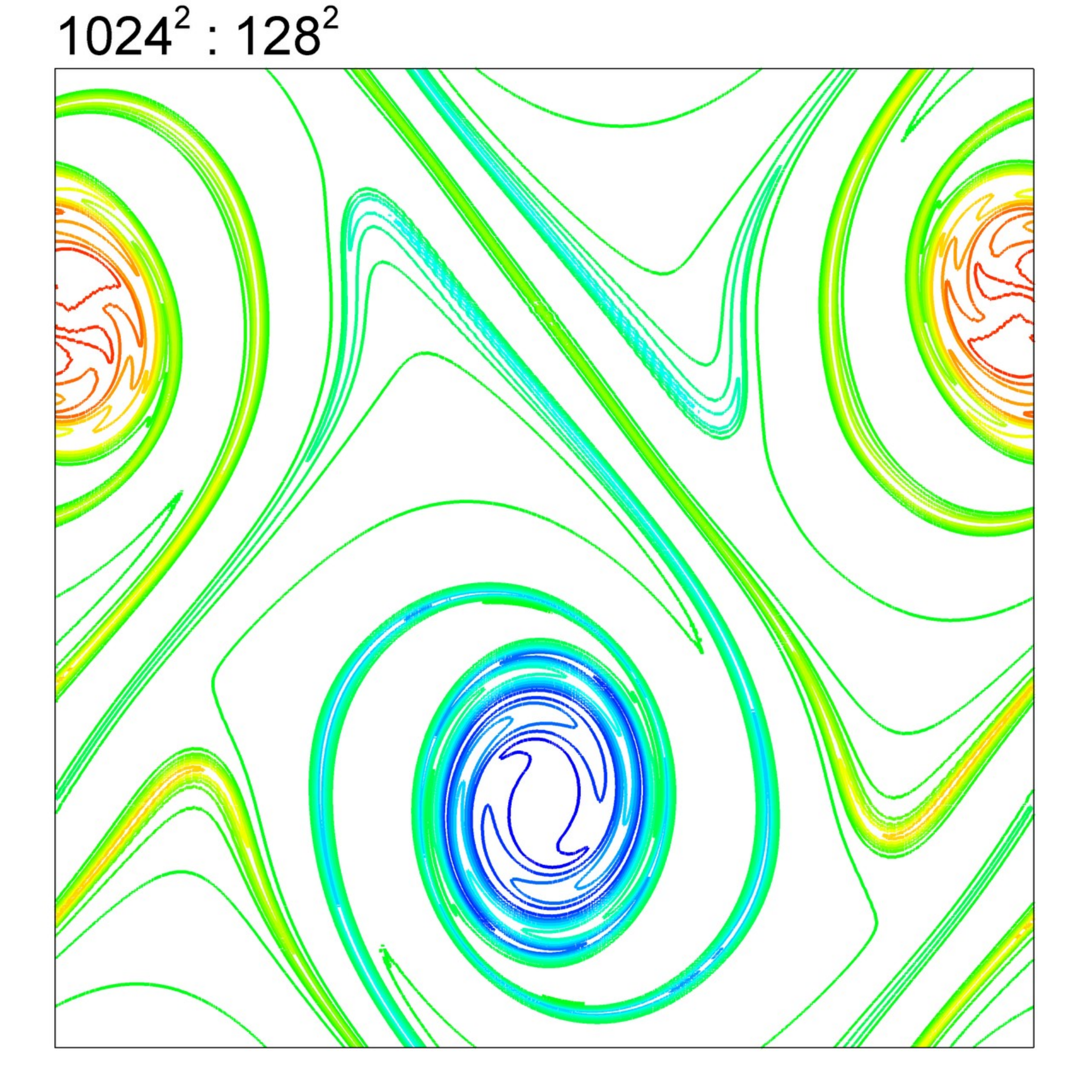}}
\subfigure{\includegraphics[width=0.33\textwidth]{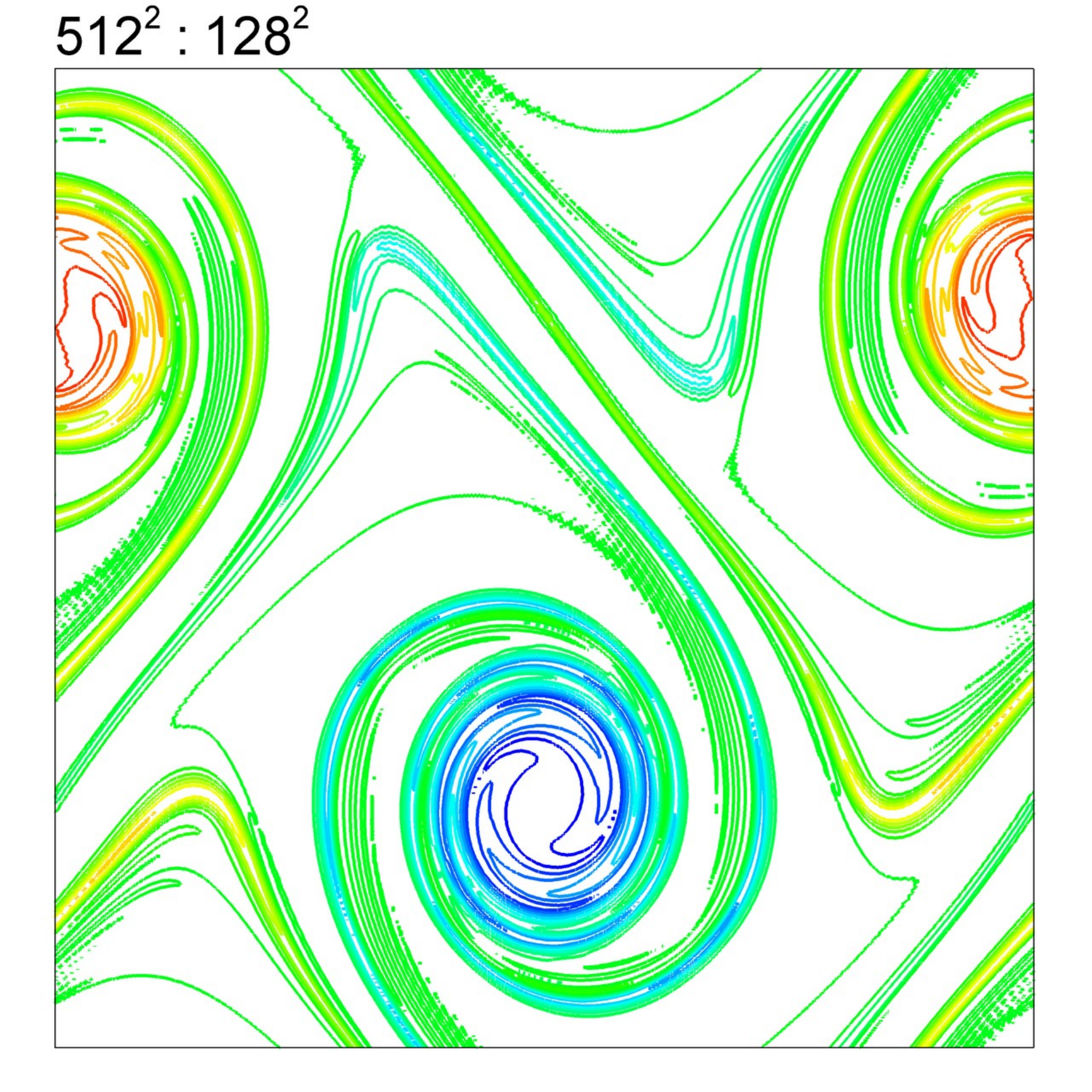}}
}
\mbox{
\subfigure{\includegraphics[width=0.33\textwidth]{dsl-512-512.pdf}}
\subfigure{\includegraphics[width=0.33\textwidth]{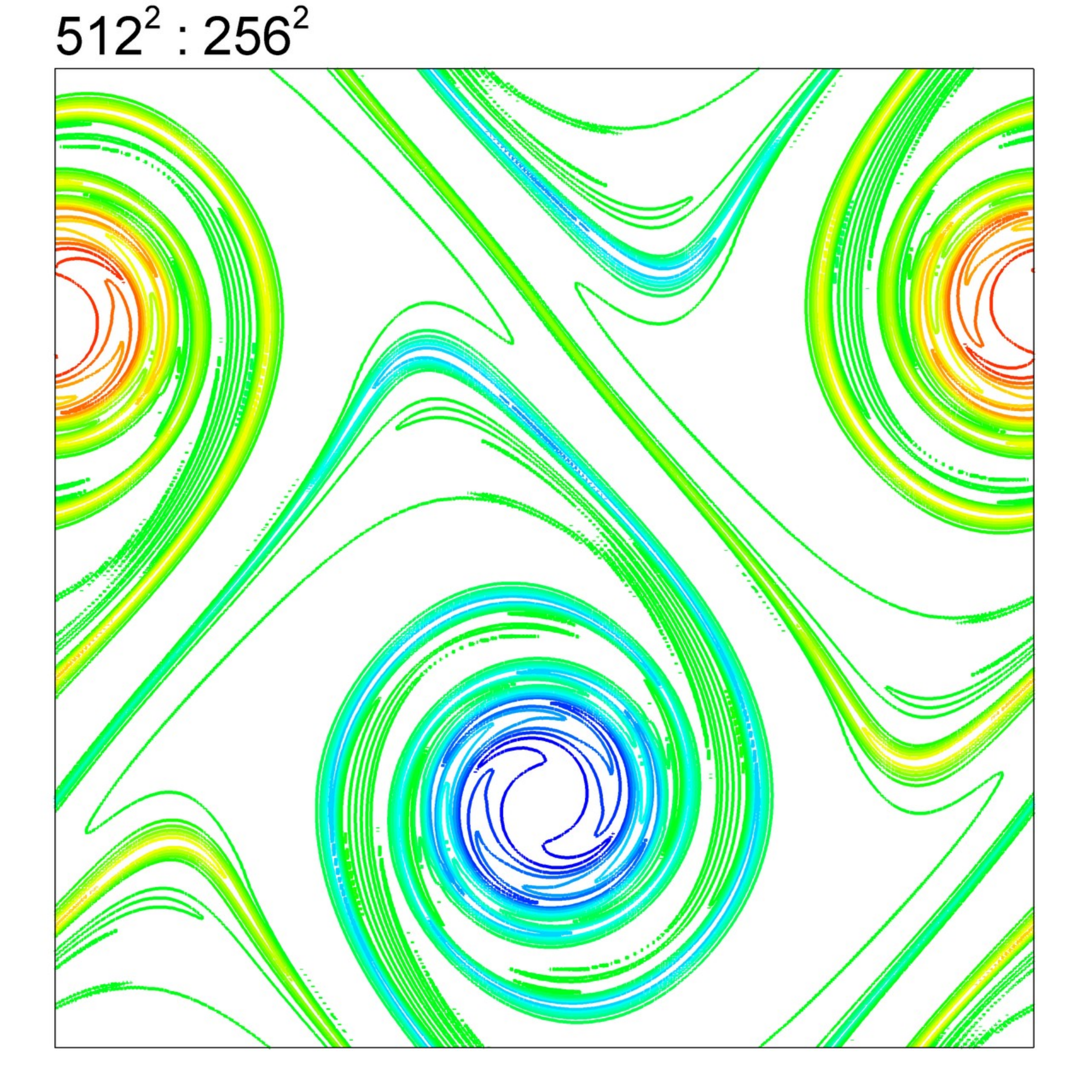}}
\subfigure{\includegraphics[width=0.33\textwidth]{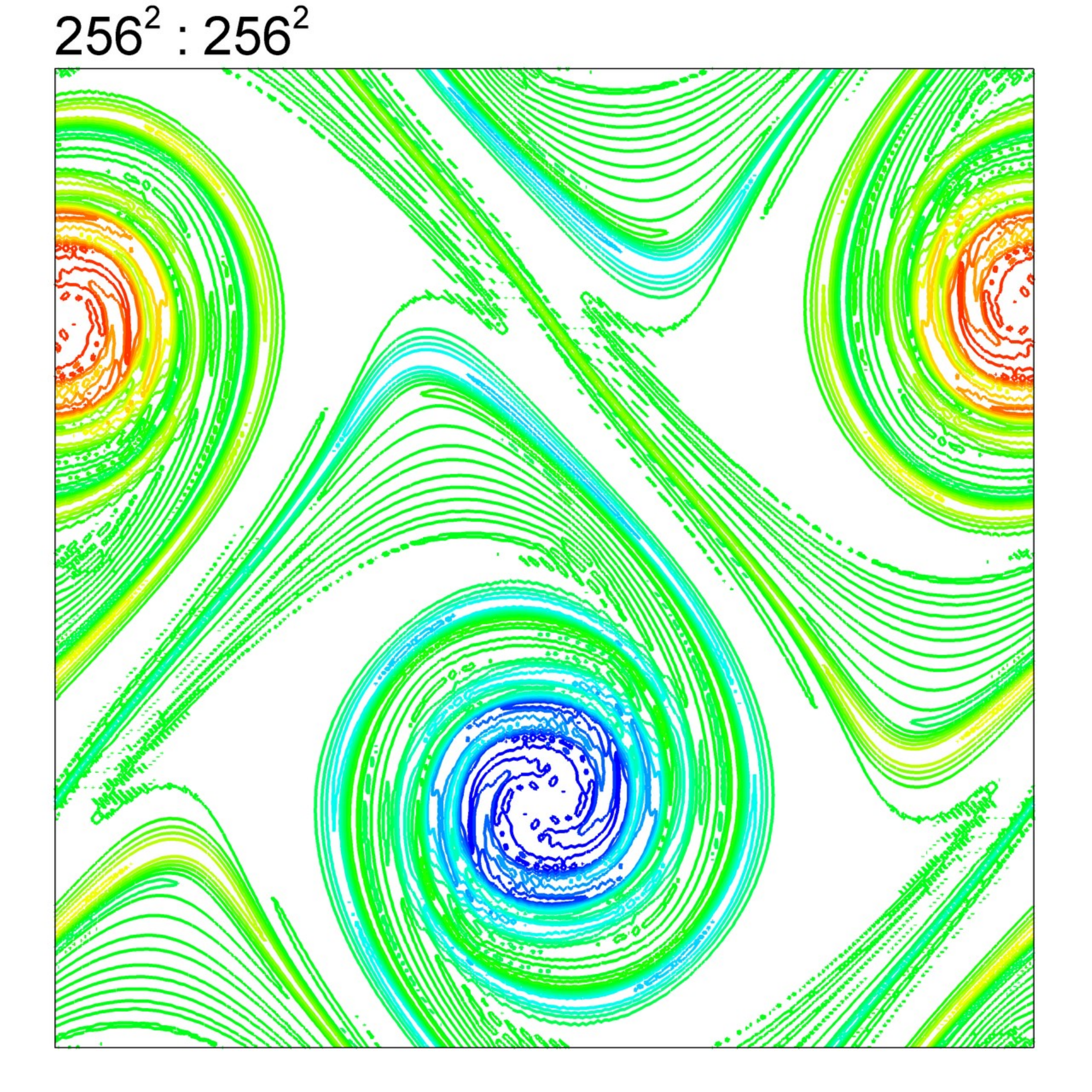}}
}
\mbox{
\subfigure{\includegraphics[width=0.33\textwidth]{dsl-256-256.pdf}}
\subfigure{\includegraphics[width=0.33\textwidth]{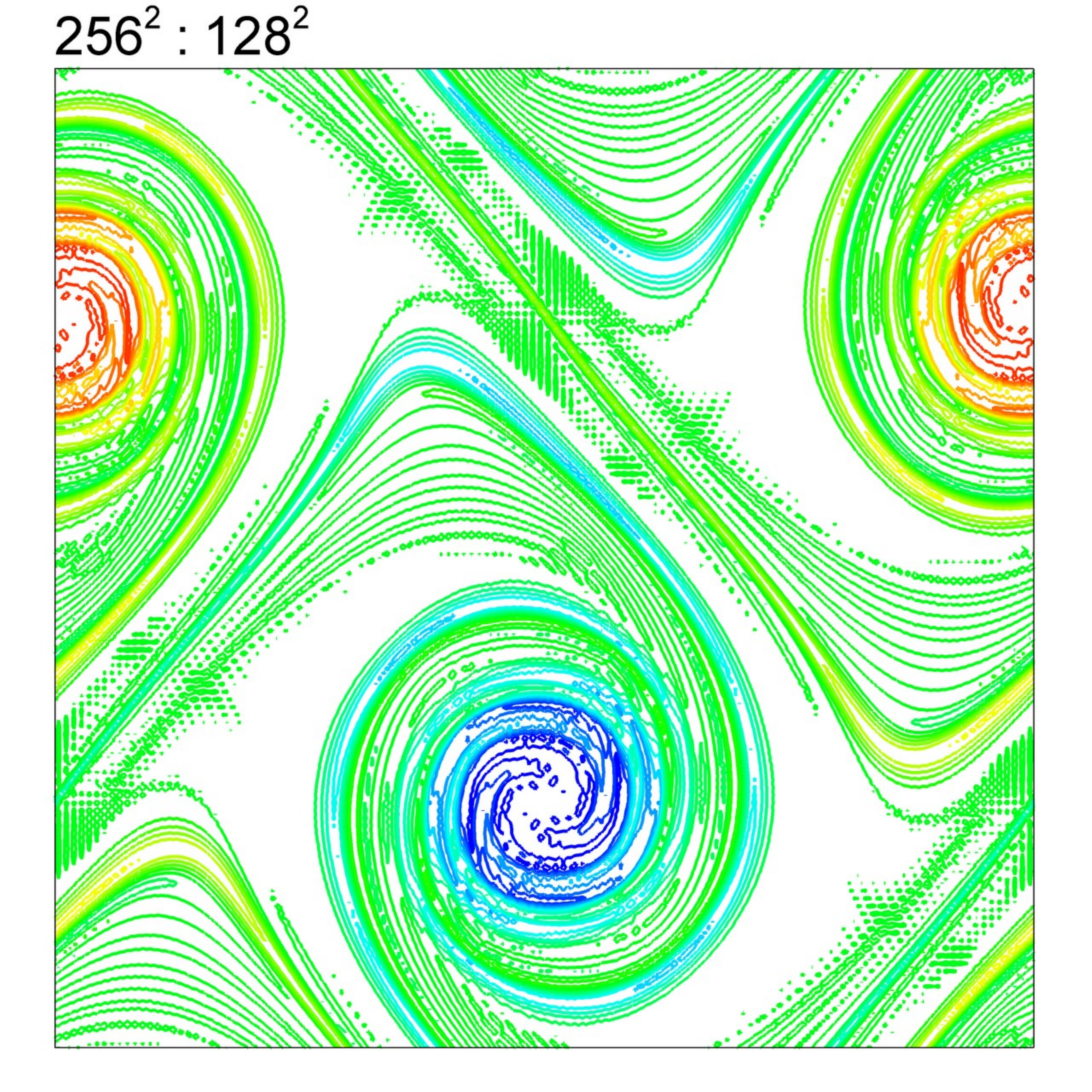}}
\subfigure{\includegraphics[width=0.33\textwidth]{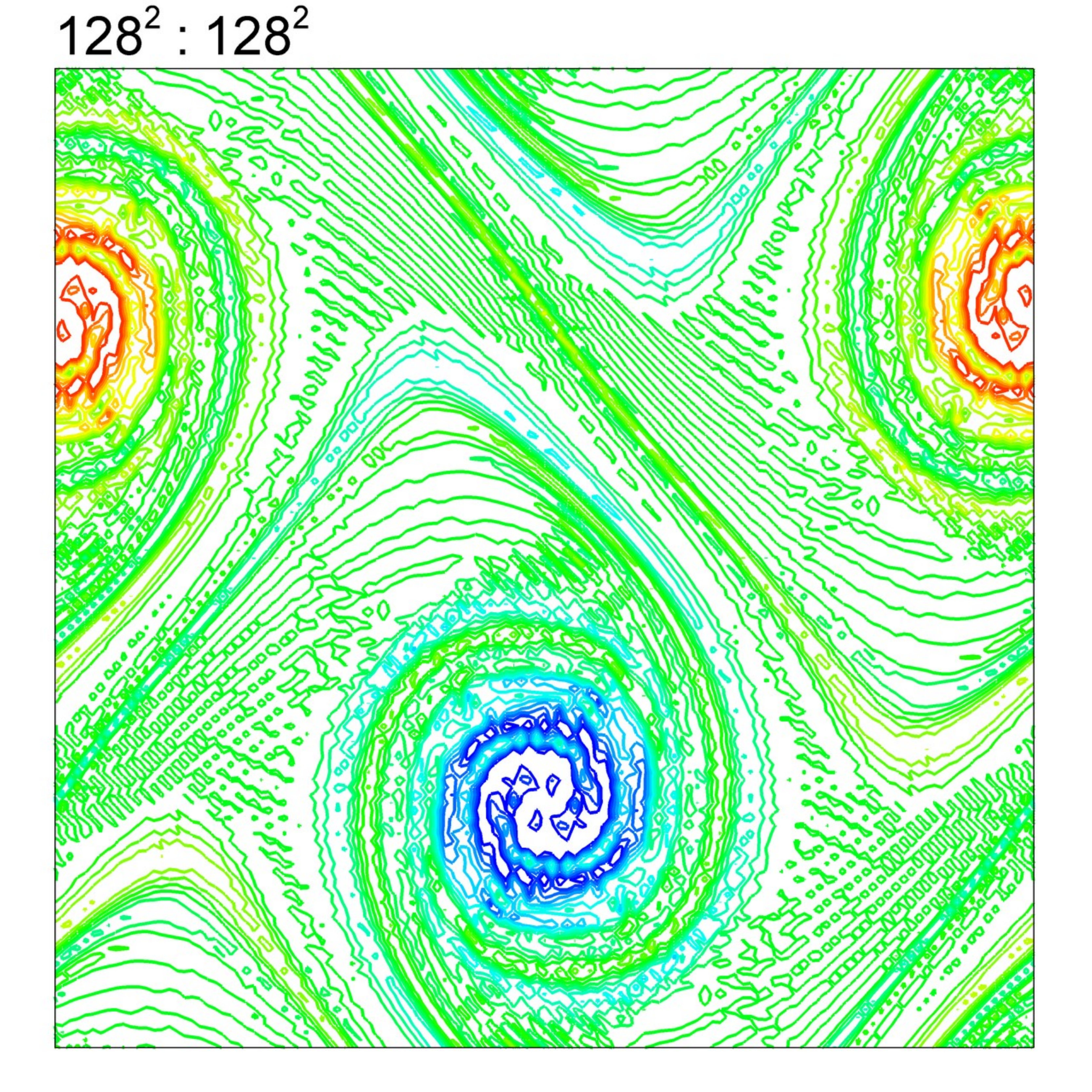}}
}
\caption{The vorticity fields for the double shear layer problem at $t=10$ obtained using the vorticity-stream function formulation. Labels shows the resolutions for both parts of the solver in the form $N^2:M^2$, where $N^2$ is the resolution for the vorticity-transport equation, and $M^2$ is the resolution for the Poisson equation.}
\label{fig:dsl-time10}
\end{figure*}

Vorticity contours at $t=10$ obtained using the vorticity-stream function formulation with and without the CGP method are shown in Fig.~\ref{fig:dsl-time10} for different spatial resolutions. These results demonstrate that high fidelity numerical simulations can be obtained using the CGP method. For example, if we consider two numerical experiments with $512^2$:$512^2$ and $512^2$:$256^2$ resolutions, they have almost the same resulting vorticity field, but the latter results (with a half-coarsened grid for the Poisson equation) were obtained almost 3 times faster than the former results (without CGP), and more importantly, the resulting field obtained using the CGP method with the $512^2$:$256^2$ resolution are better than those obtained on the $256^2$:$256^2$ grid without the CGP method. These results demonstrate that the CGP methodology can provide an accelerated method for solving flow problems with strong shear components.

The CPU times and the relative percentages for the main operational subroutines are tabulated in Table~\ref{tab:cpu}. Although we use one of the fastest (linear-cost) Poisson solver in our computations, these statistics clearly show that the computational cost of the elliptic part of the problem is significantly more expensive than that of the advection-diffusion part. The data presented in Table~\ref{tab:cpu} also demonstrates that the computational speed-up increases with increasing spatial resolution. One consequence of this is that the CGP method should be a good candidate for accelerating complex three-dimensional turbulent flow simulations. And it should be able to accelerate direct numerical simulations (DNS) of turbulence by coarsening the number of grid points for the Poisson solver without affecting the well resolved DNS data, a topic we intend to investigate in a future study.

\begin{figure*}
\centering
\mbox{
\subfigure[]{\includegraphics[width=0.5\textwidth]{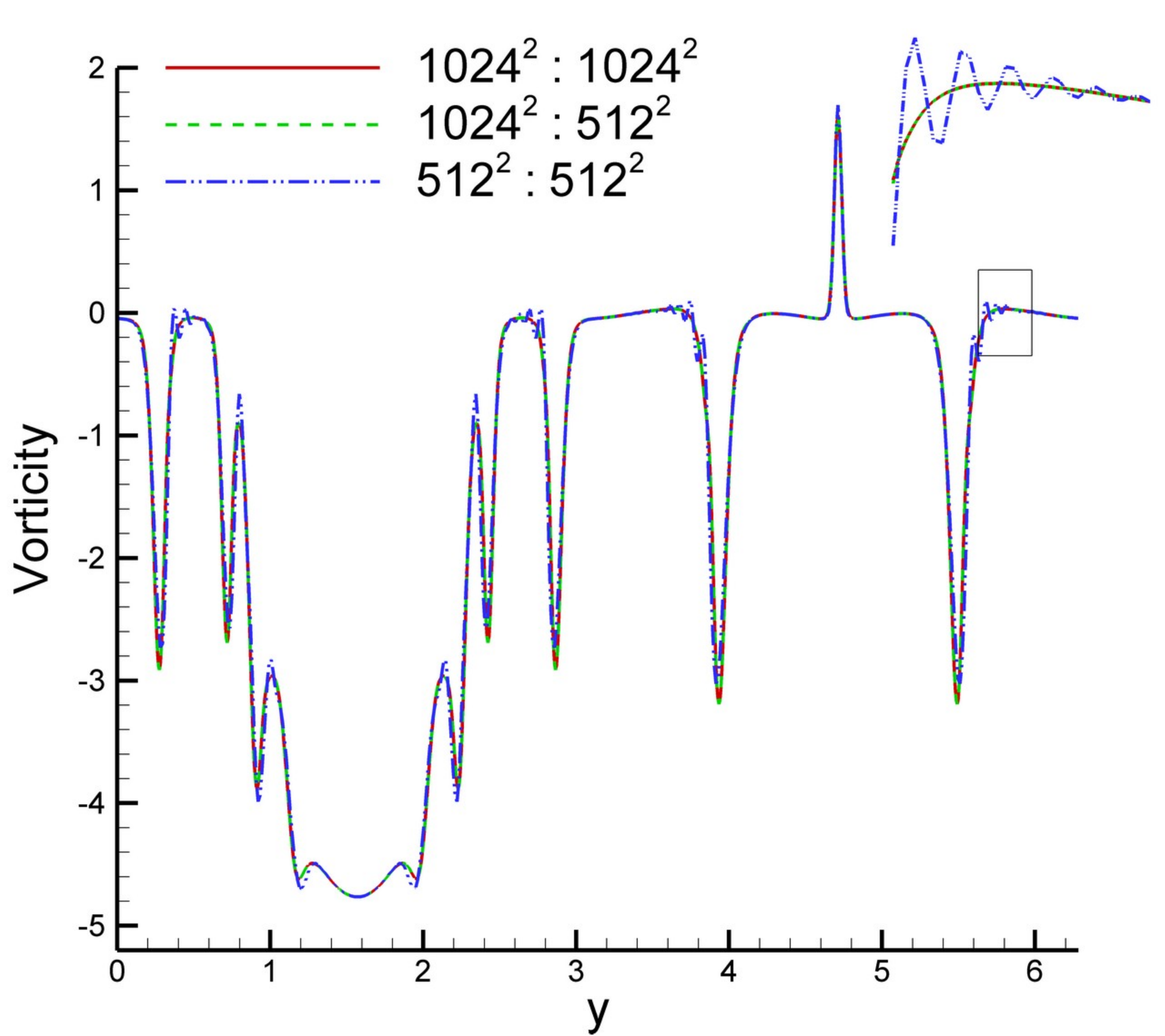}}
\subfigure[]{\includegraphics[width=0.5\textwidth]{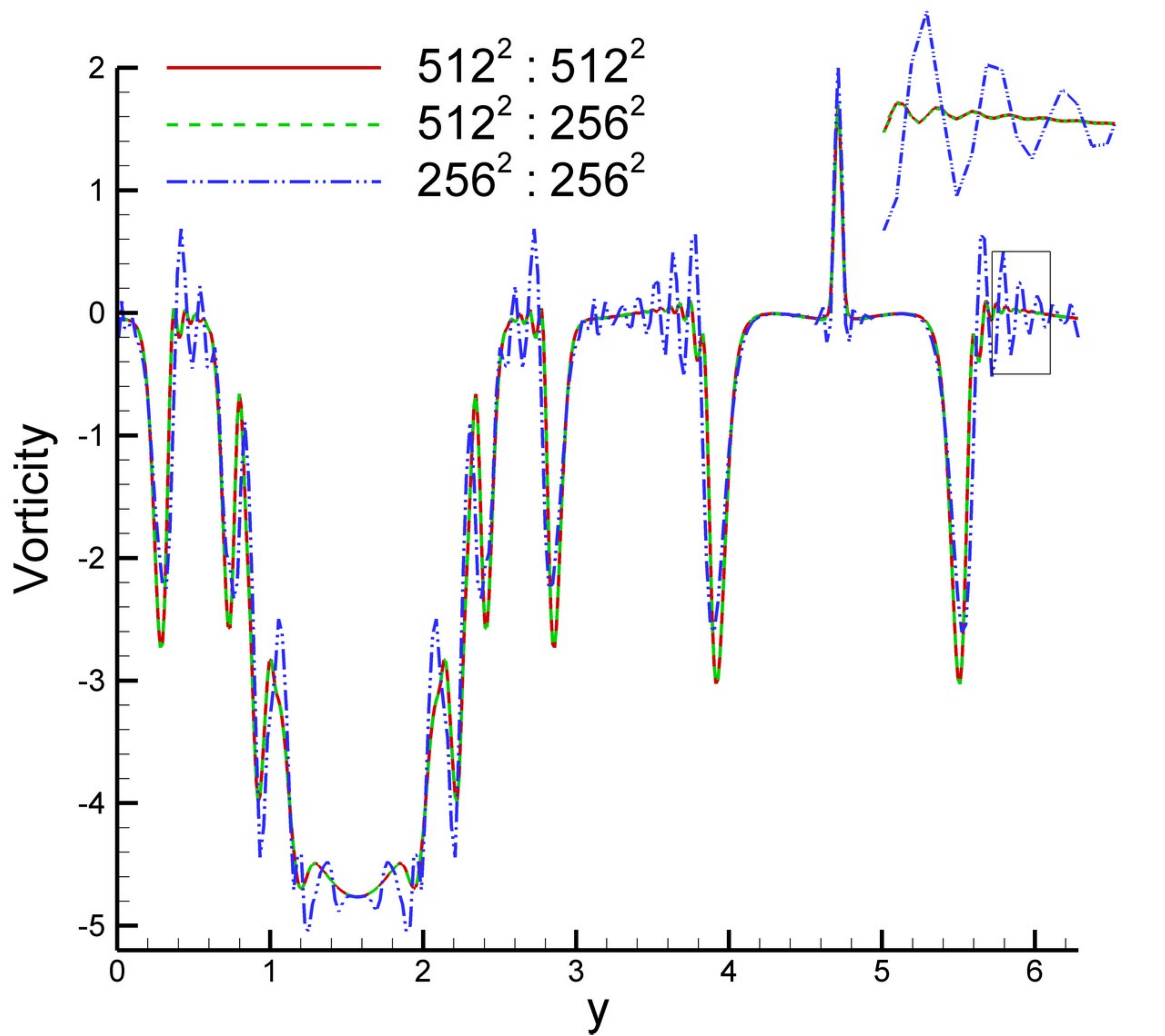}}
}
\\
\mbox{
\subfigure[]{\includegraphics[width=0.5\textwidth]{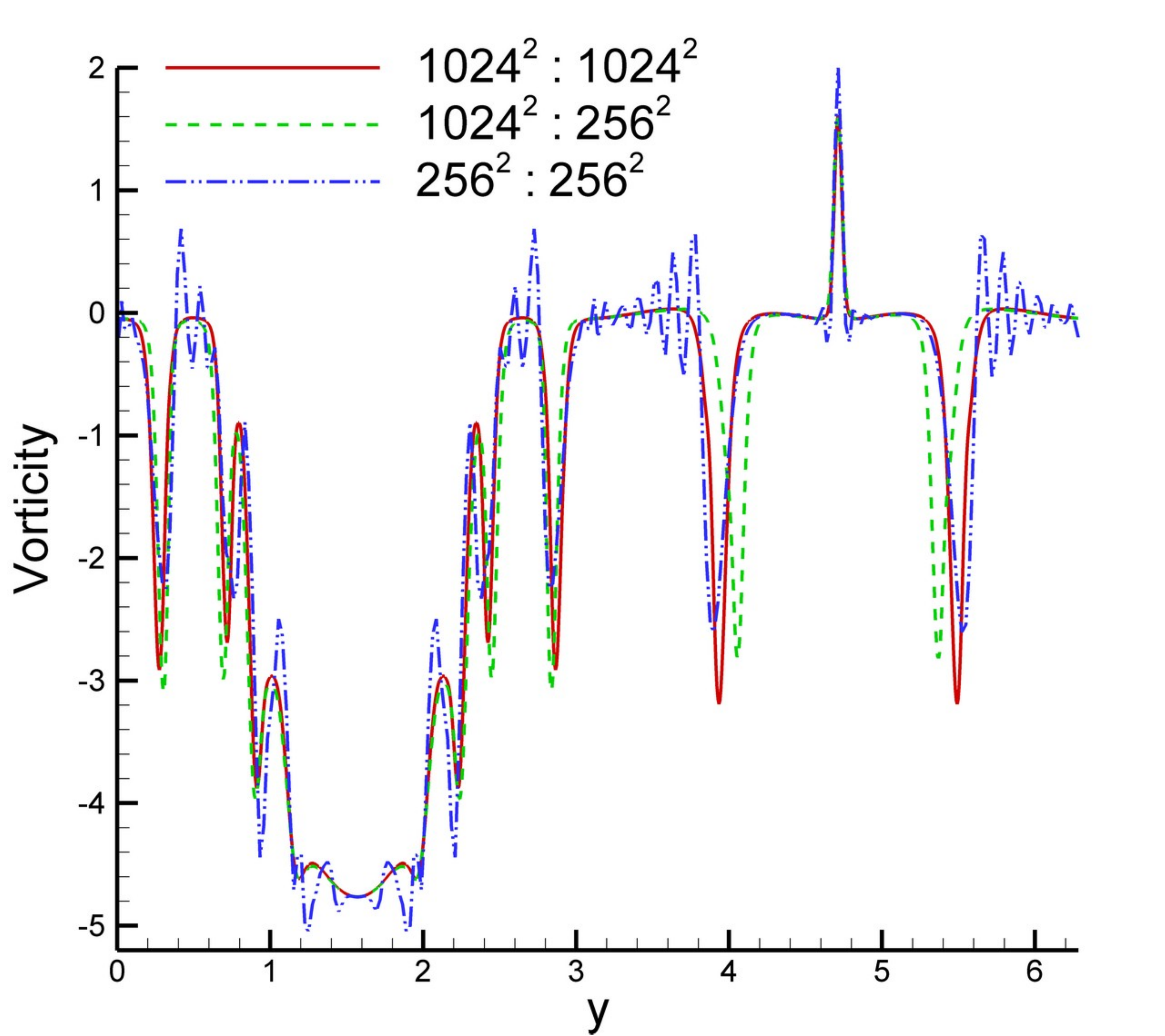}}
\subfigure[]{\includegraphics[width=0.5\textwidth]{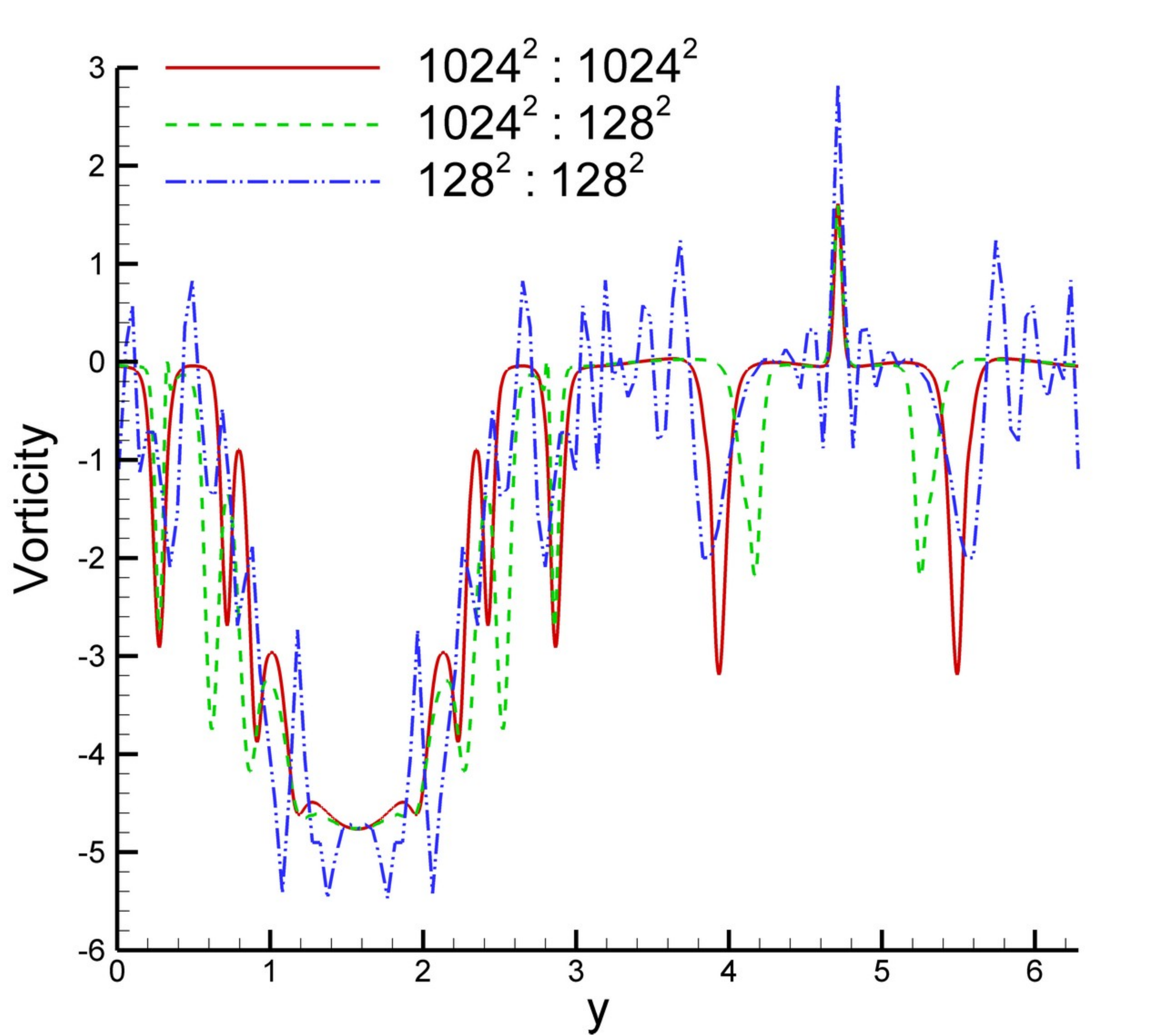}}
}
\caption{Centerline vorticity distributions for the double shear layer problem at $t=10$ obtained using the vorticity-stream function formulation; (a) comparison for one level coarsening including the standard computation (RK3) on $1024^2:1024^2$ resolution grids, the CGPRK3 method on $1024^2:512^2$ resolution grids, and the standard computation (RK3) on $512^2:512^2$ resolution grids, (b) comparison for one level coarsening including the standard computation (RK3) on $512^2:512^2$ resolution grids, the CGPRK3 method on $512^2:256^2$ resolution grids, and the standard computation (RK3) on $256^2:256^2$ resolution grids, (c) comparison for two levels of coarsening, and (d) comparison for three levels of coarsening.}
\label{fig:dsl-time10-line}
\end{figure*}

The centerline vorticity distributions along the $y$-axis at $t=10$ obtained using the vorticity-stream function formulation with and without the CGP method are shown in Fig.~\ref{fig:dsl-time10-line} for different resolutions. It can be seen clearly from the figure that we obtain the fine resolution data at a reduced computational cost using the CGP coarsening strategy, and that this data is significantly better than that obtained via the wholly coarse computation. Both Fig.~\ref{fig:dsl-time10-line}a and Fig.~\ref{fig:dsl-time10-line}b demonstrate that the one level of coarsening CGP method obtains the fine resolution computational results, but uses less computational time. This is especially true for well resolved computations. In fact, data from less well resolved simulations, as shown in Fig.~\ref{fig:dsl-time10}, for example, demonstrate that one level of coarsening does increase the error, although there is still a significant benefit to using the CGP approach in that case. This shear layer problem is particularly important in that the presence of the thinner and thinner shear layers as the flow field evolves in time is not captured by low grid resolution representations.

The Gibbs phenomenon, numerical oscillations occurring near sharp vorticity gradients, occurs for underresolved simulations in which the grid size is larger than the shear layer thickness. For example, if we look at the results obtained with a $512^2$ resolution (at a relatively high Reynolds number), it can clearly be seen that this resolution is not high enough to capture the correct physics without numerical oscillations. To be able to obtain all the small-scale physical layers, a resolution of at least $1024^2$ needs to be used (using a spatially second-order accurate scheme for the underlying equations). Here, we demonstrate that we can obtain the properly-resolved physical behavior at the computational price of underresolved simulations. Interestingly, however, coarsening the Poisson equation using the CGP method appears to eliminate the Gibbs phenomenon, and produces a similar accuracy to the finest resolved scale simulation for one level of coarsening, while accelerating the simulation with a linear reduction in computational cost. The savings are linear because a fast Poisson solver is used in the algorithm. As shown in the previous problem, the savings would be greater if a suboptimal Poisson solver were used. It should also be noted that no Gibbs phenomenon appears for further levels of coarsening, as is shown in Fig.~\ref{fig:dsl-time10-line}c for two levels of coarsening, and in Fig.~\ref{fig:dsl-time10-line}d for three levels of coarsening. This can be understood by considering the spatial and temporal discretization of the advection-diffusion part as a low-pass filter over the grid, and the Poisson solver as a pre-filtering process. It is possible that the real-frequency limit of the advection-diffusion part is lower than the grid resolution and that the CGP method is acting as a low-pass pre-filter. Accordingly, this study demonstrates that the numerical oscillations that occur in underresolved simulations with strong shear layers can be eliminated efficiently by using the CGP method.

\begin{figure*}
\centering
\mbox{
\subfigure{\includegraphics[width=0.33\textwidth]{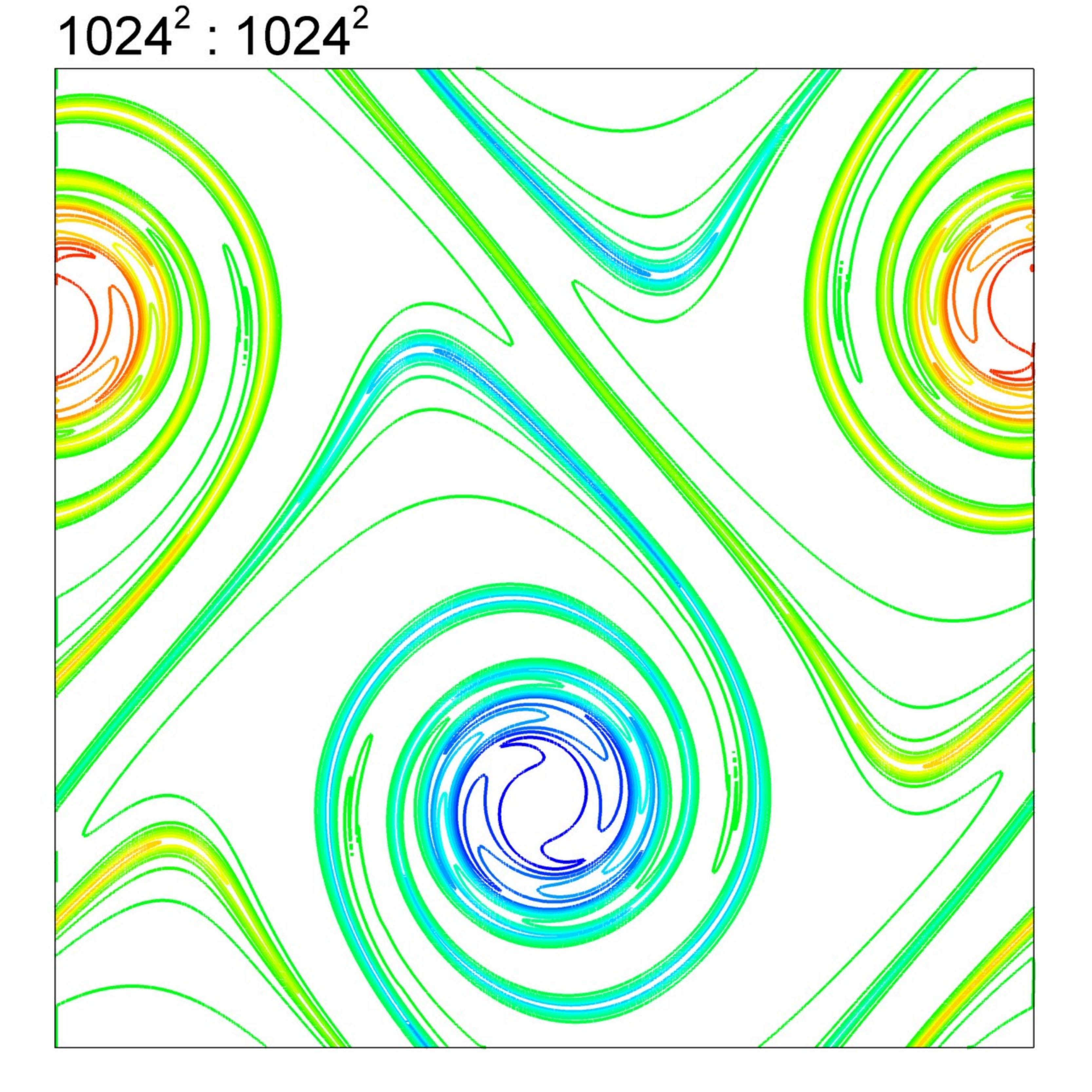}}
\subfigure{\includegraphics[width=0.33\textwidth]{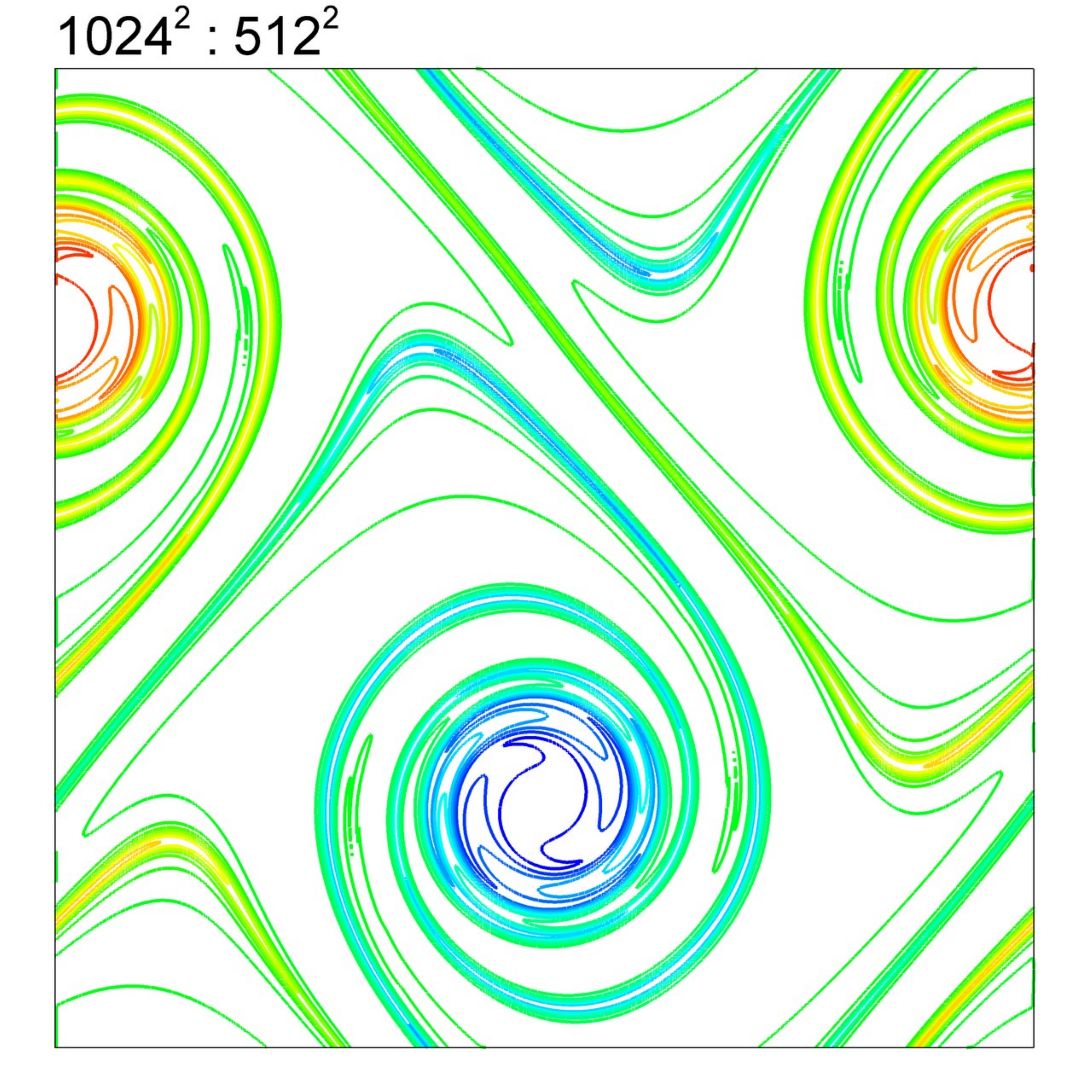}}
\subfigure{\includegraphics[width=0.33\textwidth]{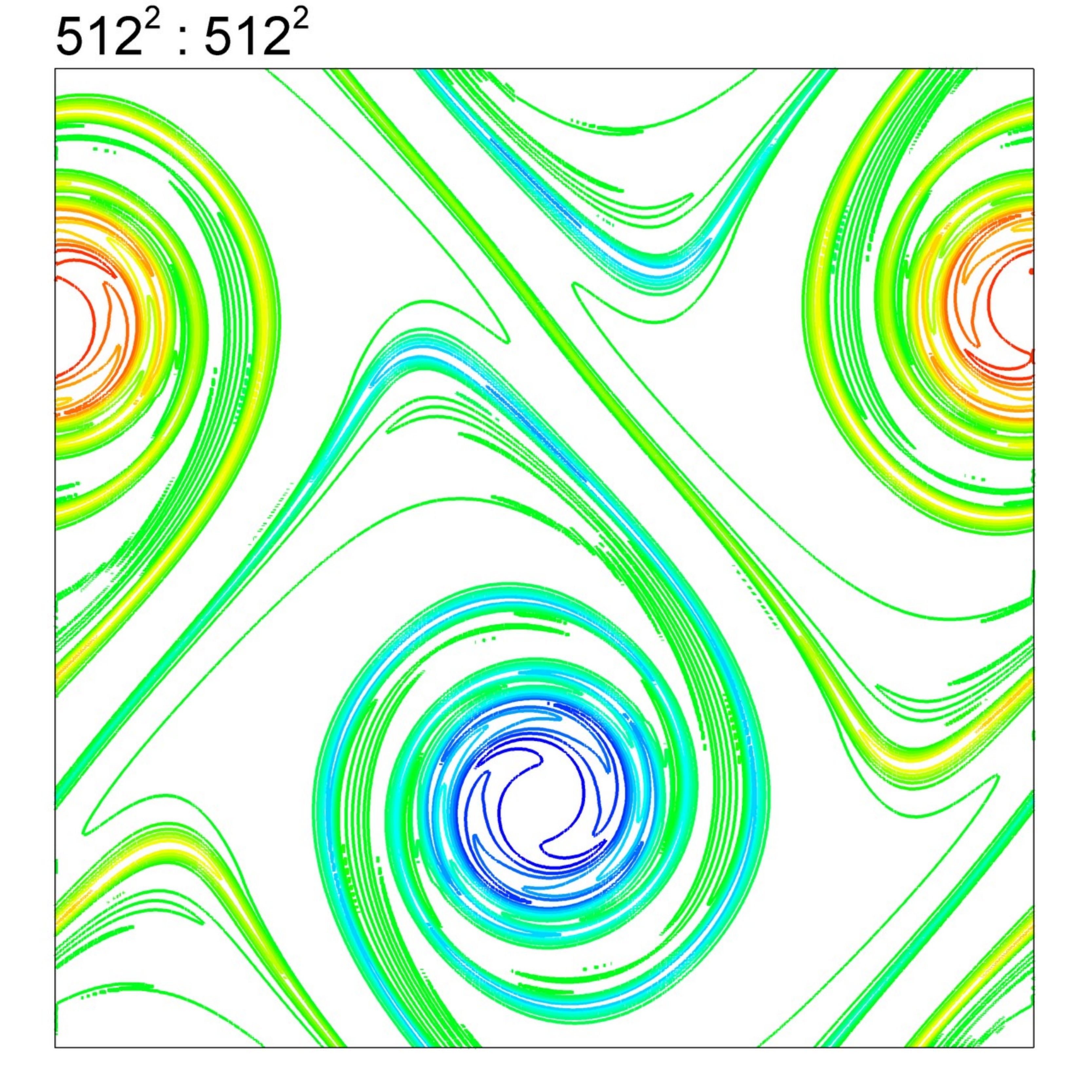}}
}
\caption{The double shear layer problem vorticity fields at $t=10$ obtained using the primitive variable fractional step formulation (CGPRK3-PV) showing the standard computation on $1024^2:1024^2$ resolution grids, the CGP method on $1024^2:512^2$ resolution grids, and the standard computation on $512^2:512^2$ resolution grids.}
\label{fig:dsl-time10p}
\end{figure*}

\begin{figure*}
\centering
\mbox{
\subfigure[]{\includegraphics[width=0.5\textwidth]{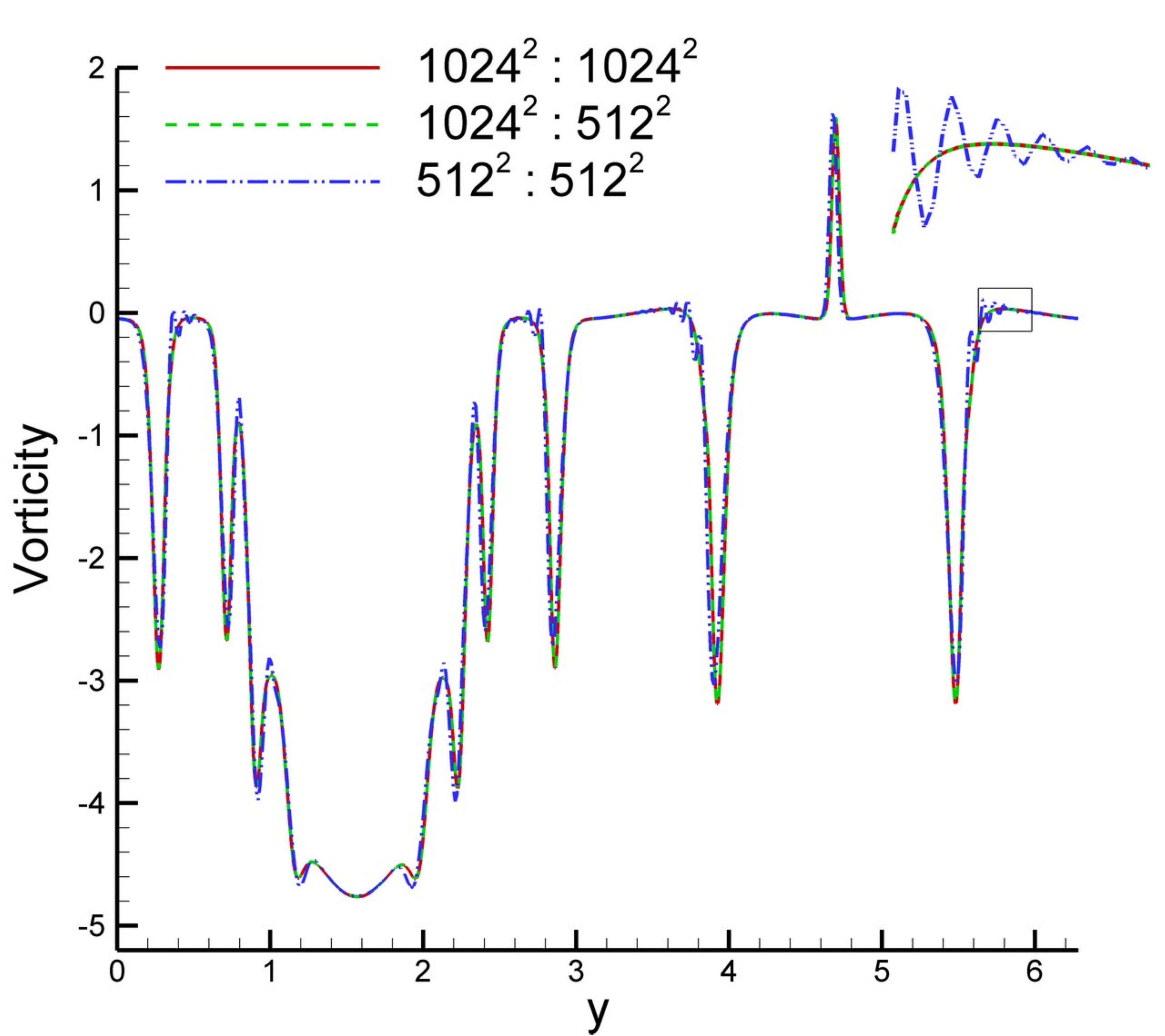}}
\subfigure[]{\includegraphics[width=0.5\textwidth]{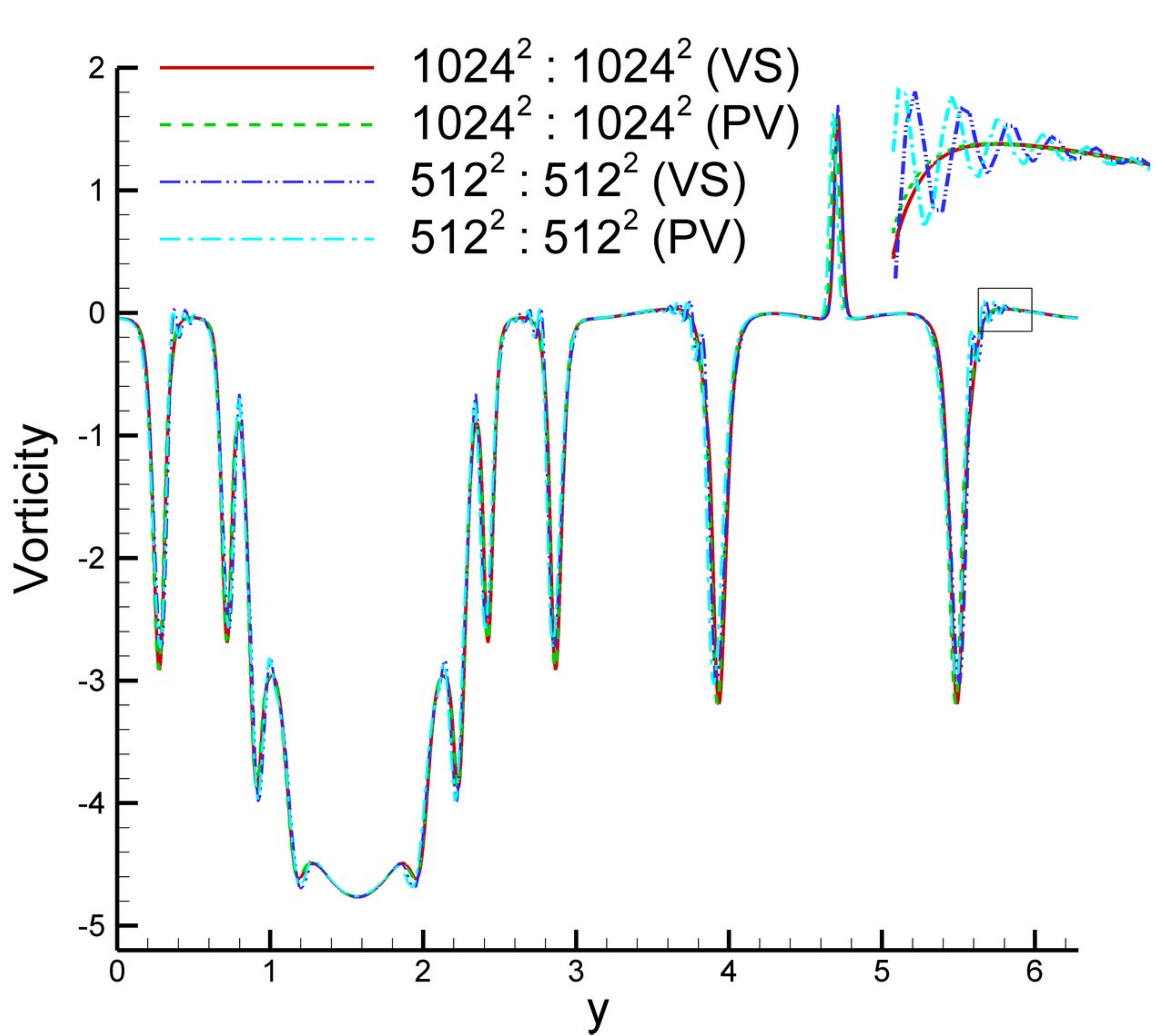}}
}
\caption{Centerline vorticity distributions at $t=10$ for the double shear layer problem: (a) Results obtained using the primitive variable fractional step formulation (CGPRK3-PV) showing a comparison of the standard computation on $1024^2:1024^2$ resolution grids, the CGP method on $1024^2:512^2$ resolution grids, and the standard computation on $512^2:512^2$ resolution grids. (b) Comparison of the primitive variable (PV) formulation and the vorticity-stream function (VS) formulation.}
\label{fig:dsl-time10p-line}
\end{figure*}

Similarly, Fig.~\ref{fig:dsl-time10p} shows the vorticity field data obtained using the primitive variable fractional step method. The corresponding centerline vorticity distributions are plotted in Fig.~\ref{fig:dsl-time10p-line}. The vorticity field data is obtained via a postprocessing procedure at the desired times. Similar speed-up is observed upon performing the CGP strategy as for the primitive variable formulation case. The results also illustrate that the CGP method provides very similar data to the full fine scale simulations for the case of primitive variable fractional step formulation. If we compare the vorticity-stream function formulation and the primitive variable formulation by looking at Fig.~\ref{fig:dsl-time10} and Fig.~\ref{fig:dsl-time10p} for vorticity field plots and by looking at Fig.~\ref{fig:dsl-time10p-line} for centerline vorticity comparisons, we see that the vorticity-stream function formulation provides slightly more accurate data than the primitive variable formulation, mainly because of a small projection error in the fractional step procedure.

\subsection{Co-rotating vortex pair}
\label{sec:corotvp}
\begin{figure*}
\centering
\mbox{
\subfigure{\includegraphics[width=0.33\textwidth]{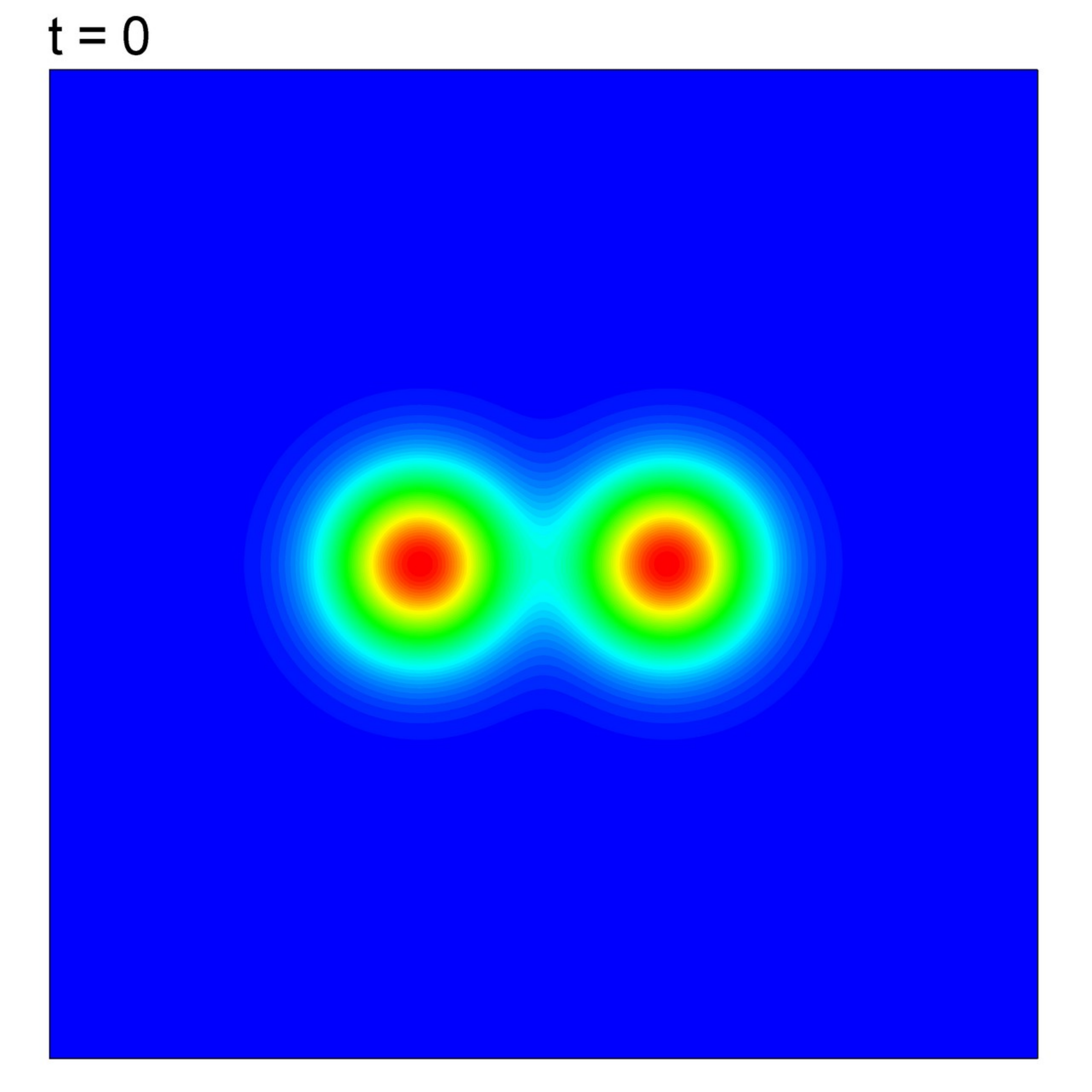}}
\subfigure{\includegraphics[width=0.33\textwidth]{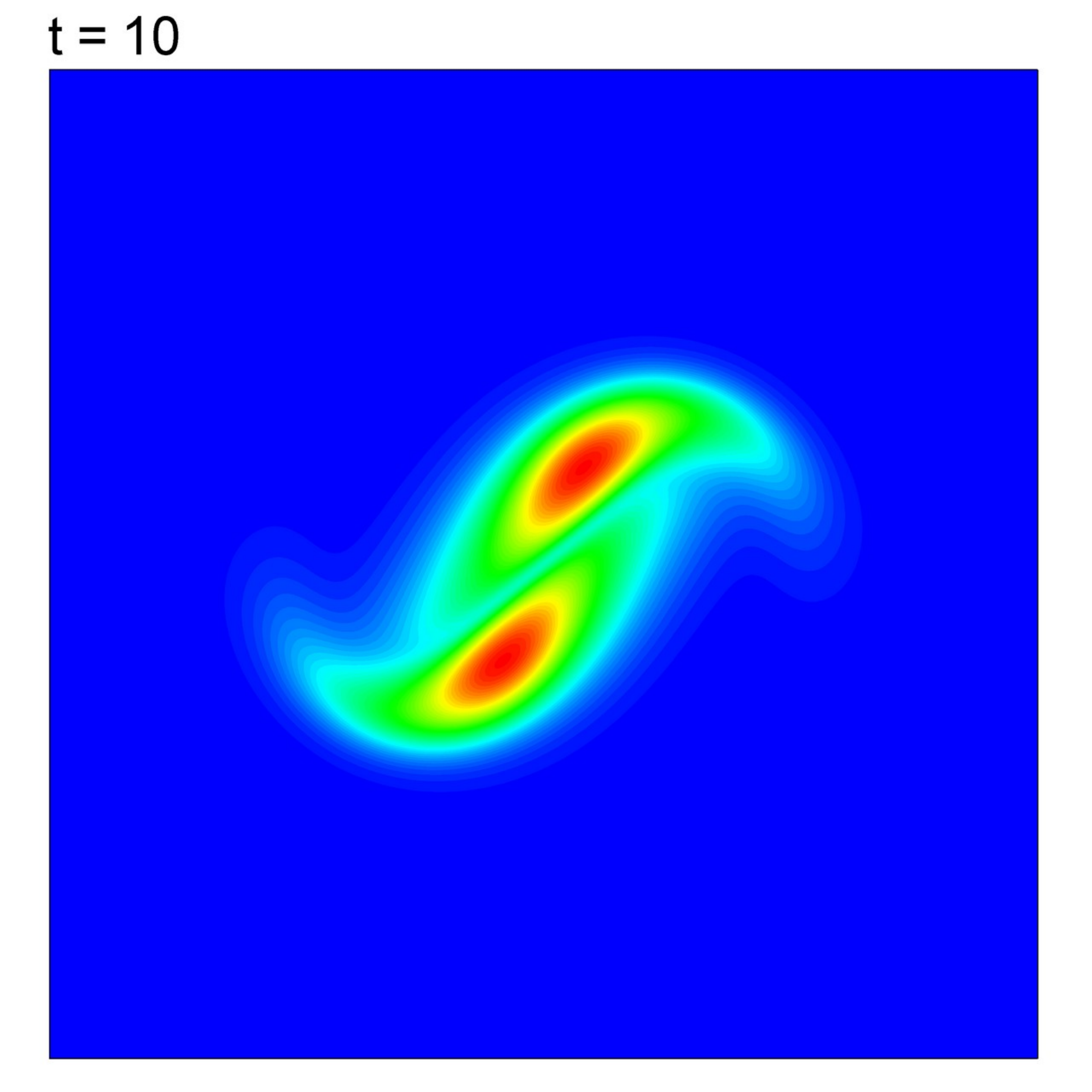}}
\subfigure{\includegraphics[width=0.33\textwidth]{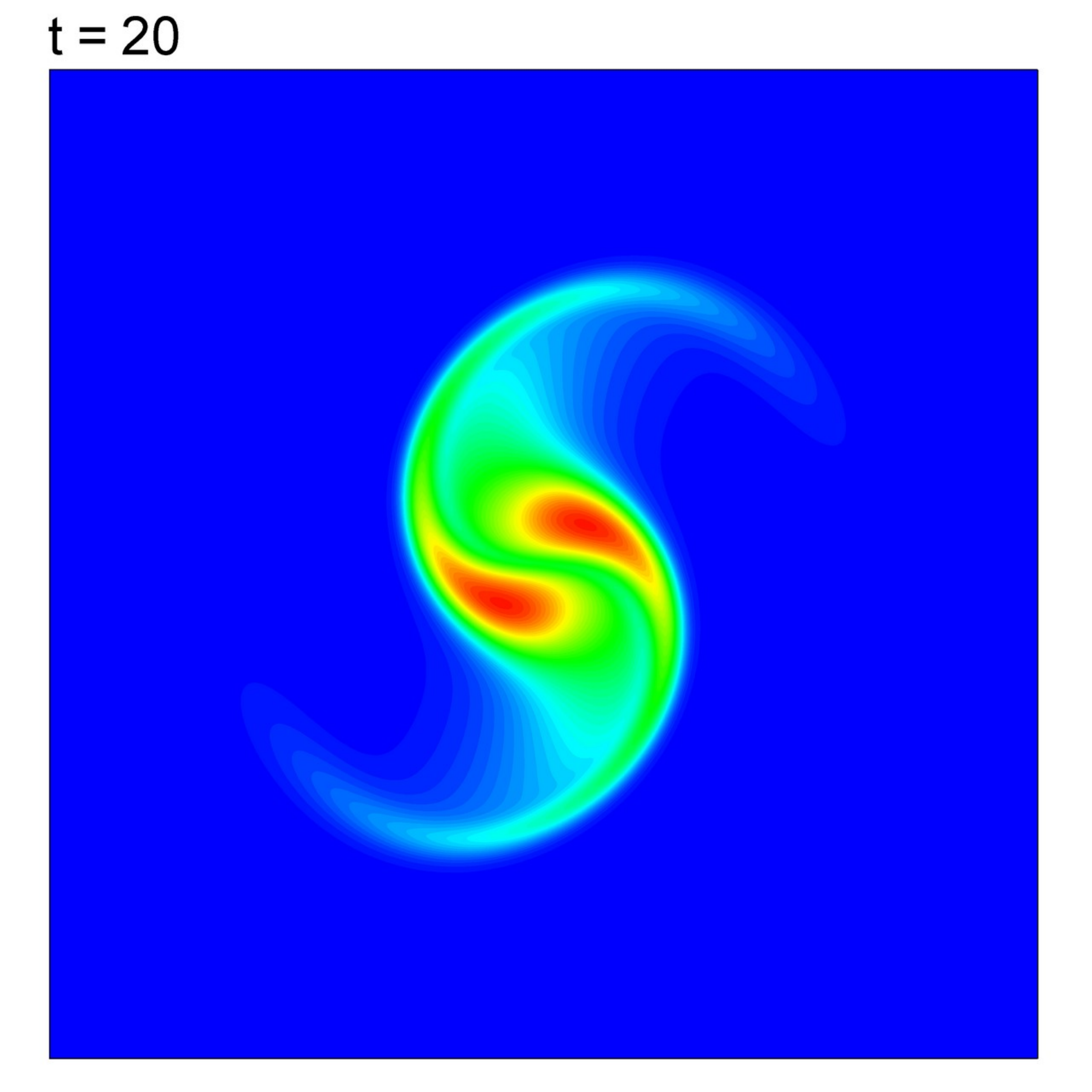}}
}
\mbox{
\subfigure{\includegraphics[width=0.33\textwidth]{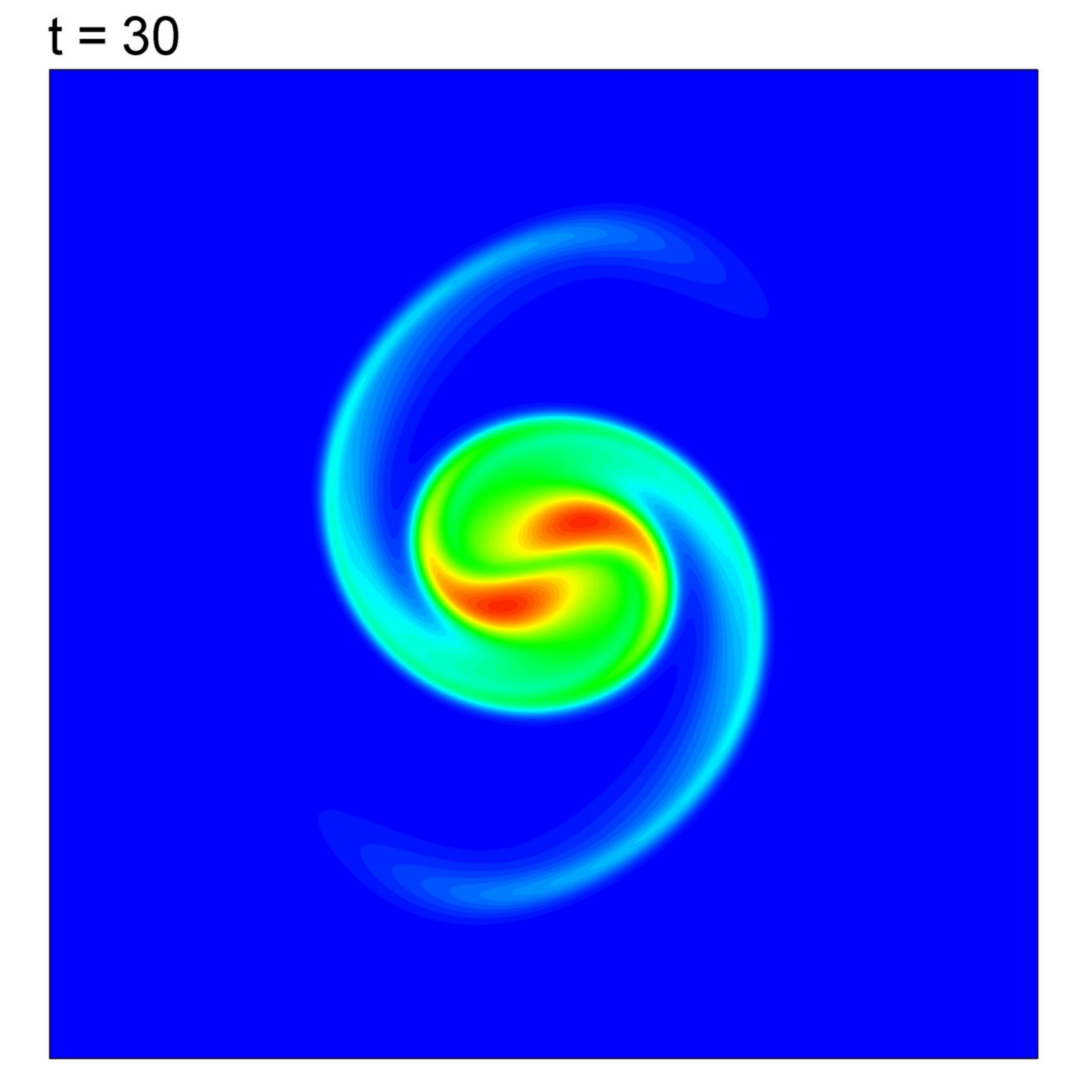}}
\subfigure{\includegraphics[width=0.33\textwidth]{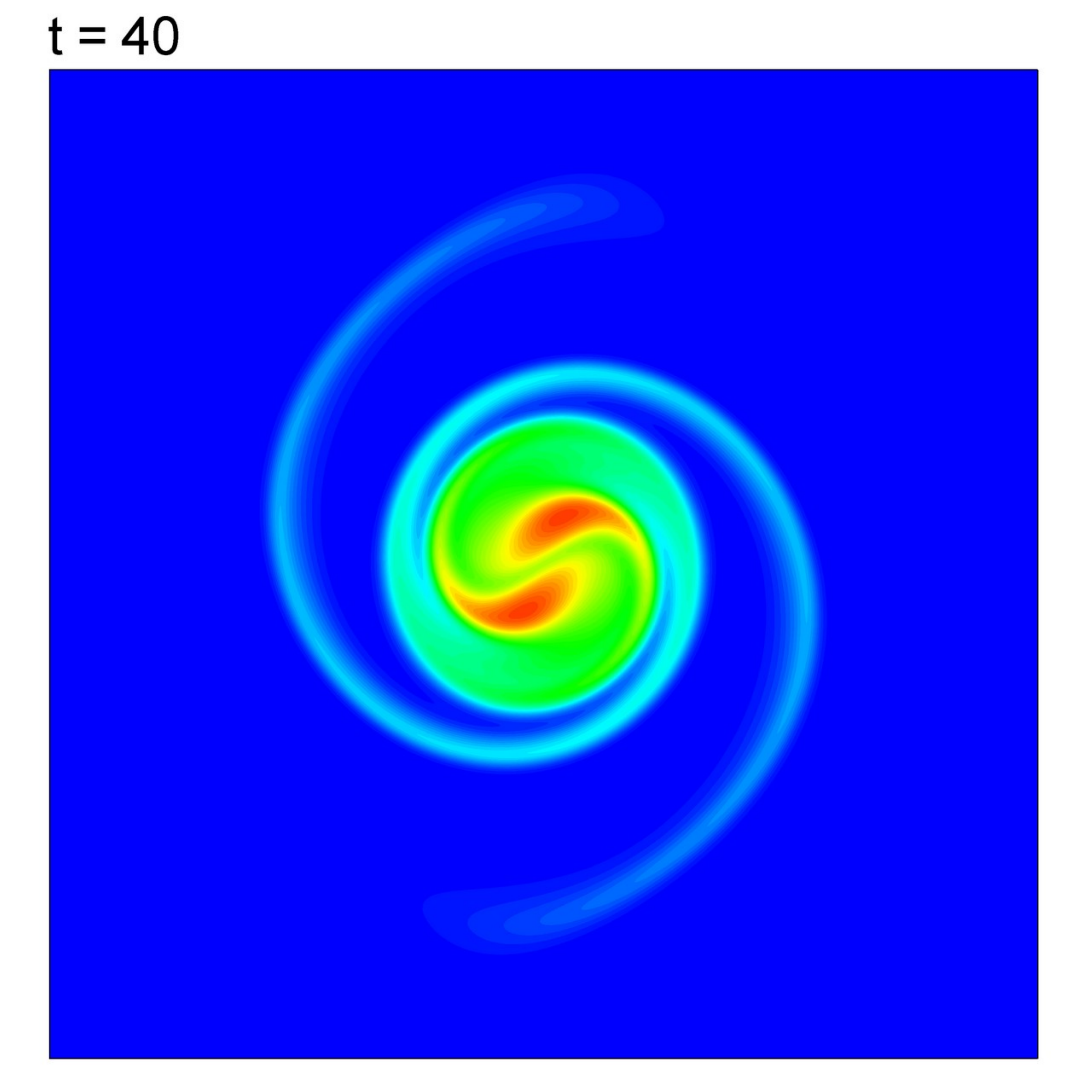}}
\subfigure{\includegraphics[width=0.33\textwidth]{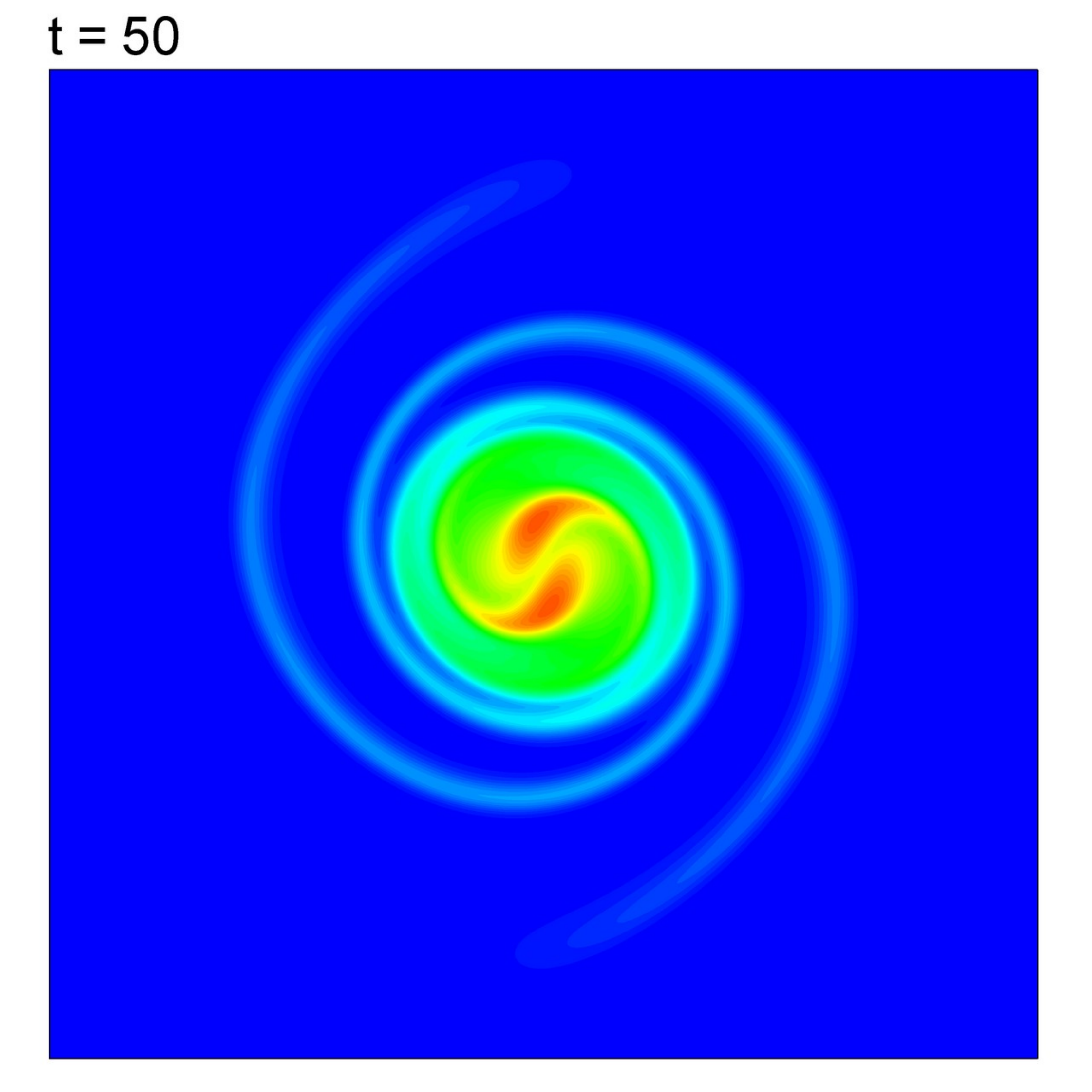}}
}
\caption{Vorticity field at different times obtained using CGPRK3 on $1024^2:512^2$ resolution grids.}
\label{fig:merger-time}
\end{figure*}

In this section, the CGP method is applied to the problem of a merging, co-rotating vortex pair \citep{buntine2006merger,von2000vortex}. The superposition of two Gaussian-distributed vortices gives the following initial vorticity field:
\begin{equation}
\omega(x,y,0)=\Gamma_1 e^{-\rho[(x-x_1)^2 + (y-y_1)^2]} + \Gamma_2 e^{-\rho[(x-x_2)^2 + (y-y_2)^2]}
\label{eq:merger}
\end{equation}
where, for our computations, the vortices have the same circulation $\Gamma_1=\Gamma_2 = 1$, the interacting constant is set to $\rho=\pi$, and the vortex centers are initially located near each other with coordinates $(x_1,y_1)=(3\pi/4,\pi)$ and $(x_2,y_2)=(5\pi/4,\pi)$. A box of side length $2\pi$ is used as the computational domain, and computations are carried out for a Reynolds number of $Re=10^4$ and a time step of $\Delta t = 10^{-3}$, with periodic boundary conditions. As is illustrated in Fig.~\ref{fig:merger-time}, as time evolves, the exterior strain field from each vortex deforms the vorticity field of the other vortex so that the vorticity fields wrap around each other.

\begin{figure*}
\centering
\mbox{
\subfigure{\includegraphics[width=0.33\textwidth]{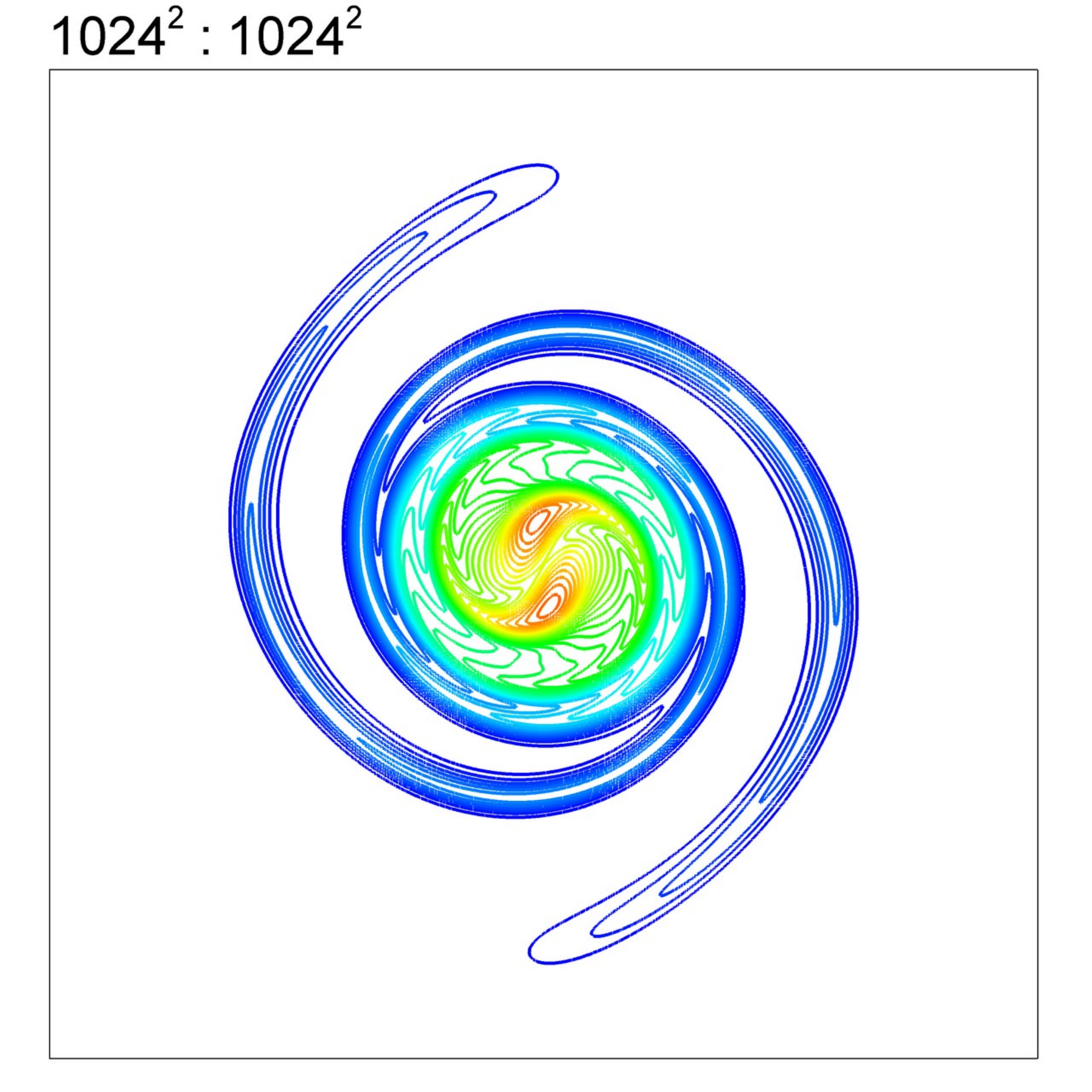}}
\subfigure{\includegraphics[width=0.33\textwidth]{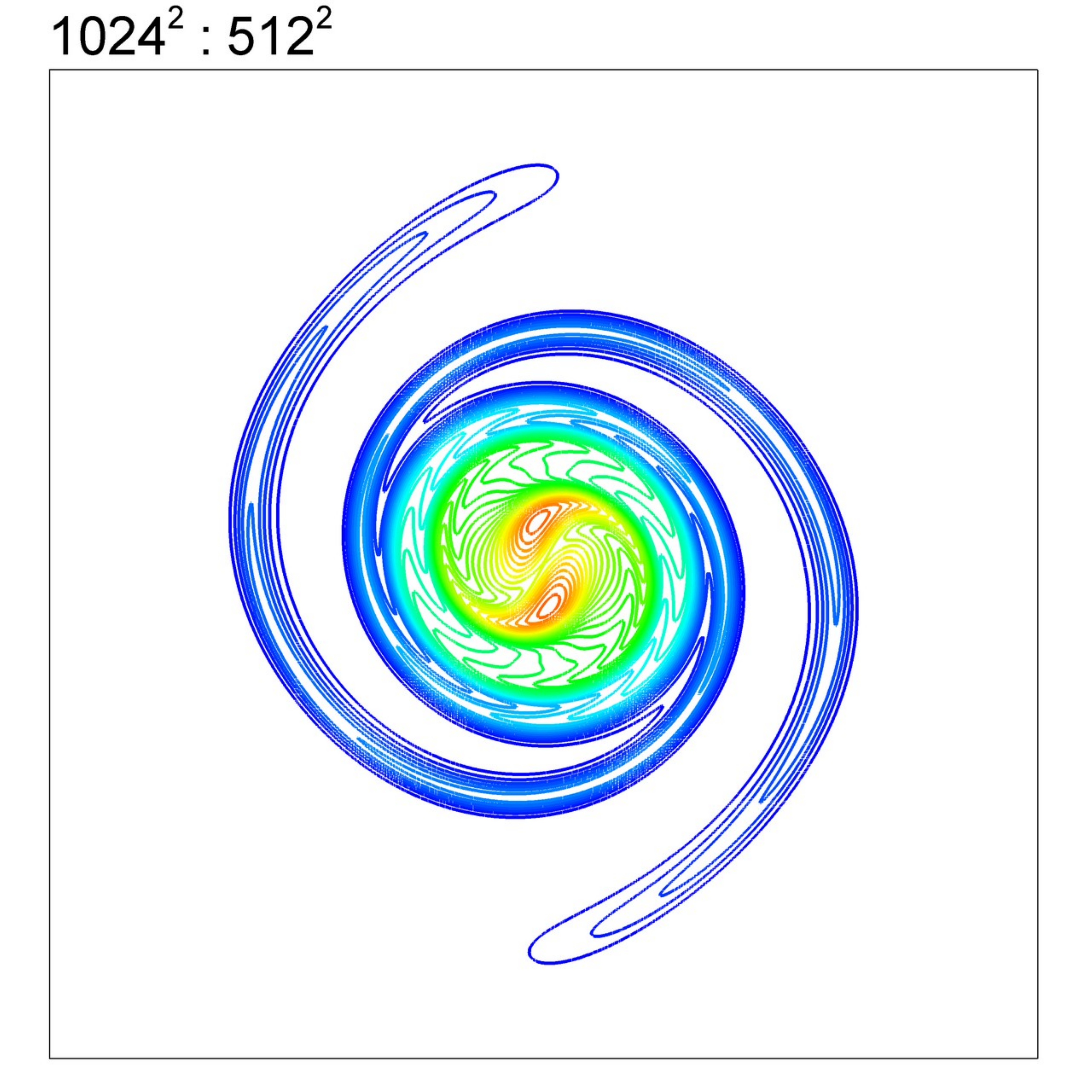}}
\subfigure{\includegraphics[width=0.33\textwidth]{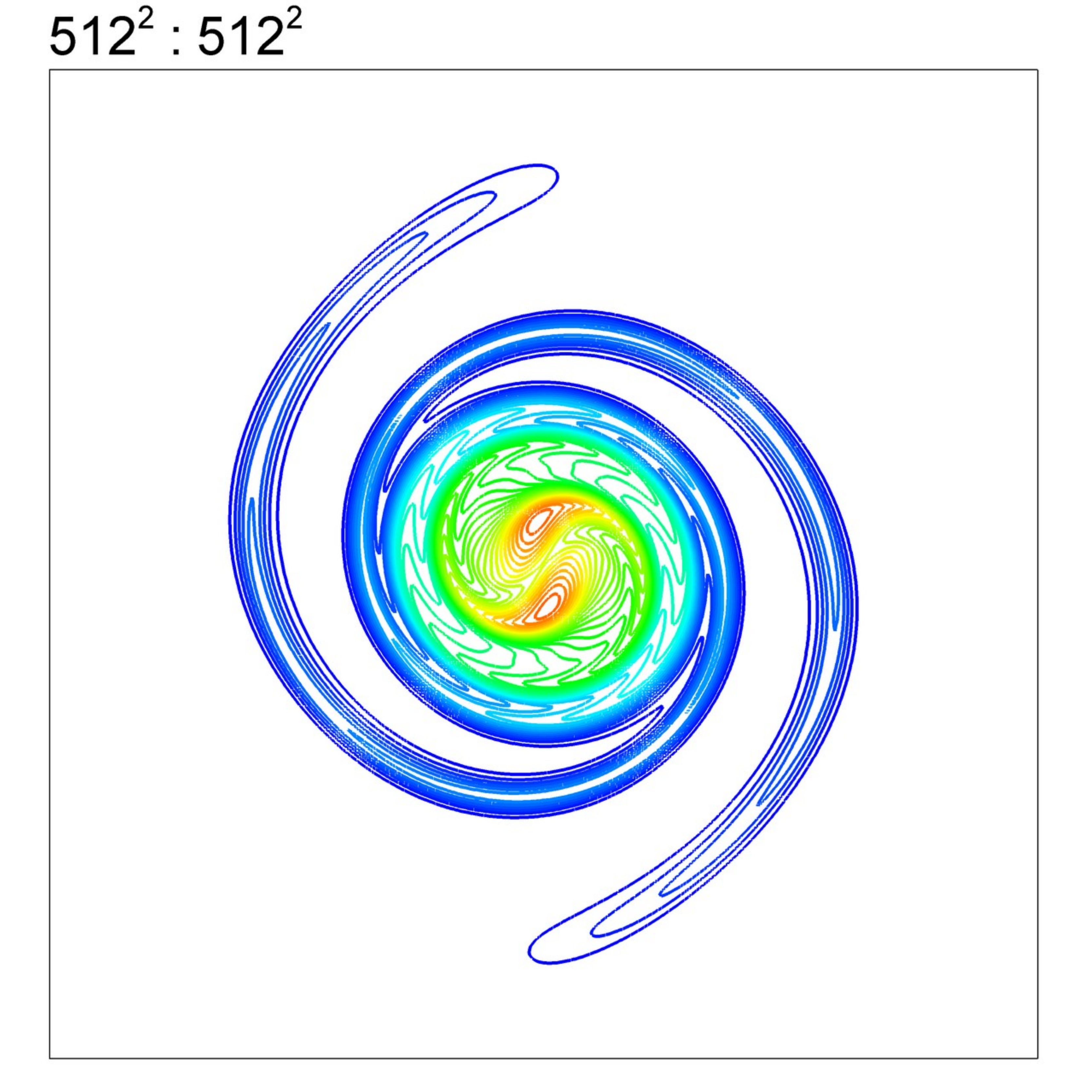}}
}
\mbox{
\subfigure{\includegraphics[width=0.33\textwidth]{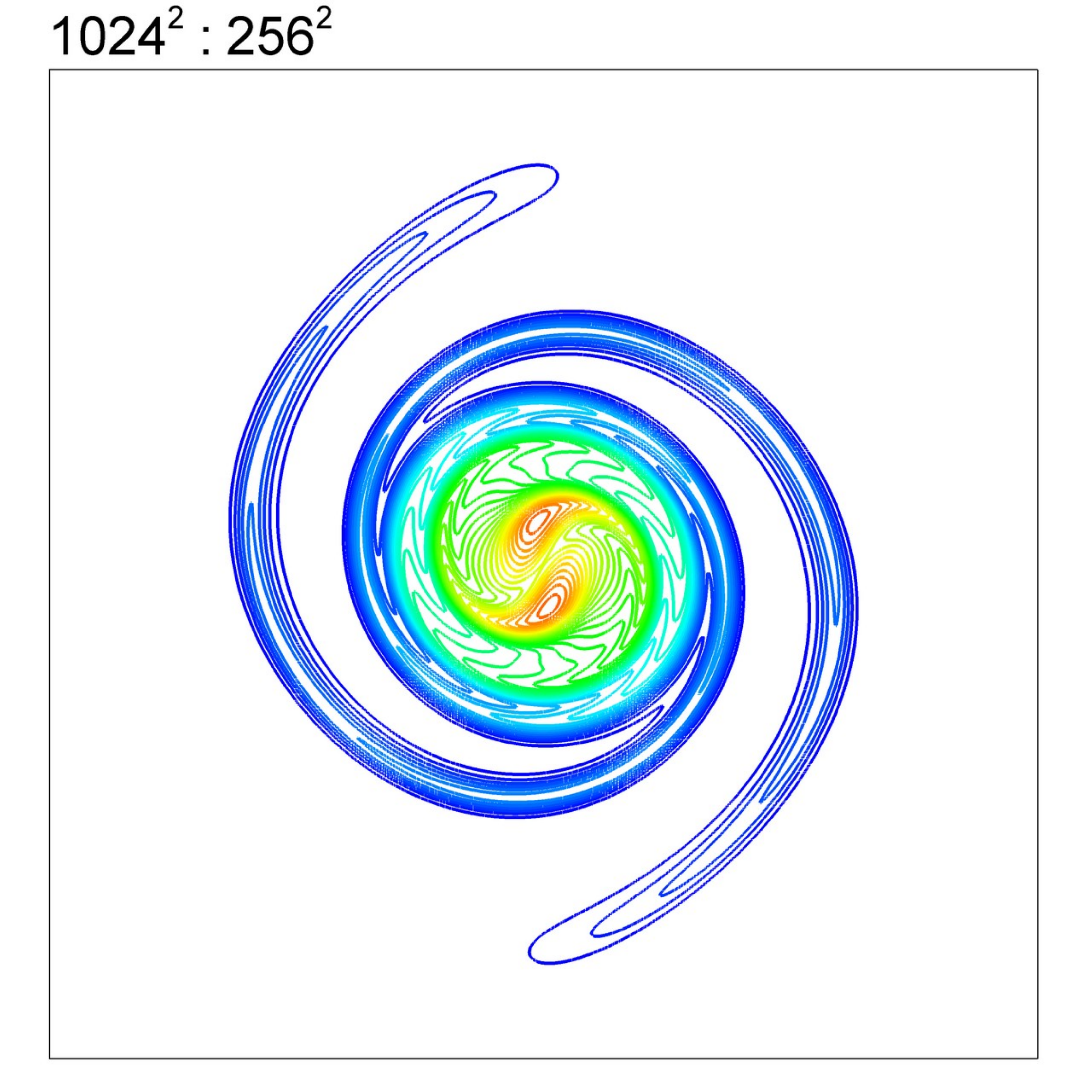}}
\subfigure{\includegraphics[width=0.33\textwidth]{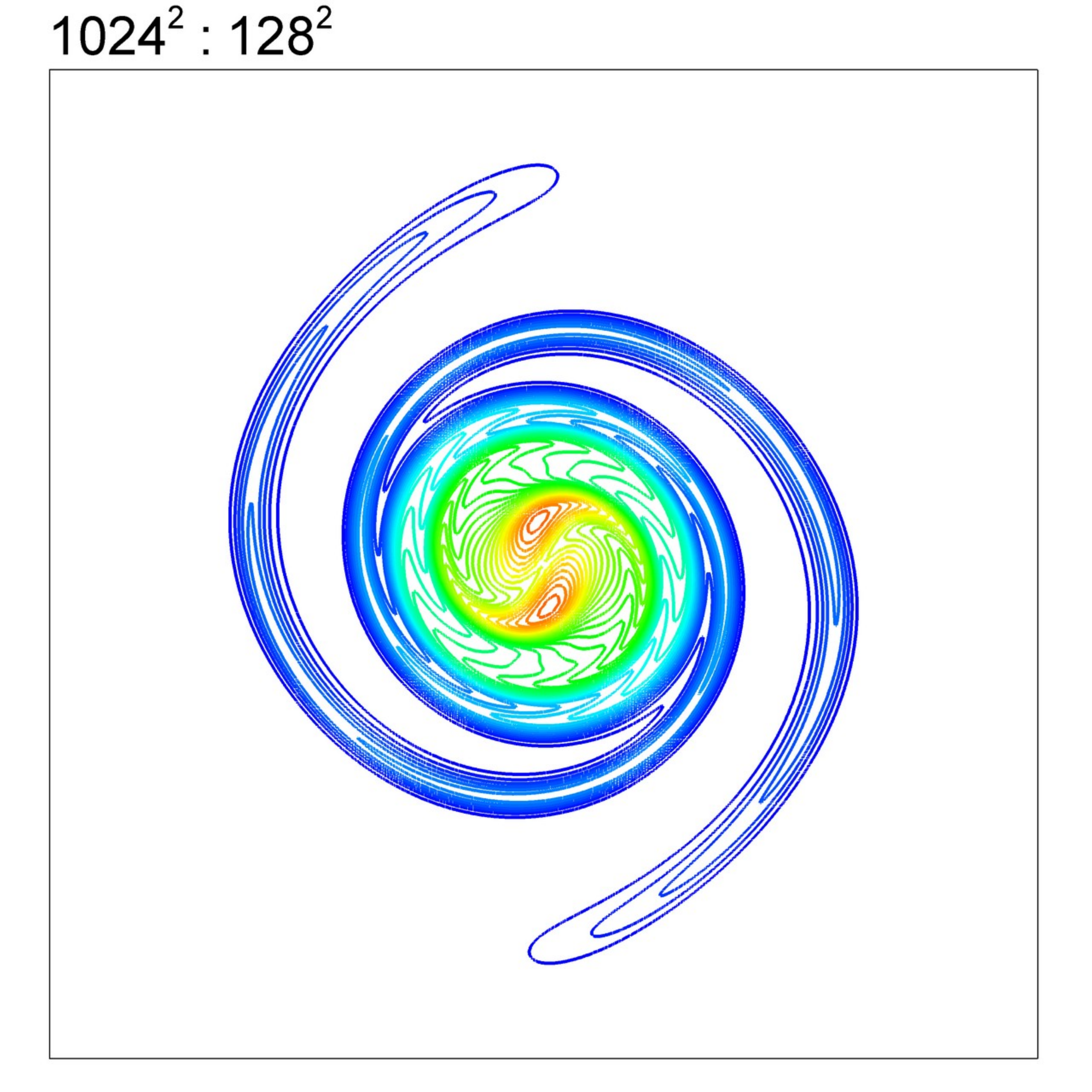}}
\subfigure{\includegraphics[width=0.33\textwidth]{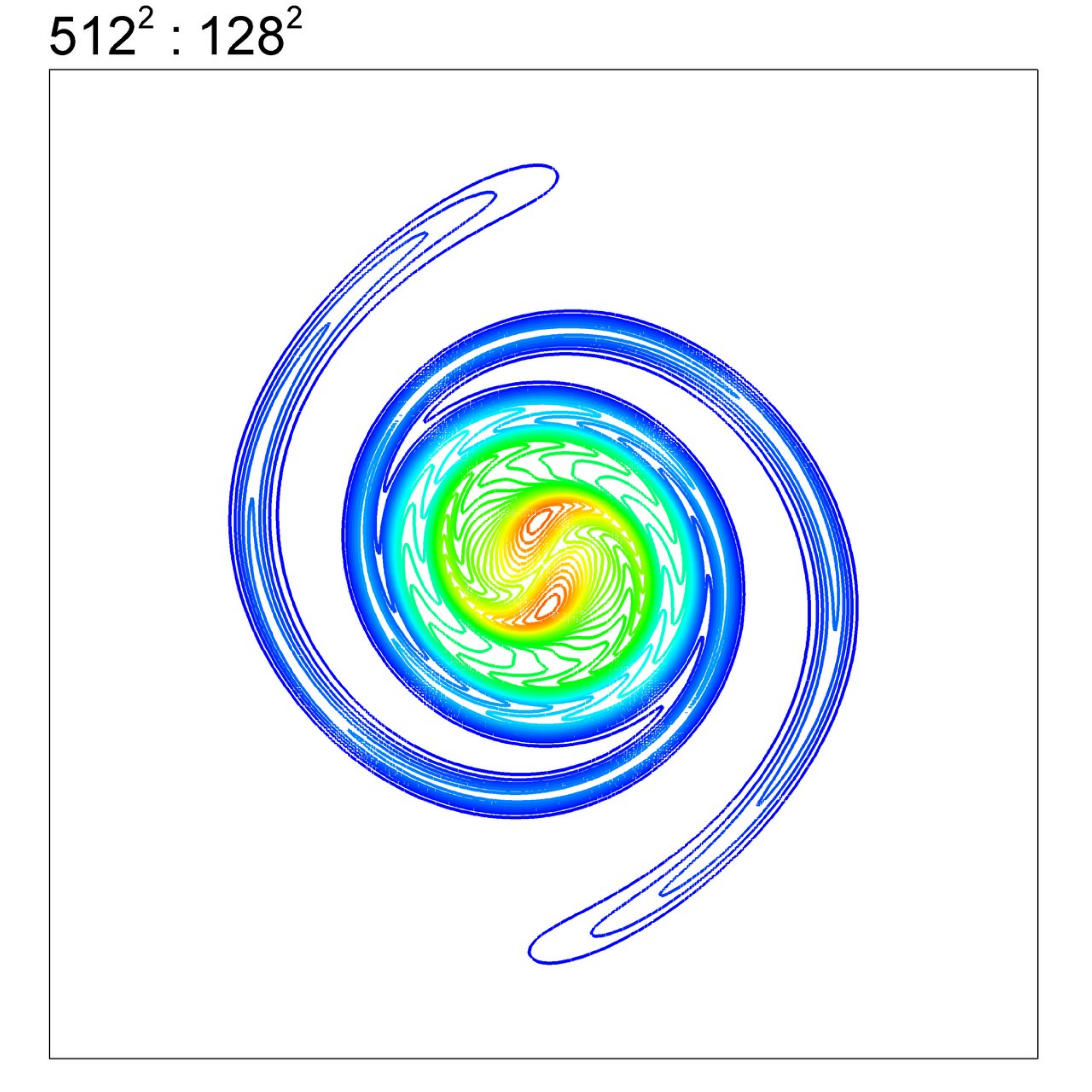}}
}
\mbox{
\subfigure{\includegraphics[width=0.33\textwidth]{merger-512-512.pdf}}
\subfigure{\includegraphics[width=0.33\textwidth]{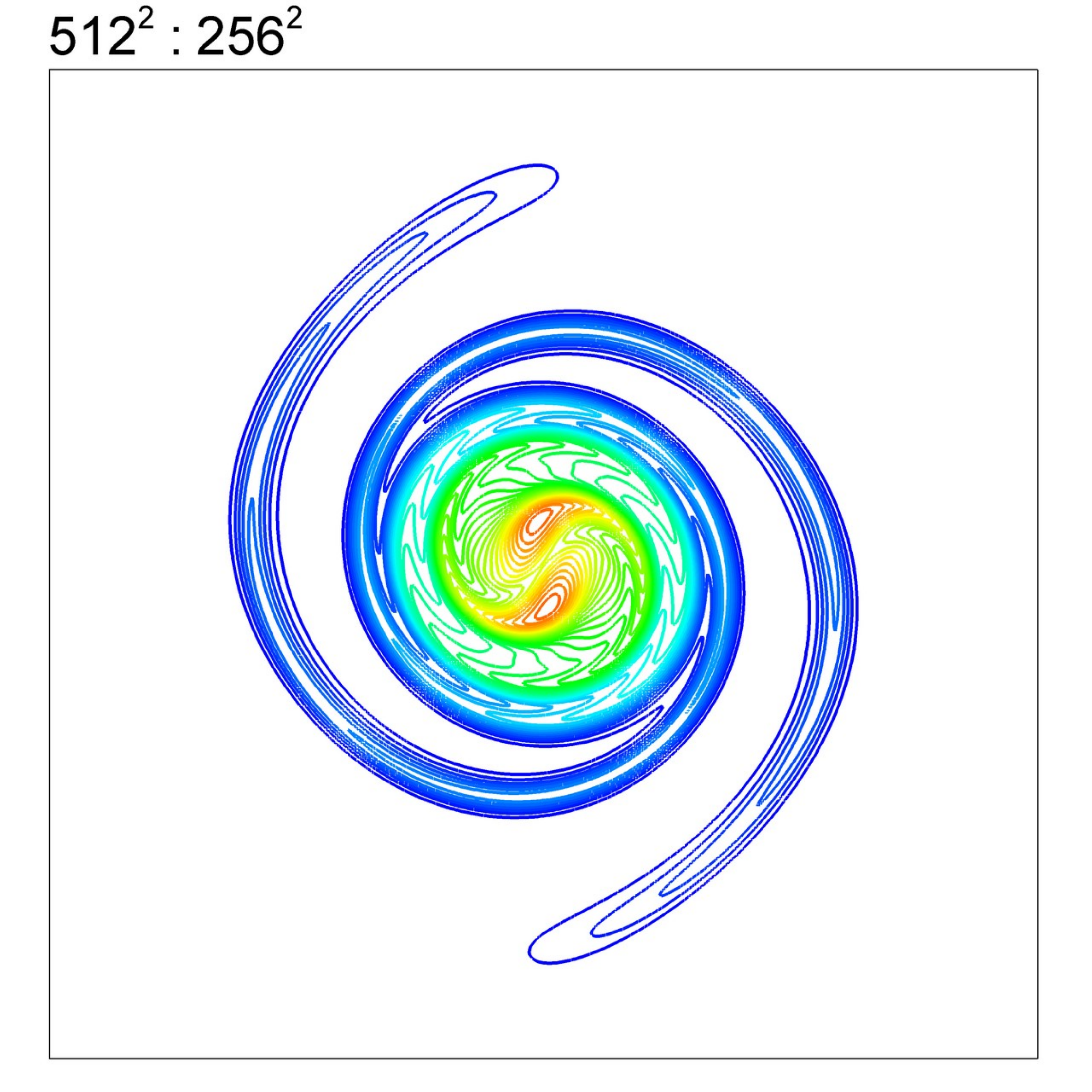}}
\subfigure{\includegraphics[width=0.33\textwidth]{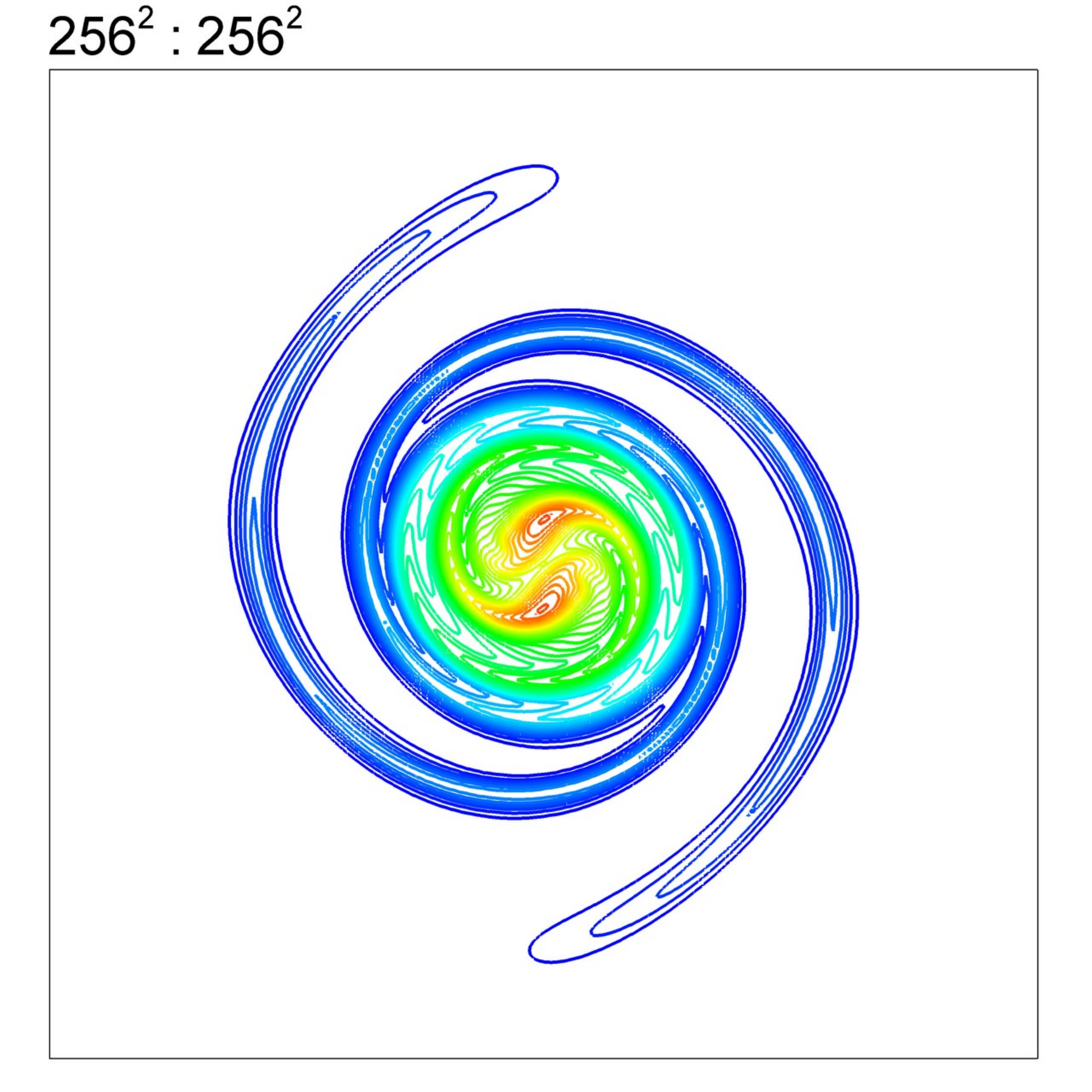}}
}
\mbox{
\subfigure{\includegraphics[width=0.33\textwidth]{merger-256-256.pdf}}
\subfigure{\includegraphics[width=0.33\textwidth]{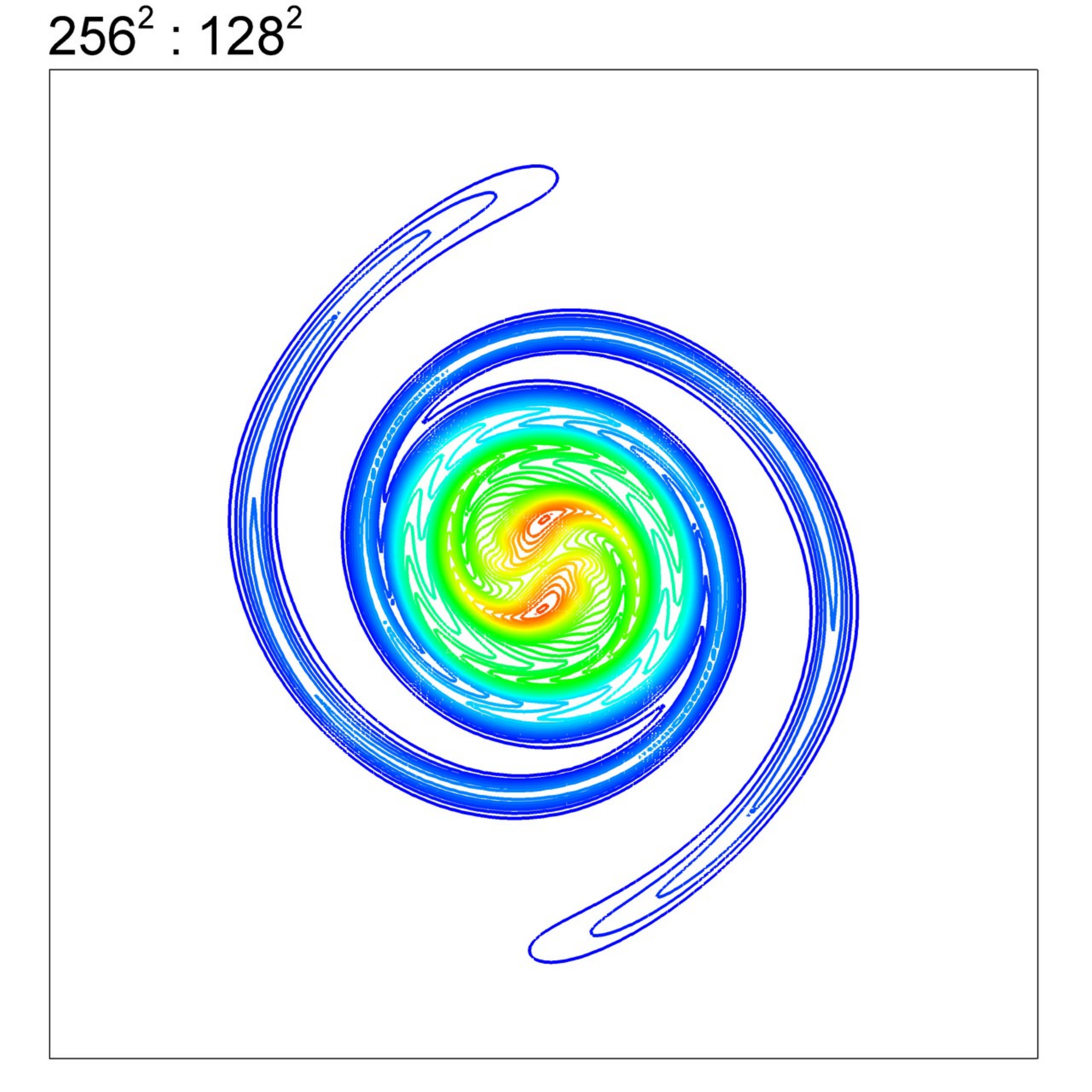}}
\subfigure{\includegraphics[width=0.33\textwidth]{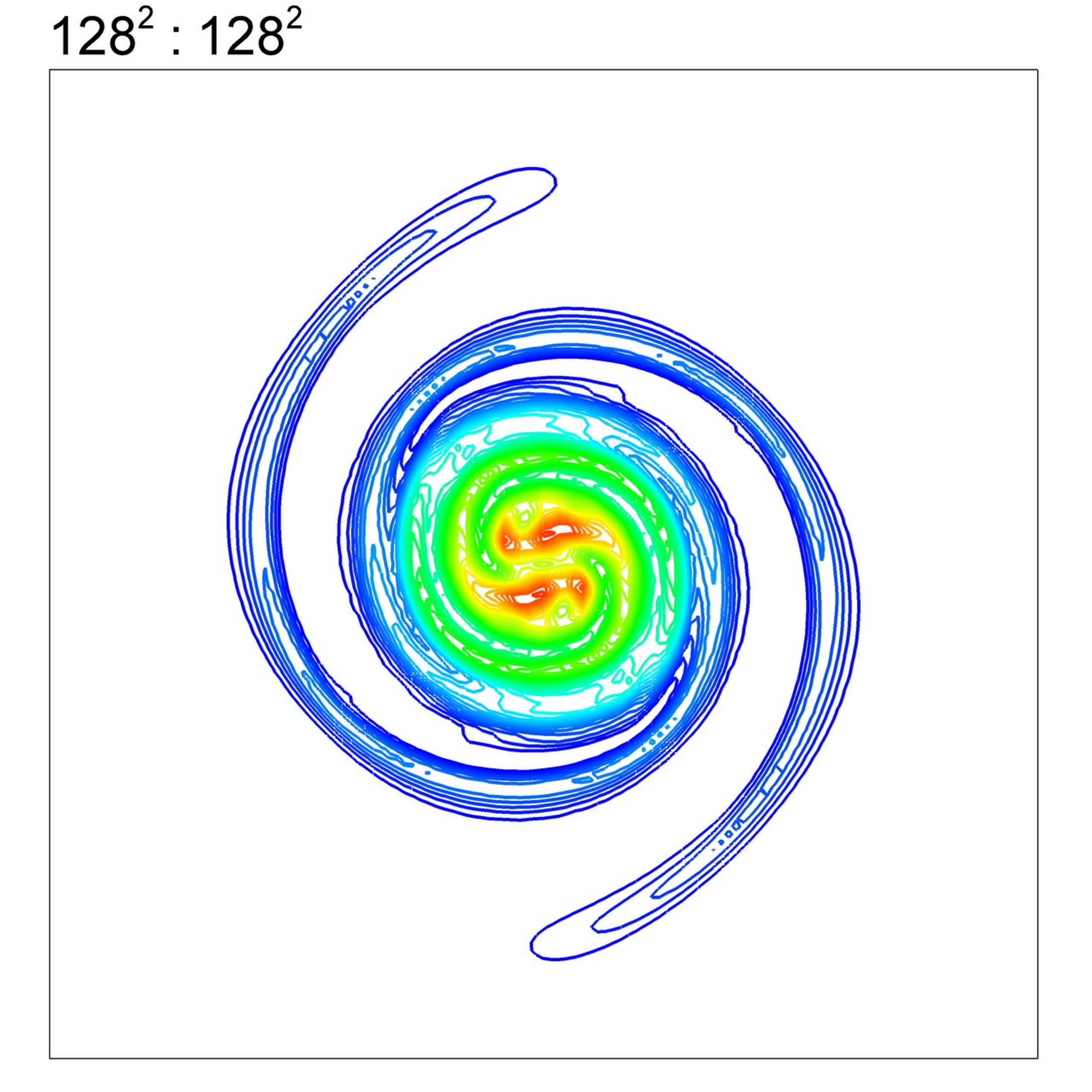}}
}
\caption{The vorticity fields for the merging co-rotating vortex pair problem at $t=50$ obtained using the vorticity-stream function formulation. Labels show the resolutions for both parts of the solver in the form $N^2:M^2$, where $N^2$ is the resolution for the vorticity-transport equation, and $M^2$ is the resolution for the Poisson equation.}
\label{fig:merger}
\end{figure*}

\begin{figure*}
\centering
\mbox{
\subfigure[]{\includegraphics[width=0.5\textwidth]{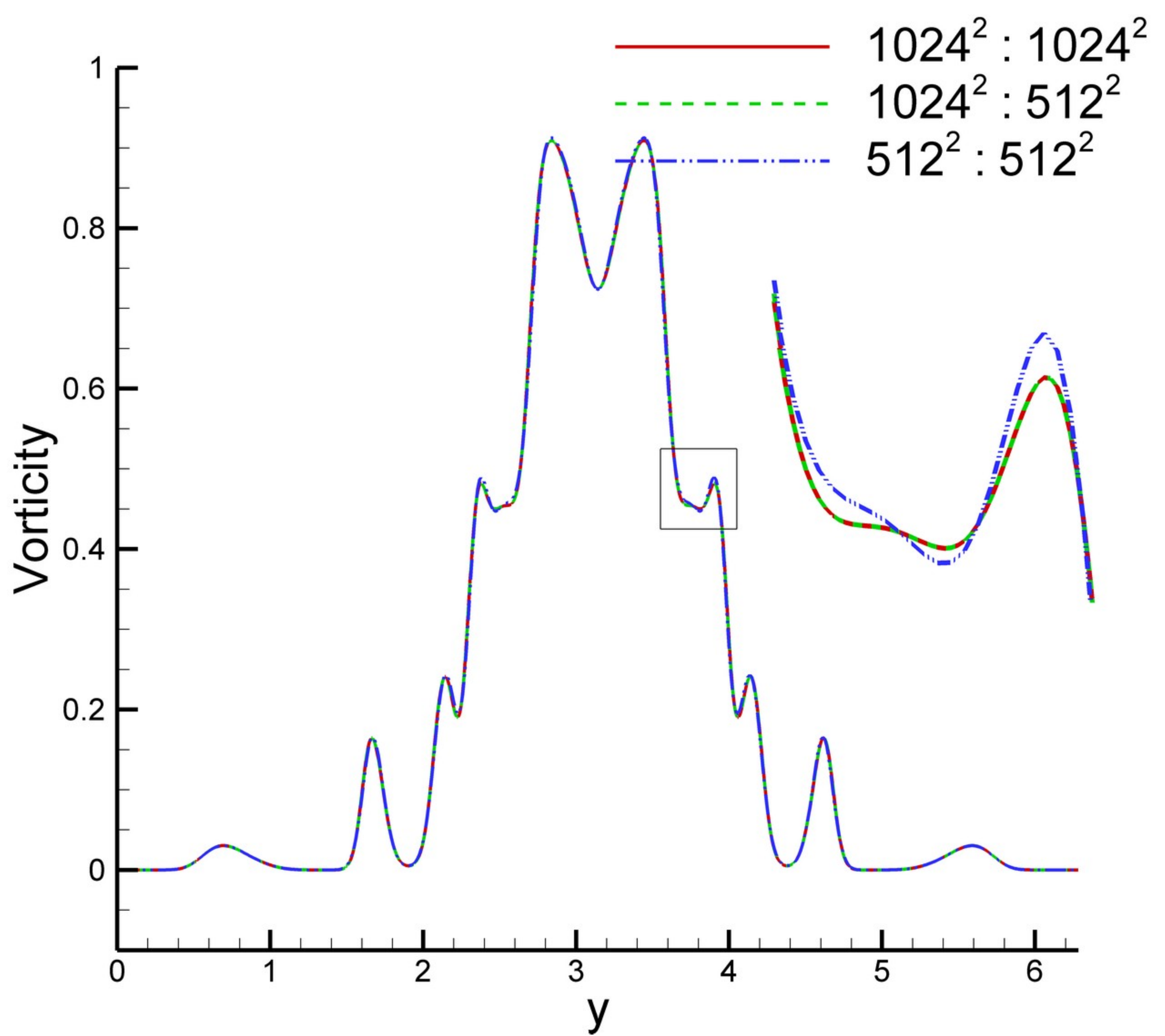}}
\subfigure[]{\includegraphics[width=0.5\textwidth]{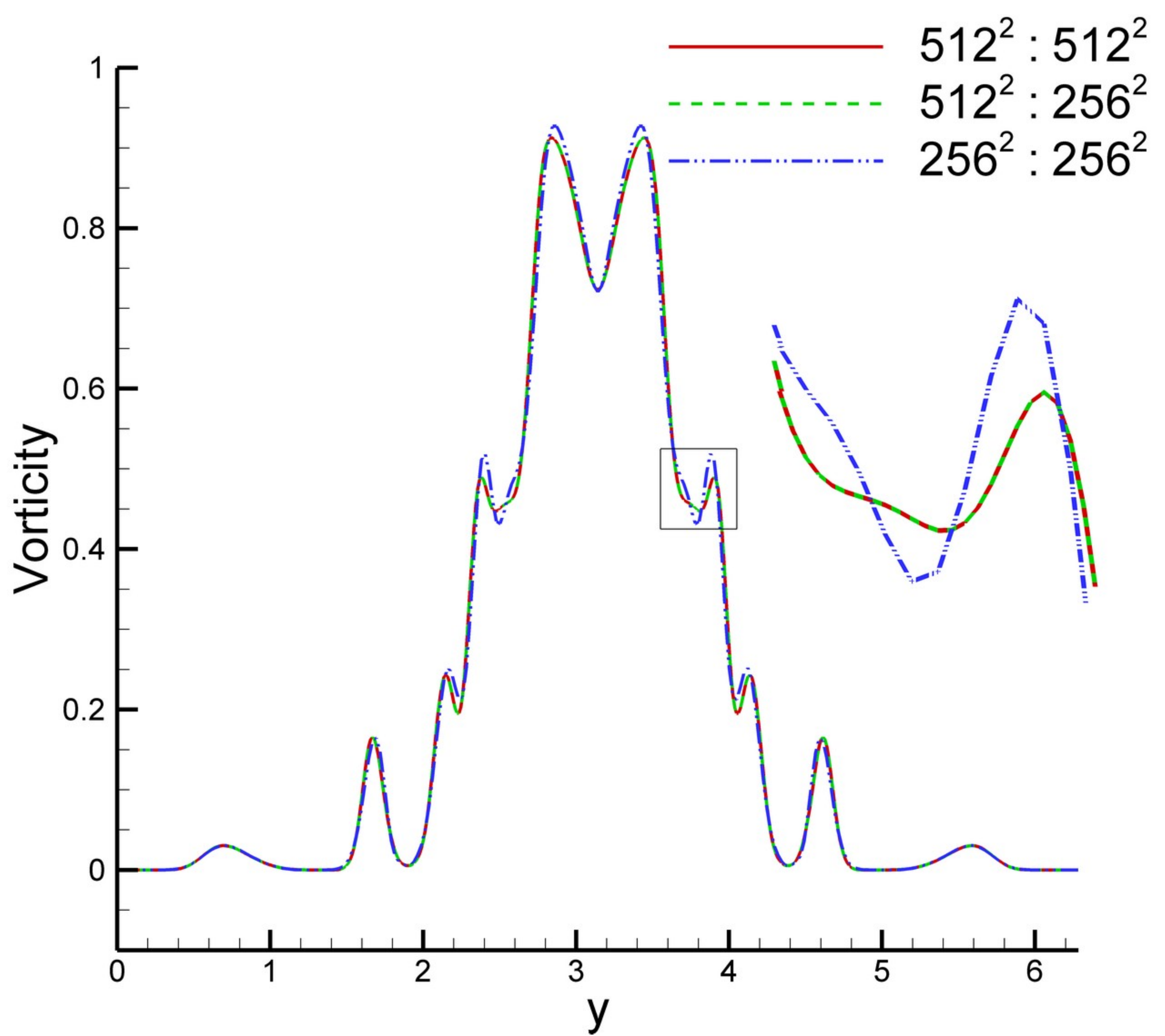}}
}
\\
\mbox{
\subfigure[]{\includegraphics[width=0.5\textwidth]{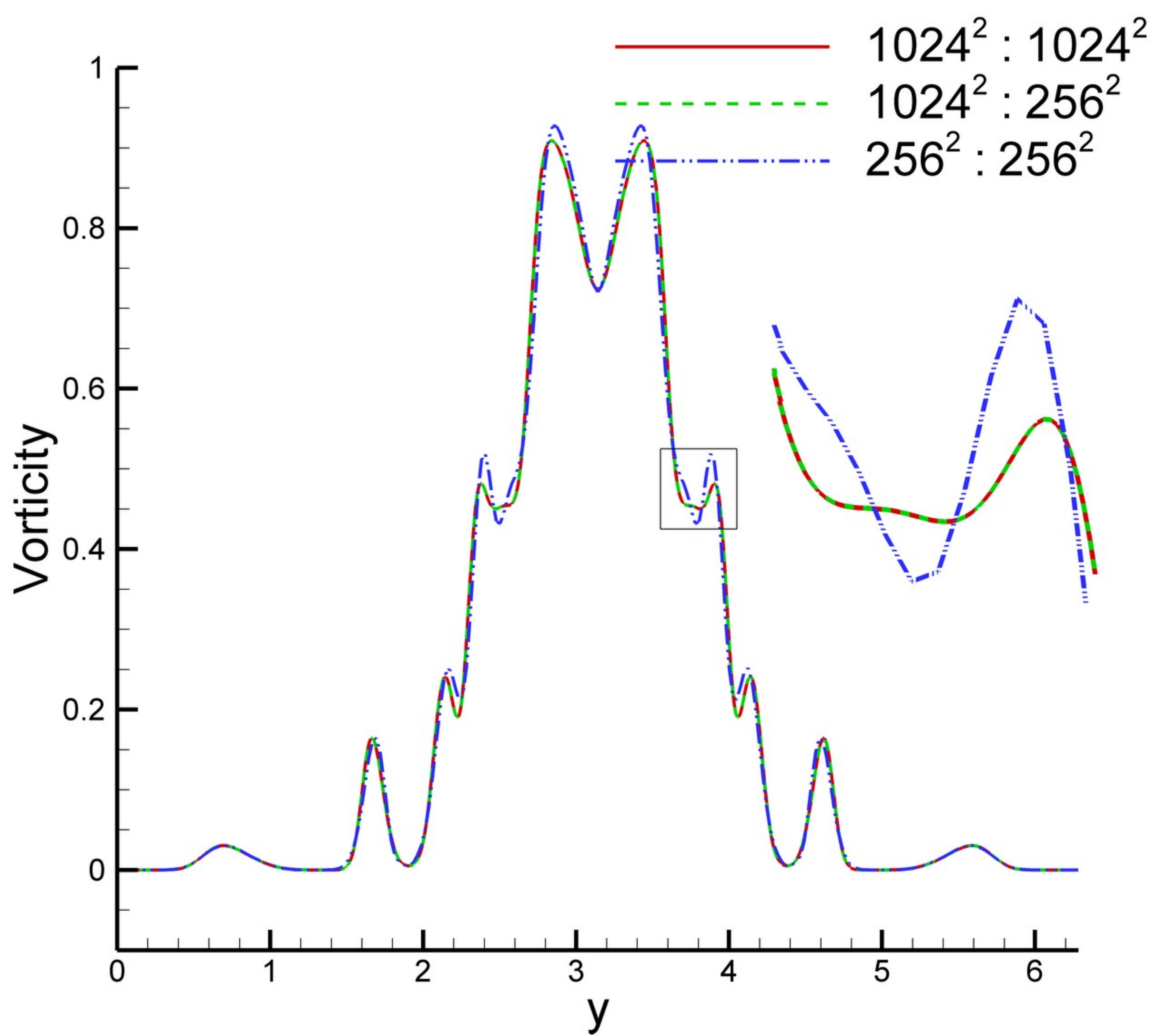}}
\subfigure[]{\includegraphics[width=0.5\textwidth]{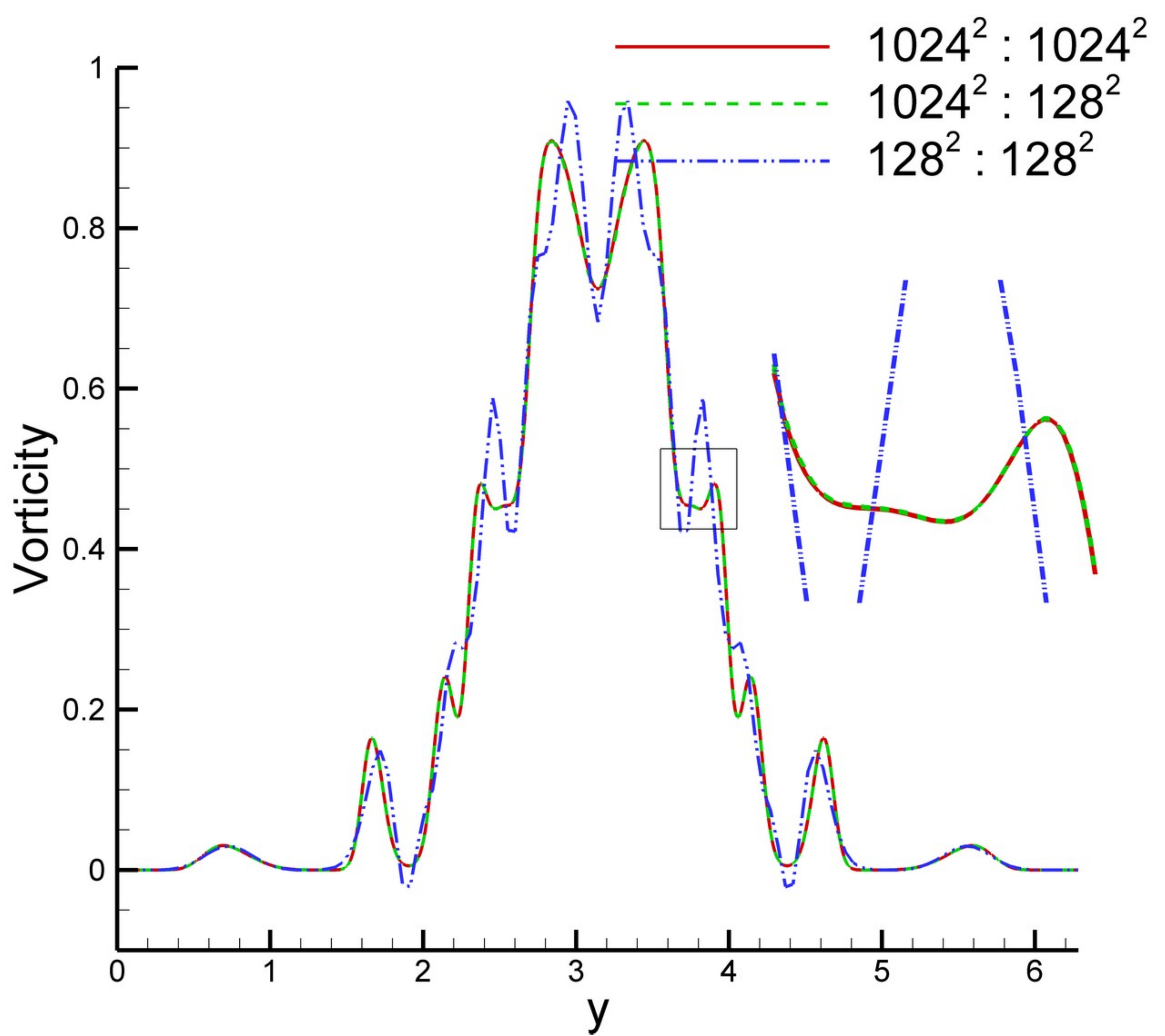}}
}
\caption{Centerline vorticity distributions for the merging co-rotating vortex pair problem at $t=50$ obtained using the vorticity-stream function formulation; (a) comparison for one level of coarsening including the standard computation (RK3) on $1024^2:1024^2$ resolution grids, the CGPRK3 method on $1024^2:512^2$ resolution grids, and the standard computation (RK3) on $512^2:512^2$ resolution grids, (b) comparison for one level of coarsening including the standard computation (RK3) on $512^2:512^2$ resolution grids, the CGPRK3 method on $512^2:256^2$ resolution grids, and the standard computation (RK3) on $256^2:256^2$ resolution grids, (c) comparison for two levels of coarsening, and (d) comparison for three levels of coarsening.}
\label{fig:merger-line}
\end{figure*}

Vorticity field contours at time $t=50$ obtained using the CGP method are plotted in Fig.~\ref{fig:merger} along with the regular fine and coarse computations. As in the double shear layer problem, the labels show the resolutions in the form of $N^2:M^2$, where $N^2$ is the resolution for the vorticity-transport equation, and $M^2$ is the resolution for the Poisson equation. The computed results with one level of coarsening are identically to those from the fine scale computations. Furthermore, results with two and three levels of coarsening also agree well with the fine scale computations. A comparison of accuracy and efficiency is also summarized in Table~\ref{tab:merger}. The absolute values of maximum amplitude of the vorticity field and corresponding $L_2$ norms are given in this table showing the applicability of the CGP method. The centerline vorticity distribution along the $y$-axis at time $t=50$ is shown in Fig.~\ref{fig:merger-line}. These results further demonstrate that both the fine scale simulation and the CGP simulations (including one level, two levels, and three levels of coarsening) provide very similar field data with negligible accuracy loss, but at a reduced computational cost.

\begin{table*}
\small
\caption{Comparison of methods for the co-rotating vortex pair merging problem for $Re=10^4$ at $t = 50$. $L_2$ norms are computed with reference to the values obtained on the finest resolution grid.}
\begin{center}
\label{tab:merger}
\begin{tabular}{llcccc}
\hline\noalign{\smallskip}
Method & Resolutions &  $ |\omega|_{max}$ & $||\omega||_{L_2}$ & CPU (hr) & Speed-up \\
\hline\noalign{\smallskip}
RK3 ($\ell=0$)     &$1024^2$ : $1024^2$ &  0.9133 & -          & 81.61 & 1.00 \\
CGPRK3 ($\ell=1$)  &$1024^2$ : $512^2$  &  0.9133 & 6.7173E-7  & 18.82 & 4.34 \\
CGPRK3 ($\ell=2$)  &$1024^2$ : $256^2$  &  0.9132 & 3.0447E-5  & 9.68  & 8.43 \\
CGPRK3 ($\ell=3$)  &$1024^2$ : $128^2$  &  0.9117 & 1.9605E-4  & 7.93  & 10.29 \\
RK3 ($\ell=0$)     &$512^2$ : $512^2$   &  0.9159 & 1.9975E-3  & 15.18 & 1.00 \\
CGPRK3 ($\ell=1$)  &$512^2$ : $256^2$   &  0.9159 & 1.9993E-3  & 4.13  & 3.68 \\
CGPRK3 ($\ell=2$)  &$512^2$ : $128^2$   &  0.9145 & 2.0611E-3  & 2.42  & 6.27 \\
RK3 ($\ell=0$)     &$256^2$ : $256^2$   &  0.9292 & 9.4755E-3  & 2.82  & 1.00 \\
CGPRK3 ($\ell=1$)  &$256^2$ : $128^2$   &  0.9291 & 9.4997E-3  & 0.95  & 2.97 \\
RK3 ($\ell=0$)     &$128^2$ : $128^2$   &  0.9588 & 2.8086E-2  & 0.61  & 1.00 \\
\hline
\end{tabular}
\end{center}
\end{table*}

\subsection{Two-dimensional decaying turbulence}
\label{sec:2dturb}
Two-dimensional homogenous decaying turbulence is an incompressible flow setting in which energy decays with time and can be thought of as the process of many random vortices merging \citep{kraichnan1967inertial,batchelor1969computation,leith1971atmospheric,brachet1988dynamics}. One of the most important reasons for studying two-dimensional turbulence is to improve our understanding of geophysical flows in the atmosphere and ocean \cite{herring1974decay,boer1983large,mcwilliams1984emergence,san2011approximate}. Recent reviews on two-dimensional turbulence have been provided by Tabeling \cite{tabeling2002two} and Boffetta and Ecke \cite{boffetta2012two}. This system has an inertial (intermediate wave number) range in the energy spectrum that is proportional to $k^{\alpha}$, where $\alpha=3$ and $k$ is the wave number, in the inviscid limit according to Kraichnan's theory of two-dimensional turbulence \cite{kraichnan1967inertial}. The computational domain we use to solve this problem is a square box whose edge has a length of $2\pi$. Periodic boundary conditions are used. The initial energy spectrum in Fourier space is given by \citep{ishiko2009implicit}:
\begin{equation}
E(k) = \frac{a_s}{2}\frac{1}{k_p}\left(\frac{k}{k_p}\right)^{2s+1}\mbox{exp} \bigg[-\left(s+\frac{1}{2}\right)\left(\frac{k}{k_p}\right)^{2} \bigg]
\label{eq:initE}
\end{equation}
where $k= |\textbf{k}|=\sqrt{k_x^2 + k_y^2}$. The maximum value of the initial energy spectrum occurs at a wavenumber of $k_p$. We use $k_p=12$ in this study. The coefficient $a_s$ normalizes the initial kinetic energy and is given by:
\begin{equation}
a_s=\frac{(2s+1)^{s+1}}{2^ss!}
\label{eq:err}
\end{equation}
where $s$ is a shape parameter. In this study, we take $s=3$. The magnitude of the vorticity Fourier coefficients related to the assumed energy spectrum becomes:
\begin{equation}
|\tilde{\omega}(\textbf{k})|=\sqrt{\frac{k}{\pi}E(k)}
\label{eq:err}
\end{equation}
and the initial vorticity distribution in Fourier space is then obtained by introducing a random phase:
\begin{equation}
\tilde{\omega}(\textbf{k})=\sqrt{\frac{k}{\pi}E(k)} \ \ e^{\emph{\textbf{i}}\zeta(\textbf{k})}
\label{eq:err}
\end{equation}
where the phase function is given by $\zeta(\textbf{k})=\xi(\textbf{k})+\eta(\textbf{k})$, where $\xi(\textbf{k})$ and $\eta(\textbf{k})$ are independent random values chosen in [$0,2\pi$] at each coordinate point in the first quadrant of the $k_x$-$k_y$ plane. Following \citep{ishiko2009implicit}, the conjugate relations for other quadrants are:
\begin{eqnarray}
\xi(-k_x,k_y)&=& -\xi(k_x,k_y) \nonumber \\
\xi(-k_x,-k_y)&=&-\xi(k_x,k_y) \nonumber \\
\xi(k_x,-k_y)&=&\xi(k_x,k_y) \nonumber \\
\eta(-k_x,k_y)&=&\eta(k_x,k_y)\nonumber \\
\eta(-k_x,-k_y)&=&-\eta(k_x,k_y)\nonumber \\
\eta(k_x,-k_y)&=&-\eta(k_x,k_y).
\label{eq:err}
\end{eqnarray}

\begin{figure*}
\centering
\mbox{
\subfigure{\includegraphics[width=0.33\textwidth]{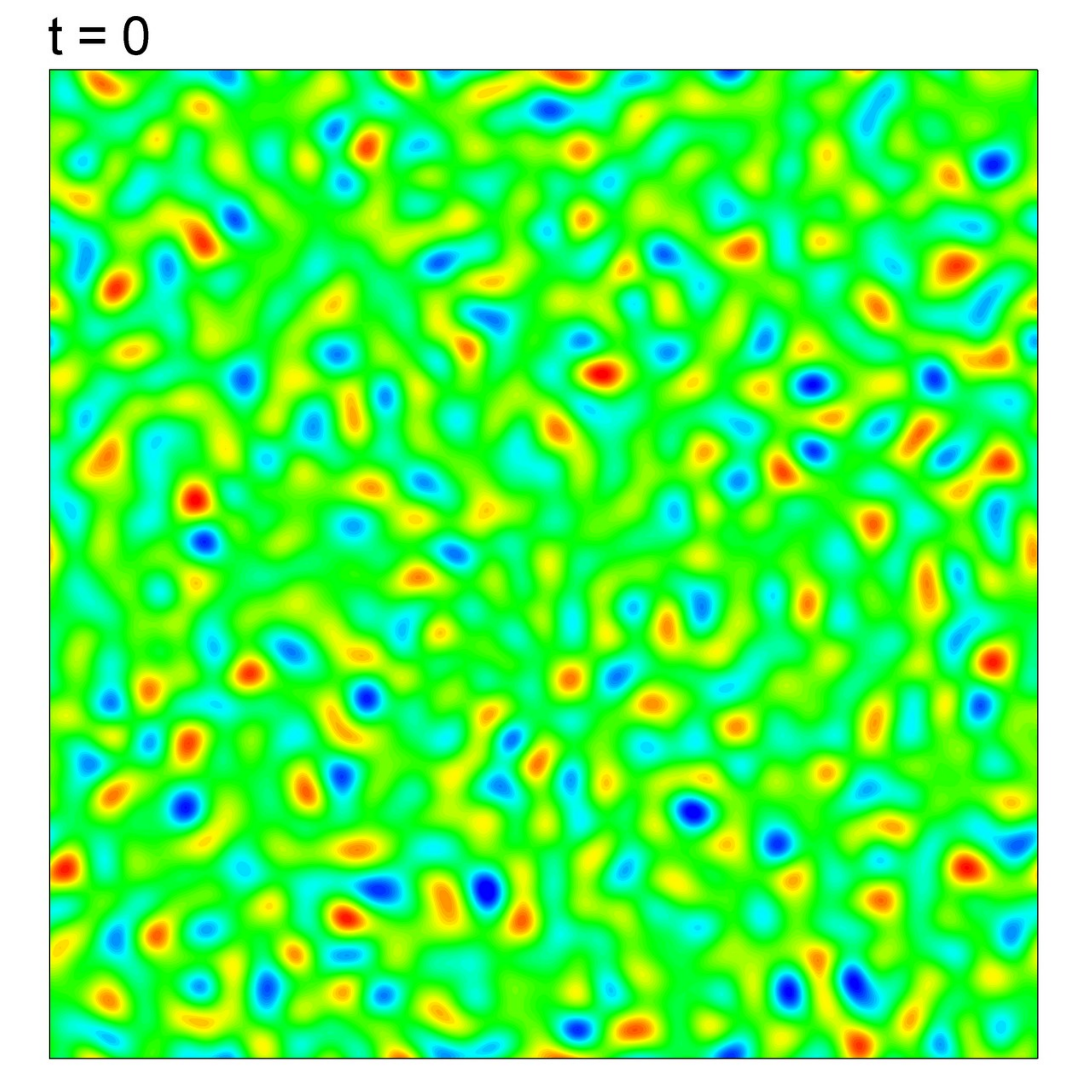}}
\subfigure{\includegraphics[width=0.33\textwidth]{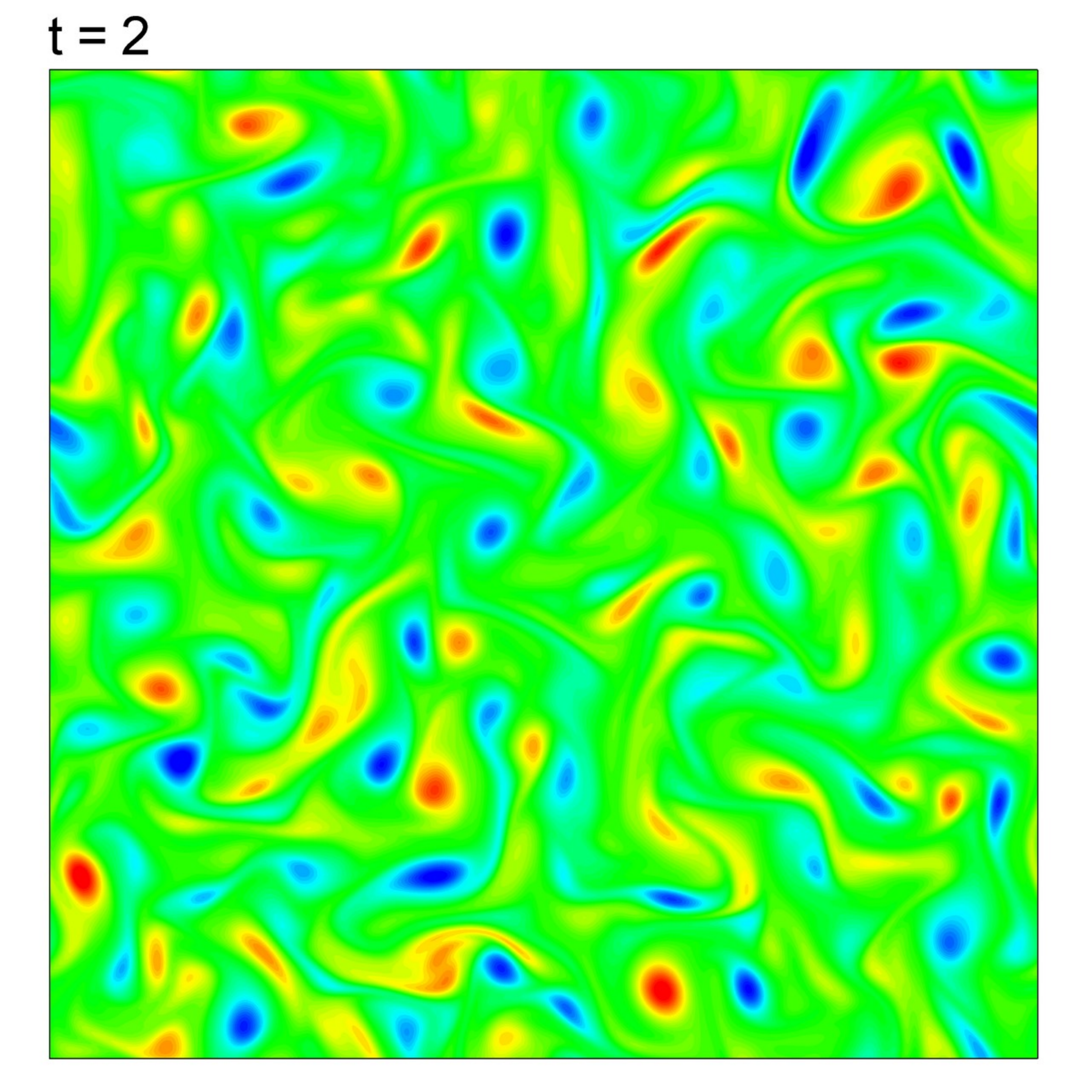}}
\subfigure{\includegraphics[width=0.33\textwidth]{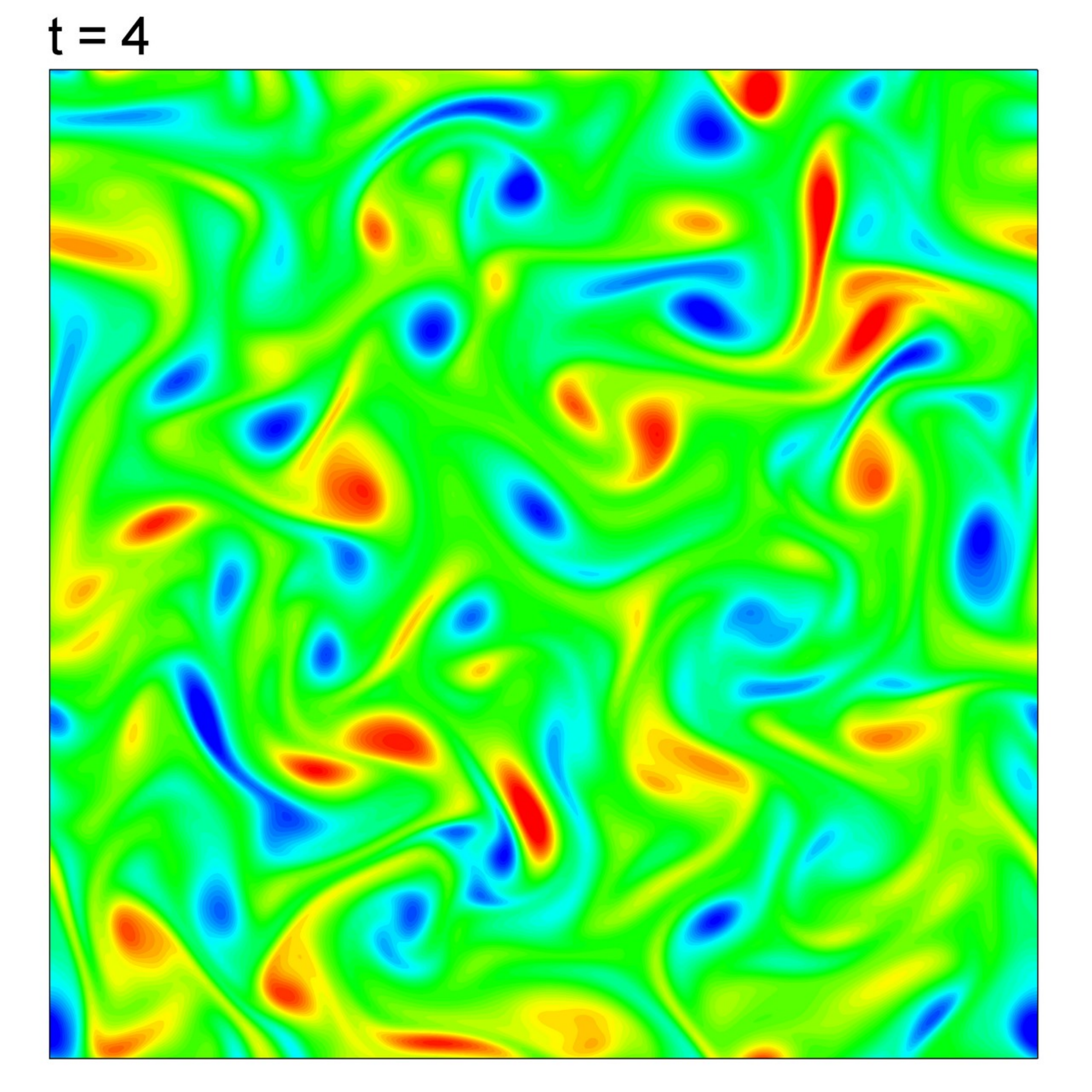}}
}
\mbox{
\subfigure{\includegraphics[width=0.33\textwidth]{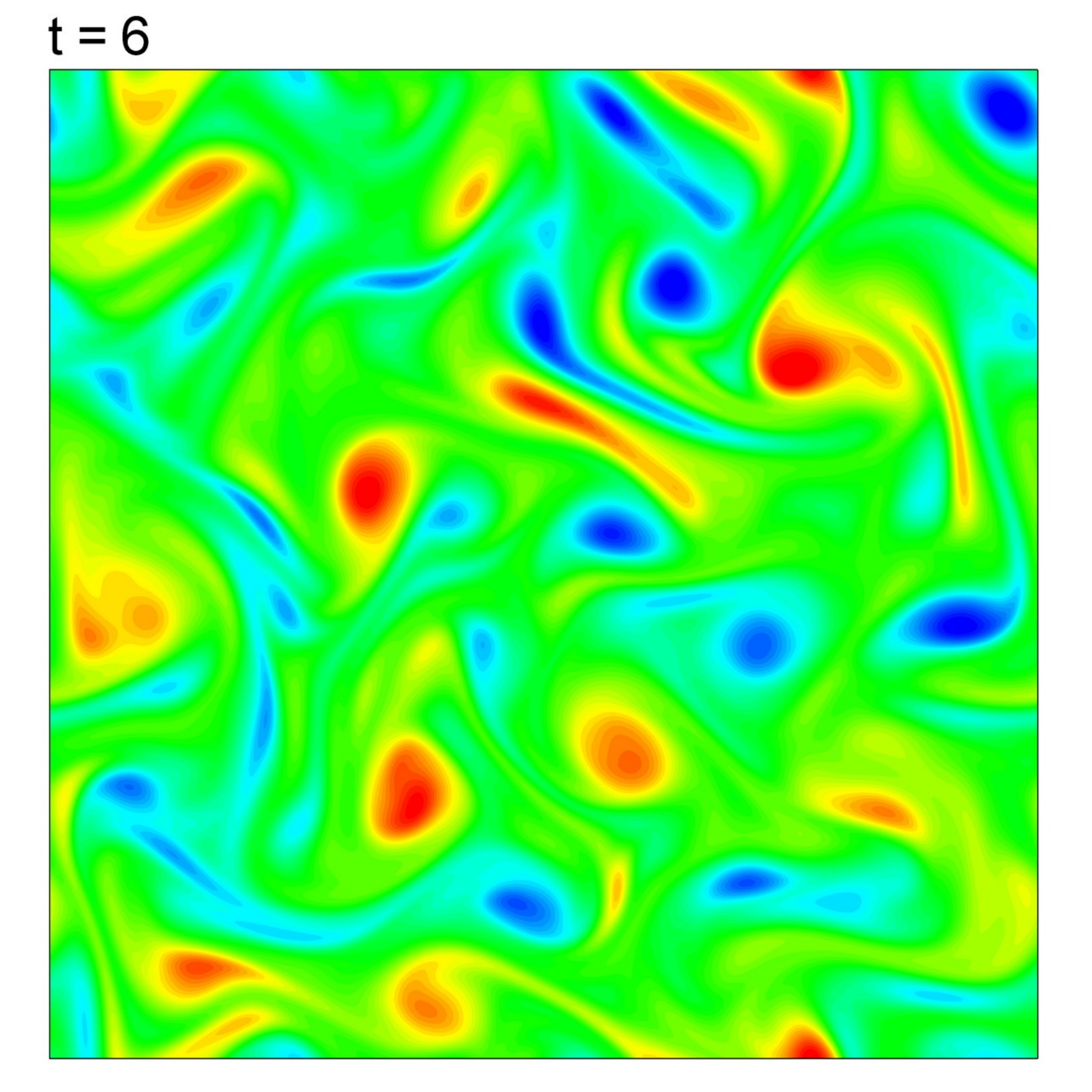}}
\subfigure{\includegraphics[width=0.33\textwidth]{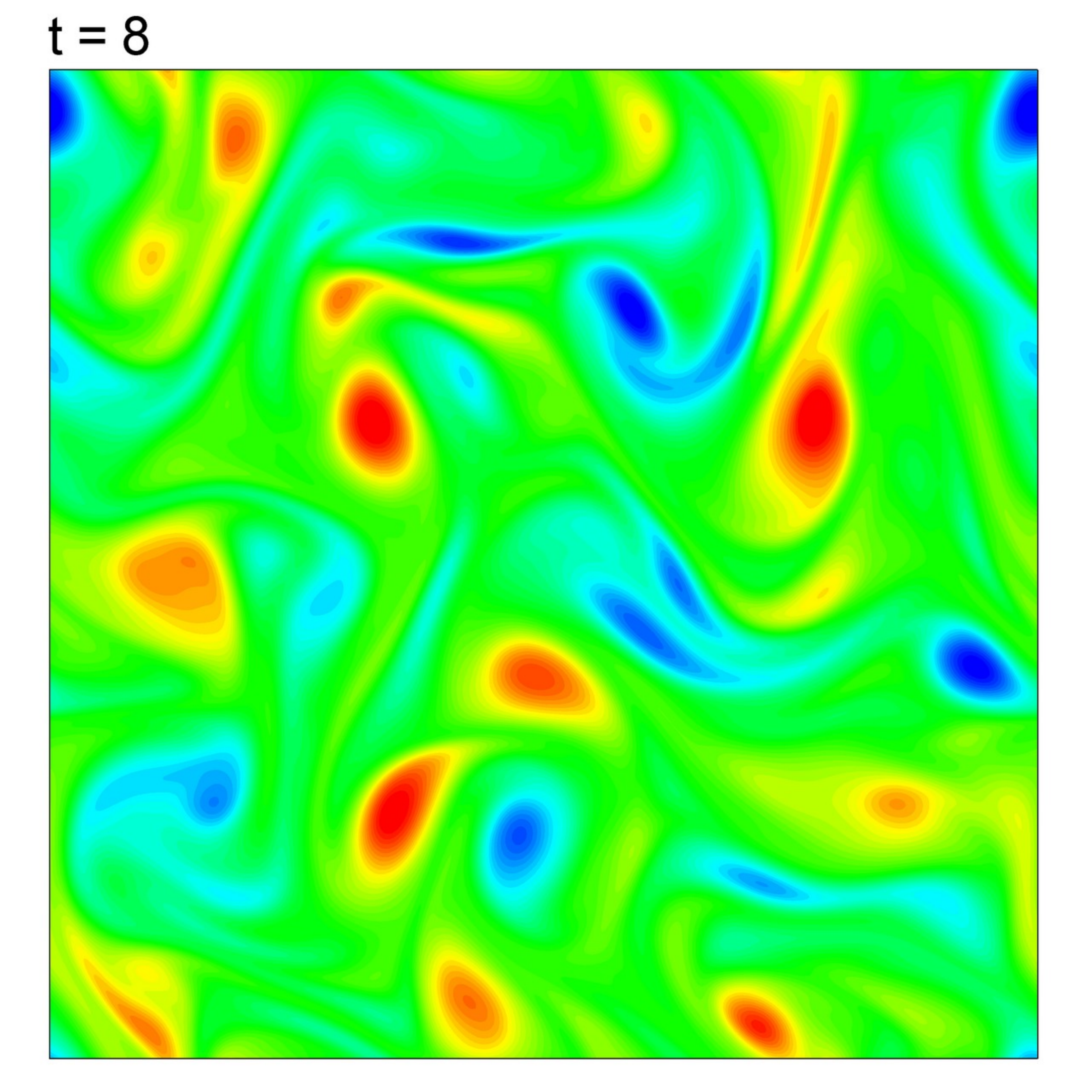}}
\subfigure{\includegraphics[width=0.33\textwidth]{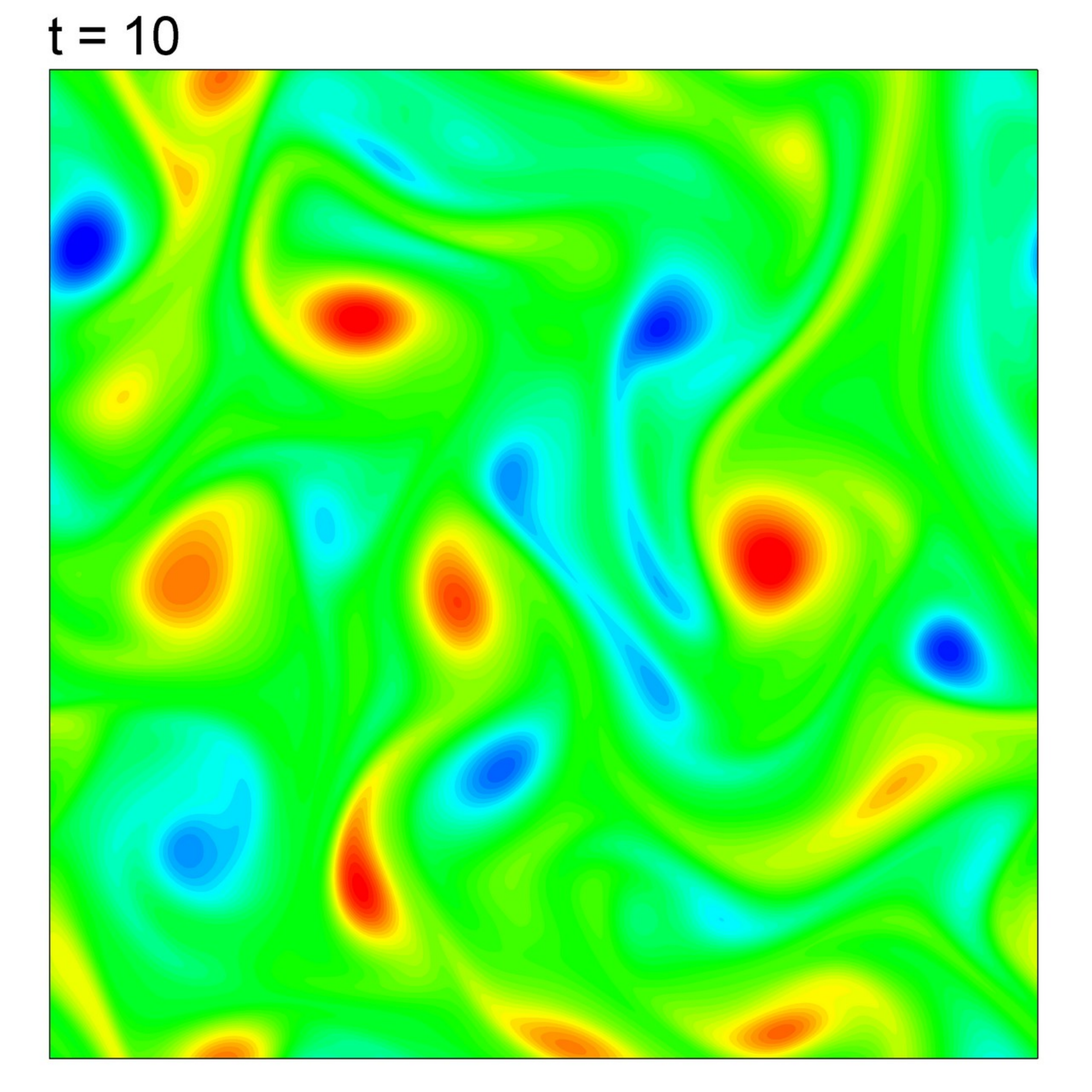}}
}
\caption{The vorticity field at different times for two-dimensional decaying turbulence obtained using the CGP method on $1024^2:512^2$ resolution grids for $Re=500$.}
\label{fig:turb-time}
\end{figure*}

After the randomization process described above for the phases, the initial vorticity field in physical space is obtained by performing an inverse FFT. The resulting initial vorticity field contour plot is illustrated in Fig.~\ref{fig:turb-time}(a).

In this section, first, two different flow field with Reynolds numbers $Re=250$ and $Re=500$ are computed using the CGP method and compared to results using the standard method without the CGP methodology. The time step is chosen to be $\Delta t=2\times10^{-4}$, which leads to a solution that is well resolved in time. The numerical experiments are performed until a final time of $t=10$, and the resulting vorticity contour plots are shown in Fig.~\ref{fig:turb-time} for $Re=500$.

\begin{figure*}
\centering
\mbox{
\subfigure{\includegraphics[width=0.33\textwidth]{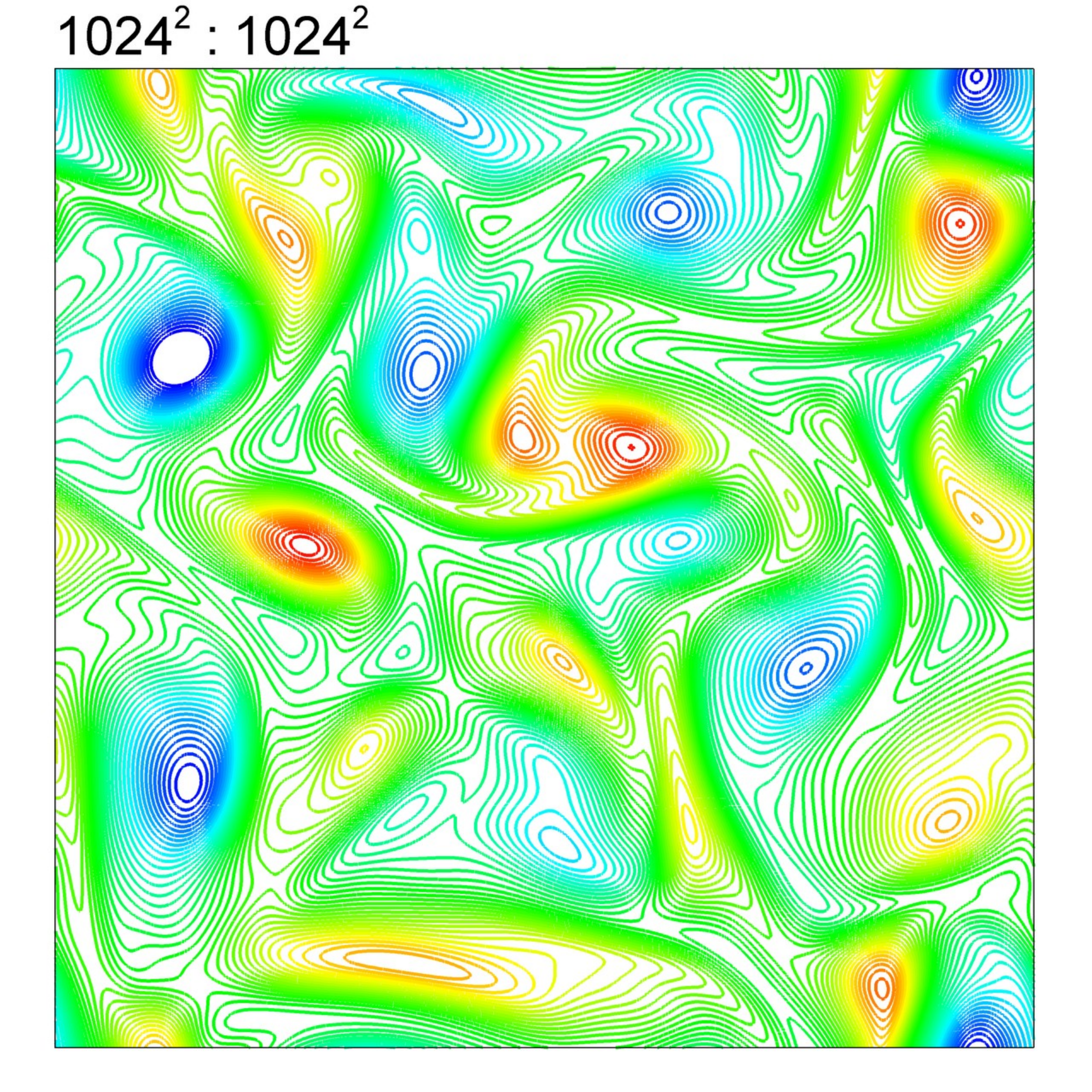}}
\subfigure{\includegraphics[width=0.33\textwidth]{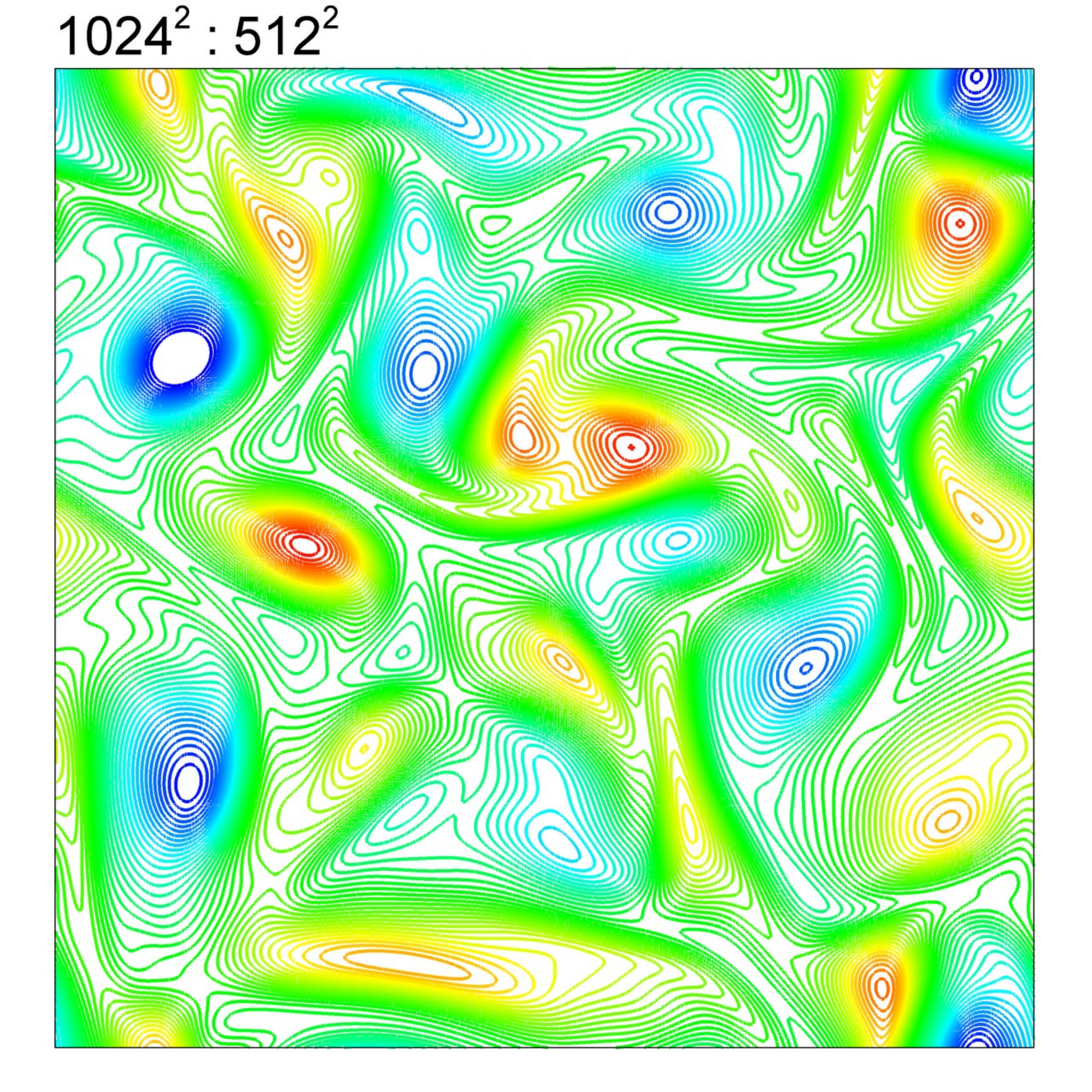}}
\subfigure{\includegraphics[width=0.33\textwidth]{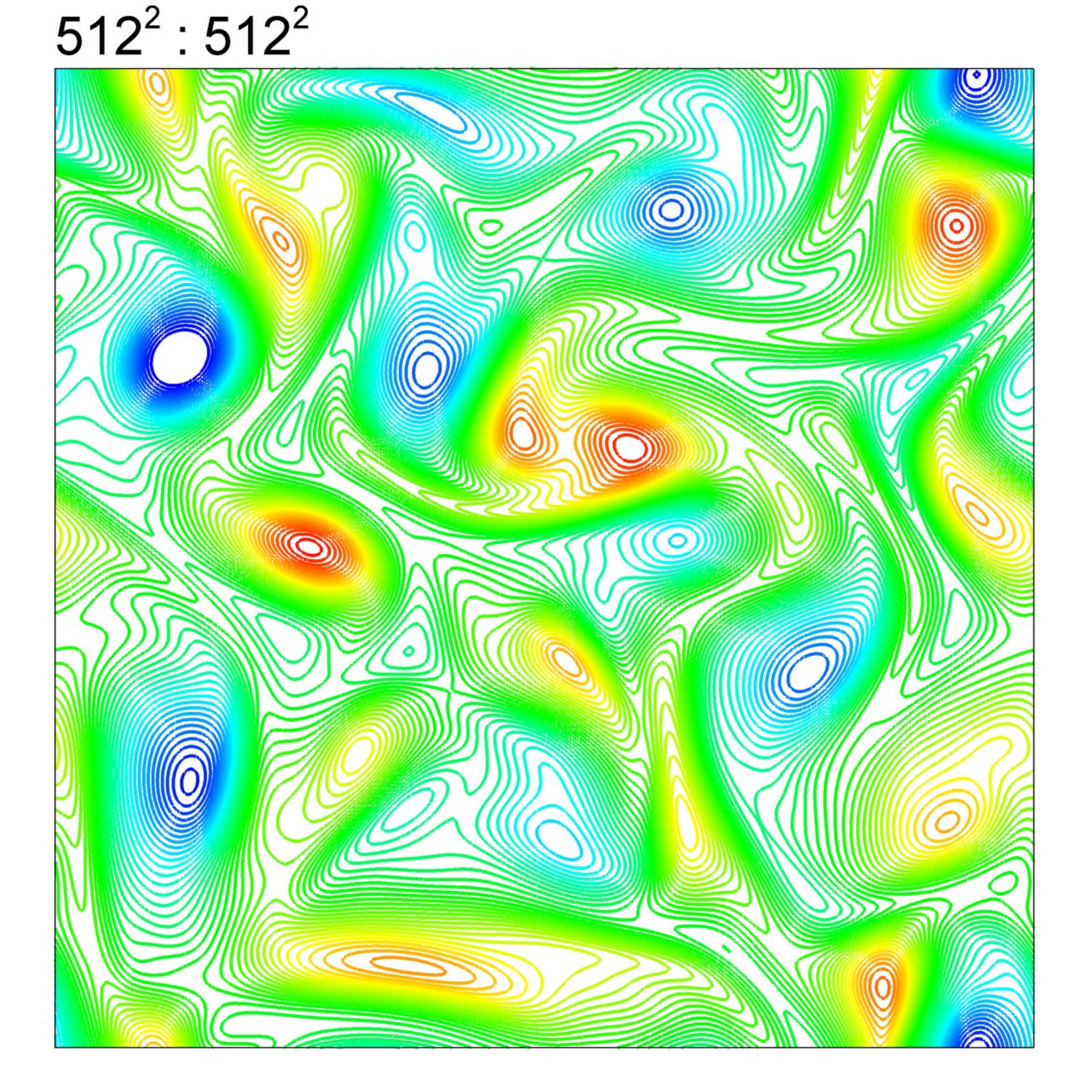}}
}
\mbox{
\subfigure{\includegraphics[width=0.33\textwidth]{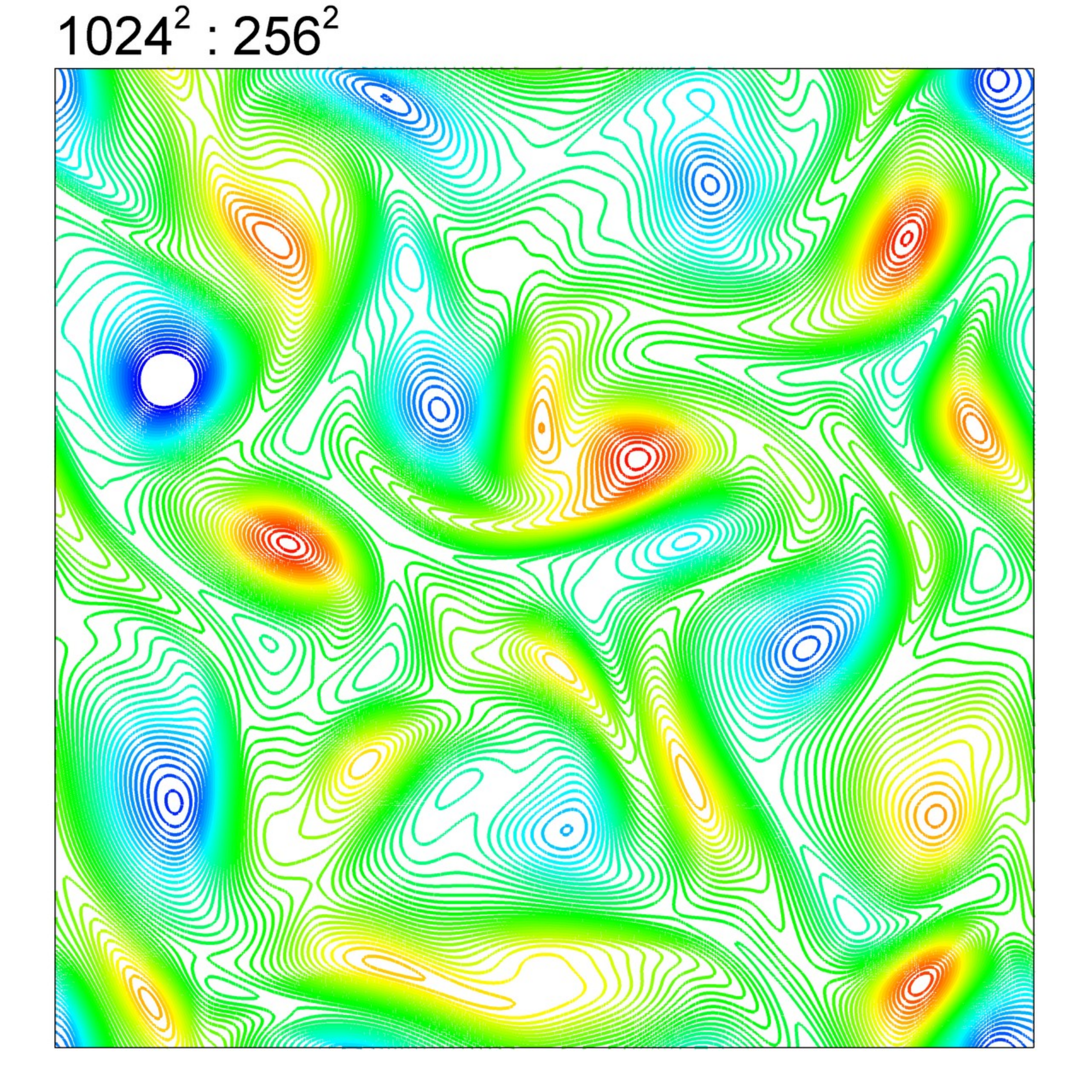}}
\subfigure{\includegraphics[width=0.33\textwidth]{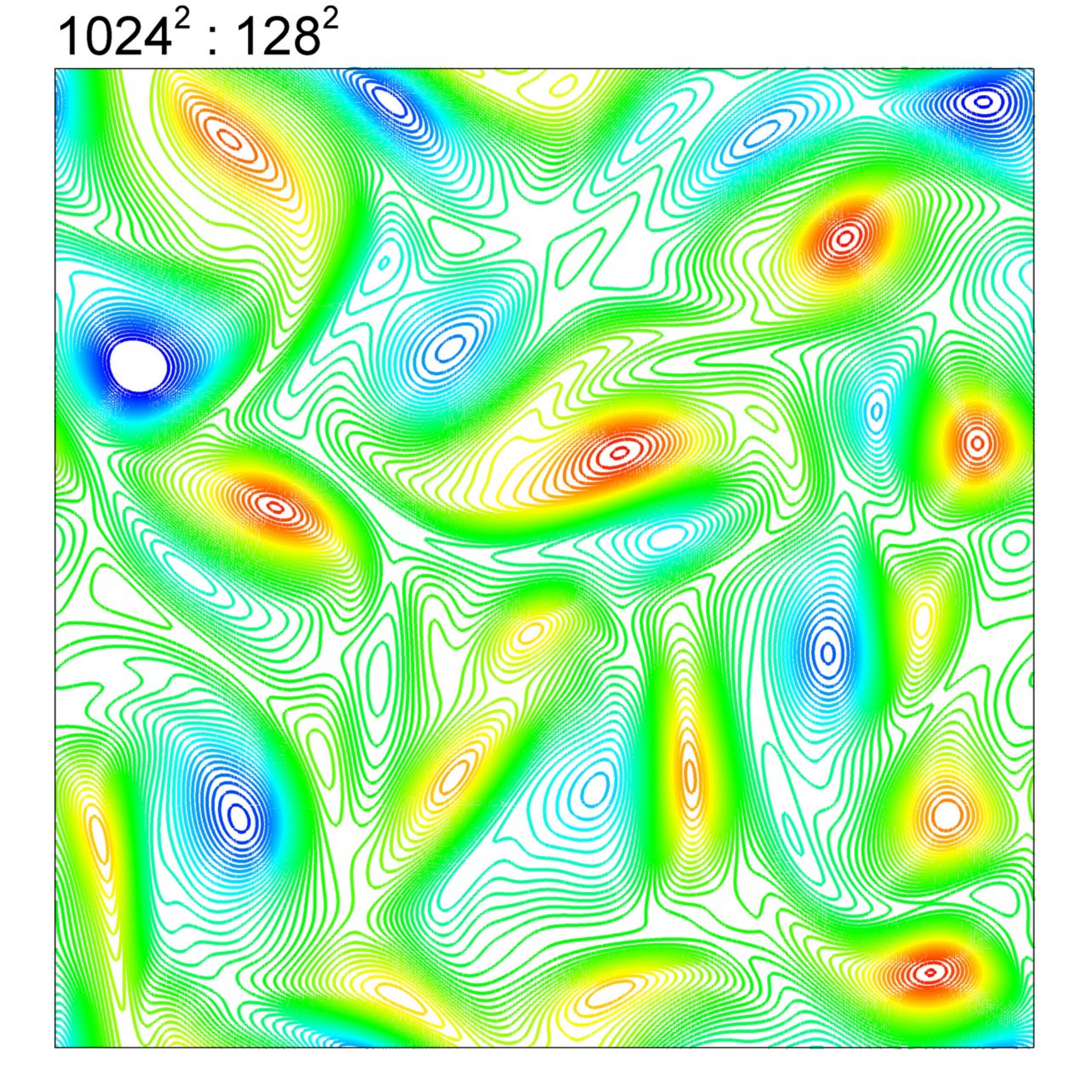}}
\subfigure{\includegraphics[width=0.33\textwidth]{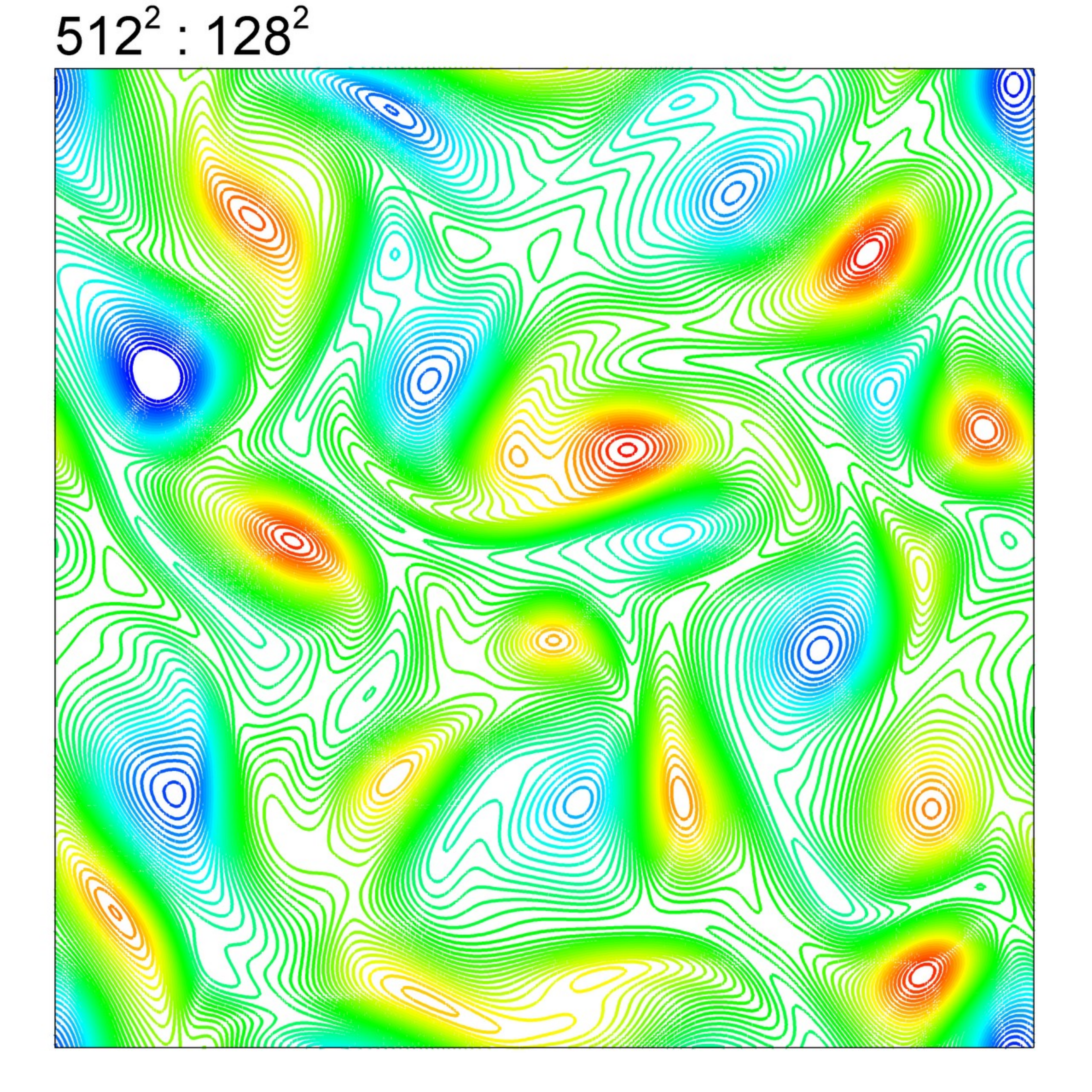}}
}
\mbox{
\subfigure{\includegraphics[width=0.33\textwidth]{turb2d250-512-512.pdf}}
\subfigure{\includegraphics[width=0.33\textwidth]{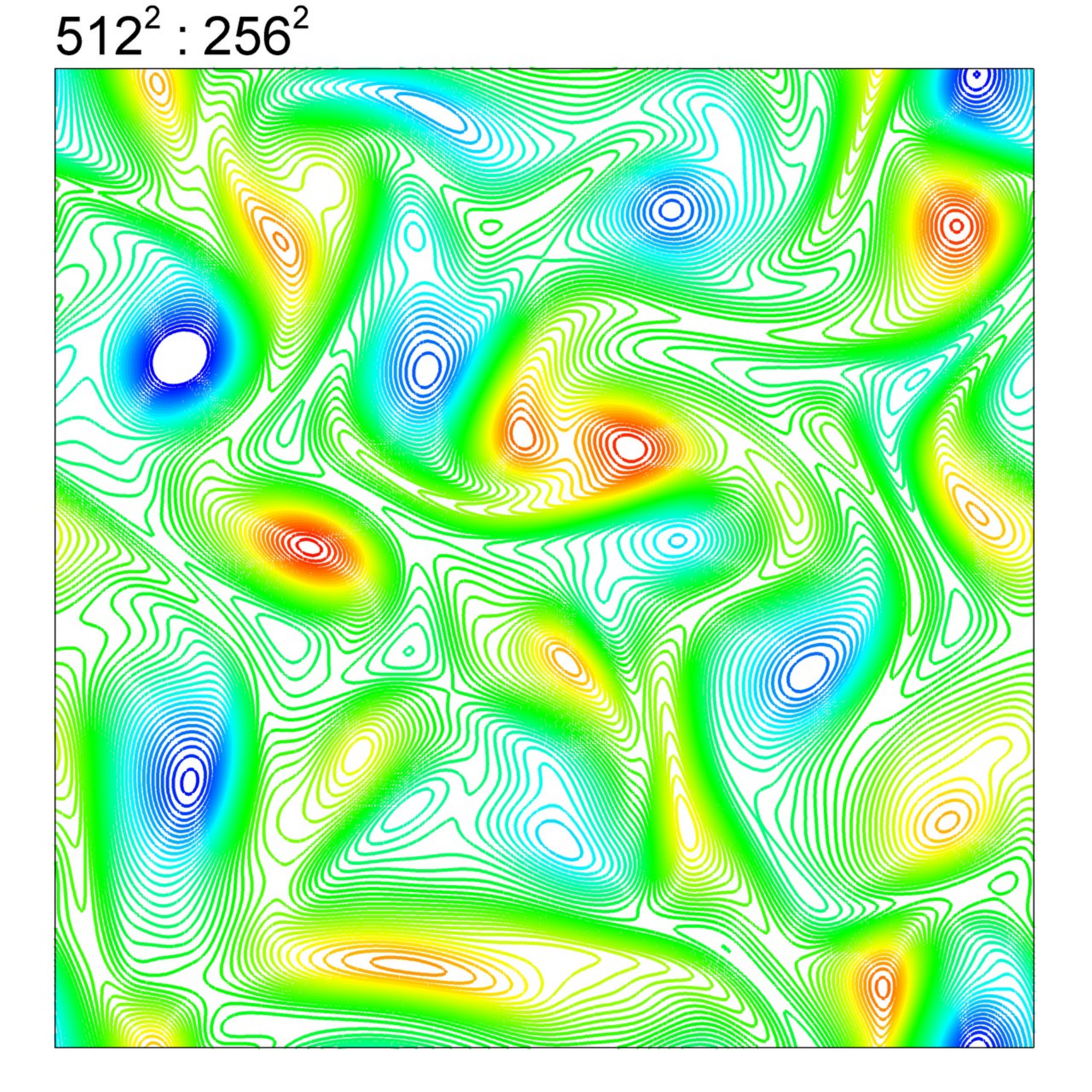}}
\subfigure{\includegraphics[width=0.33\textwidth]{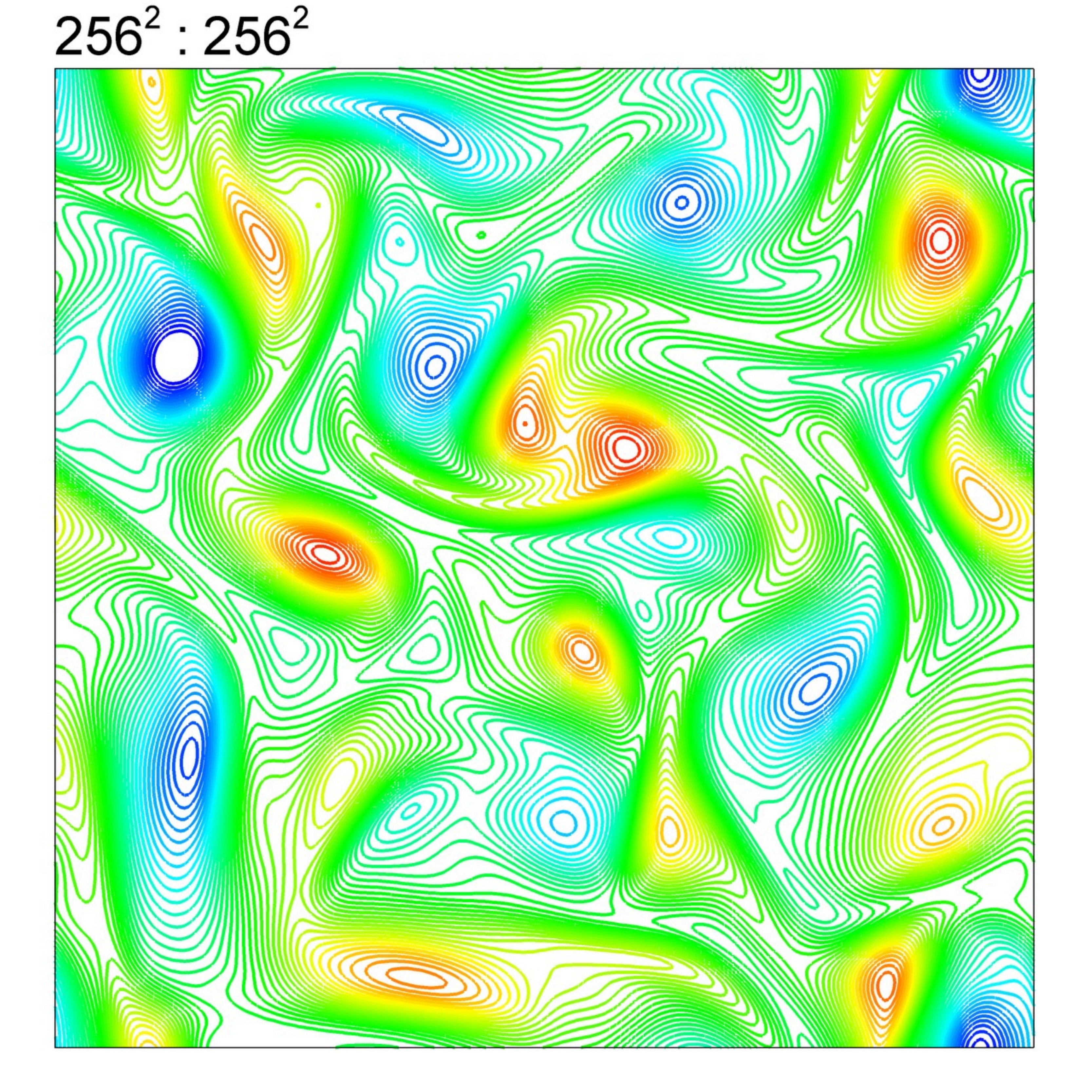}}
}
\mbox{
\subfigure{\includegraphics[width=0.33\textwidth]{turb2d250-256-256.pdf}}
\subfigure{\includegraphics[width=0.33\textwidth]{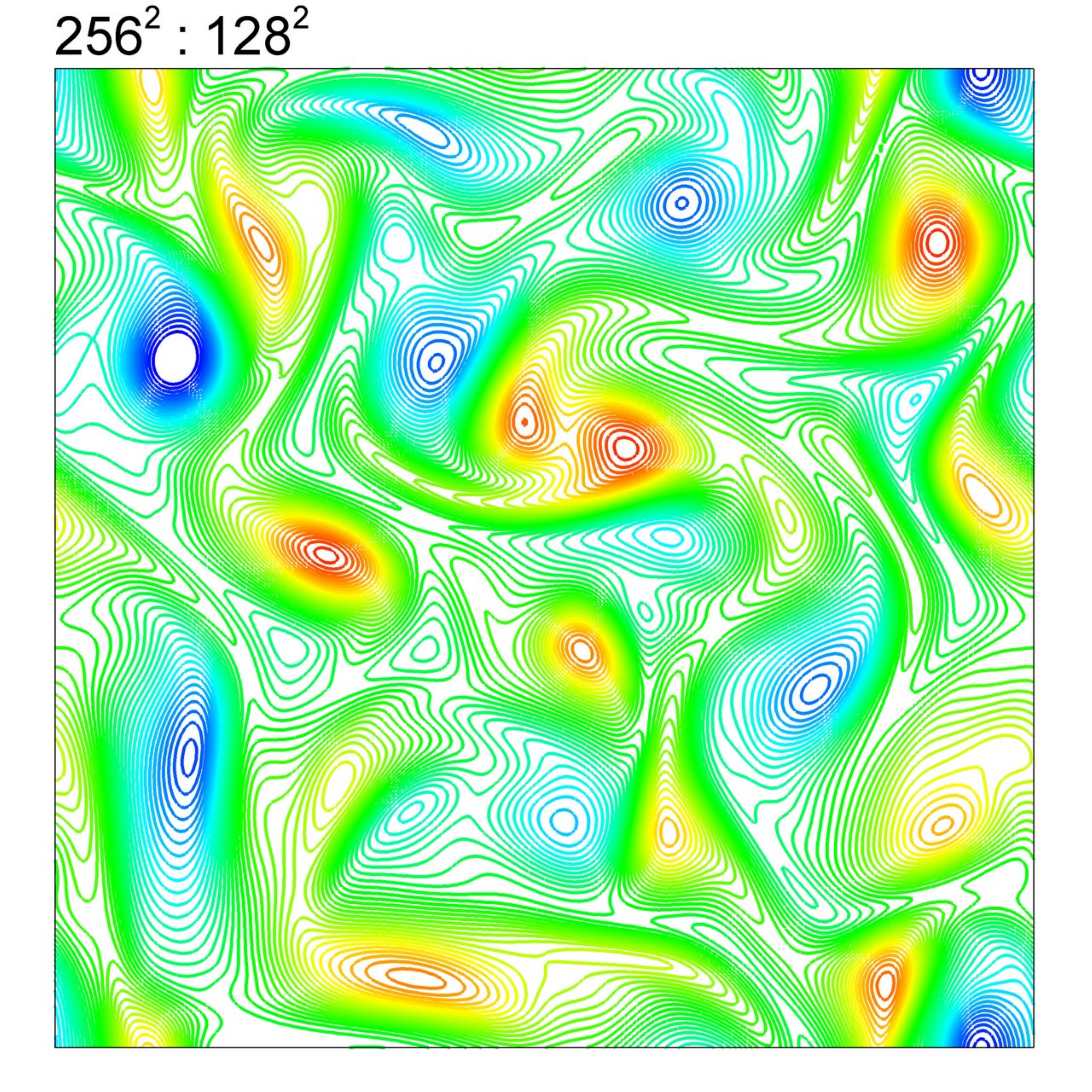}}
\subfigure{\includegraphics[width=0.33\textwidth]{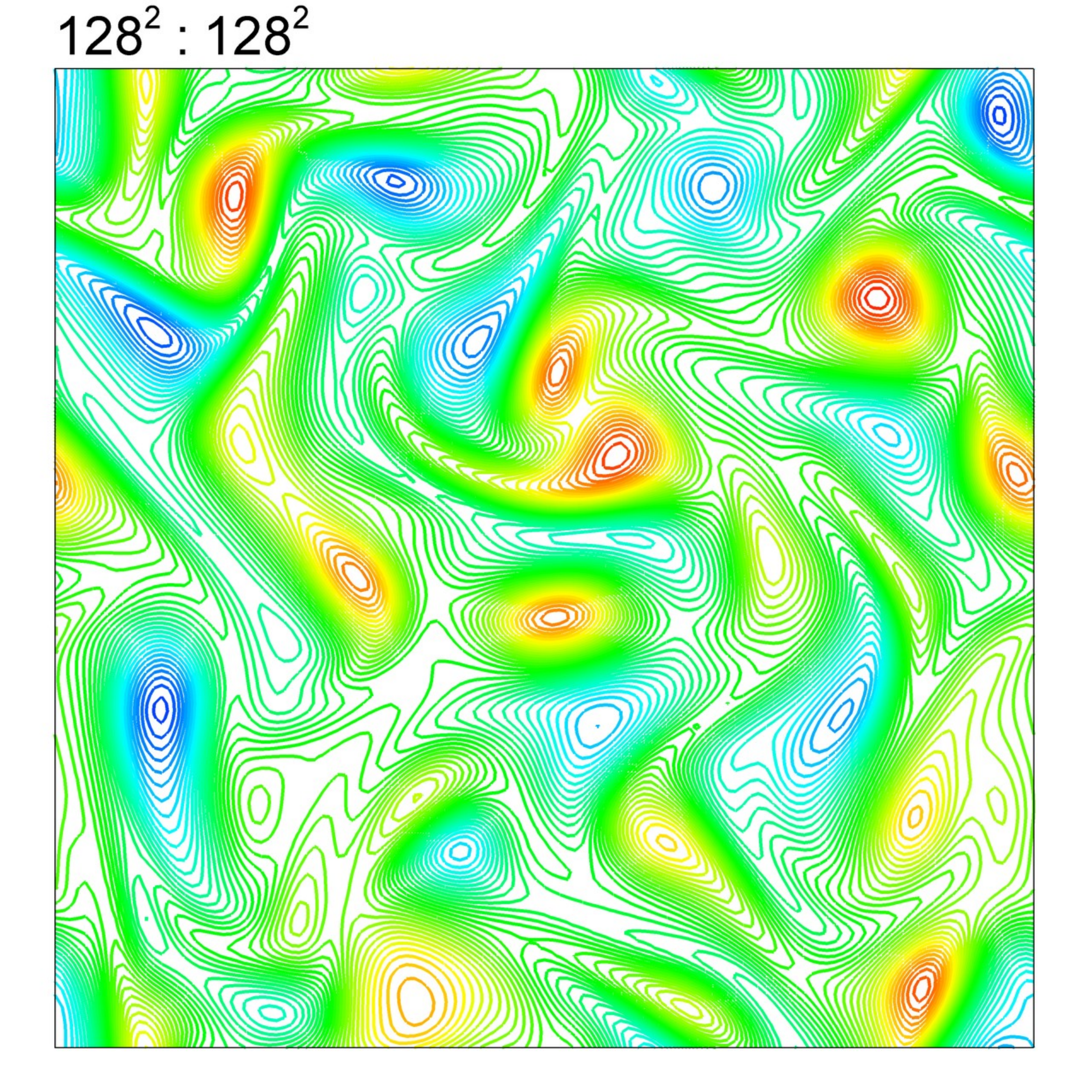}}
}
\caption{Vorticity fields for the decaying turbulence problem at $t=10$ obtained using the vorticity-stream function formulation for $Re=250$. Labels show the resolutions for both parts of the solver in the form $N^2:M^2$, where $N^2$ is the resolution for the vorticity-transport equation, and $M^2$ is the resolution for the Poisson equation. The same equidistant contour levels are used in all cases.}
\label{fig:turb2d-250}
\end{figure*}

\begin{figure*}
\centering
\mbox{
\subfigure{\includegraphics[width=0.33\textwidth]{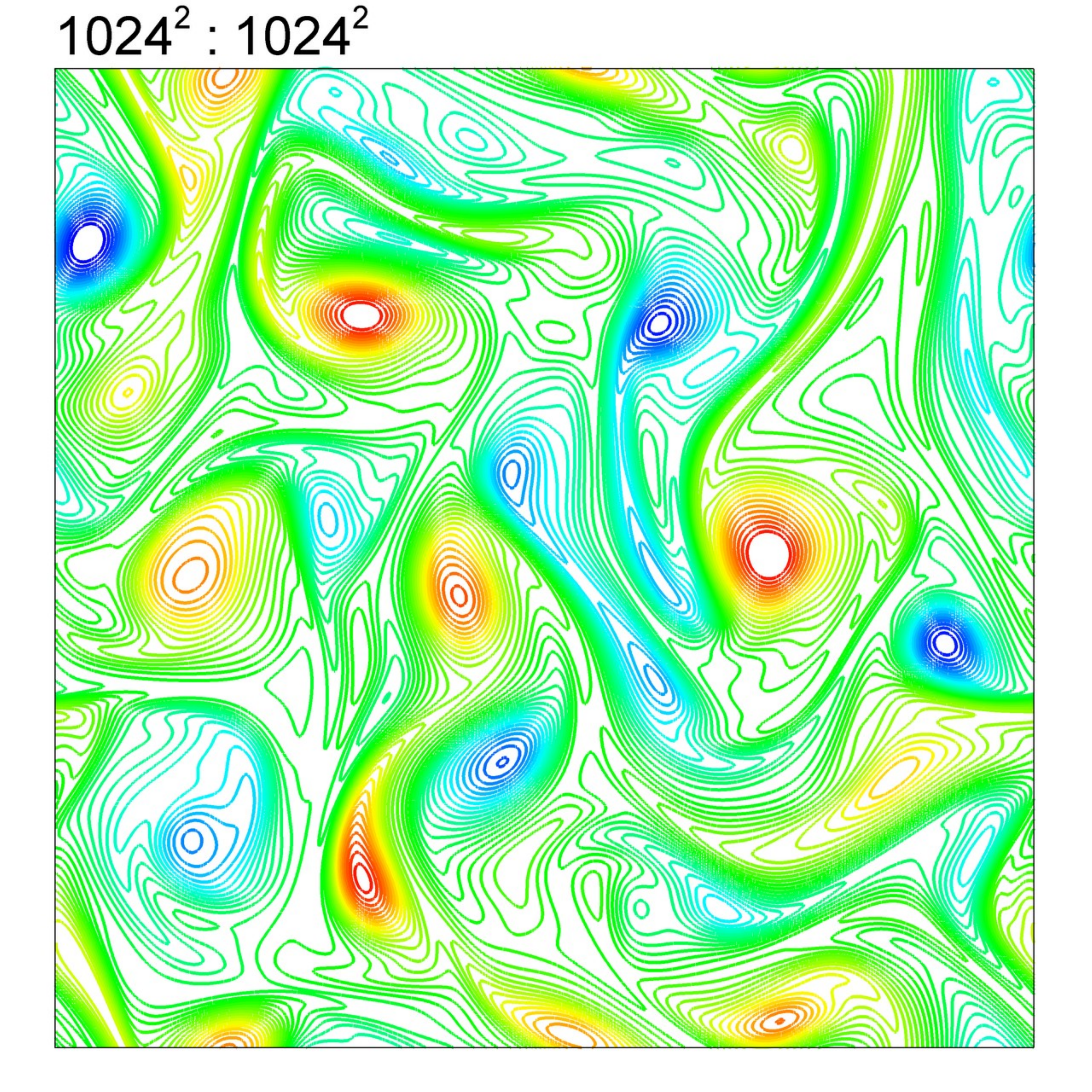}}
\subfigure{\includegraphics[width=0.33\textwidth]{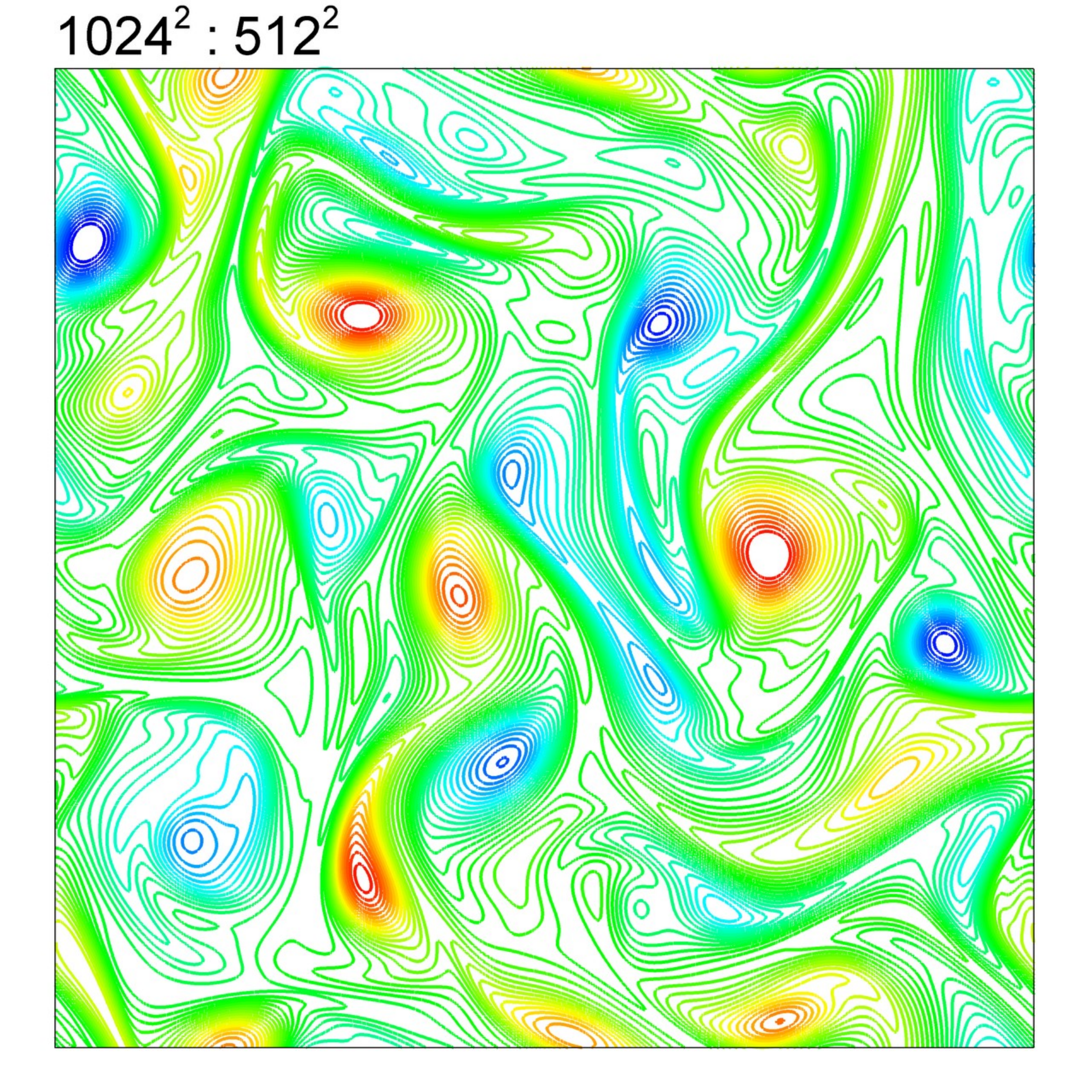}}
\subfigure{\includegraphics[width=0.33\textwidth]{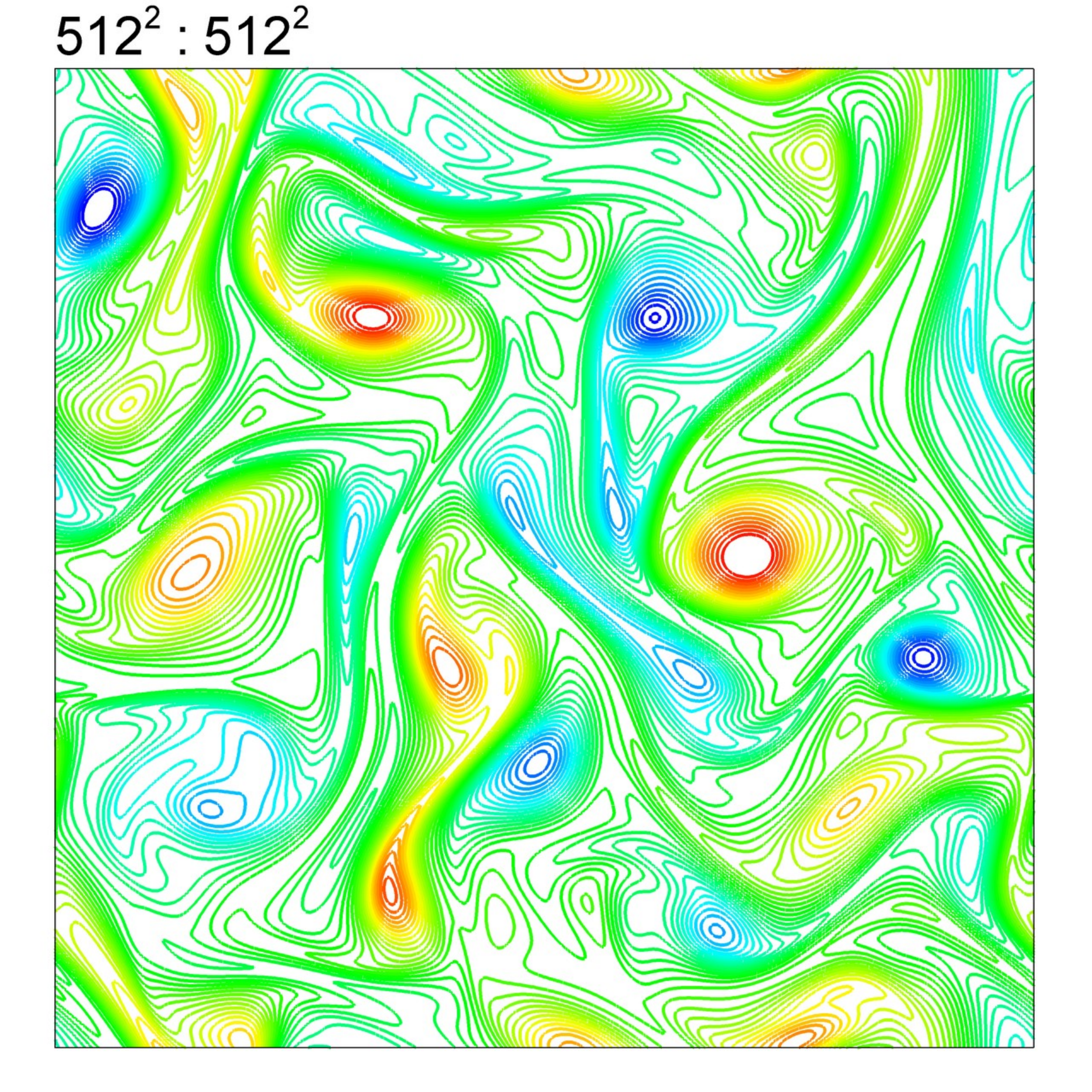}}
}
\mbox{
\subfigure{\includegraphics[width=0.33\textwidth]{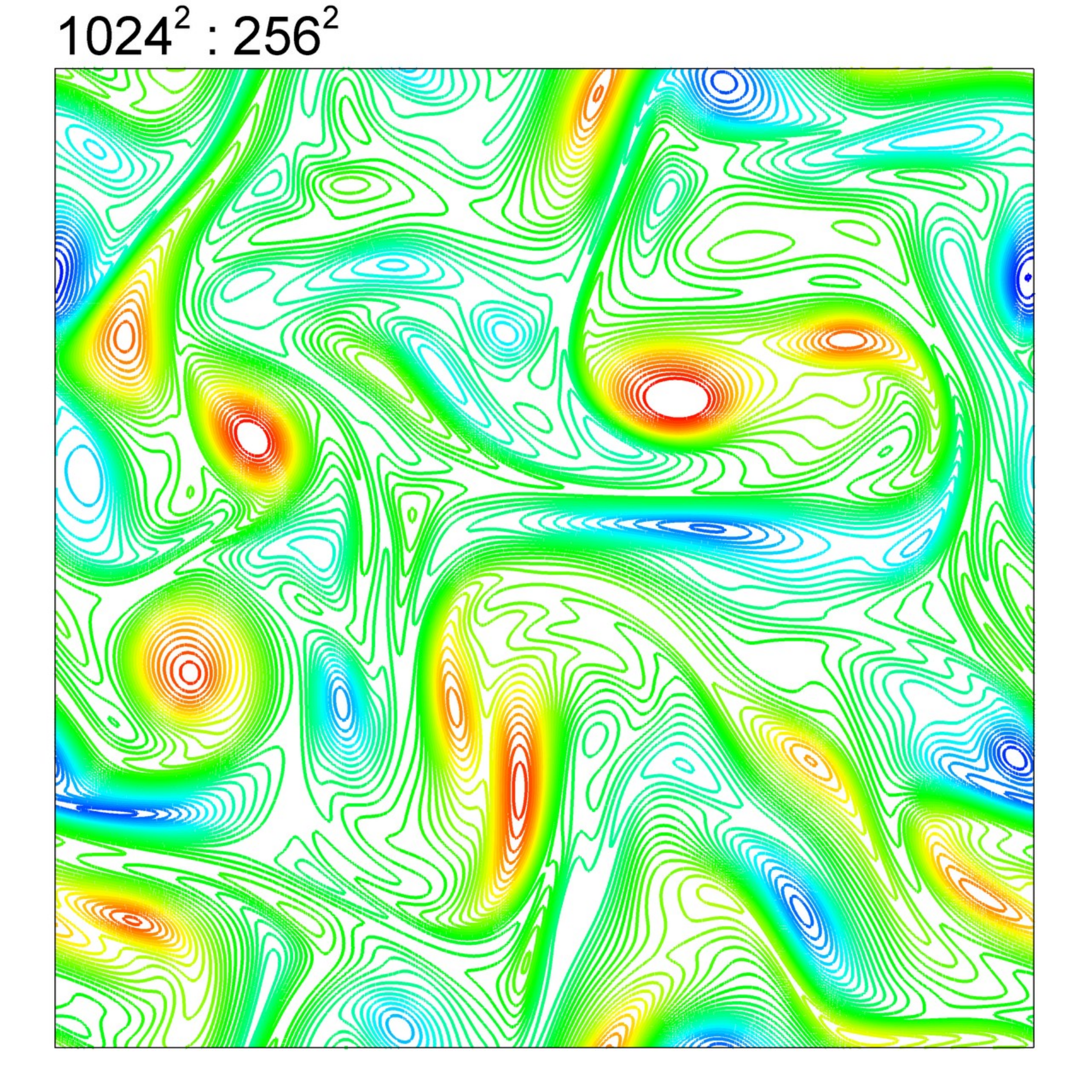}}
\subfigure{\includegraphics[width=0.33\textwidth]{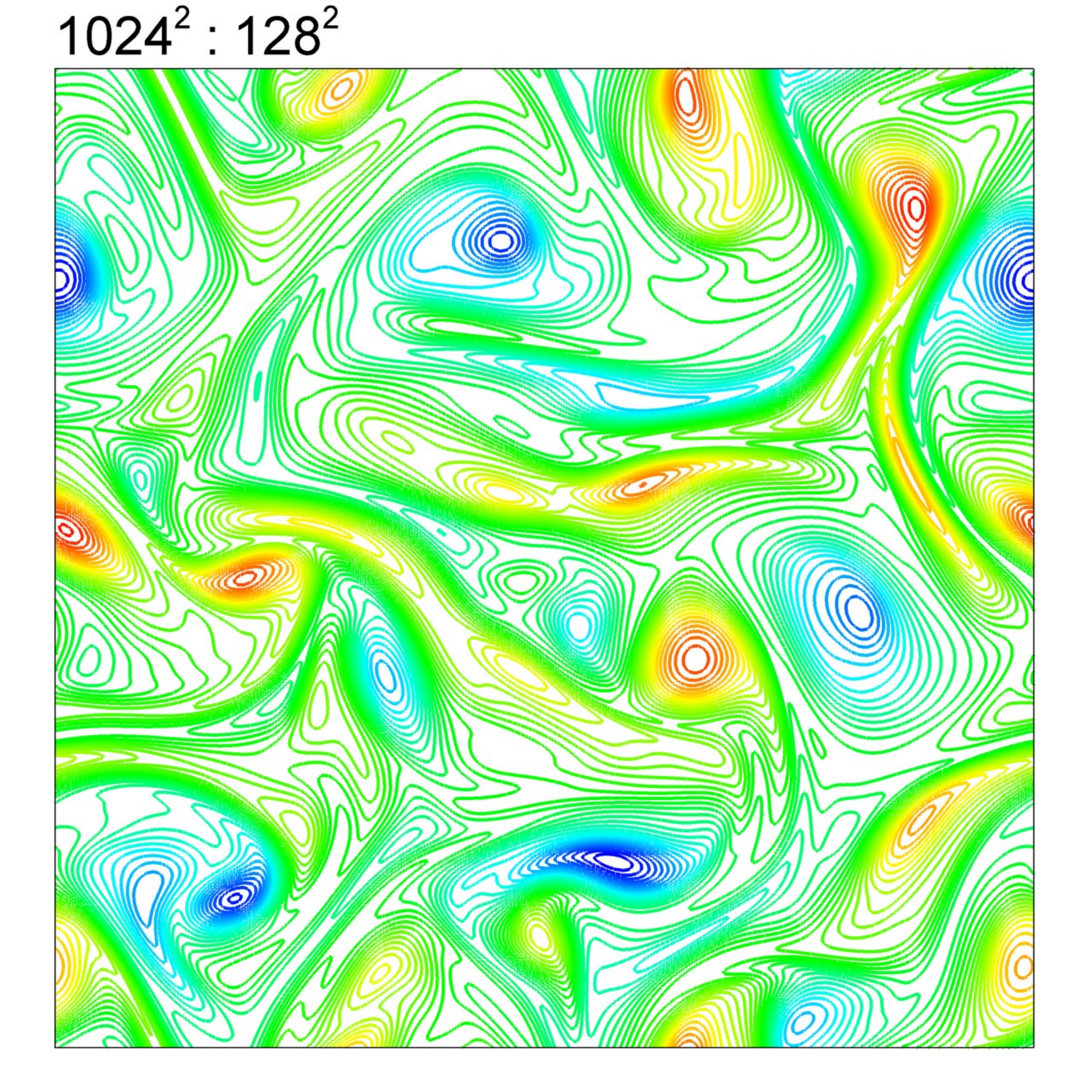}}
\subfigure{\includegraphics[width=0.33\textwidth]{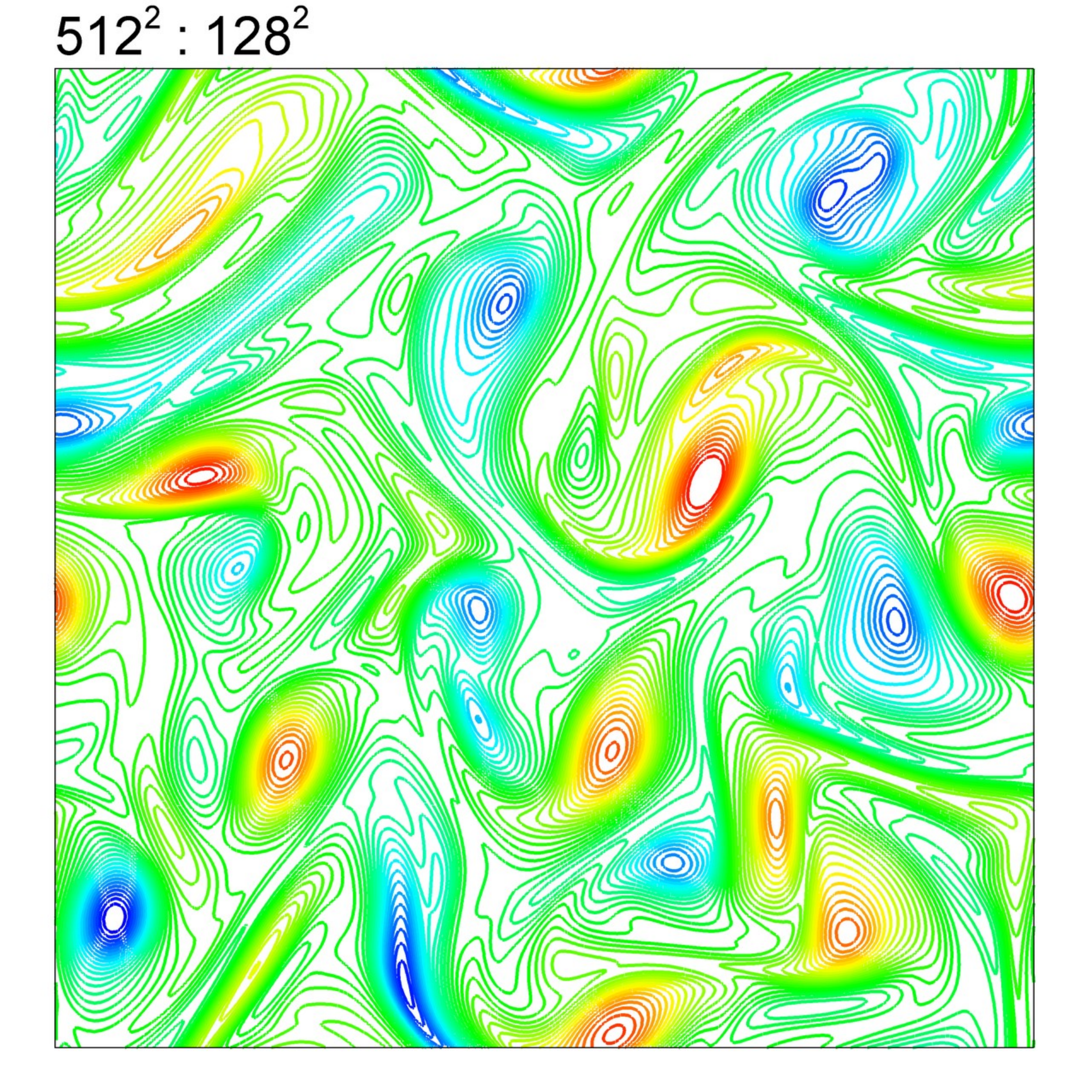}}
}
\mbox{
\subfigure{\includegraphics[width=0.33\textwidth]{turb2d-512-512.pdf}}
\subfigure{\includegraphics[width=0.33\textwidth]{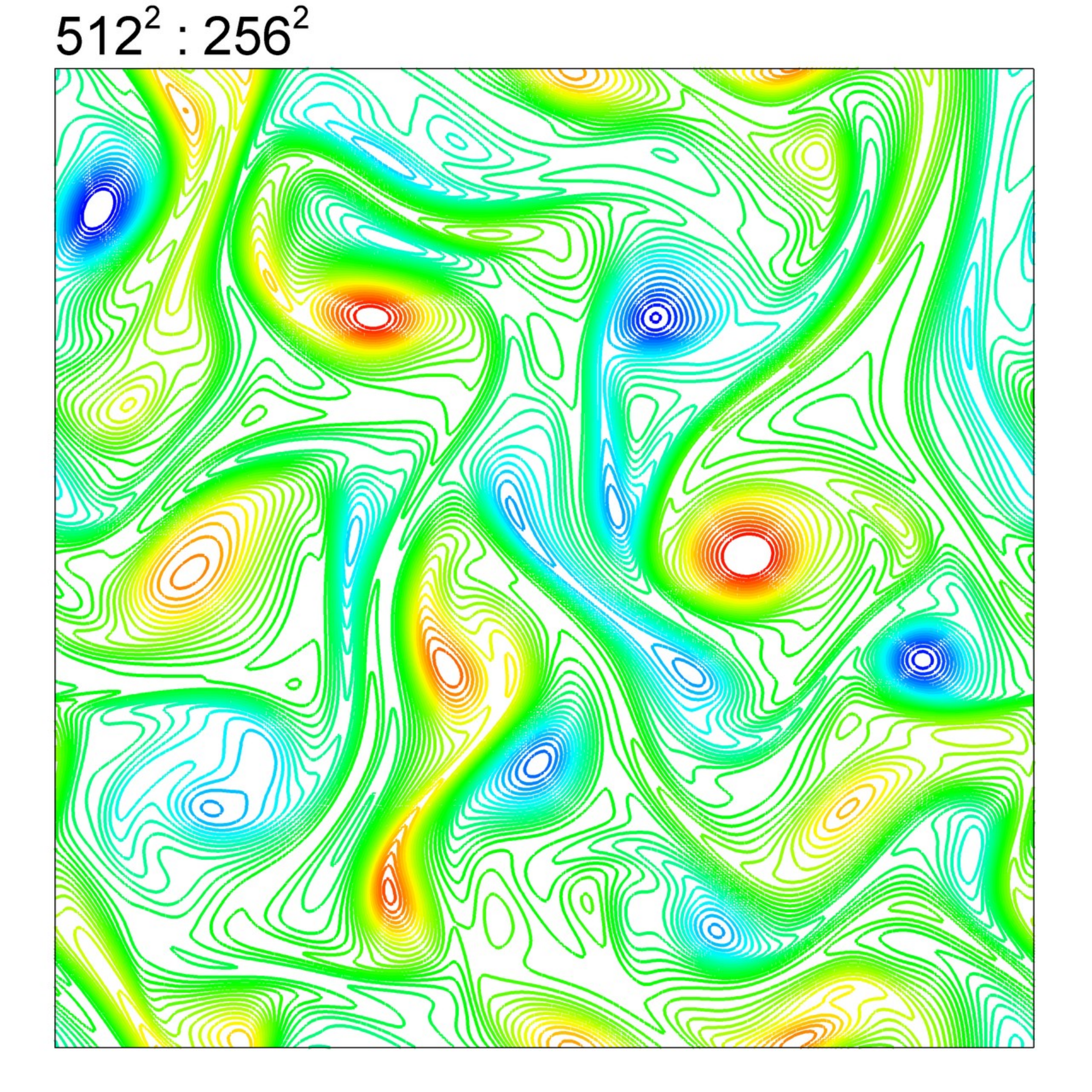}}
\subfigure{\includegraphics[width=0.33\textwidth]{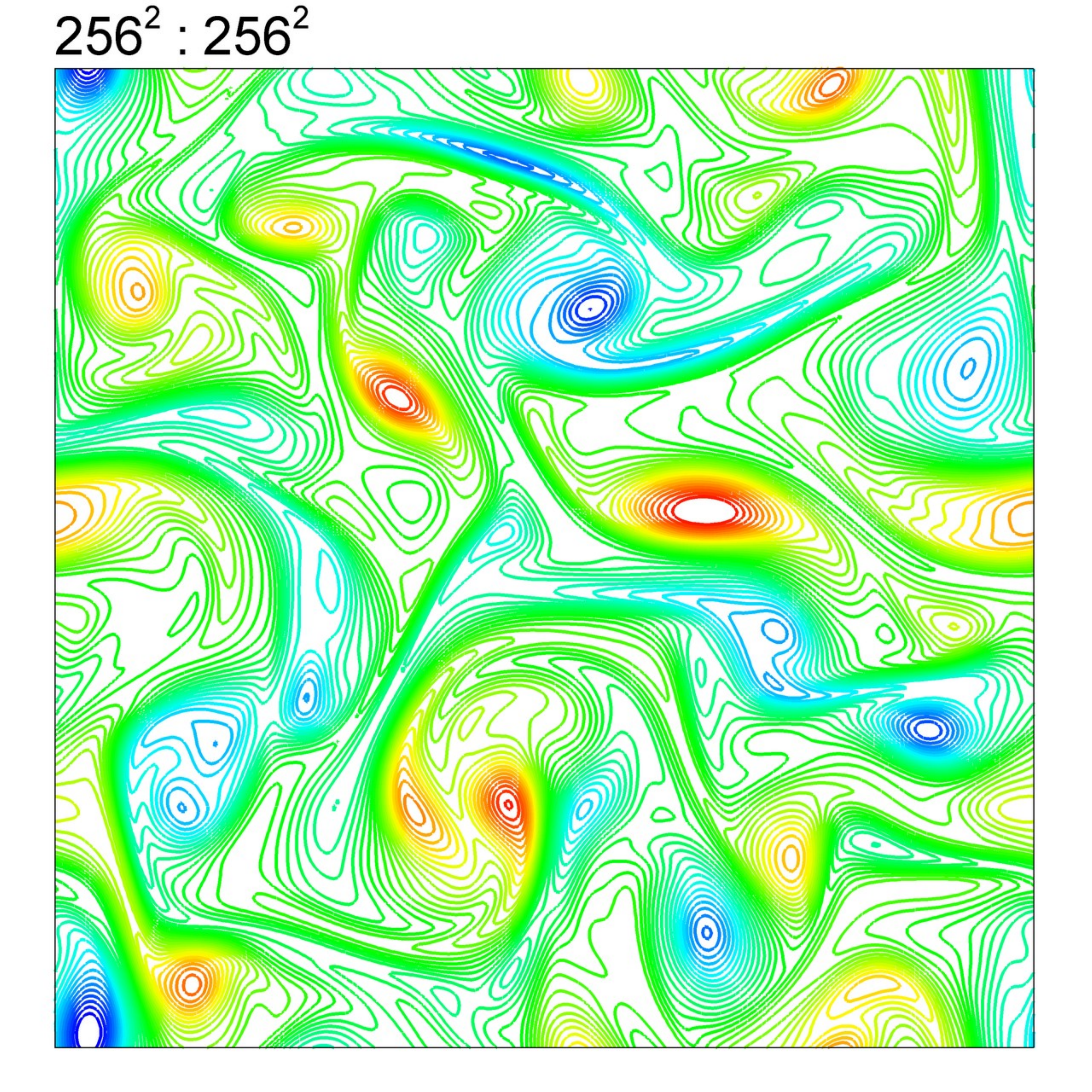}}
}
\mbox{
\subfigure{\includegraphics[width=0.33\textwidth]{turb2d-256-256.pdf}}
\subfigure{\includegraphics[width=0.33\textwidth]{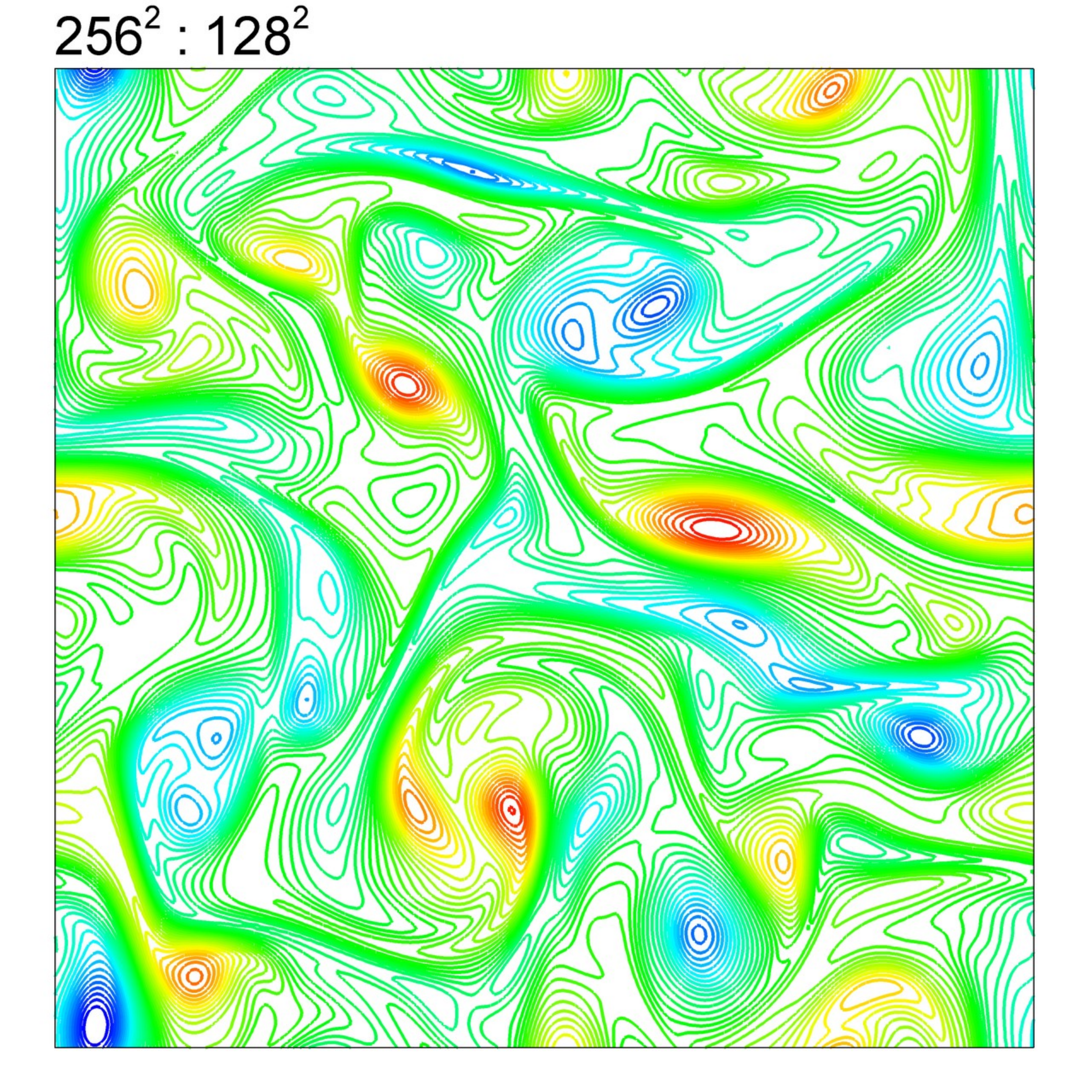}}
\subfigure{\includegraphics[width=0.33\textwidth]{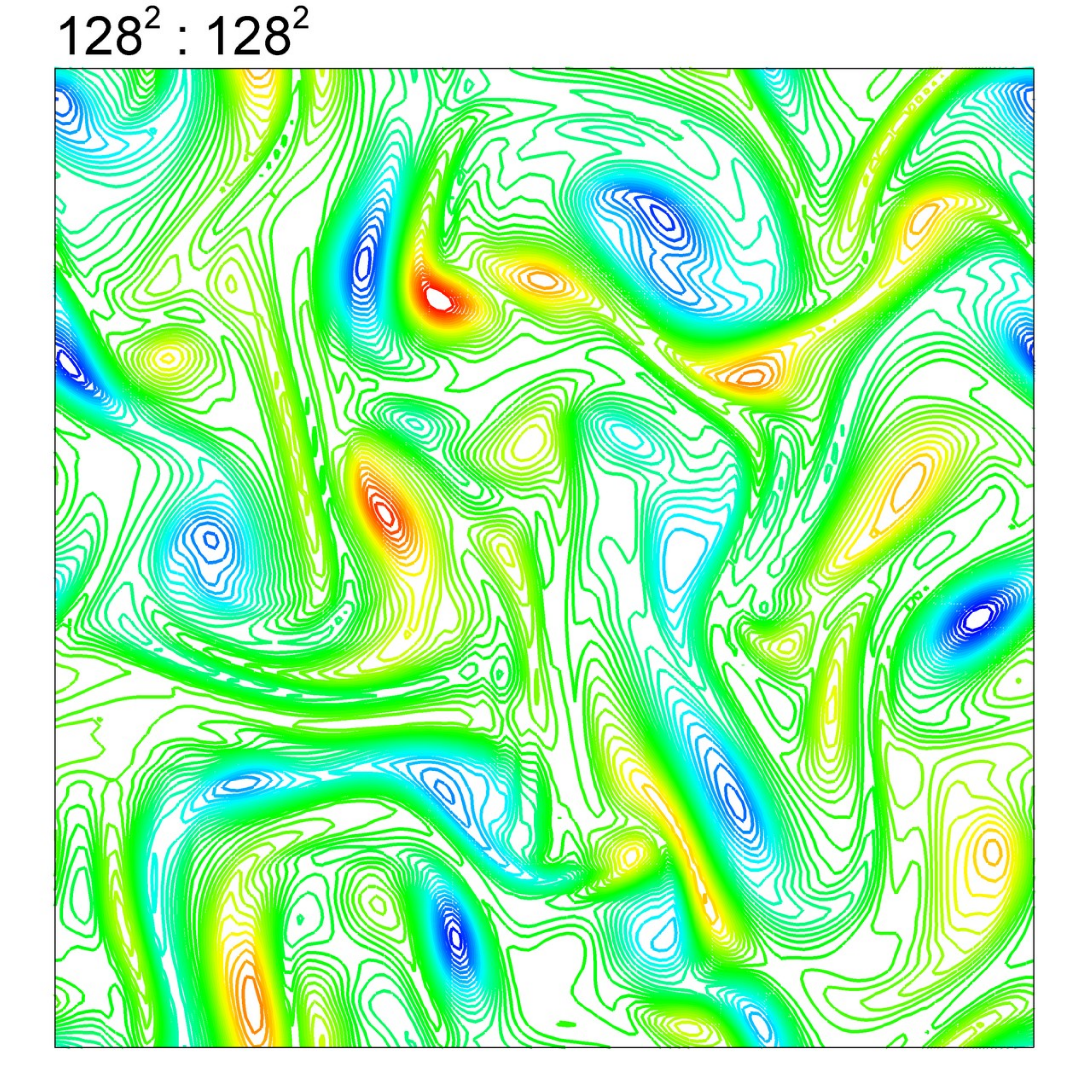}}
}
\caption{Vorticity fields for the decaying turbulence problem at $t=10$ obtained using the vorticity-stream function formulation for $Re=500$. Labels show the resolutions for both parts of the solver in the form $N^2:M^2$, where $N^2$ is the resolution for the vorticity-transport equation, and $M^2$ is the resolution for the Poisson equation. The same equidistant contour levels are used in all cases.}
\label{fig:turb2d-500}
\end{figure*}

\begin{figure*}
\centering
\mbox{
\subfigure[]{\includegraphics[width=0.5\textwidth]{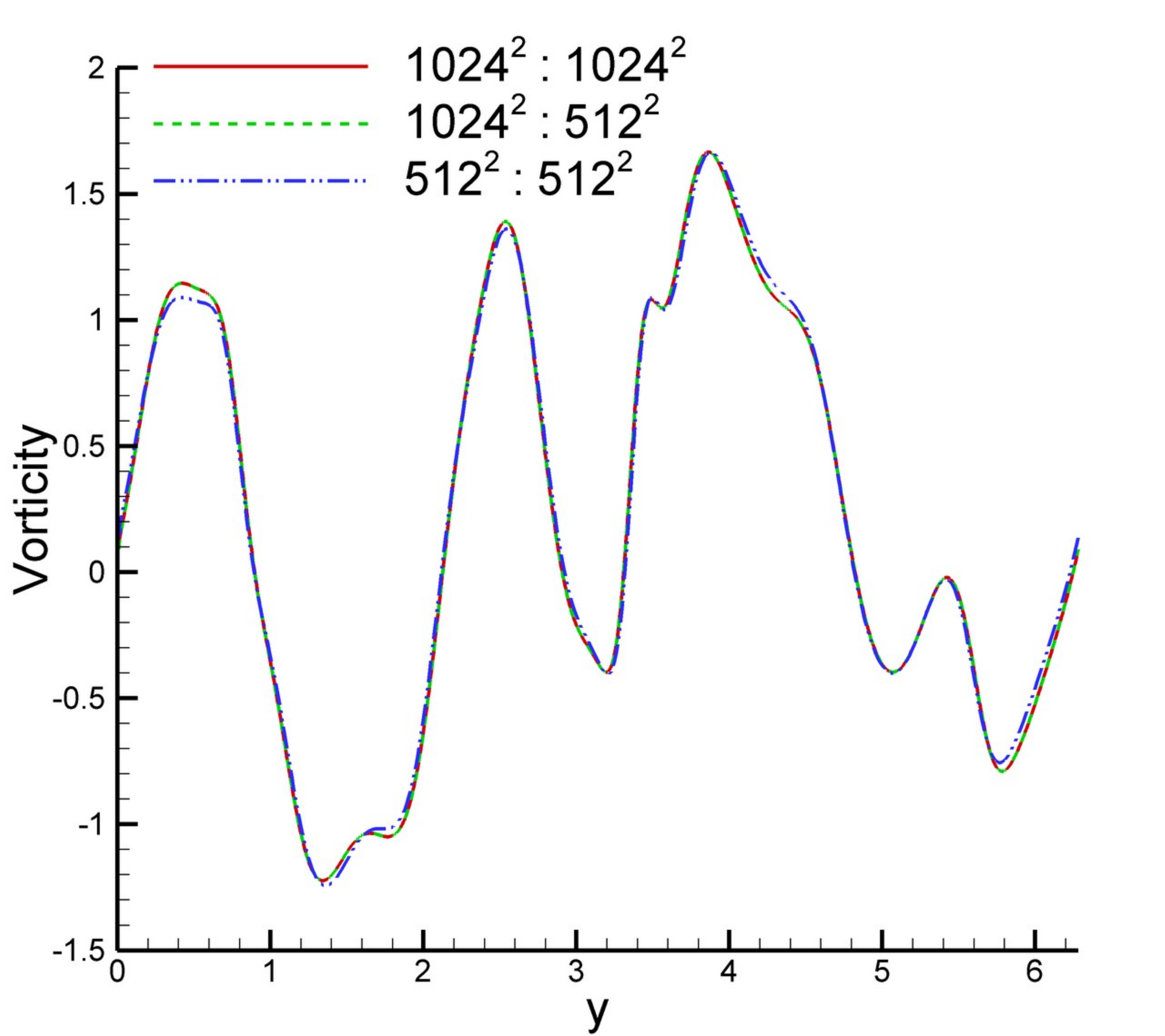}}
\subfigure[]{\includegraphics[width=0.5\textwidth]{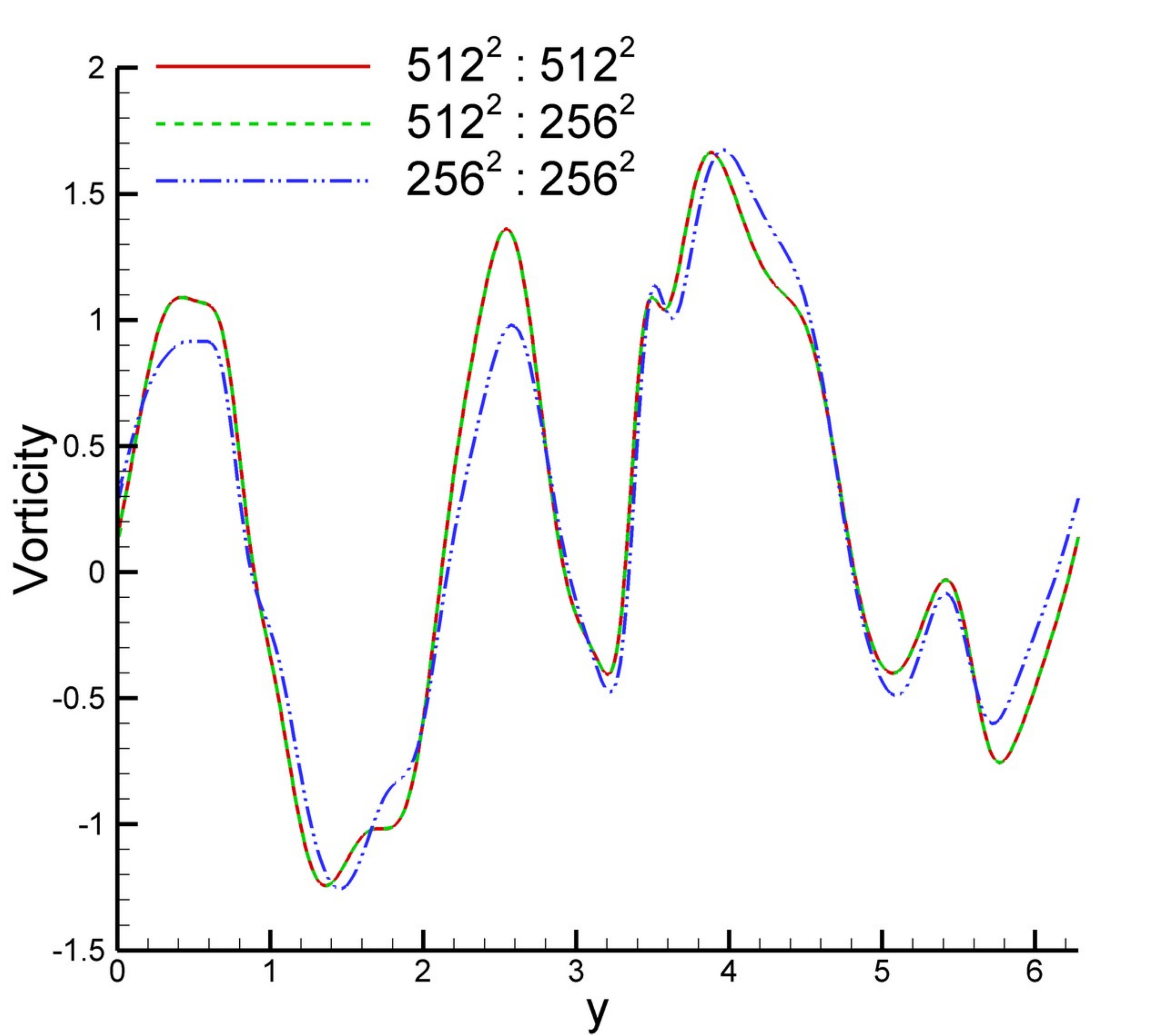}}
}
\caption{Centerline vorticity distributions for the two-dimensional decaying turbulence problem at $t=10$ for $Re=250$; (a) comparison for the standard computation (RK3) on $1024^2:1024^2$ resolution grids, the CGPRK3 method on $1024^2:512^2$ resolution grids, and the standard computation (RK3) on $512^2:512^2$ resolution grids, (b) comparison for the standard computation (RK3) on $512^2:512^2$ resolution grids, the CGPRK3 method on $512^2:256^2$ resolution grids, and the standard computation (RK3) on $256^2:256^2$ resolution grids.}
\label{fig:turb2d250-line}
\end{figure*}

\begin{figure*}
\centering
\mbox{
\subfigure[]{\includegraphics[width=0.5\textwidth]{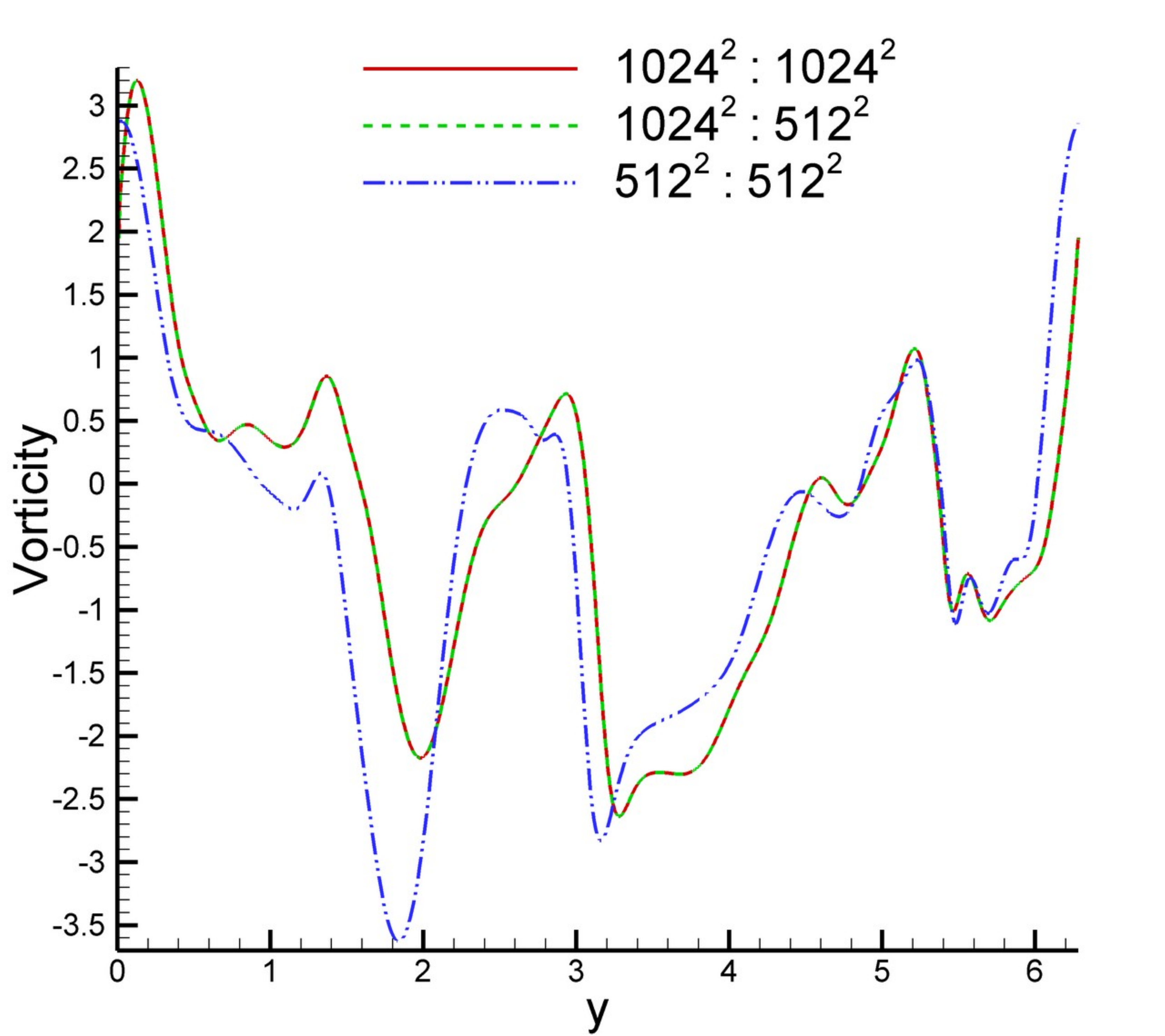}}
\subfigure[]{\includegraphics[width=0.5\textwidth]{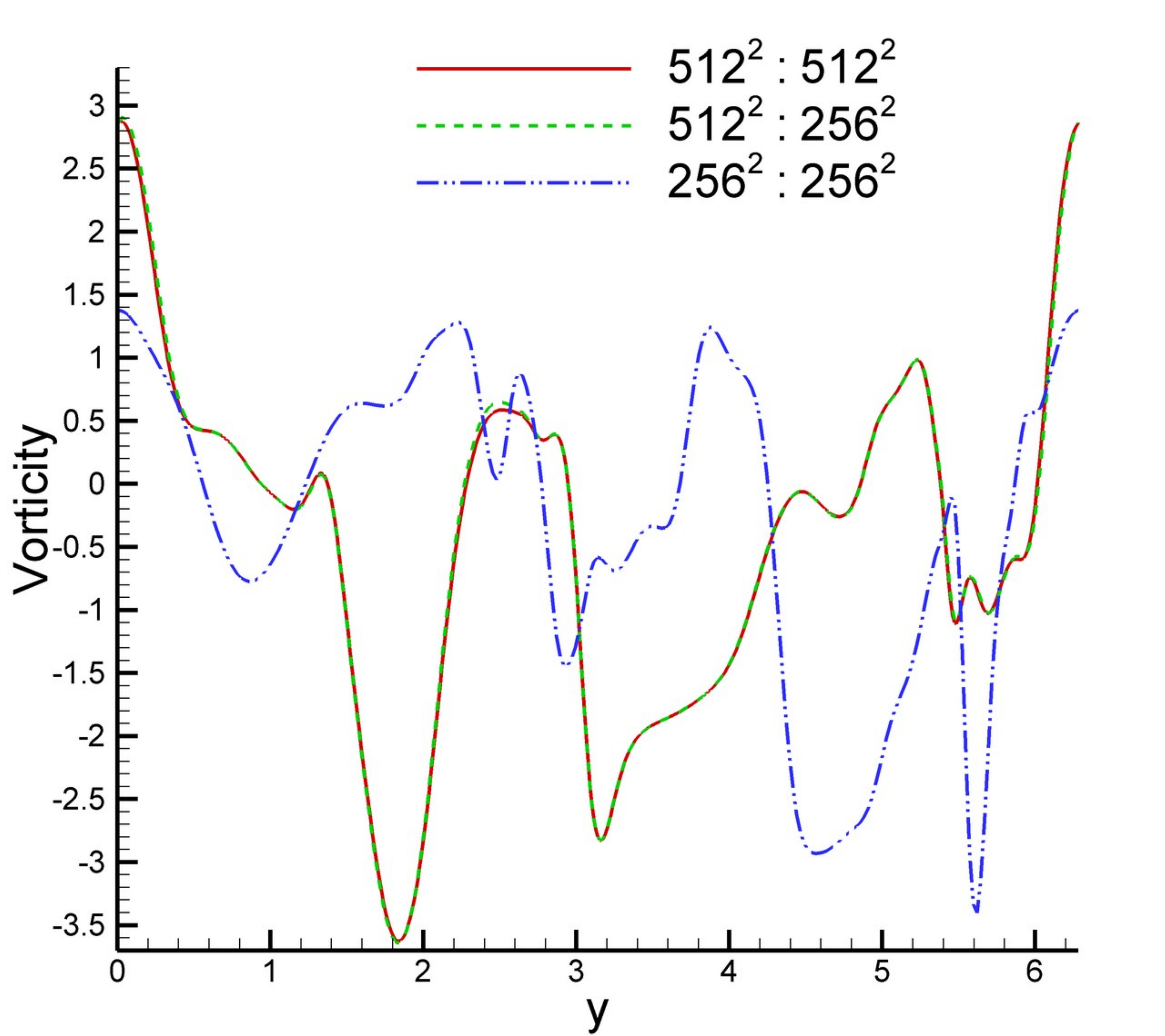}}
}
\caption{Centerline vorticity distributions for the two-dimensional decaying turbulence problem at $t=10$ for $Re=500$; (a) comparison for the standard computation (RK3) on $1024^2:1024^2$ resolution grids, the CGPRK3 method on $1024^2:512^2$ resolution grids, and the standard computation (RK3) on $512^2:512^2$ resolution grids, (b) comparison for the standard computation (RK3) on $512^2:512^2$ resolution grids, the CGPRK3 method on $512^2:256^2$ resolution grids, and the standard computation (RK3) on $256^2:256^2$ resolution grids.}
\label{fig:turb2d500-line}
\end{figure*}

\begin{figure*}
\centering
\includegraphics[width=1.0\textwidth]{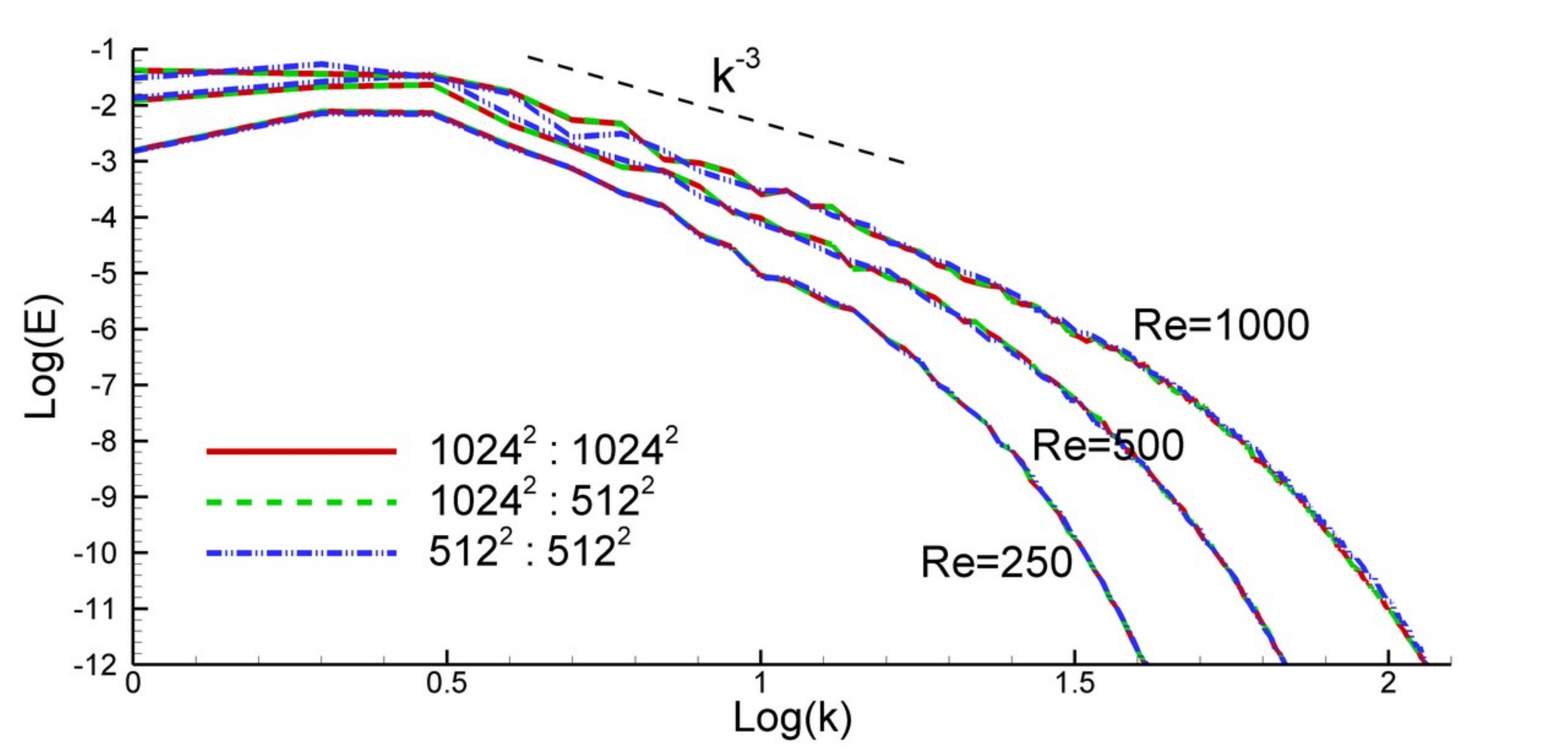}
\caption{Comparisons of the angle averaged energy spectra at time $t=10$ for different Reynolds numbers obtained by standard computations (RK3) on $1024^2:1024^2$ resolution grids, the CGPRK3 method on $1024^2:512^2$ resolution grids, and standard computations (RK3) on $512^2:512^2$ resolution grids. Angle averaged energy spectra asymptotically approach $k^{-3}$ scaling in the inertial range with increasing $Re$.}
\label{fig:spec}
\end{figure*}

Fig.~\ref{fig:turb2d-250} shows vorticity contours obtained using the CGP method as well as standard fine and coarse scale computations for $Re=250$. The first resolution number in the legend corresponds to the grid resolution for computing the advection-diffusion part of the problem, and the second one corresponds to the grid resolution for the Poisson equation. We chose this Reynolds number in particular to show the effect of the cell Reynolds number, defined as:
\begin{equation}
Re_{cell}=Re \frac{2\pi}{N_x}
\label{eq:cell}
\end{equation}
where $N_x=N_y$ is the grid resolution along one direction. $Re_{cell}$ should be smaller than 2 in order to accurately capture viscous effects for a particular resolution. For $Re=250$, the cell Reynolds numbers are 1.53 and 3.07 for resolutions of $1024^2$ and $512^2$, respectively. Since the cell Reynolds numbers are smaller than 2, both the fine and coarse simulations produce similar results (i.e., grid independent solutions). The CGP method provides data identical with that of the pure fine scale simulations.

Similar comparisons are made for $Re=500$ in Fig.~\ref{fig:turb2d-500}. The cell Reynolds numbers are now 3.07 and 6.14 for resolutions of $1024^2$ and $512^2$, respectively. The results of the CGP and the fine resolution computations are very close and highly superior to those from the coarse resolution computation. The CGP method captures the same flow details as the fine scale simulation at a lower computational price when we use one level of coarsening. We see that the accuracy of the CGP method also depends on the resolution. For well resolved cases, the difference between the standard fine resolution computation and the coarse-grid projection computation is negligible, although the error is not that small for lower resolution cases. However, there is still strong benefit to using the coarse-grid projection approach due to the significant improvement of the accuracy of the results compared to the standard low resolution computations. The total computational times for these simulations are: 87.4 hr for the standard computation on a fine resolution ($1024^2:1024^2$) grid, 22.9 hr for the computation using the CGP method on a $1024^2:512^2$ resolution grid, and 13.1 hr for the standard computation on a grid of resolution $512^2:512^2$. Vorticity distributions along the vertical centerline at time $t=10$ for $Re=250$ and $Re=500$ are shown in Fig.~\ref{fig:turb2d250-line} and Fig.~\ref{fig:turb2d500-line}, respectively. As discussed earlier, the speed up in computations would be even greater if we used a suboptimal Poisson solver in the computations.

Finally, we perform a comparison of turbulence statistics. Fig.~\ref{fig:spec} demonstrates the angle averaged energy spectrum for different Reynolds numbers. The energy spectrum is defined as:
\begin{equation}
\hat{E}(\textbf{k},t)=\frac{1}{2}k^2|\tilde{\psi}(\textbf{k},t)|^2
\label{eq:esp}
\end{equation}
and the angle averaged energy spectrum is:
\begin{equation}
E(k,t)= \sum_{k\leq|\acute{k}|\leq k+1} \hat{E}(\acute{\textbf{k}},t).
\label{eq:esp}
\end{equation}
The angle averaged energy spectra obtained by CGP and standard methods are shown in Fig.~\ref{fig:spec}, and demonstrate that the energy spectra converge to $k^{-3}$ scaling in the inertial range for increasing $Re$. Our primary goal in this study is to investigate the behavior of the CGP method in long time integration turbulence simulations. Therefore, we do not perform any further computations at larger Reynolds numbers. This decaying turbulence test case primarily shows that the CGP method can be used to obtain highly accurate simulations of complex flow fields at greatly accelerated rates, with a speed-up factor of 3 obtained for this particular Poisson solver. Again, we expect this gain to be even more pronounced with the choice any other suboptimal Poisson solver.

\subsection{Taylor-Green vortex on a distorted grid}
\begin{figure}[h]
\centering
\mbox{
\subfigure{\includegraphics[width=0.4\textwidth]{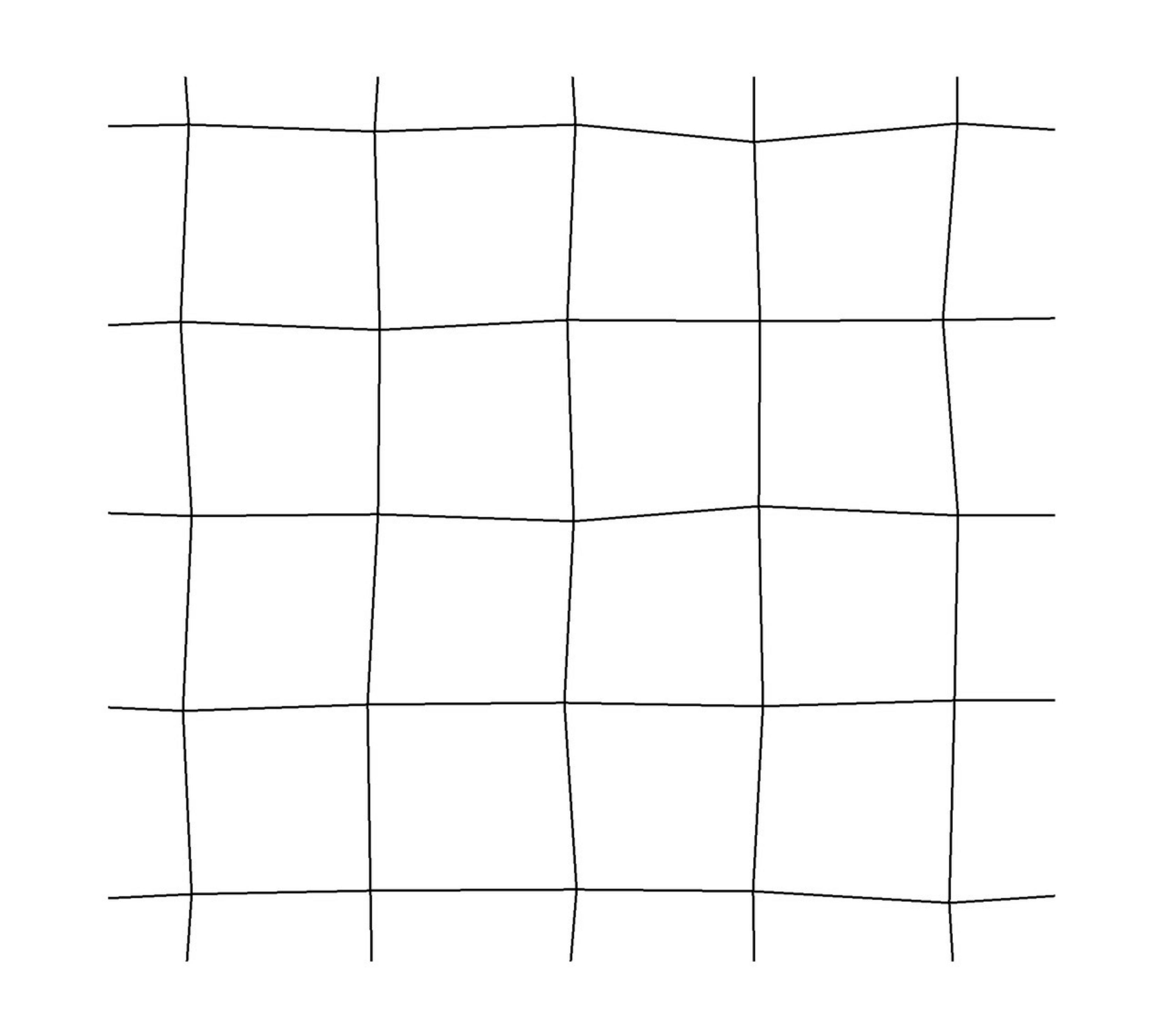}}
\subfigure{\includegraphics[width=0.4\textwidth]{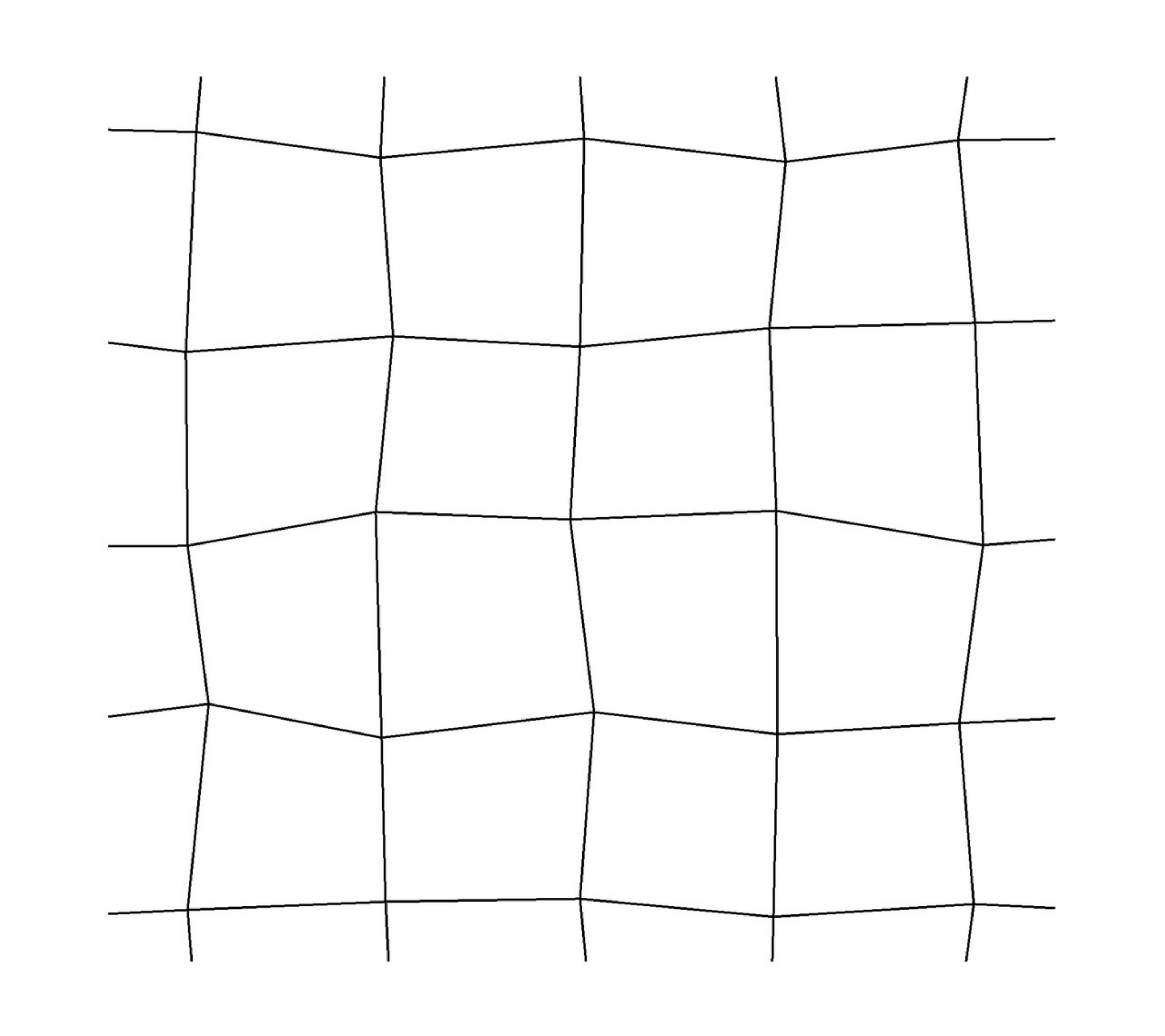}}
}
\caption{Distorted grids for the Taylor-Green vortex problem for distortion parameters $\tau=0.1$ (left), and $\tau=0.2$ (right).}
\label{fig:dist-grid}
\end{figure}

All the previous test cases were computed on Cartesian grids. Since the method is intended to be generally applicable, we analyze here the performance of the CGP method on a non-Cartesian, randomly distorted grid. Specifically, we perform quantitative error analyzes for the Taylor-Green vortex problem on the distorted grid. The grid distortion is accomplished using a random number generator returning a real number between zero and a defined distortion parameter, $\tau$, at each query. For $\tau=0$, we obtain a regular Cartesian grid. Fig.~\ref{fig:dist-grid} shows the distorted grids with $\tau=0.1$ and $\tau=0.2$, that are used in our computations. First we transform the vorticity-stream function formulation of the governing equations to generalized curvilinear coordinates. The metrics and Jacobian are computed again via finite differences \cite{tannehill1997computational}. Here, we apply the CGP method to the vorticity-stream function formulation in curvilinear coordinates. We apply the previous full weighting restriction and bilinear prolongation operators before and after solving the elliptic constraint equation to reduce the computational time. Since we transformed our equations to generalized coordinates, it is not at all straightforward to use FFT-based elliptic Poisson solvers for this case. For that reason, we use another linear-cost efficient Poisson solver, the V-cycle multigrid method \cite{gupta1997comparison}. Both the $L_{\infty}$ and $L_2$ norms, are computed according to the exact analytical solution given in Eq.~\ref{eq:exw}.
\begin{table*}
\small
\caption{Computed error norms for the Taylor-Green vortex problem on distorted grids for $Re=10$ and $\Delta t=2.5\times10^{-4}$ at time $t=1$. The first resolution number is the number of grid points used to solve the advection-diffusion equation (vorticity-transport equation), and the second resolution number is that used to solve the Poisson equation (the V-cycle multigrid Poisson solver is used). The speed-up ratio is defined as the ratio of the CPU times for the computations performed without using the CGP procedure (using standard methods) and with the CGP procedure.}
\begin{center}
\label{tab:dist-grid}
\begin{tabular}{llcccc}
\hline\noalign{\smallskip}
Method & Resolutions &  $||\omega||_{\infty}$ & $||\omega||_{L_2}$ & CPU (s) & Speed-up \\
\hline\noalign{\smallskip}
RK3 ($\tau=0$)      &$256^2$ : $256^2$     &  1.6439E-5 & 8.1561E-6 & 3171.09 & 1.00 \\
CGPRK3 ($\tau=0$)   &$256^2$ : $128^2$     &  1.6439E-5 & 8.1568E-6 & 1234.72 & 2.56 \\
RK3 ($\tau=0$)      &$128^2$ : $128^2$     &  6.5756E-5 & 3.3133E-5 & 1076.88 & 1.00 \\
CGPRK3 ($\tau=0$)   &$128^2$ : $64^2$      &  6.5757E-5 & 3.3176E-5 & 550.73  & 1.96 \\
\hline\noalign{\smallskip}
RK3 ($\tau=0.1$)    &$256^2$ : $256^2$     &  6.3111E-4 & 1.1422E-4 & 9566.87 & 1.00 \\
CGPRK3 ($\tau=0.1$) &$256^2$ : $128^2$     &  6.3253E-4 & 1.2802E-4 & 1137.14 & 8.41 \\
RK3 ($\tau=0.1$)    &$128^2$ : $128^2$     &  1.0933E-3 & 1.8148E-4 & 1886.91 & 1.00 \\
CGPRK3 ($\tau=0.1$) &$128^2$ : $64^2$      &  1.0946E-3 & 1.9874E-4 & 482.37  & 3.91 \\
\hline\noalign{\smallskip}
RK3 ($\tau=0.2$)    &$256^2$ : $256^2$     &  1.4867E-3 & 3.0031E-4 & 88109.04 & 1.00 \\
CGPRK3 ($\tau=0.2$) &$256^2$ : $128^2$     &  1.4836E-3 & 3.4331E-4 & 2080.43  & 42.35 \\
RK3 ($\tau=0.2$)    &$128^2$ : $128^2$     &  2.4964E-3 & 4.2399E-4 & 5642.90  & 1.00 \\
CGPRK3 ($\tau=0.2$) &$128^2$ : $64^2$      &  2.4696E-3 & 5.0689E-4 & 969.37   & 5.82 \\
\hline
\end{tabular}
\end{center}
\end{table*}

A comparison of the efficiency and accuracy of the CGP method and the other methods is shown in Table~\ref{tab:dist-grid}, with three different grid distortion ratios used. These results clearly demonstrate the applicability of the CGP method to problems solved on generalized non-Cartesian grids. As we expect, for no distortion ($\tau=0$) the results are nearly the same as those found in Table~\ref{tab:norm1} and verify that the method is independent of the choice of Poisson solver. The measured CPU times are different due to the different Poisson solver as well as the curvilinear transform, but the speed-up ratios are identical. It can be seen that the errors due to using coarser grids is due mainly due to the restriction and prolongation operators rather than solution of the Poisson equation. Increasing the distortion ratio results an increased CPU times for the simulation. We compare the CPU times obtained by the CGPRK3 algorithm with the results obtained with the regular solver (RK3) for the same resolutions. We find that increasing the distortion ratio results in more and more computational efficiency when using the CGP method. For $\tau=0.2$ the simulation was 42 times faster when performed with the CGP method, without an accompanying degeneracy in the accuracy of the results.

\subsection{Laminar flow over a circular cylinder}
\begin{figure}[h]
\centering
\includegraphics[width=0.9\textwidth]{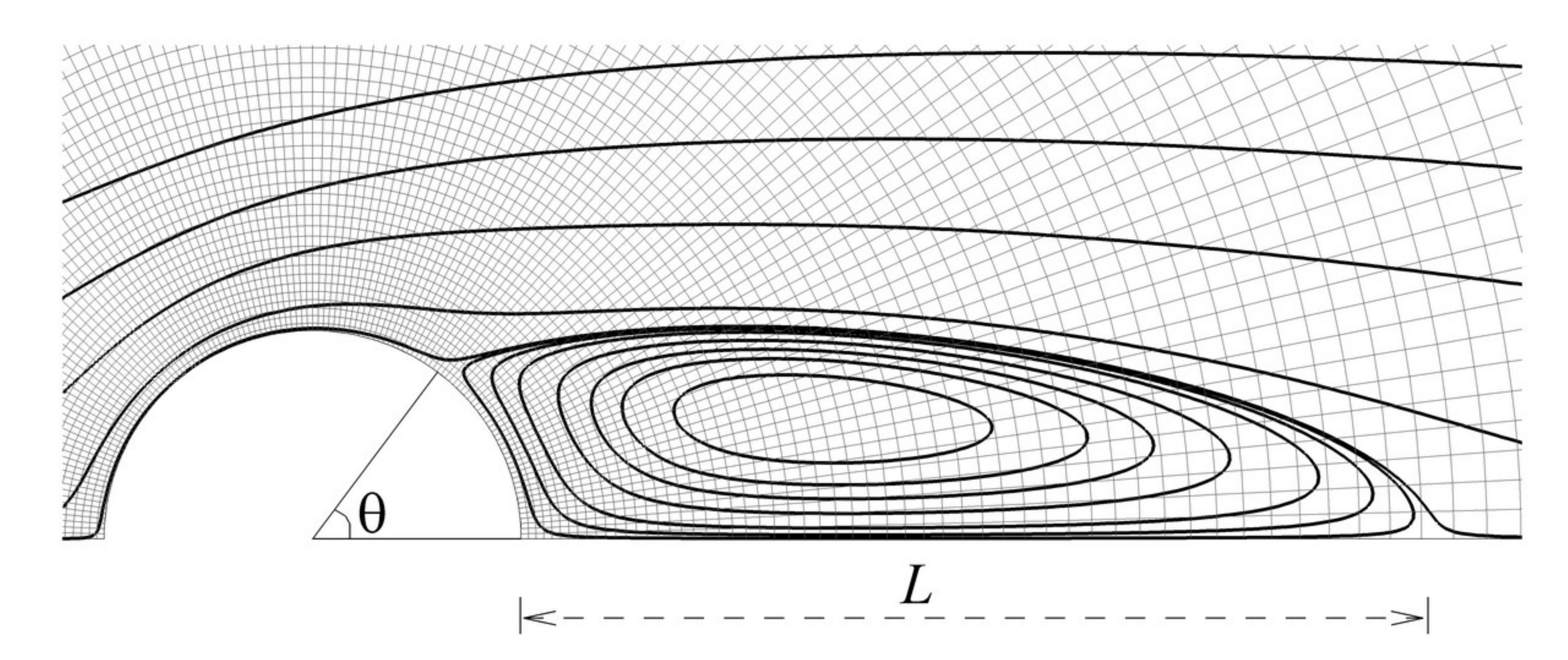}
\caption{Computational mesh for a cylinder with 128 radial and 128 circumferential nodes using the stretching parameter $\beta=1.04$ with the definitions of separation angle, $\theta$, and separation bubble length, $L$.}
\label{fig:def-cyl}
\end{figure}

The laminar flow over a cylinder was computed to demonstrate that the coarse-grid projection method also is applicable to external flows. Flow over a cylinder is a fundamental fluid mechanics problem of practical importance. The cylinder is represented in two dimensions by a circle, and a flow domain surrounds the circle, for which we set the outer diameter to be 30 times bigger than the cylinder diameter. The flow field over the cylinder is symmetric at low values of the Reynolds number. As the Reynolds number increases, the flow begins to separate behind the cylinder causing vortex shedding, an unsteady phenomenon. Since we present our results for $Re=40$ based on the free stream velocity and cylinder diameter, a relatively low Reynolds number, we only compute the flow in the upper half of the domain. In this case, a stationary separation bubble forms behind the cylinder and the flow field is symmetric about the $x$-axis. In order to compare the results with those available in the literature, we have computed the separation bubble length (reattachment length), $L$, and separation angle, $\theta$, as shown in Fig.~\ref{fig:def-cyl}. This canonical flow problem has been used as a benchmark case in many computational studies \cite{choi2007immersed,dennis1970numerical,galbraith2011implicit,linnick2005high,russell2003cartesian}.

We used the vorticity-stream function formulation in generalized curvilinear coordinates. The metrics and Jacobian of the transformation were computed analytically according to the following transformation:
\begin{eqnarray}
x &=& \left( \frac{D-1}{2}\kappa(\eta) + \frac{1}{2} \right)\mbox{cos}\left(\pi(1-\xi)\right)\\
y &=& \left( \frac{D-1}{2}\kappa(\eta) + \frac{1}{2} \right)\mbox{sin}\left(\pi(1-\xi)\right)
\label{eq:transform}
\end{eqnarray}
where $D$ is the outer diameter, and $\kappa(\eta)$ is the logarithmic stretching function, which is given as \cite{hoffmann2000computational}:
\begin{equation}
\kappa(\eta) = \frac{(\beta+1)-(\beta-1)\left(\frac{\beta+1}{\beta-1}\right)^{1-\eta}}{1+ \left( \frac{\beta+1}{\beta-1}\right)^{1-\eta}}
\label{eq:kappa}
\end{equation}
where $\beta$ is the stretching parameter. Here, we set $\beta=1.04$ and $D=30$. The computational mesh shown in Fig.~\ref{fig:def-cyl} consists of an O-grid with 128 radial and 128 circumferential nodes.

\begin{figure*}
\centering
\mbox{
\subfigure{\includegraphics[width=0.33\textwidth]{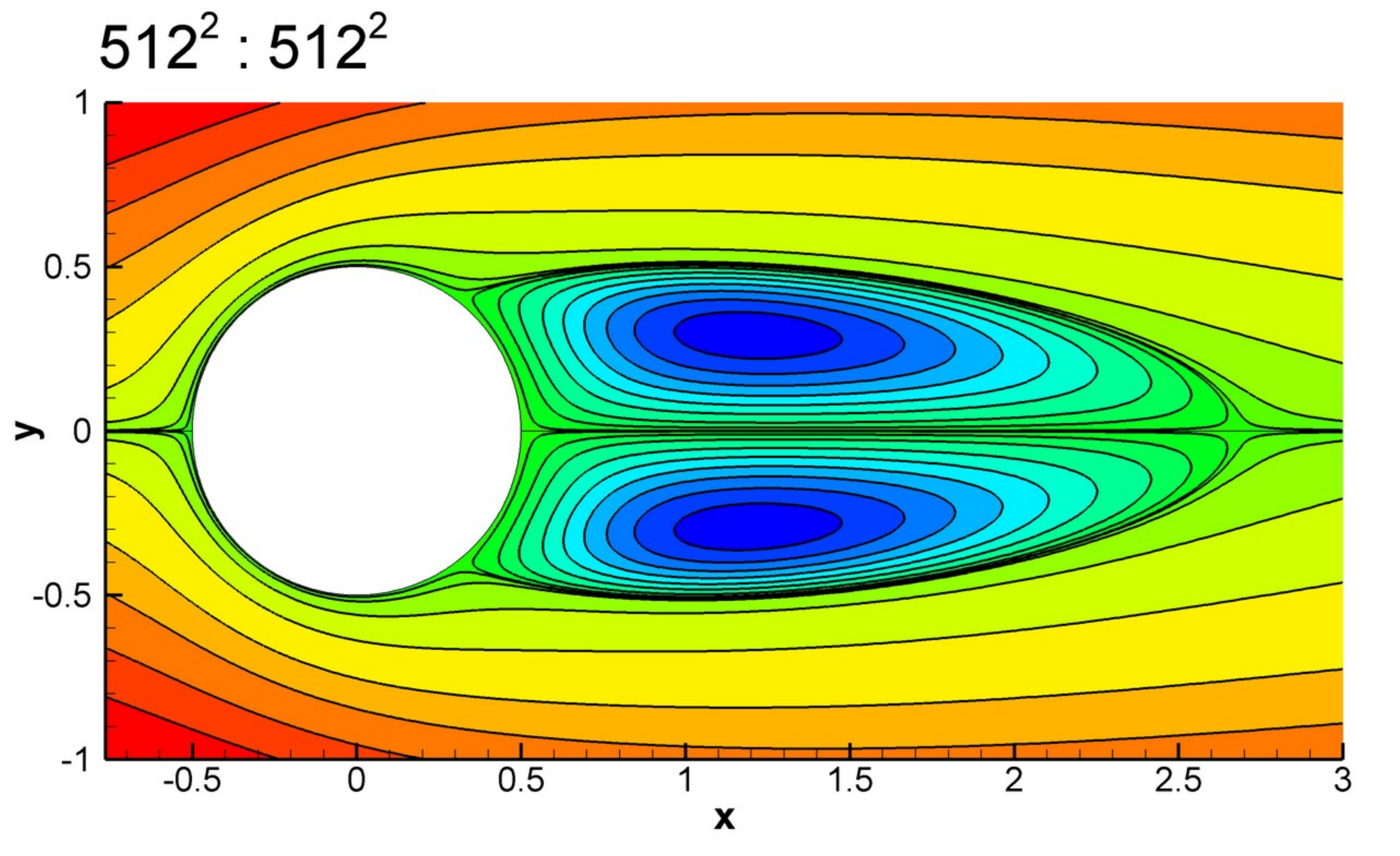}}
\subfigure{\includegraphics[width=0.33\textwidth]{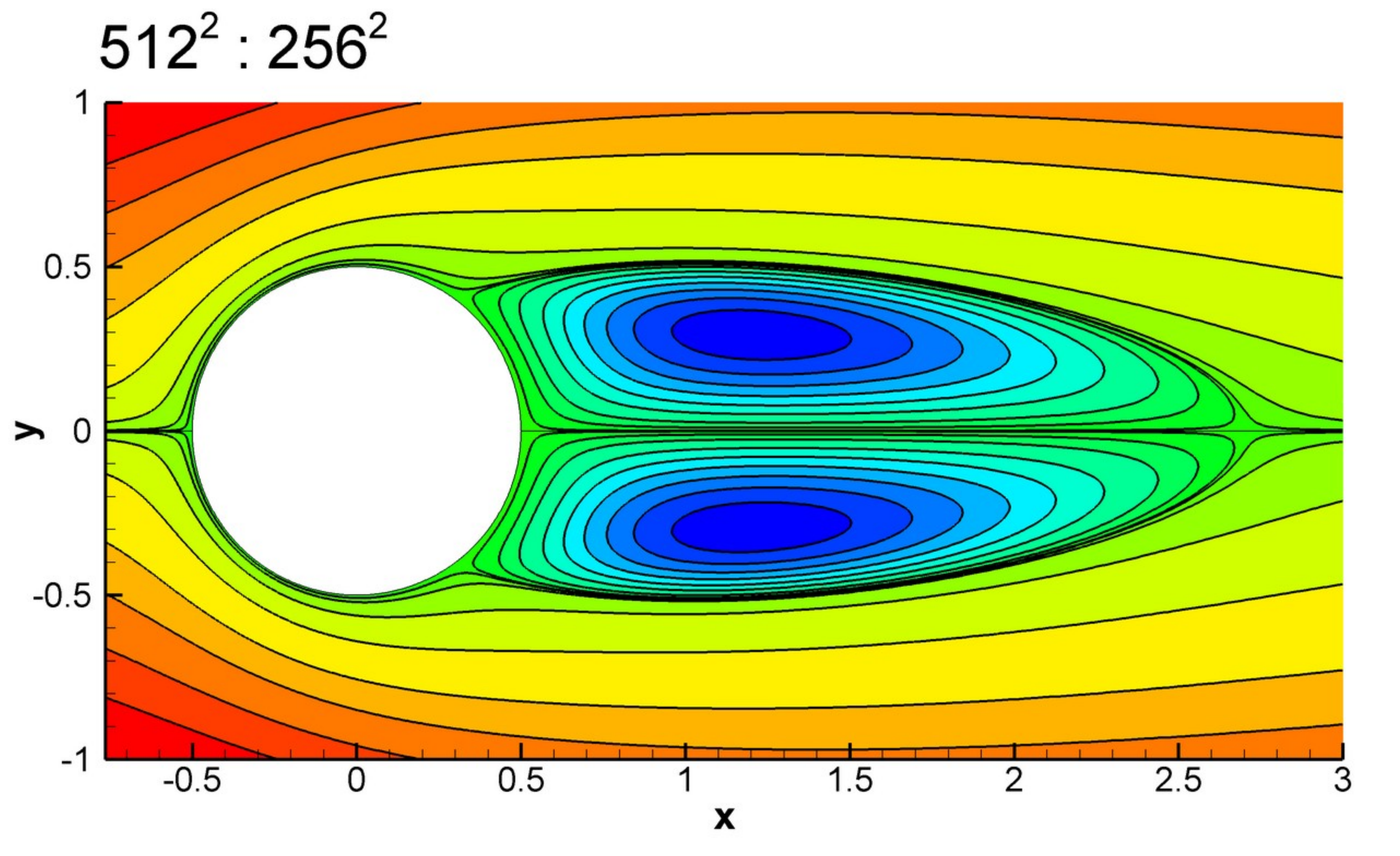}}
\subfigure{\includegraphics[width=0.33\textwidth]{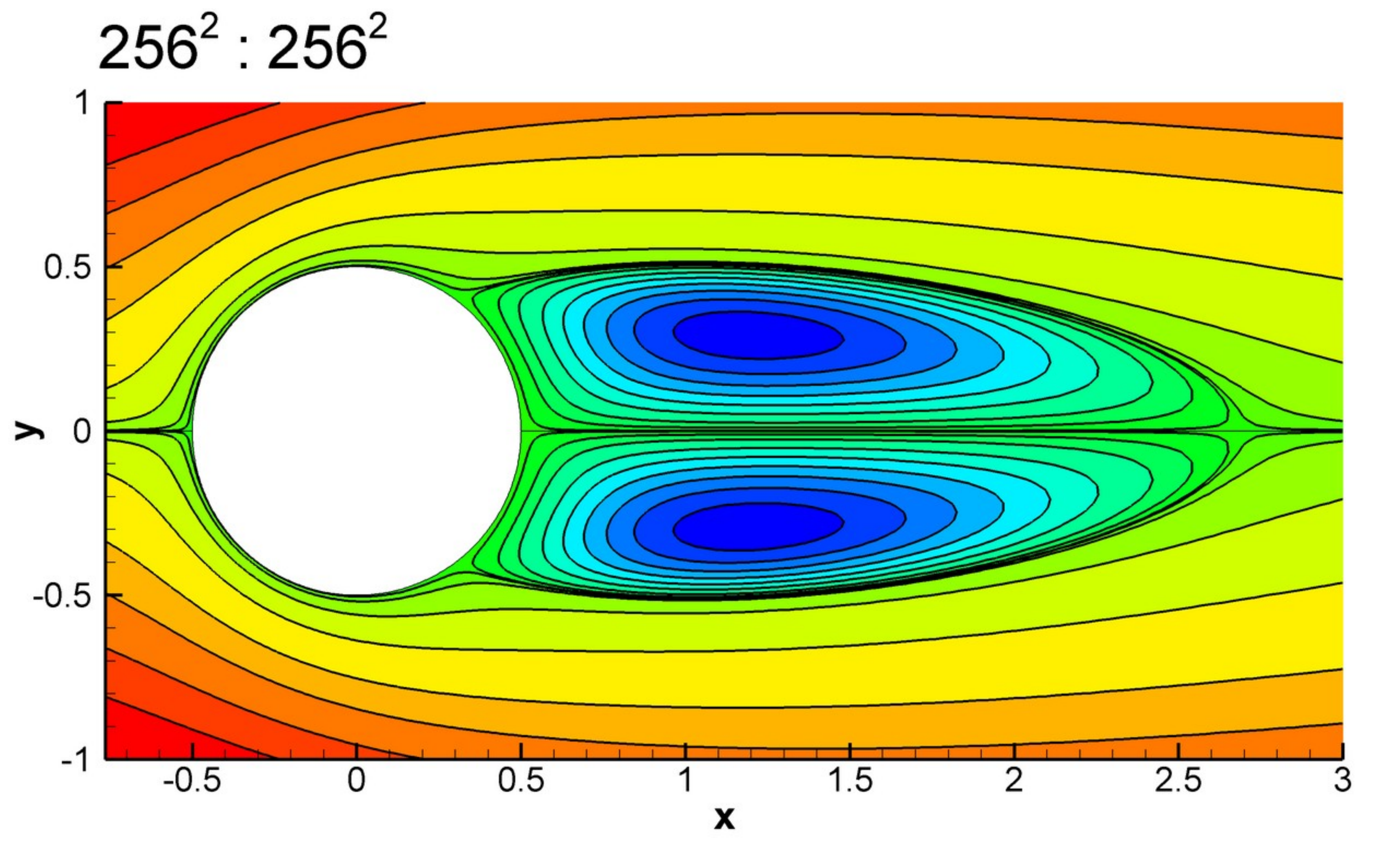}}
}
\mbox{
\subfigure{\includegraphics[width=0.33\textwidth]{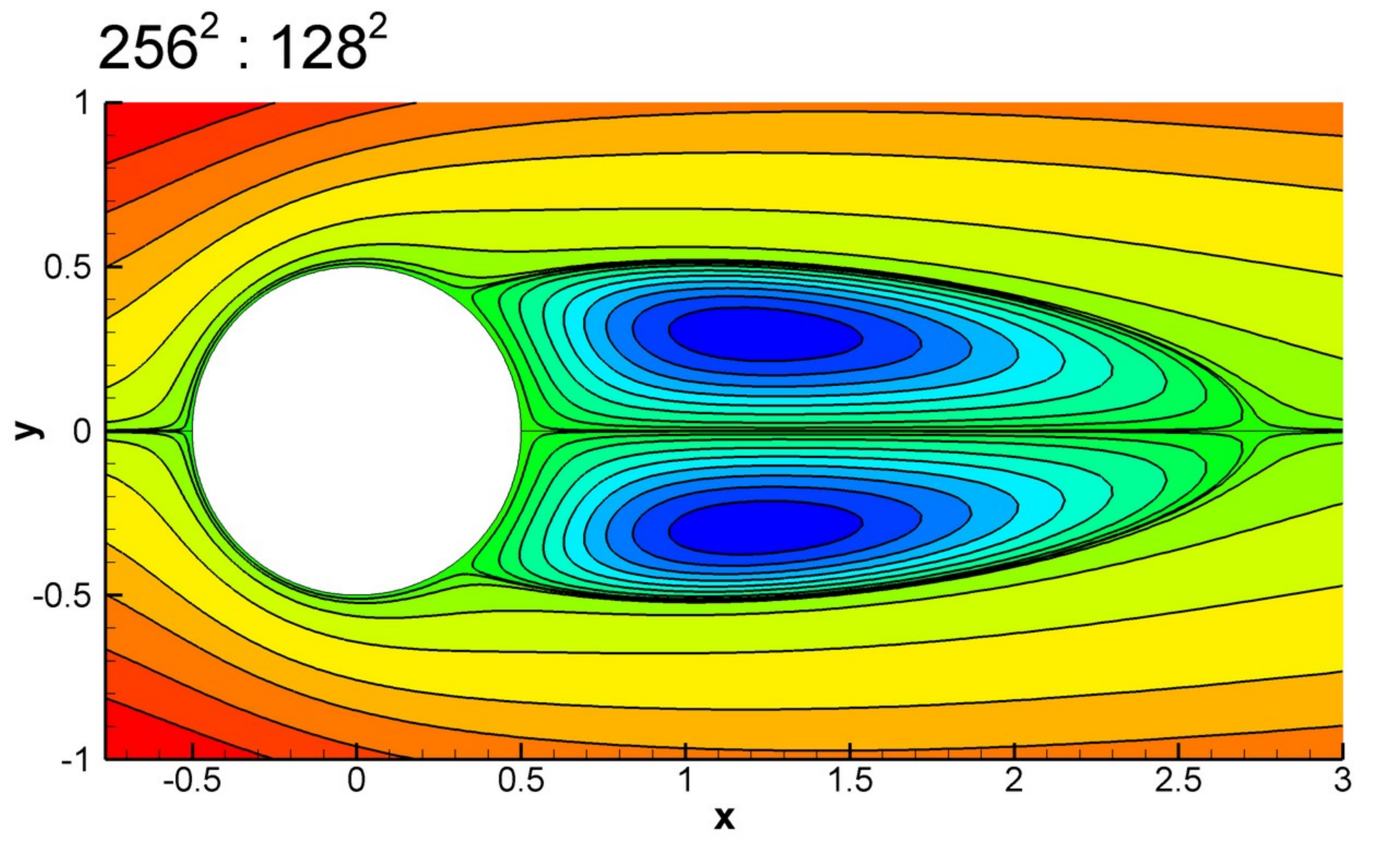}}
\subfigure{\includegraphics[width=0.33\textwidth]{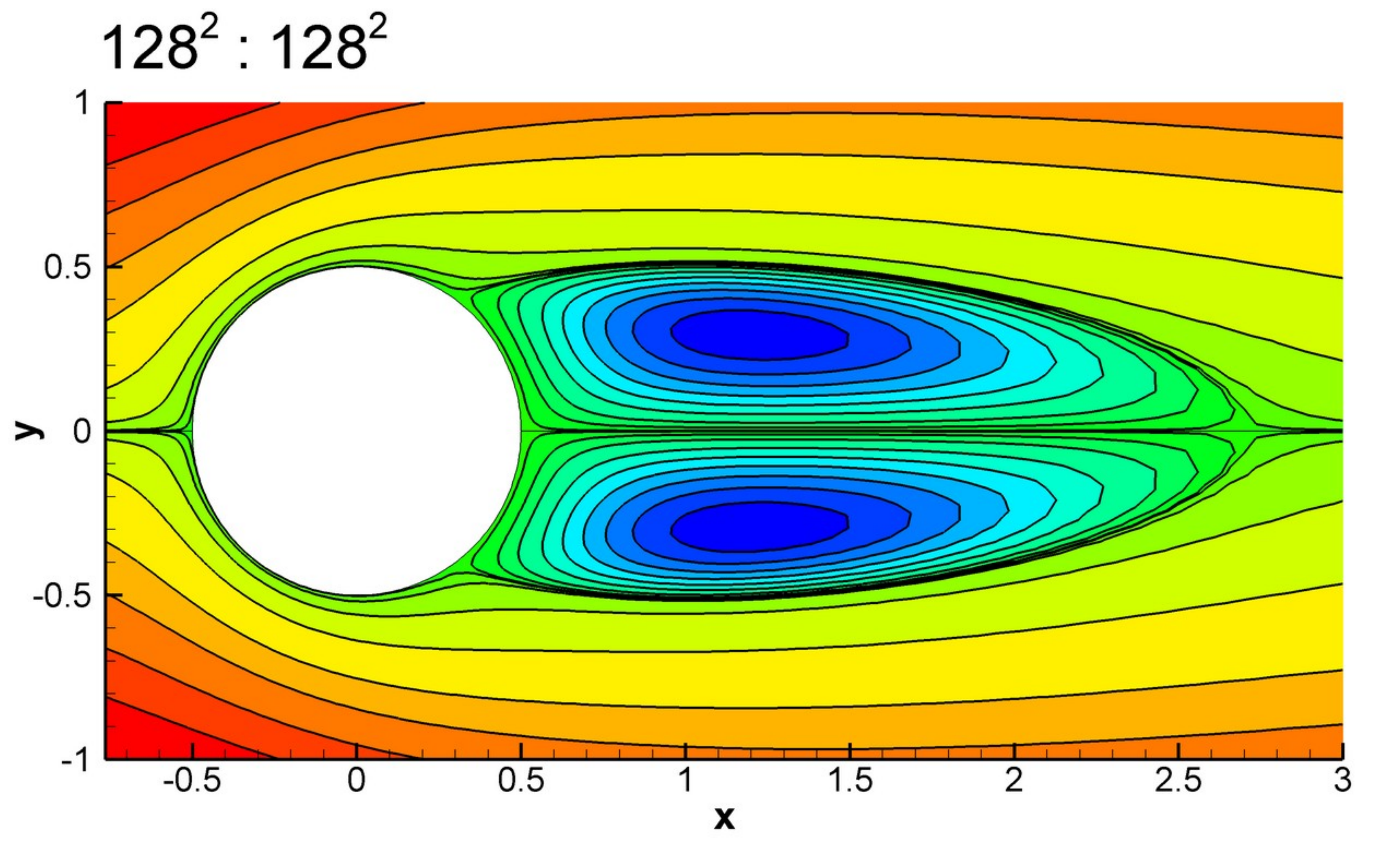}}
\subfigure{\includegraphics[width=0.33\textwidth]{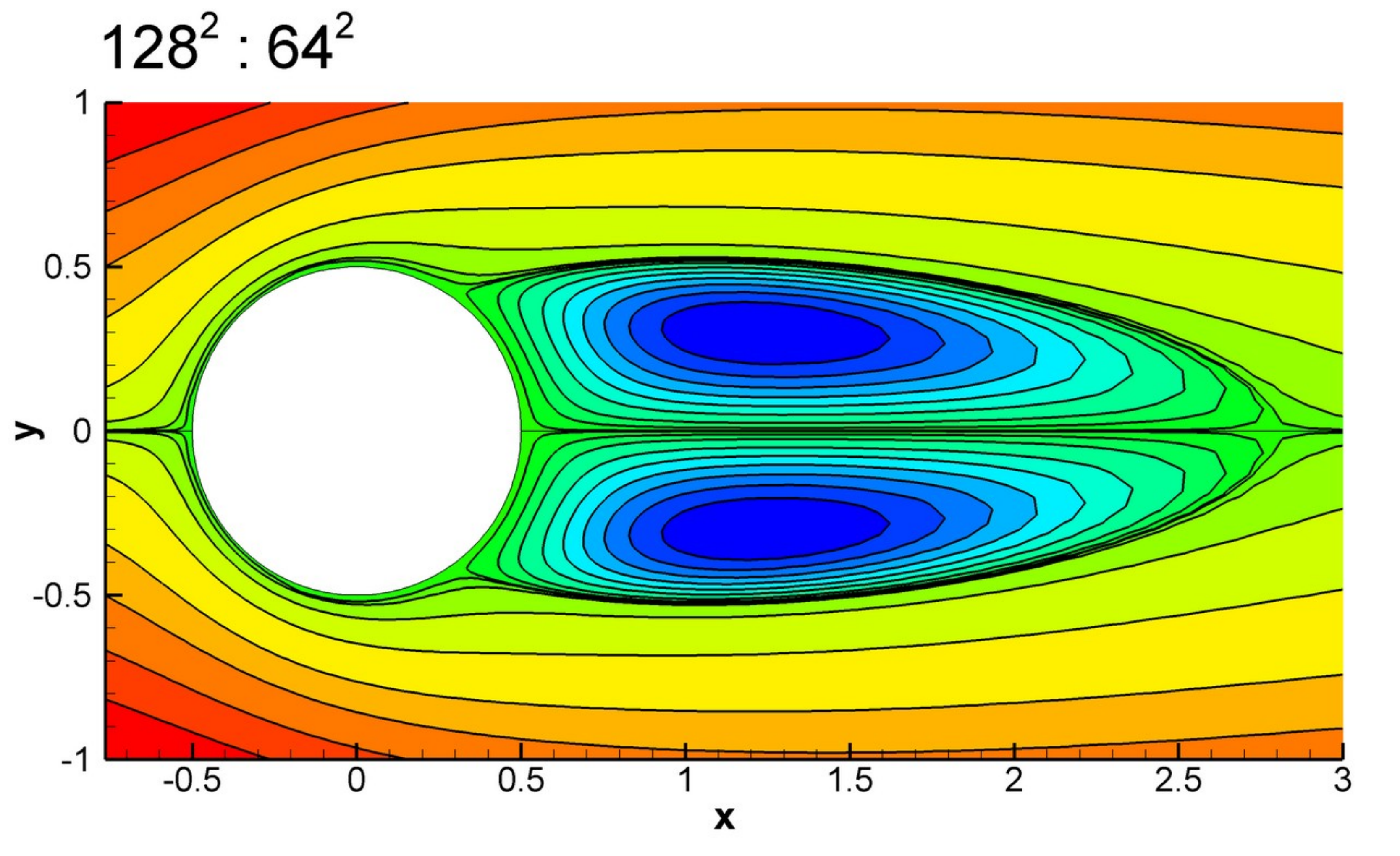}}
}
\mbox{
\subfigure{\includegraphics[width=0.33\textwidth]{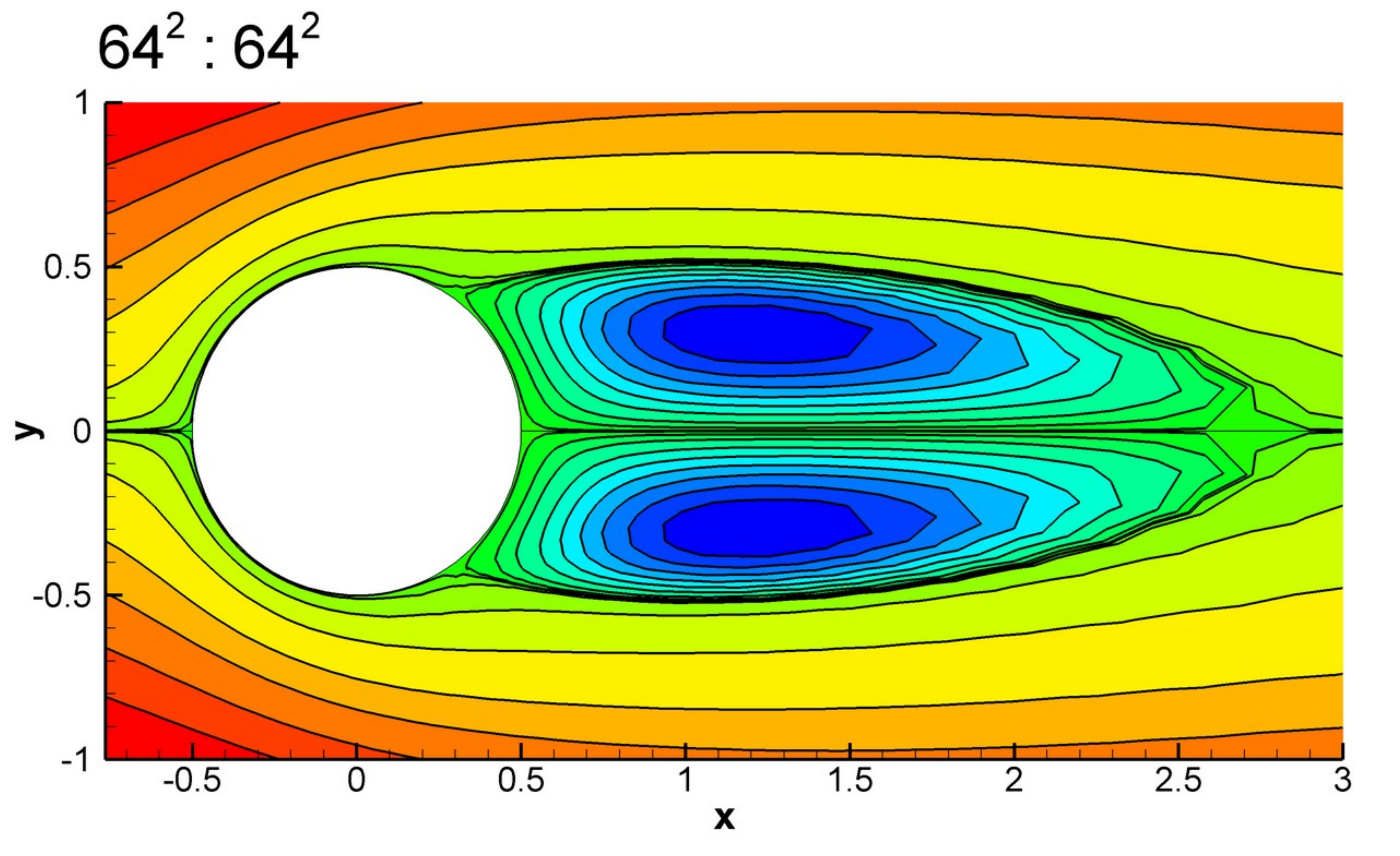}}
\subfigure{\includegraphics[width=0.33\textwidth]{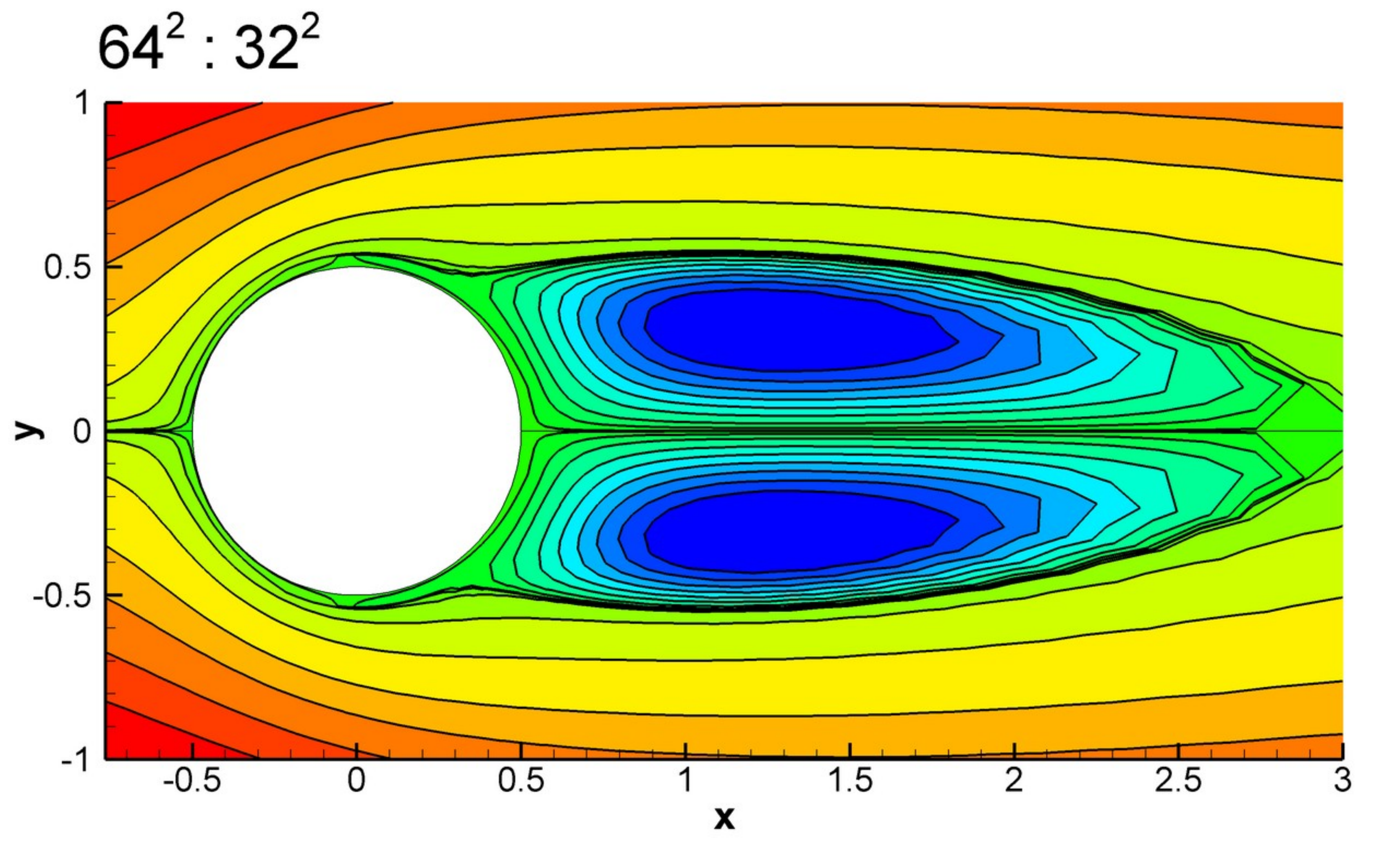}}
\subfigure{\includegraphics[width=0.33\textwidth]{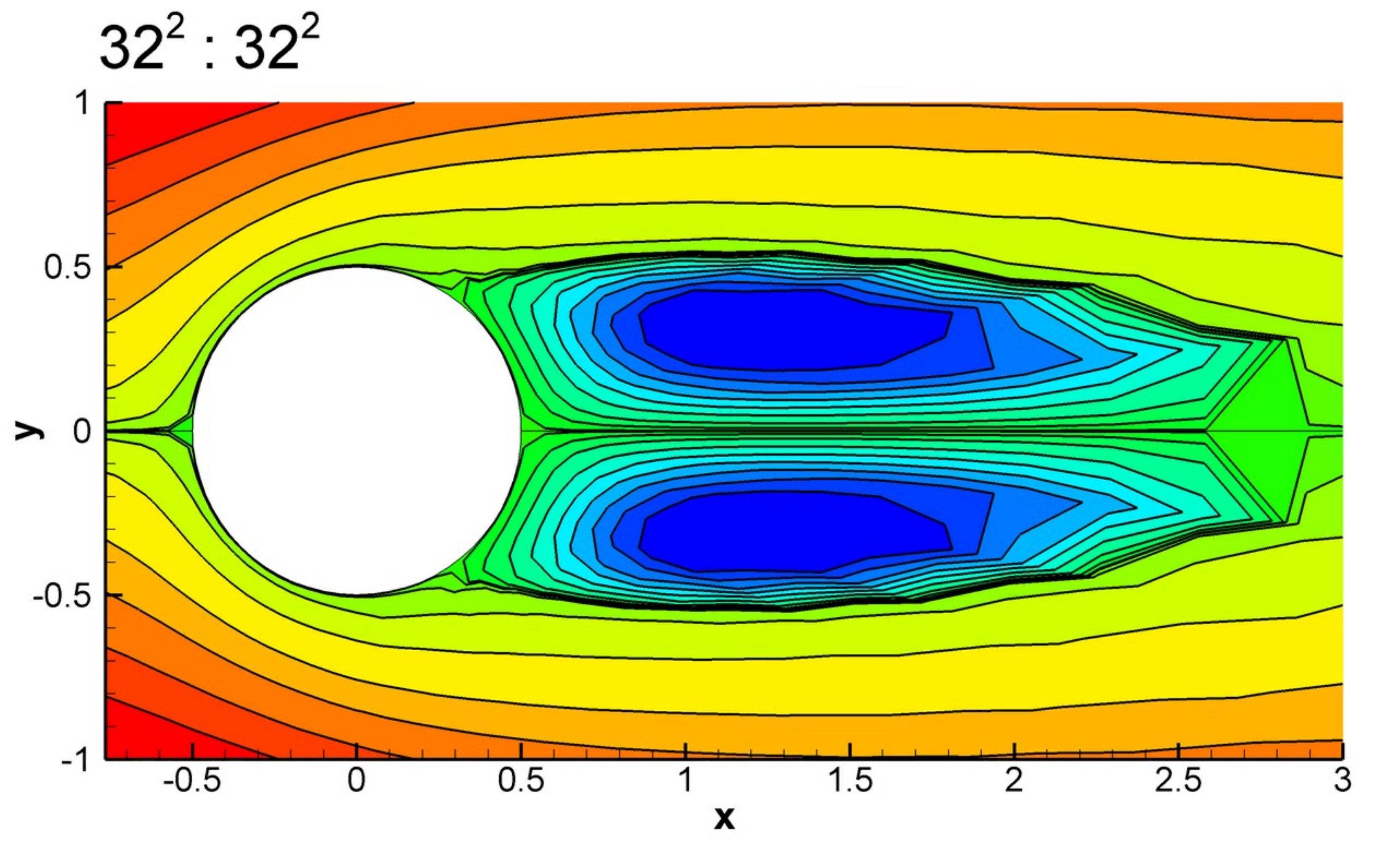}}
}
\caption{Stream function contours for laminar flow over a cylinder at $Re=40$. Labels include the resolutions for both parts of the solver in the form $N^2:M^2$, where $N^2$ is the resolution for the vorticity-transport equation, and $M^2$ is the resolution for the Poisson equation. The contour interval layouts are identical in all cases.}
\label{fig:sf-cyl}
\end{figure*}

\begin{figure*}
\centering
\mbox{
\subfigure{\includegraphics[width=0.33\textwidth]{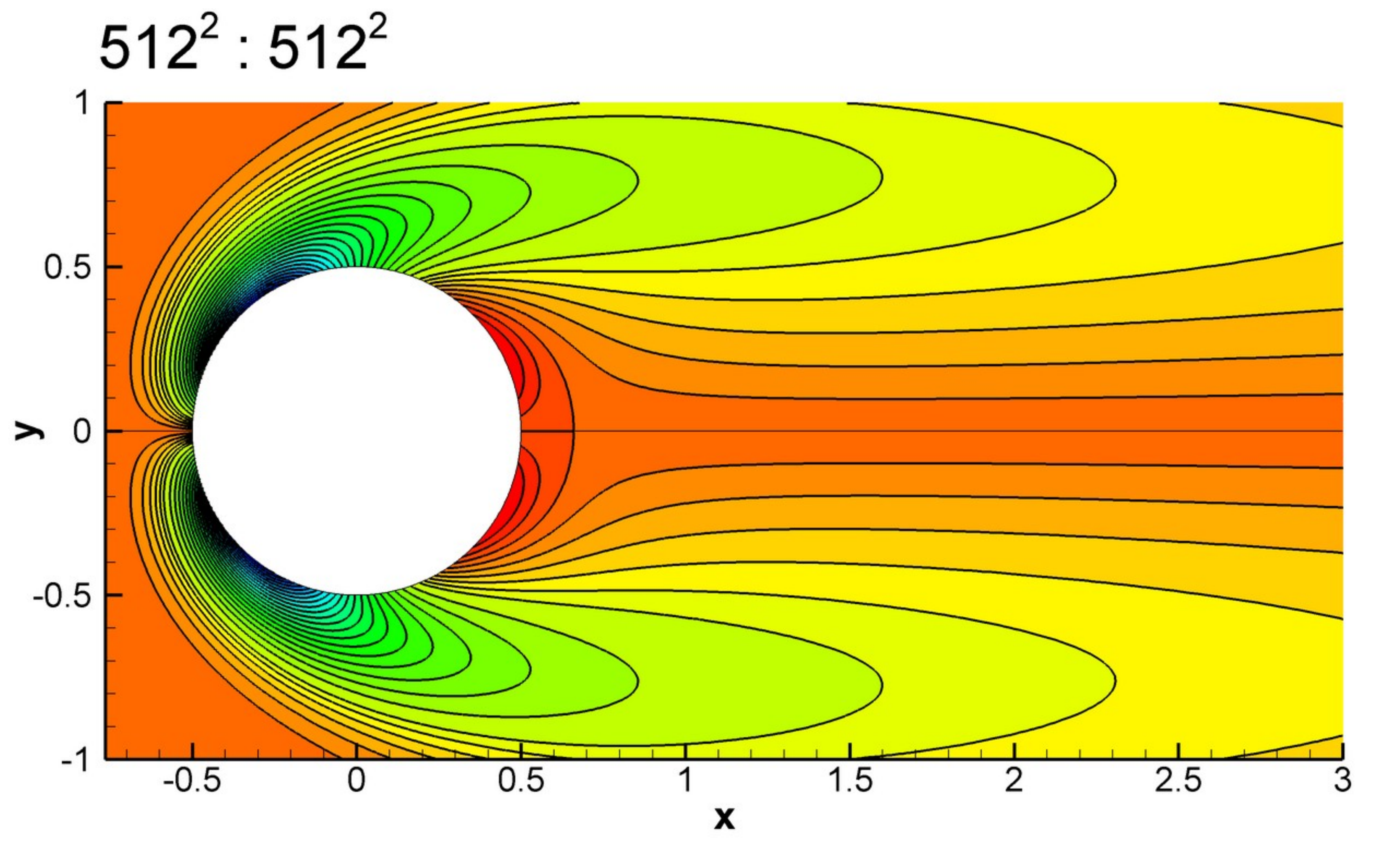}}
\subfigure{\includegraphics[width=0.33\textwidth]{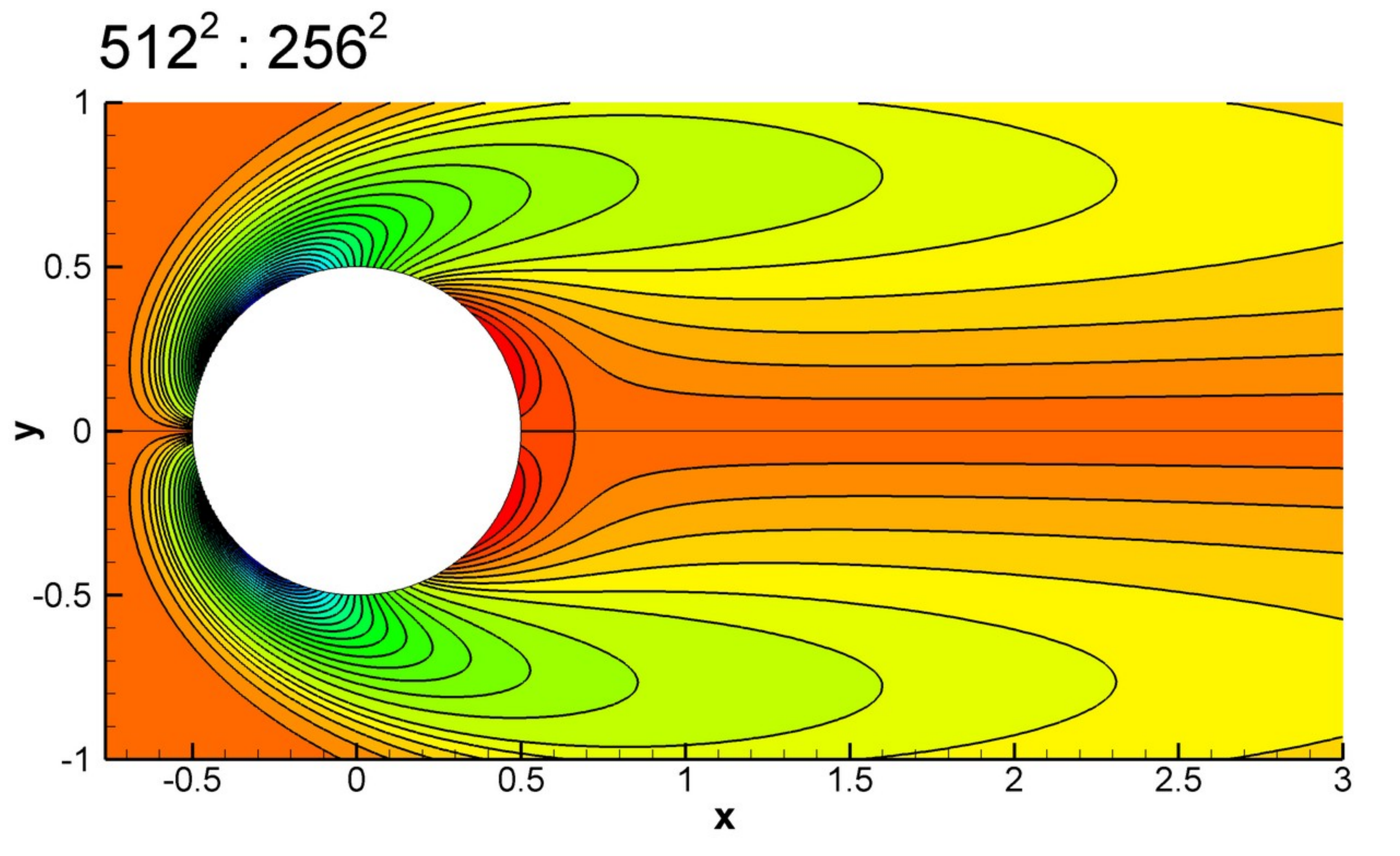}}
\subfigure{\includegraphics[width=0.33\textwidth]{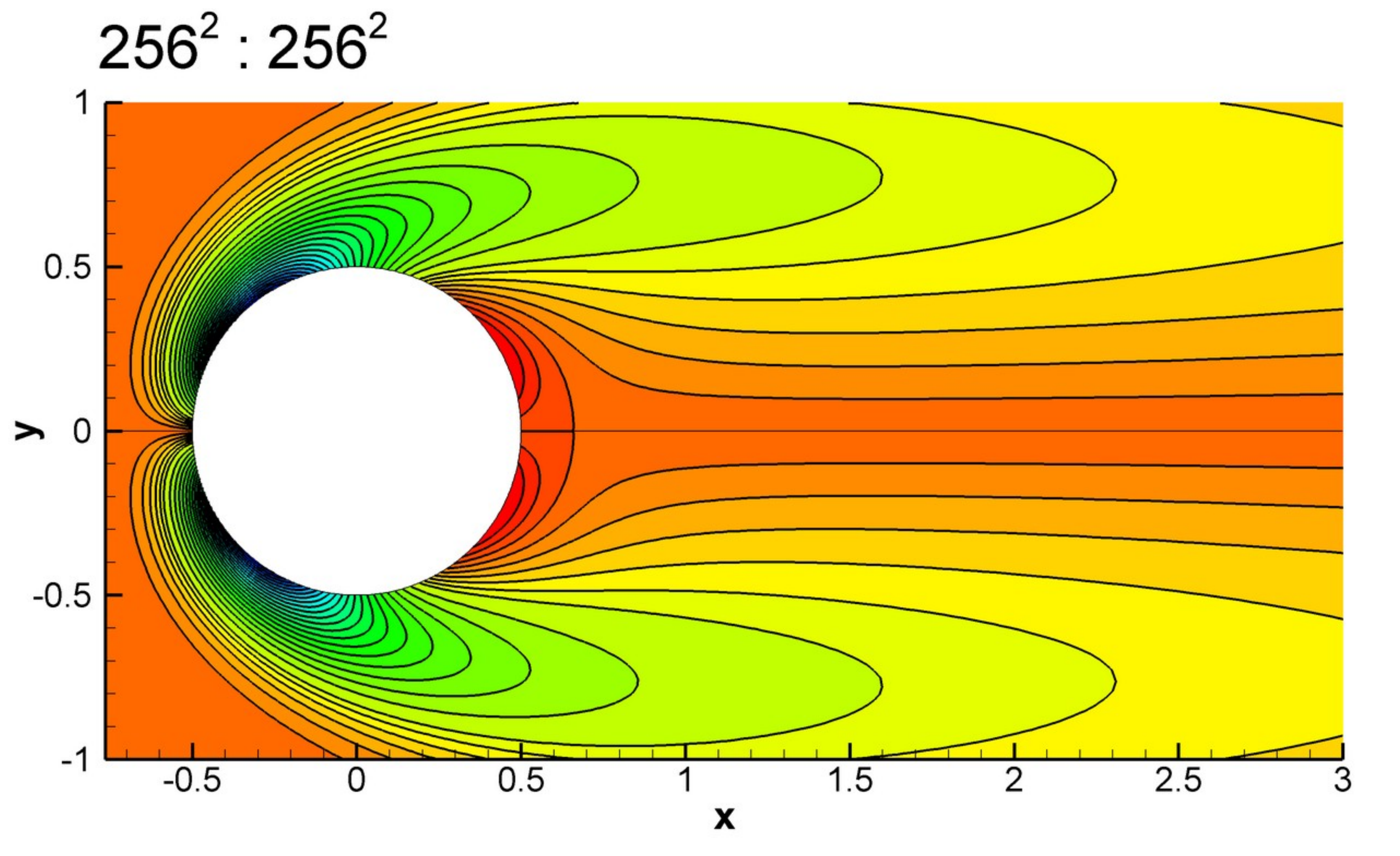}}
}
\mbox{
\subfigure{\includegraphics[width=0.33\textwidth]{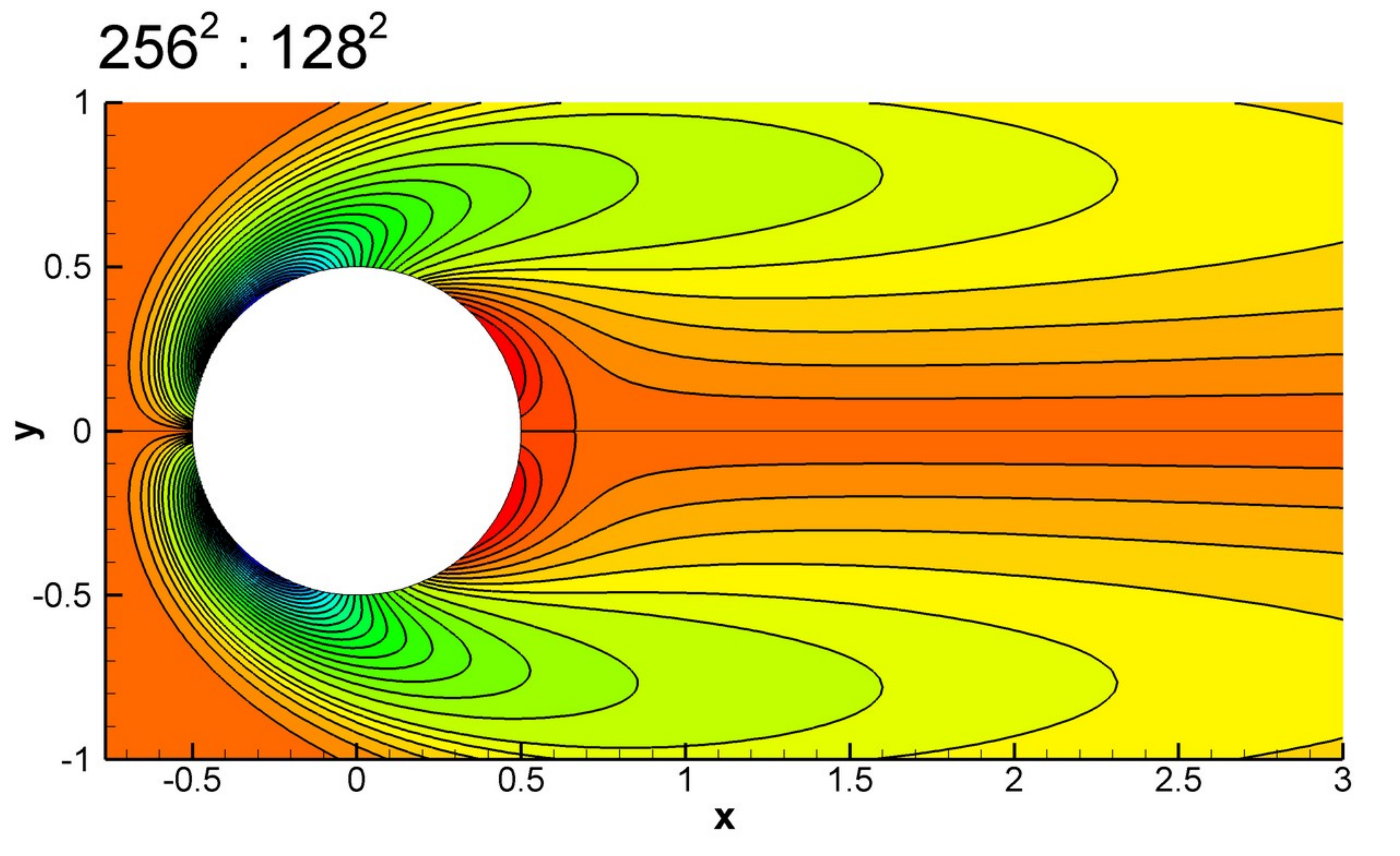}}
\subfigure{\includegraphics[width=0.33\textwidth]{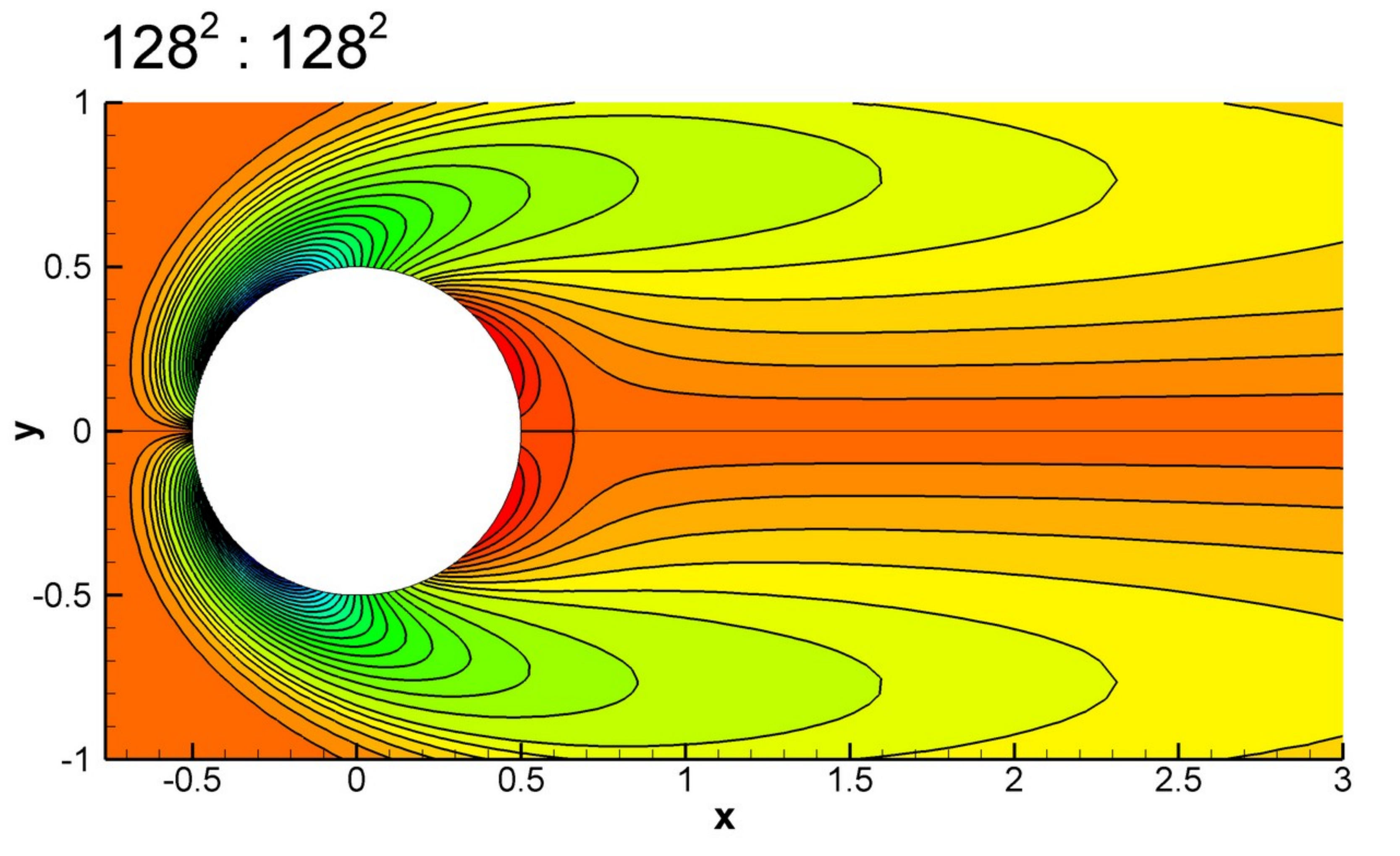}}
\subfigure{\includegraphics[width=0.33\textwidth]{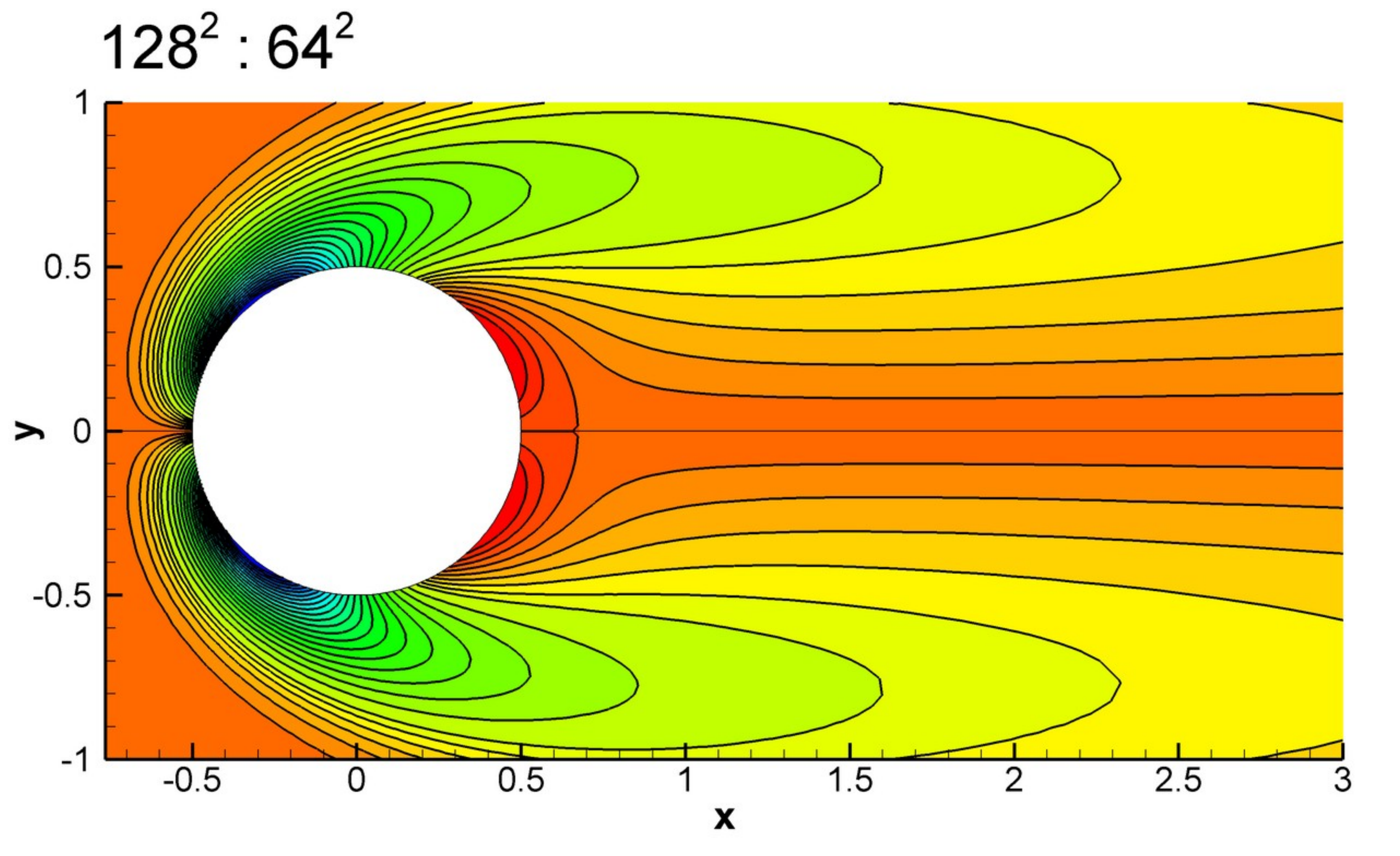}}
}
\mbox{
\subfigure{\includegraphics[width=0.33\textwidth]{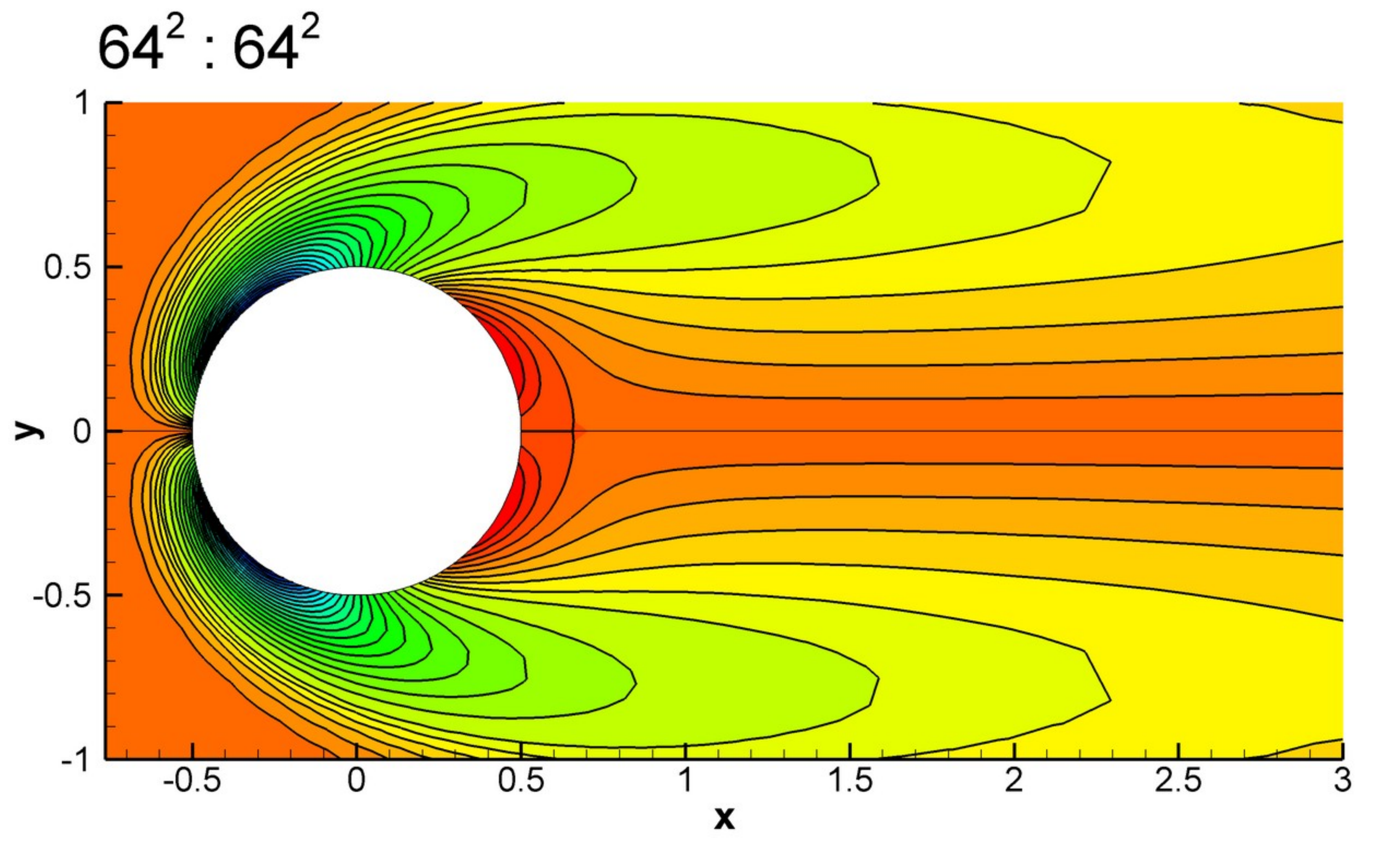}}
\subfigure{\includegraphics[width=0.33\textwidth]{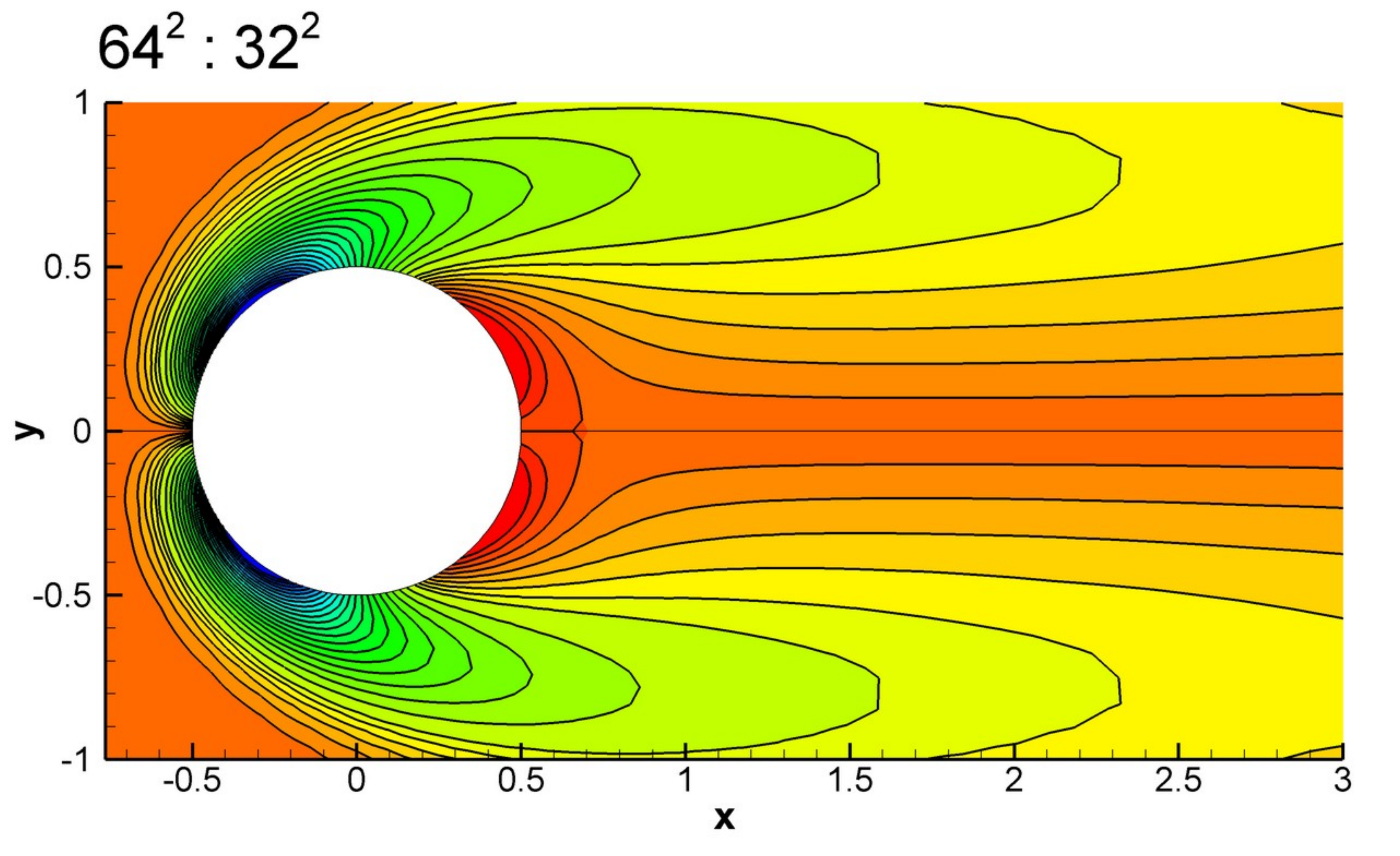}}
\subfigure{\includegraphics[width=0.33\textwidth]{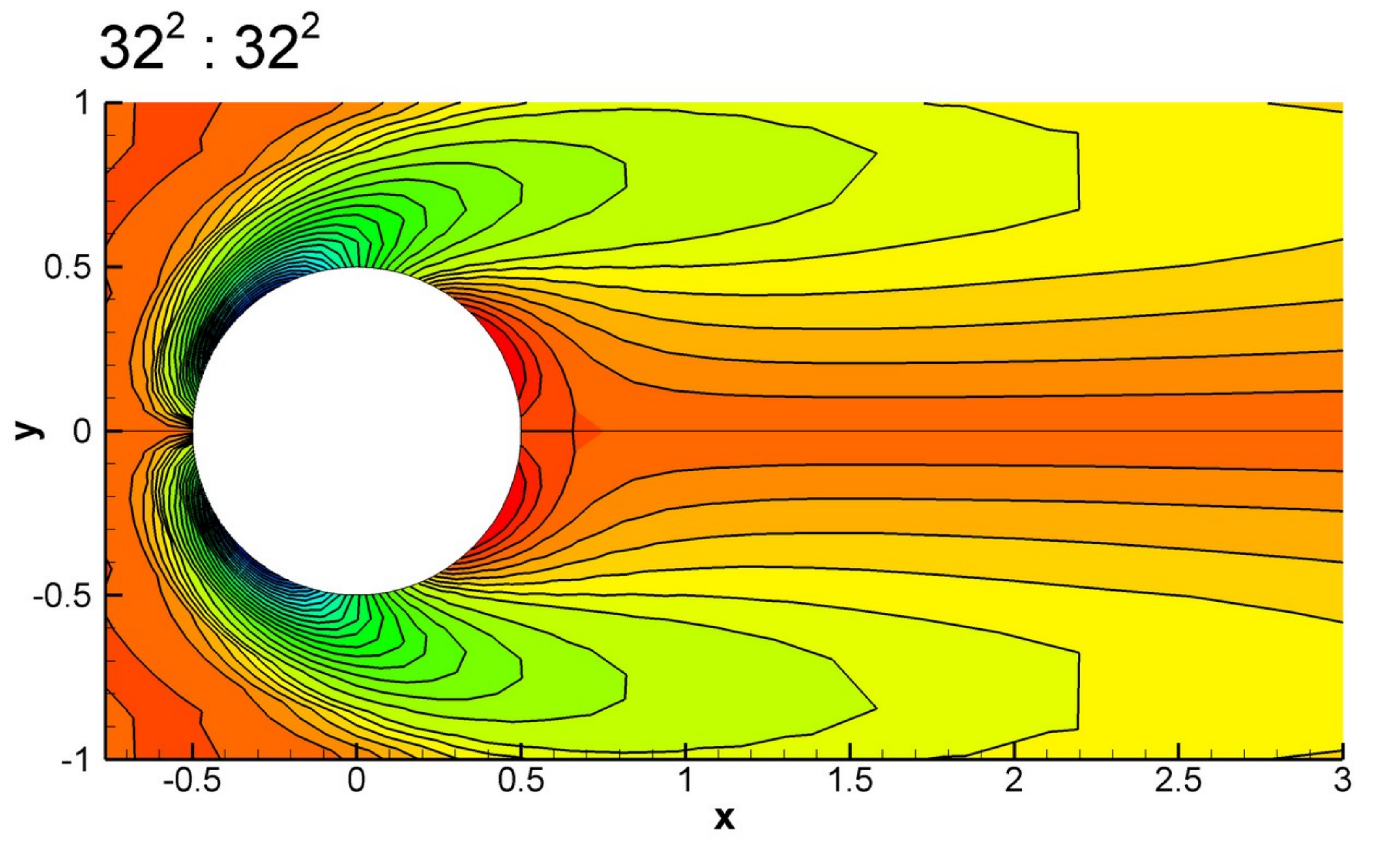}}
}
\caption{Vorticity contours for laminar flow over a cylinder at $Re=40$. Labels include the resolutions for both parts of the solver in the form $N^2:M^2$, where $N^2$ is the resolution for the vorticity-transport equation, and $M^2$ is the resolution for the Poisson equation. The contour interval layouts are identical in all cases.}
\label{fig:vor-cyl}
\end{figure*}

As in the Taylor-Green problem on a distorted grid, transforming the physical space ($x$-$y$) to a Cartesian computational space ($\xi$-$\eta$) allows one to utilize the coarse-grid projection methodology for the flow around a circular cylinder. The full weighting restriction operator and bilinear interpolation are used for data transfer between the vorticity-transport equation and the elliptic equation in transformed coordinates. Here, we utilize a fast sine transform linear-cost elliptic solver (FFTs in the circumfrential direction and the Thomas algorithm in the radial direction \cite{moin2001fundamentals}), and obtain a speed up of about 2, as was obtained in the previous cases for one level of coarsening. As in the previous cases, the speed up would be greater if we used other, suboptimal elliptic solvers. Starting with zero initial conditions, we integrated until the flow field reached stationarity and recirculating eddies were obtained. Stream function and vorticity field contours are shown in Fig.~\ref{fig:sf-cyl} and Fig.~\ref{fig:vor-cyl}, respectively. The various resolutions used in the computations are written in the labels of the subfigures. The labels include the resolution for the advection-diffusion part and the resolution for the elliptic part, as described in the captions. As was the case in previous sections, we compare the CGPRK3 computations with standard RK3 computations. For example, the flow field obtained by a standard computation with $128^2:128^2$ agrees well with that obtained by the CGP method with $128^2:64^2$, and better than that of the standard computation with $64^2:64^2$. Similar observations holds for higher and lower resolution computations as well. It can clearly be seen that the CGP approach provides results at the same level of accuracy, but at a reduced computational cost compared to the standard, fine resolution simulations, by reducing the resolution of the elliptic sub-problem. According to our findings, as shown in Table~\ref{tab:cylinder}, the separation bubble recirculation length is $L=2.2$ and the separation angle is $\theta=53.8$ which agree well with the experimental results of Coutanceau and Bouard \cite{coutanceau1977experimental}, as well as the computational studies presented in \cite{choi2007immersed,dennis1970numerical,galbraith2011implicit,linnick2005high,russell2003cartesian}.

\begin{table}[h]
\small
\caption{Separation angle and separation bubble length for a cylinder at $Re=40$.}
\begin{center}
\label{tab:cylinder}
\begin{tabular}{lcc}
\hline\noalign{\smallskip}
Study &  $\theta^{\circ}$ & $L$ \\
\hline\noalign{\smallskip}
Coutanceau and Bouard \cite{coutanceau1977experimental} & 53.5& 2.13 \\
Choi et al. \cite{choi2007immersed} & 53.6 & 2.21 \\
Dennis and Chang \cite{dennis1970numerical} & 53.8 & 2.35 \\
Galbraith and Abdallah \cite{galbraith2011implicit} & 54.0 & 2.23 \\
Linnick and Fasel \cite{linnick2005high} & 53.6 & 2.28 \\
Russell and Wang \cite{russell2003cartesian} & 53.1 & 2.29 \\
CGPRK3 & 53.8& 2.20 \\
\hline
\end{tabular}
\end{center}
\end{table}

\subsection{Three-dimensional Taylor-Green vortex}
Here, the coarse-grid projection method is applied to the three-dimensional Taylor-Green vortex flow problem \citep{taylor1937mechanism,brachet1983small}, which is perhaps the simplest system in which to study the generation of increasingly smaller scale motions and the resulting turbulence. The fundamental mechanism involved in homogeneous three-dimensional turbulent flows is the enhancement of vorticity by vortex stretching and consequent production of small eddies. Energy is transferred forward in spectral space, from large scales to smaller scales, via the vortex stretching mechanism. This process controls the turbulent energy dynamics and hence the global structure of the evolution of the turbulent flow. A prototype of this process is given by the generalized Taylor-Green vortex problem \cite{brachet1983small}, which is a three-dimensional flow that develops from the single mode initial condition:
\begin{eqnarray}
u(x,y,z,t=0)&=&\frac{2}{\sqrt{3}} \mbox{sin}\left(\theta+\frac{2 \pi}{3}\right) \mbox{sin}(x)\mbox{cos}(y)\mbox{cos}(z)\\
v(x,y,z,t=0)&=&\frac{2}{\sqrt{3}} \mbox{sin}\left(\theta-\frac{2 \pi}{3}\right) \mbox{cos}(x)\mbox{sin}(y)\mbox{cos}(z)\\
w(x,y,z,t=0)&=&\frac{2}{\sqrt{3}} \mbox{sin}\left(\theta \right) \mbox{cos}(x)\mbox{cos}(y)\mbox{sin}(z)
\label{eq:tgv3}
\end{eqnarray}
where $u$, $v$, and $w$ are the Cartesian velocity components. The computational domain used is a cubic box whose edge has a length of $2\pi$. Periodic boundary conditions are used in all directions and we set $\theta=0$. In this case, the initial flow has two-dimensional streamlines, but the flow is three-dimensional for all $t>0$. We performed numerical simulations for $Re=200$ using the vorticity-velocity formulation of the governing equations given in Section \ref{sec:vv} with $\Delta t=4\times10^{-3}$.

\begin{table*}
\small
\caption{Comparison of methods for the three-dimensional Taylor-Green vortex problem for $Re=200$ at $t = 10$. Percent errors are computed with respect to the values computed at the finest resolution.}
\begin{center}
\label{tab:tgv3}
\begin{tabular}{llcccc}
\hline\noalign{\smallskip}
Method & Resolutions &  $ |\omega_x|_{max}$ & \% Error & CPU (hr) & Speed-up \\
\hline\noalign{\smallskip}
RK3 ($\ell=0$)     &$256^3$ : $256^3$ &  4.6039 & -     & 206.84 & 1.00 \\
CGPRK3 ($\ell=1$)  &$256^3$ : $128^3$ &  4.8667 & 5.70  & 37.91  & 5.46 \\
CGPRK3 ($\ell=2$)  &$256^3$ : $64^3$  &  5.1513 & 11.89 & 26.41  & 7.83 \\
RK3 ($\ell=0$)     &$128^3$ : $128^3$ &  5.2037 & 13.03 & 21.29  & 1.00 \\
CGPRK3 ($\ell=1$)  &$128^3$ : $64^3$  &  5.6668 & 23.08 & 4.61   & 4.62 \\
RK3 ($\ell=0$)     &$64^3$ : $64^3$   &  8.5696 & 86.14 & 1.56   & 1.00 \\
\hline
\end{tabular}
\end{center}
\end{table*}

\begin{figure*}
\centering
\mbox{
\subfigure{\includegraphics[width=0.33\textwidth]{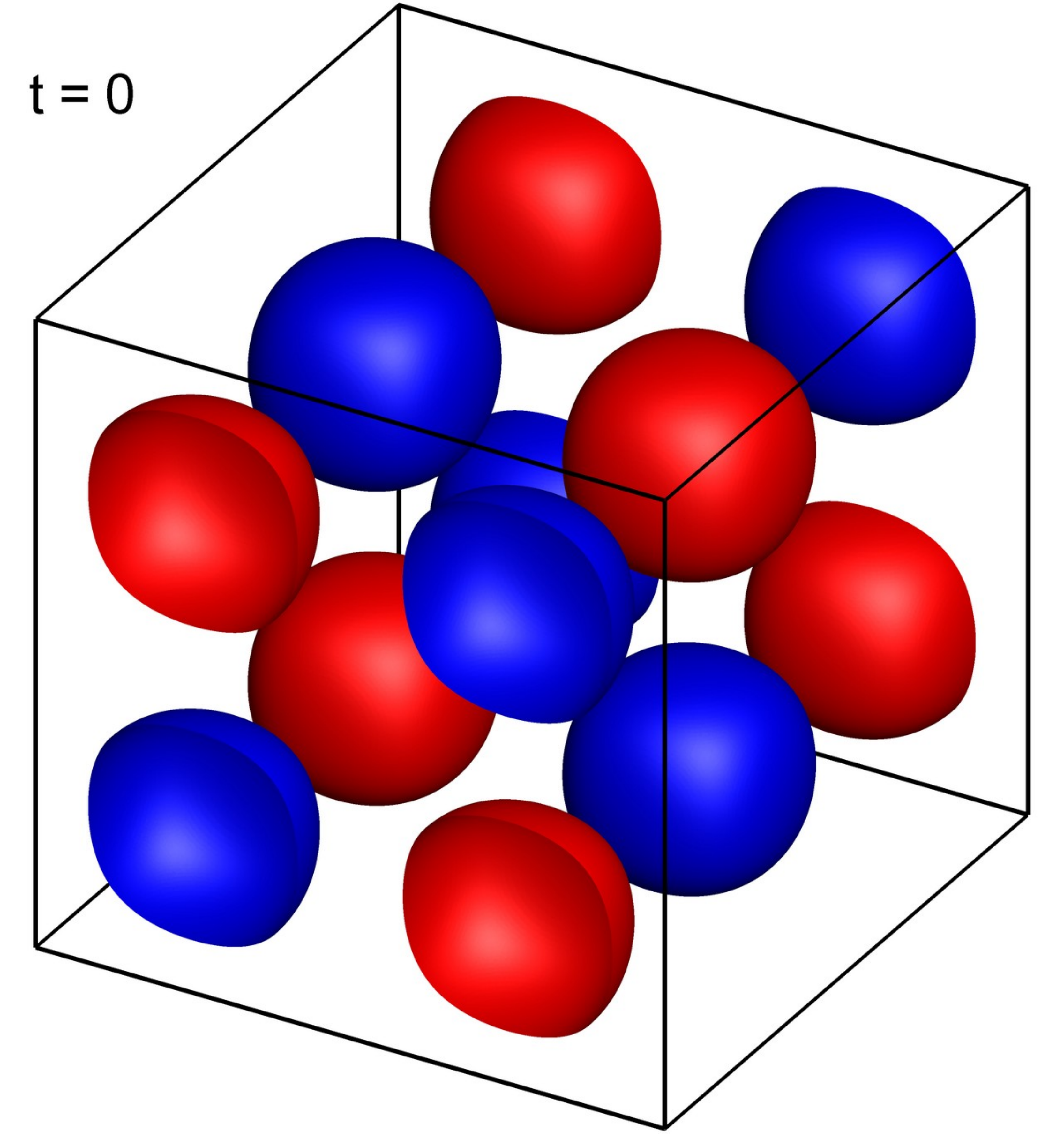}}
\subfigure{\includegraphics[width=0.33\textwidth]{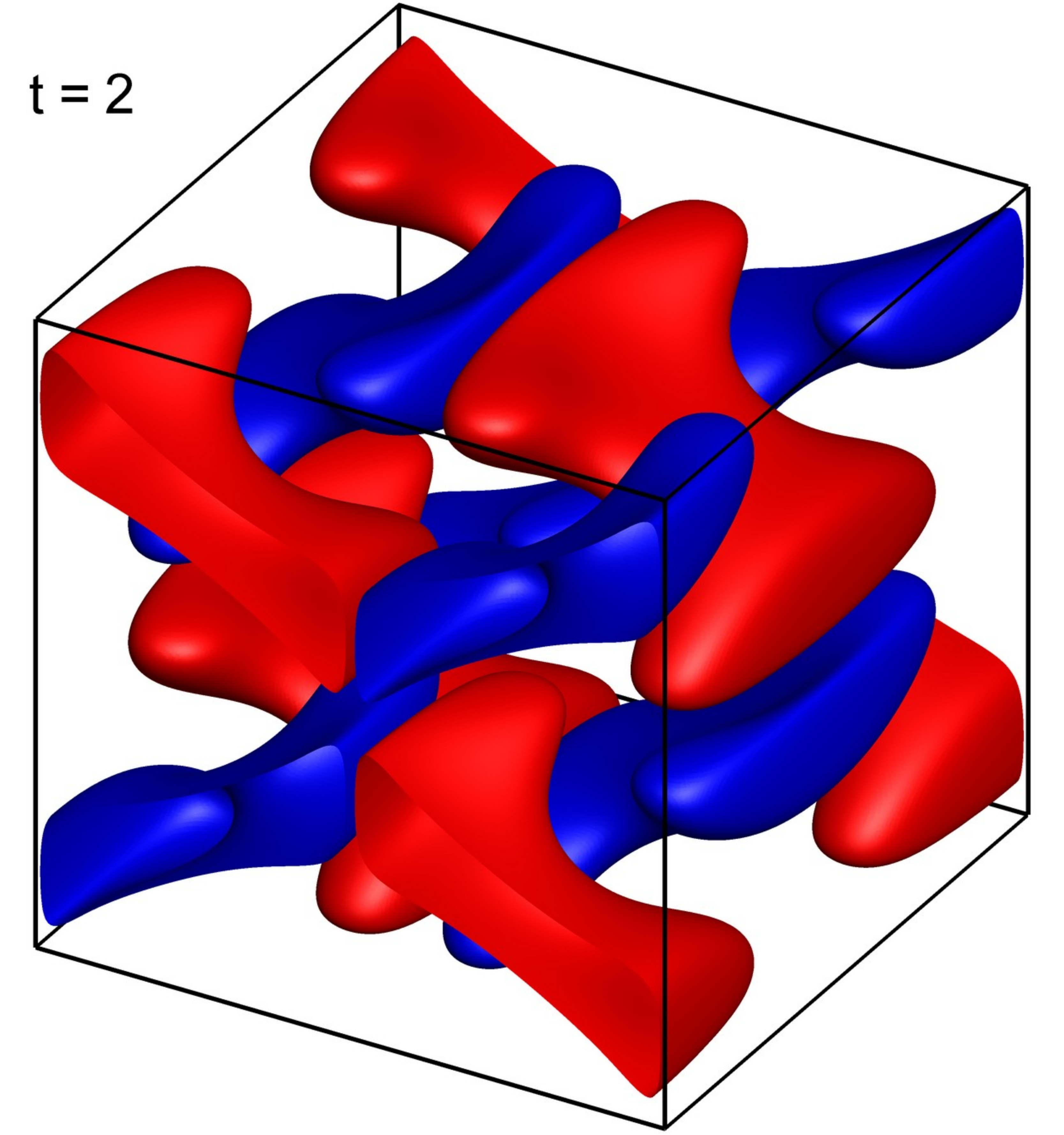}}
\subfigure{\includegraphics[width=0.33\textwidth]{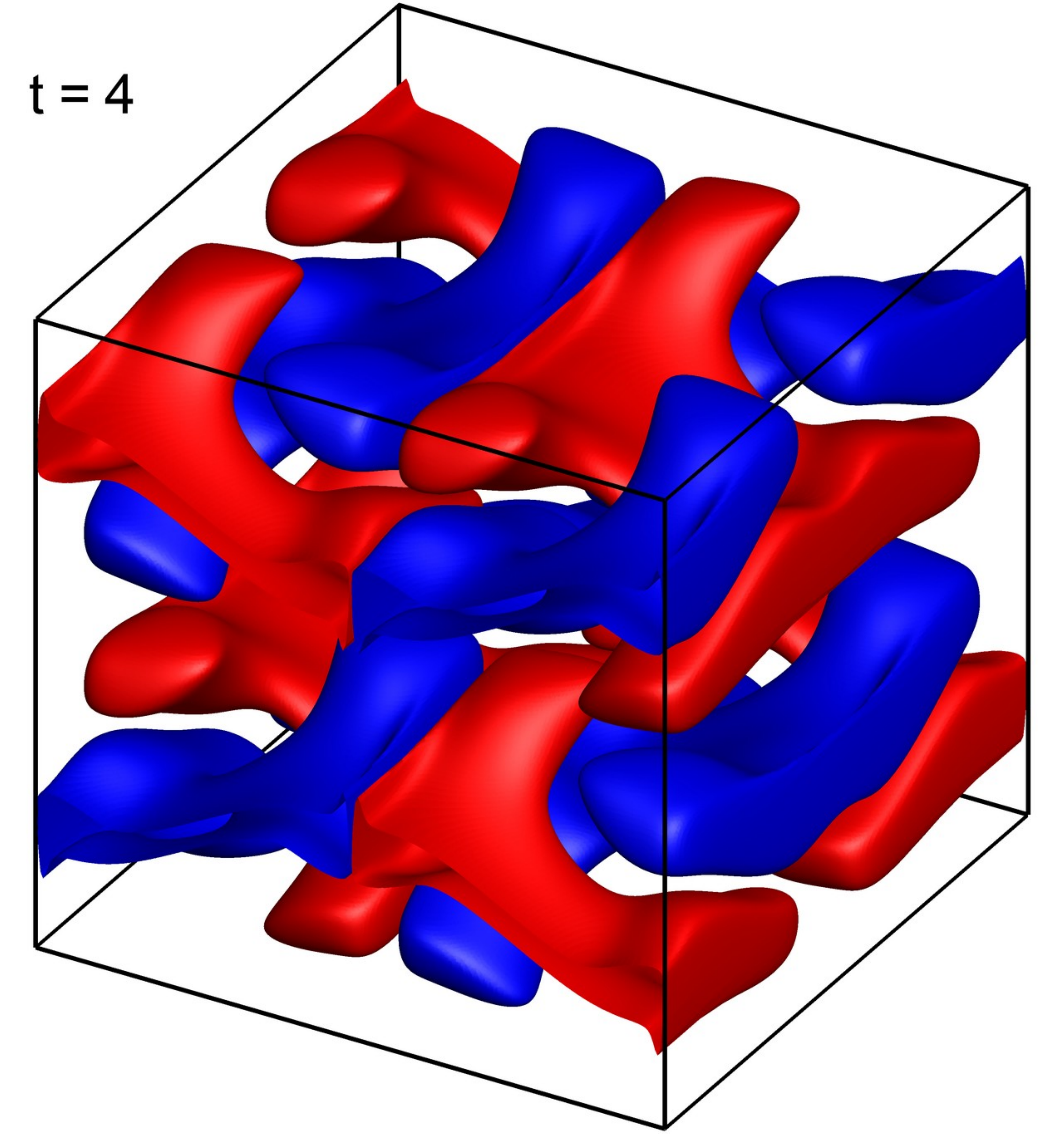}}
}
\mbox{
\subfigure{\includegraphics[width=0.33\textwidth]{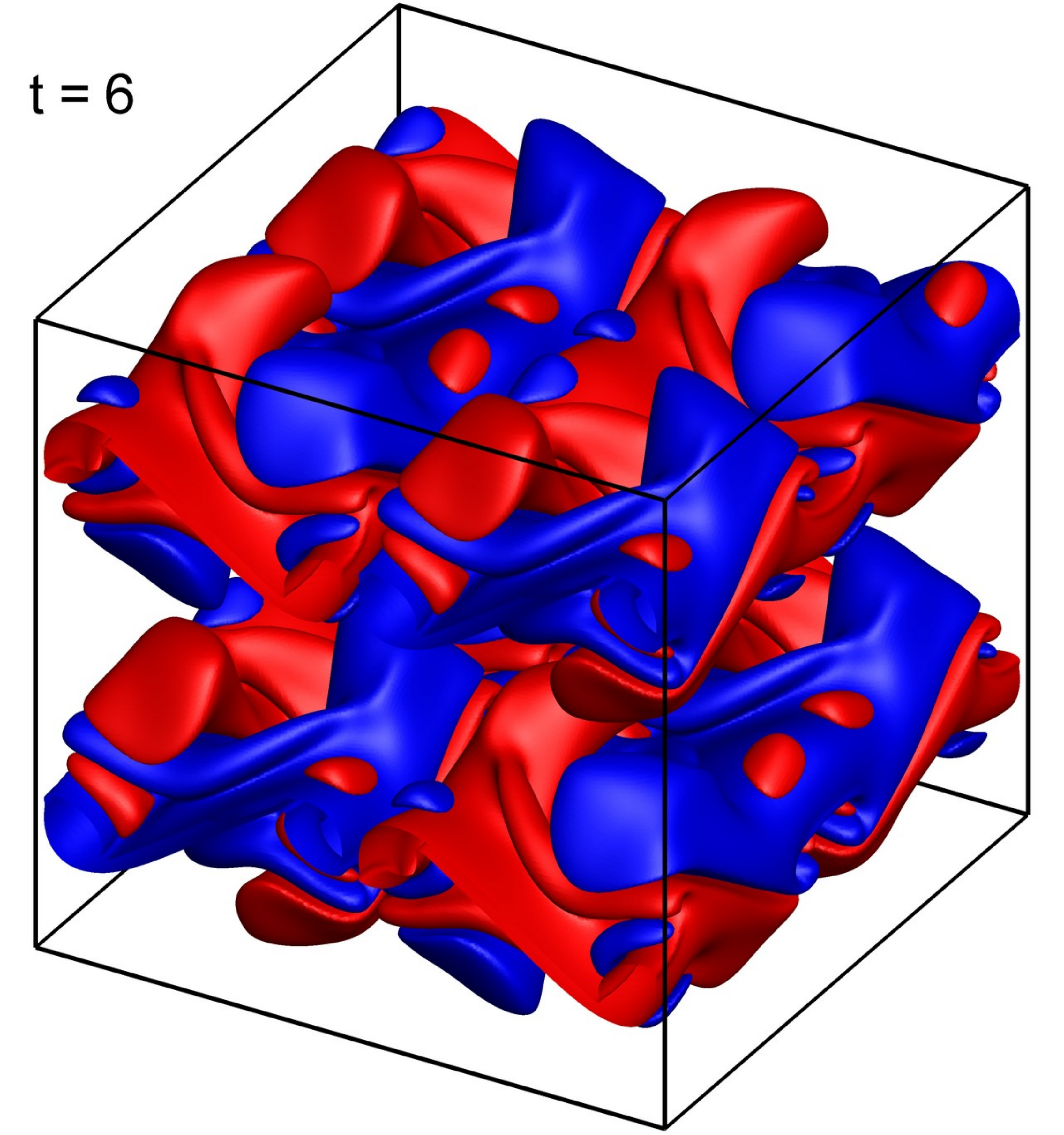}}
\subfigure{\includegraphics[width=0.33\textwidth]{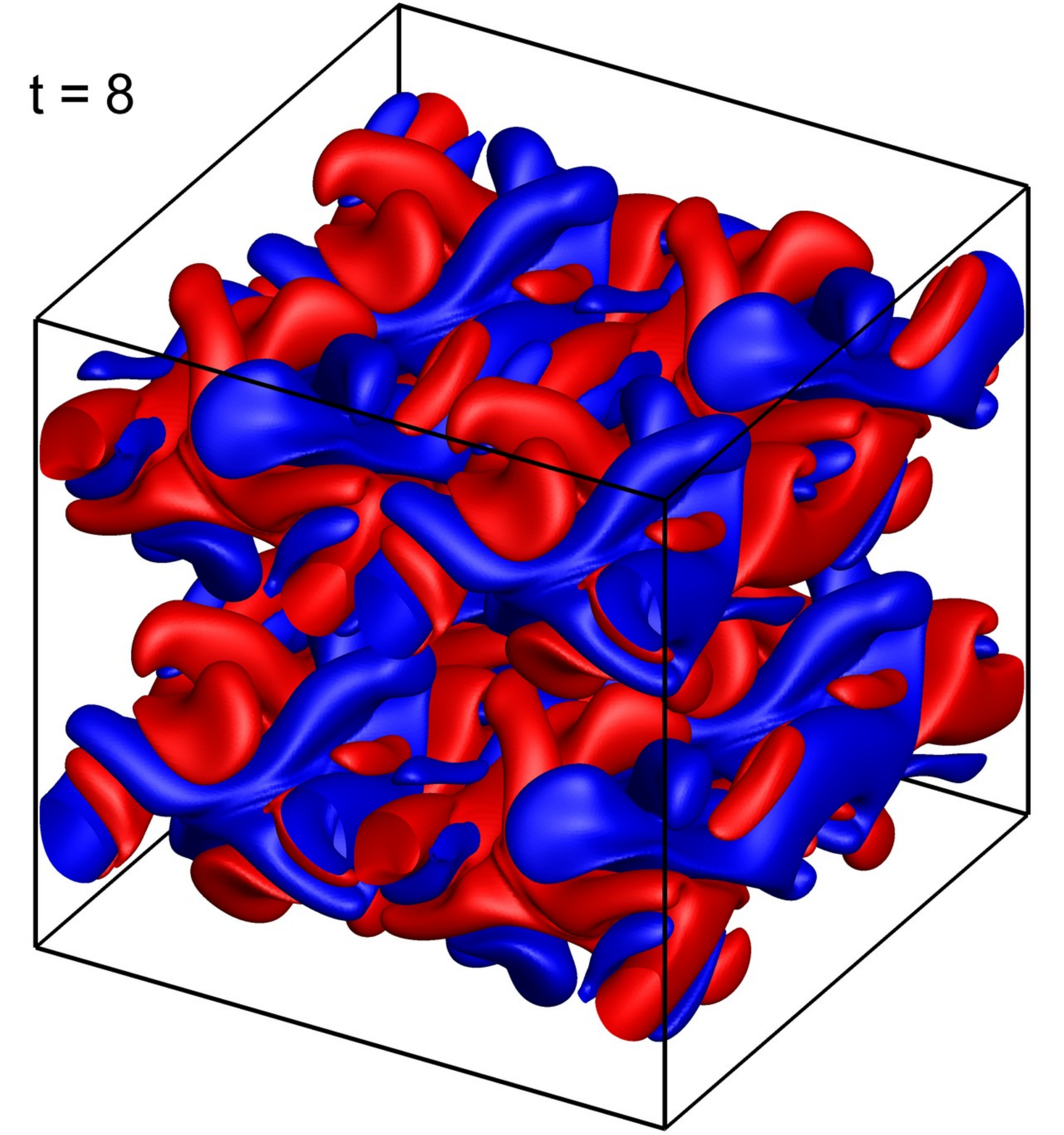}}
\subfigure{\includegraphics[width=0.33\textwidth]{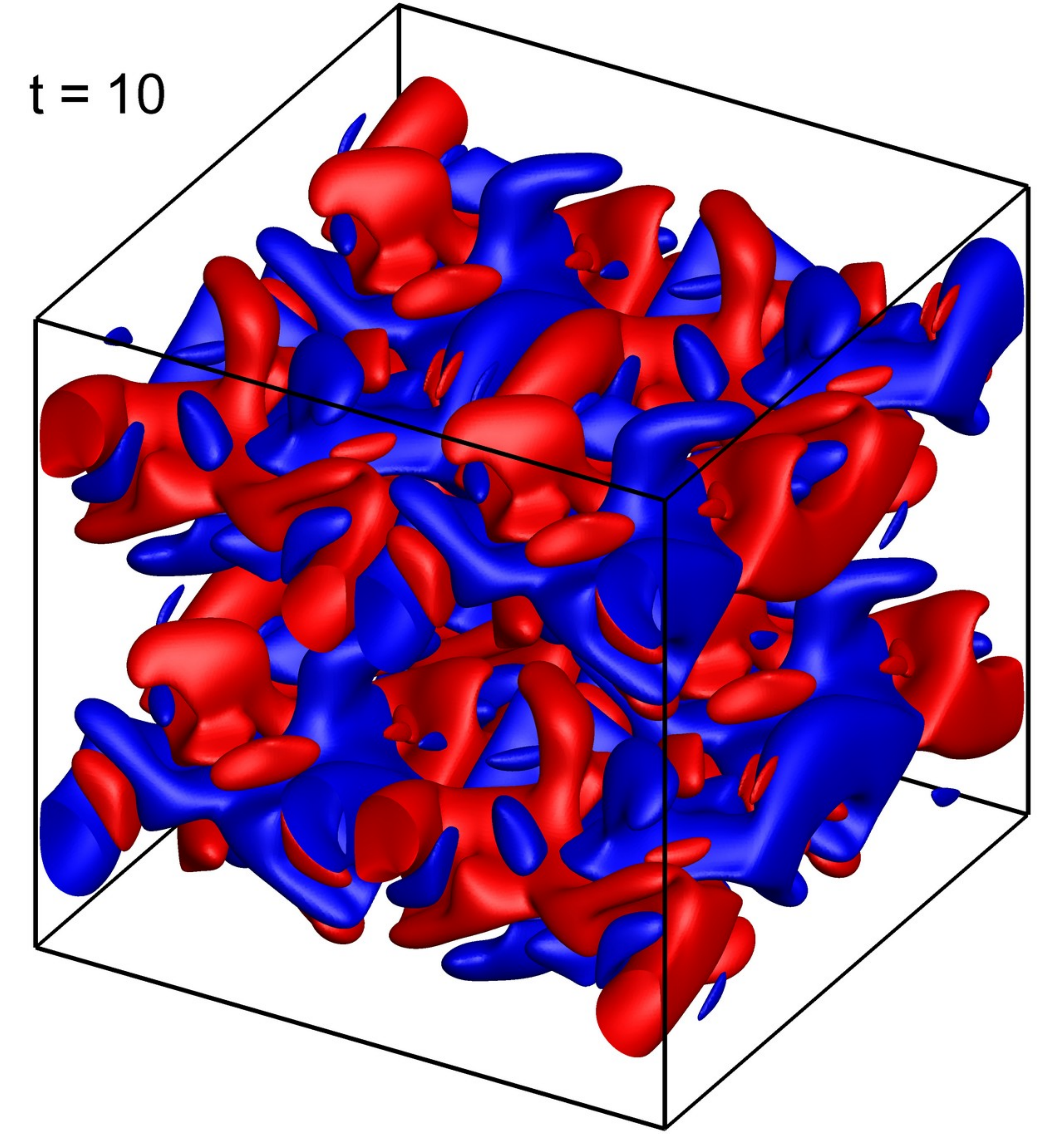}}
}
\caption{Evolution of the $x$-component of the vorticity obtained using the CGP method on a $256^3:128^3$ resolution grid for $Re=200$. Iso-surfaces of $\omega_x=-0.5$ (blue) and  $\omega_x=0.5$ (red) are shown.}
\label{fig:time-tgv3}
\end{figure*}

\begin{figure*}
\centering
\mbox{
\subfigure{\includegraphics[width=0.33\textwidth]{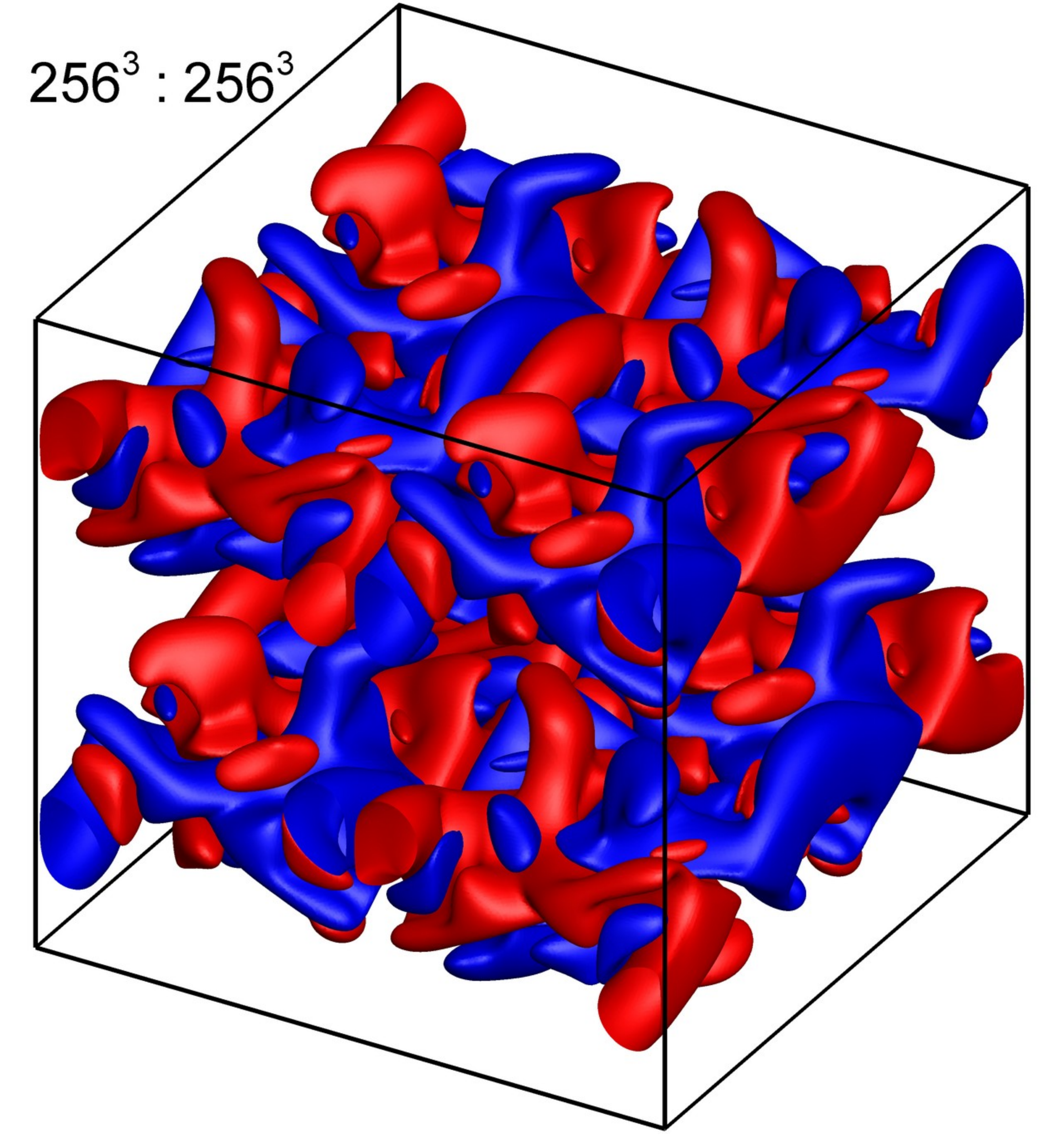}}
\subfigure{\includegraphics[width=0.33\textwidth]{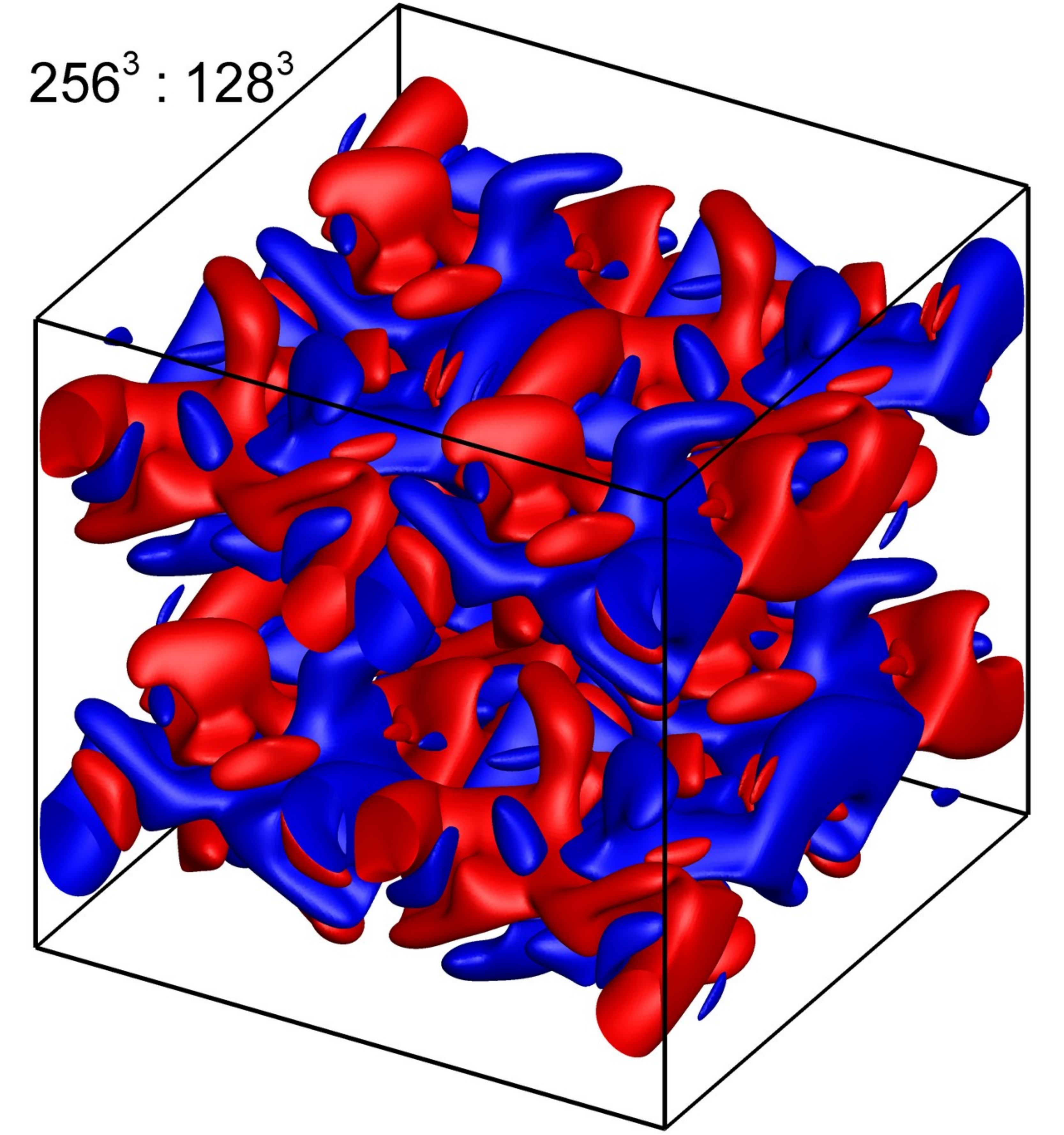}}
\subfigure{\includegraphics[width=0.33\textwidth]{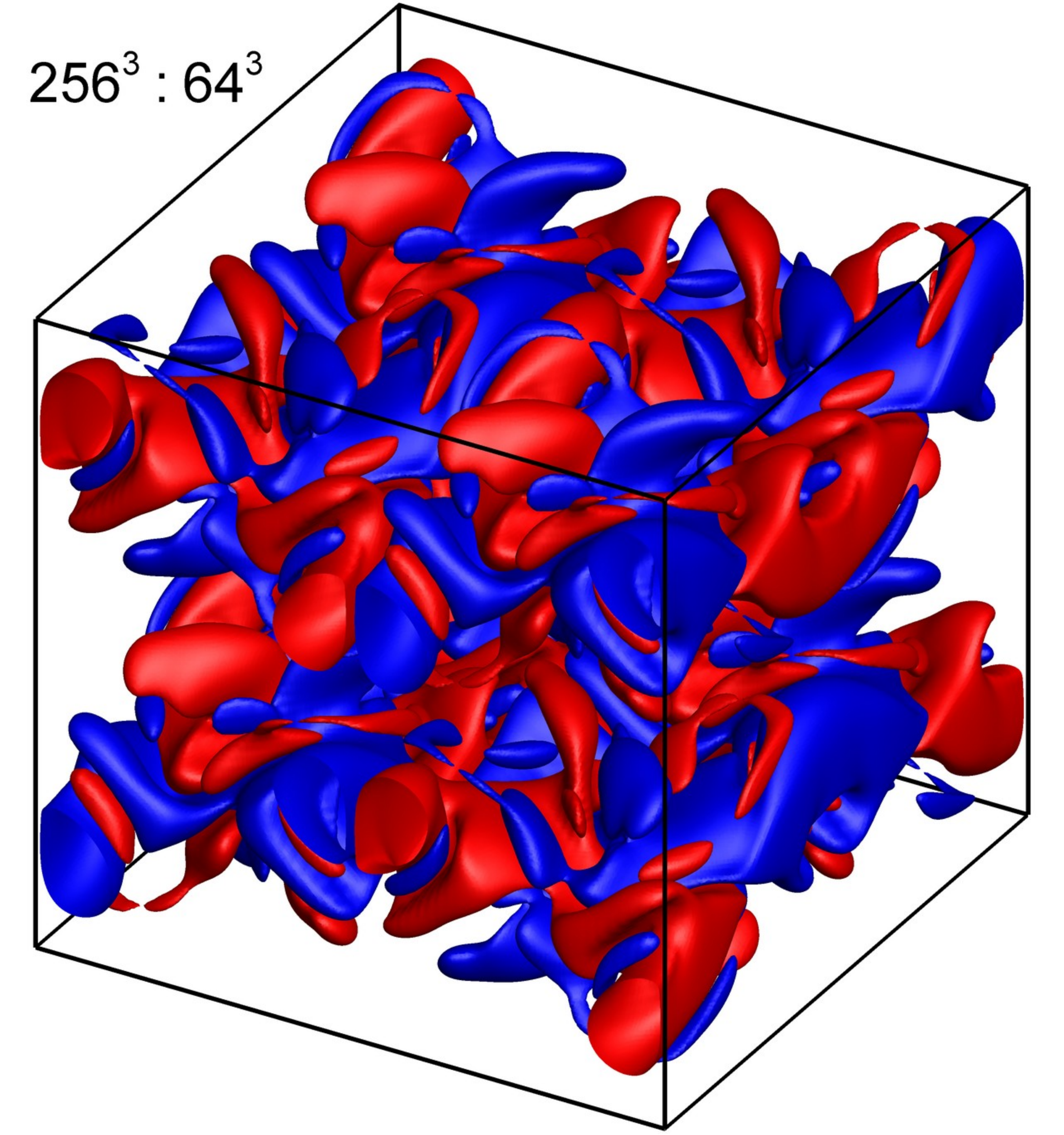}}
}
\\
\mbox{
\subfigure{\includegraphics[width=0.33\textwidth]{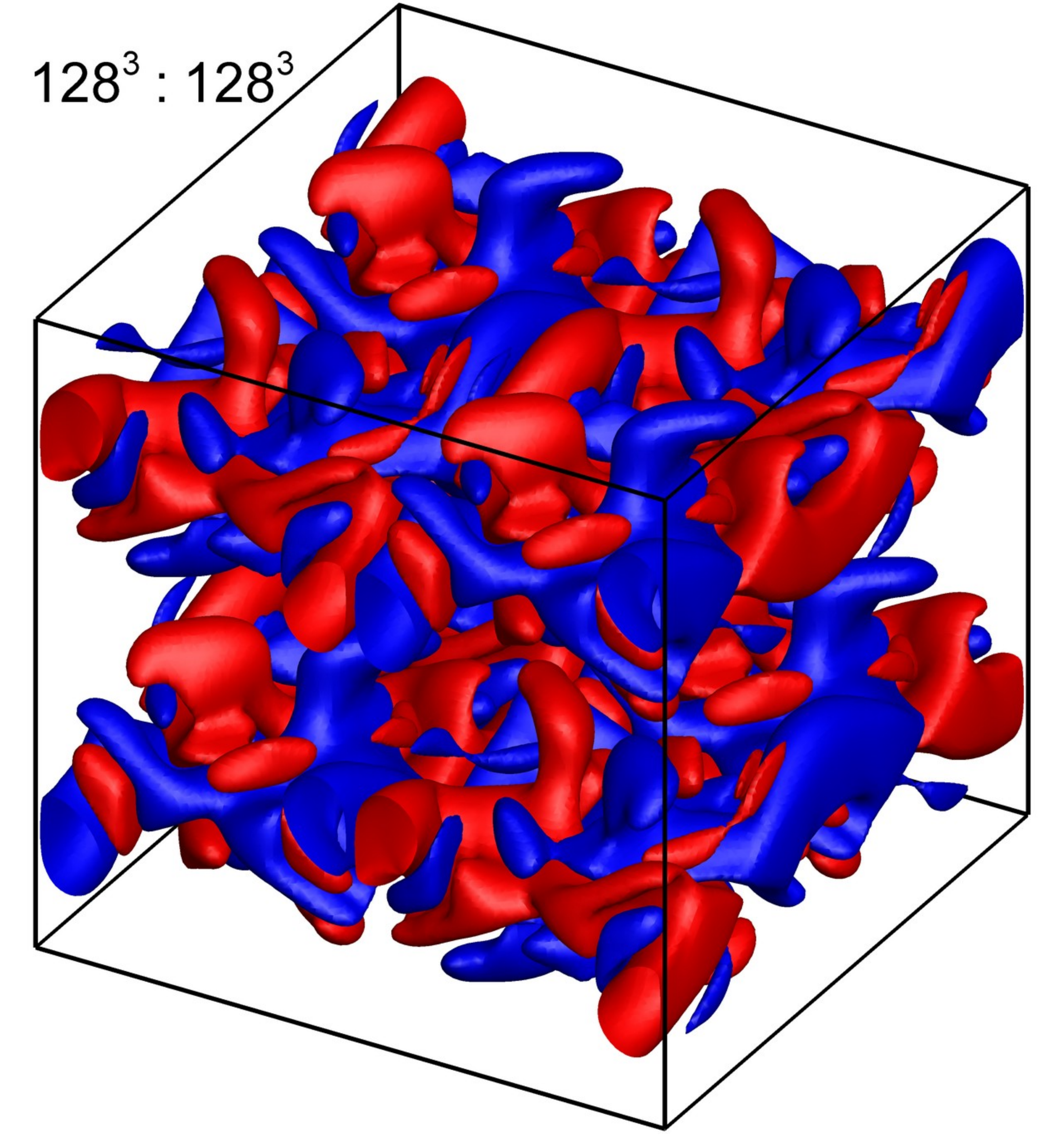}}
\subfigure{\includegraphics[width=0.33\textwidth]{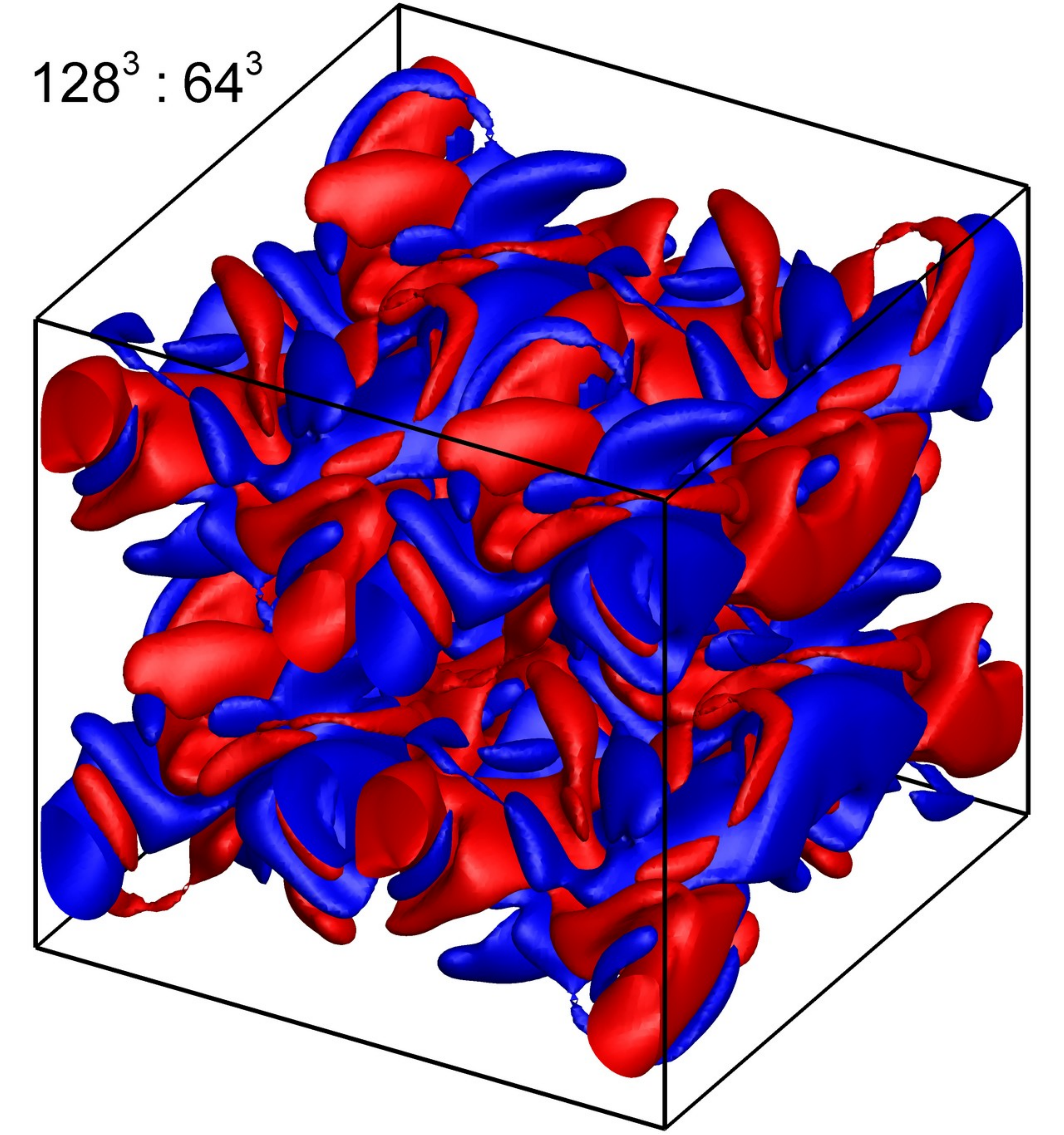}}
\subfigure{\includegraphics[width=0.33\textwidth]{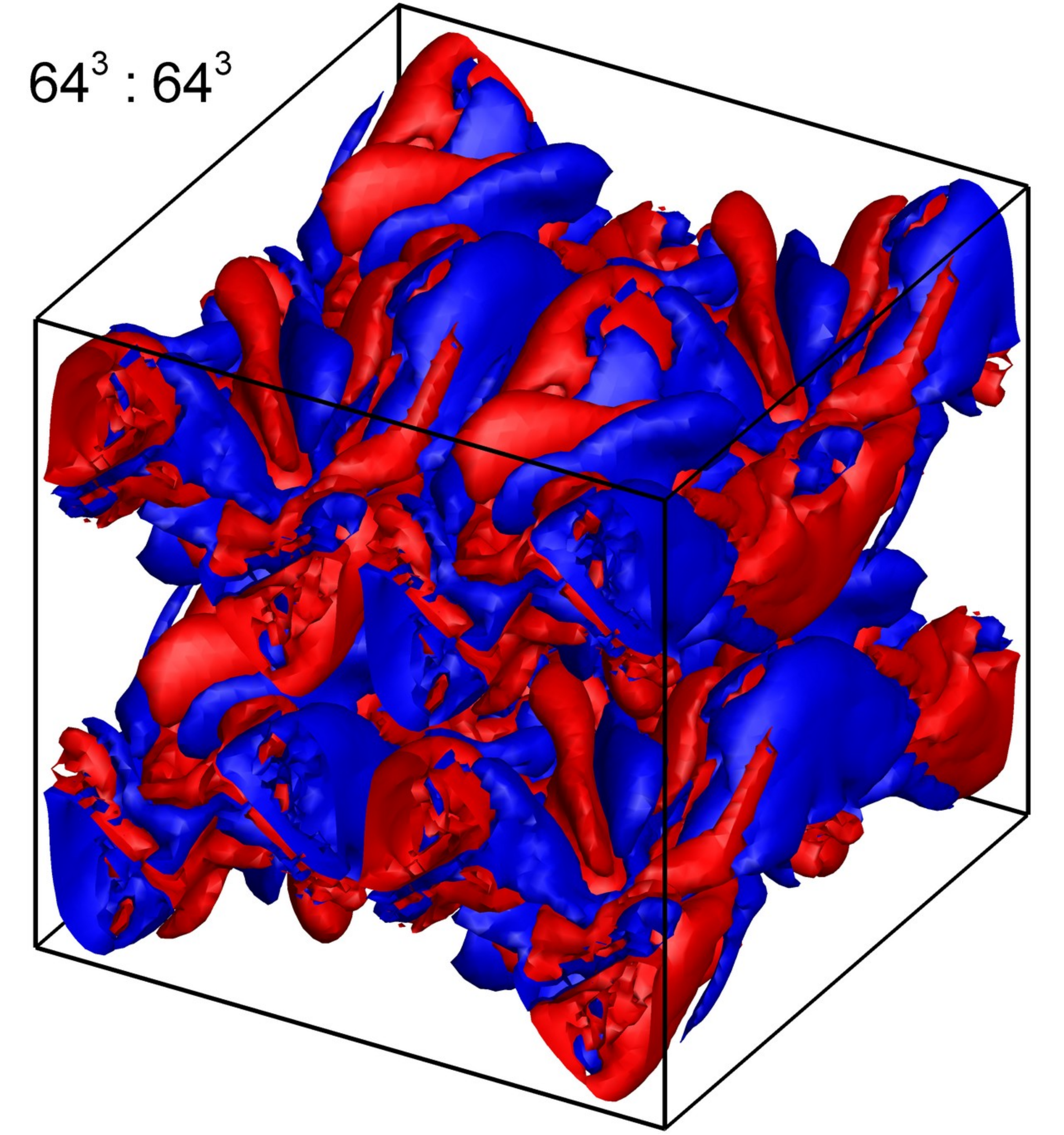}}
}
\caption{Comparison of the $x$-component of the vorticity at $t=10$ for $Re=200$. Iso-surfaces of $\omega_x =-0.5$ (blue) and  $\omega_x = 0.5$ (red) are shown. Labels include the resolutions for both parts of the solver in the form $N^3 : M^3$, where $N^3$ is the resolution for the vorticity-transport equations, and $M^3$ is the resolution for the elliptic sub-problems.}
\label{fig:wx-05}
\end{figure*}

\begin{figure*}
\centering
\mbox{
\subfigure{\includegraphics[width=0.33\textwidth]{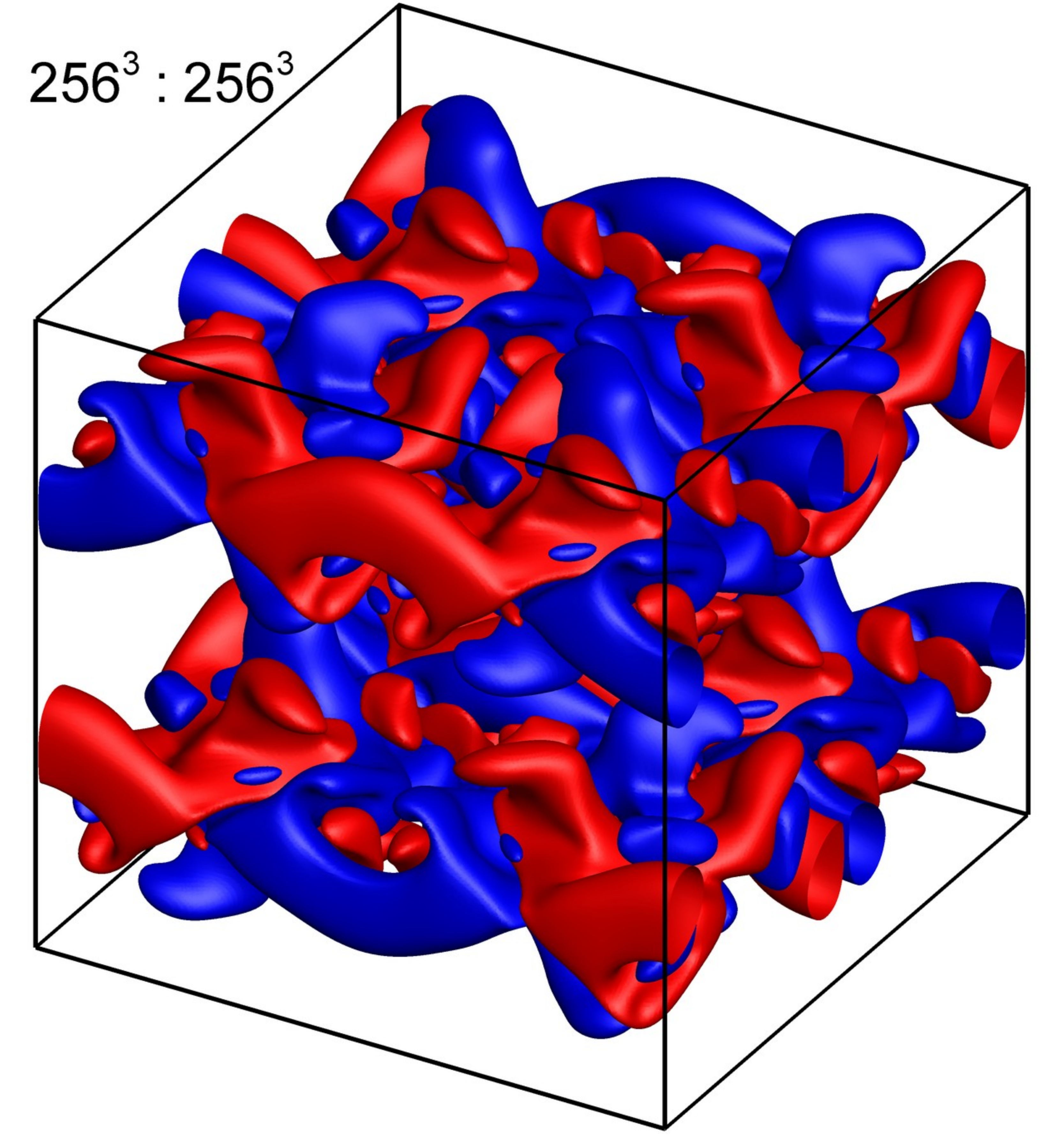}}
\subfigure{\includegraphics[width=0.33\textwidth]{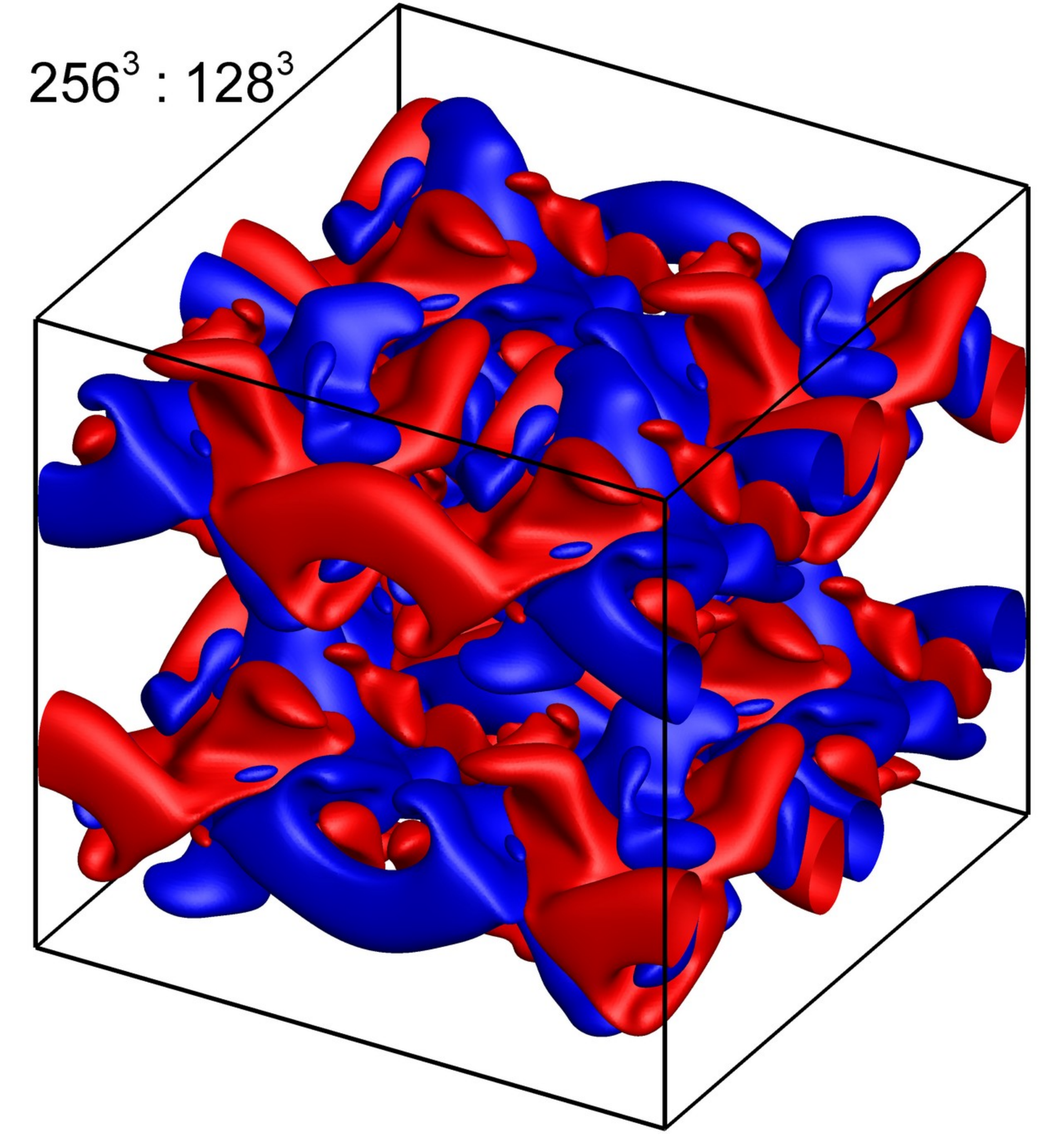}}
\subfigure{\includegraphics[width=0.33\textwidth]{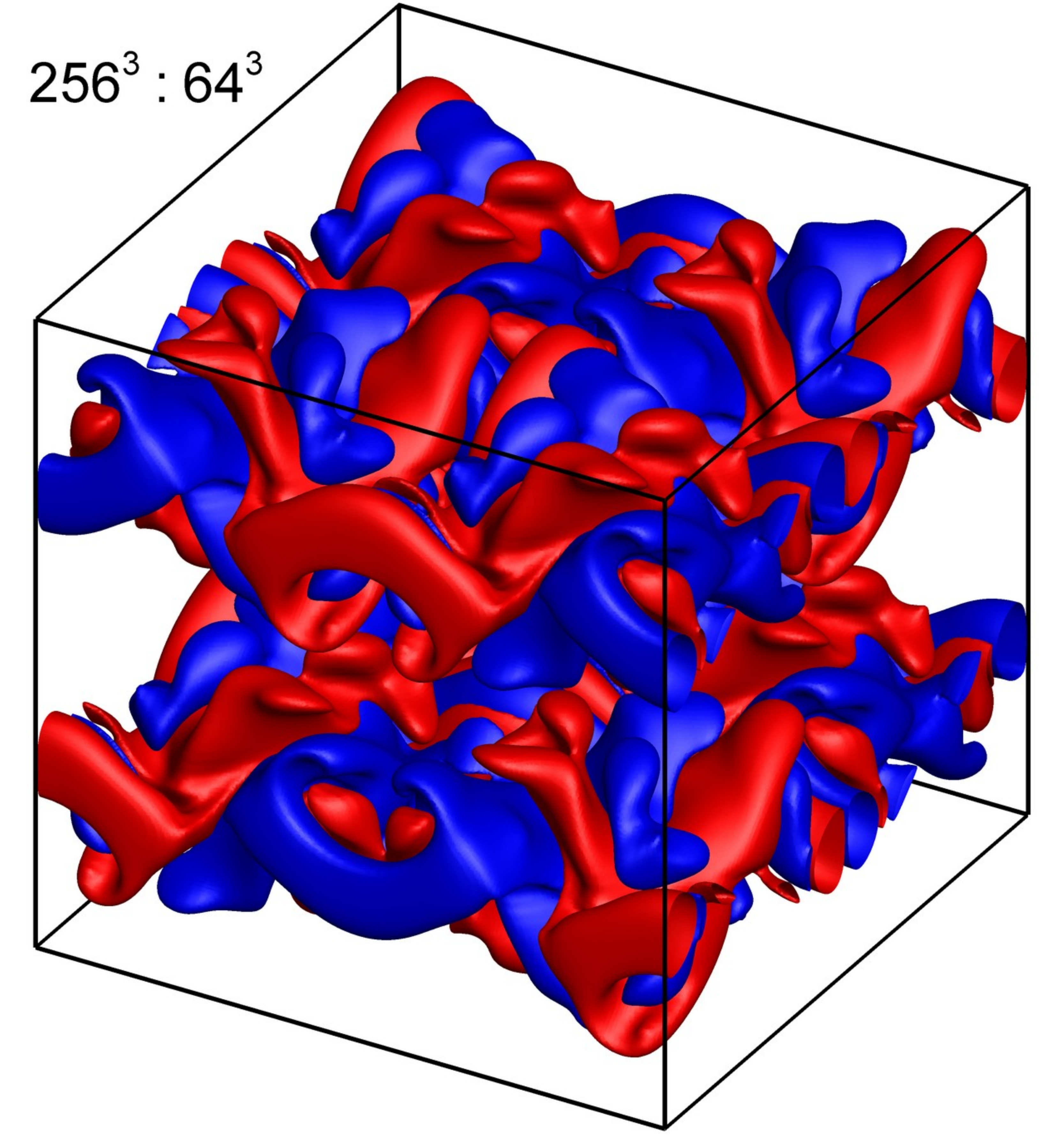}}
}
\\
\mbox{
\subfigure{\includegraphics[width=0.33\textwidth]{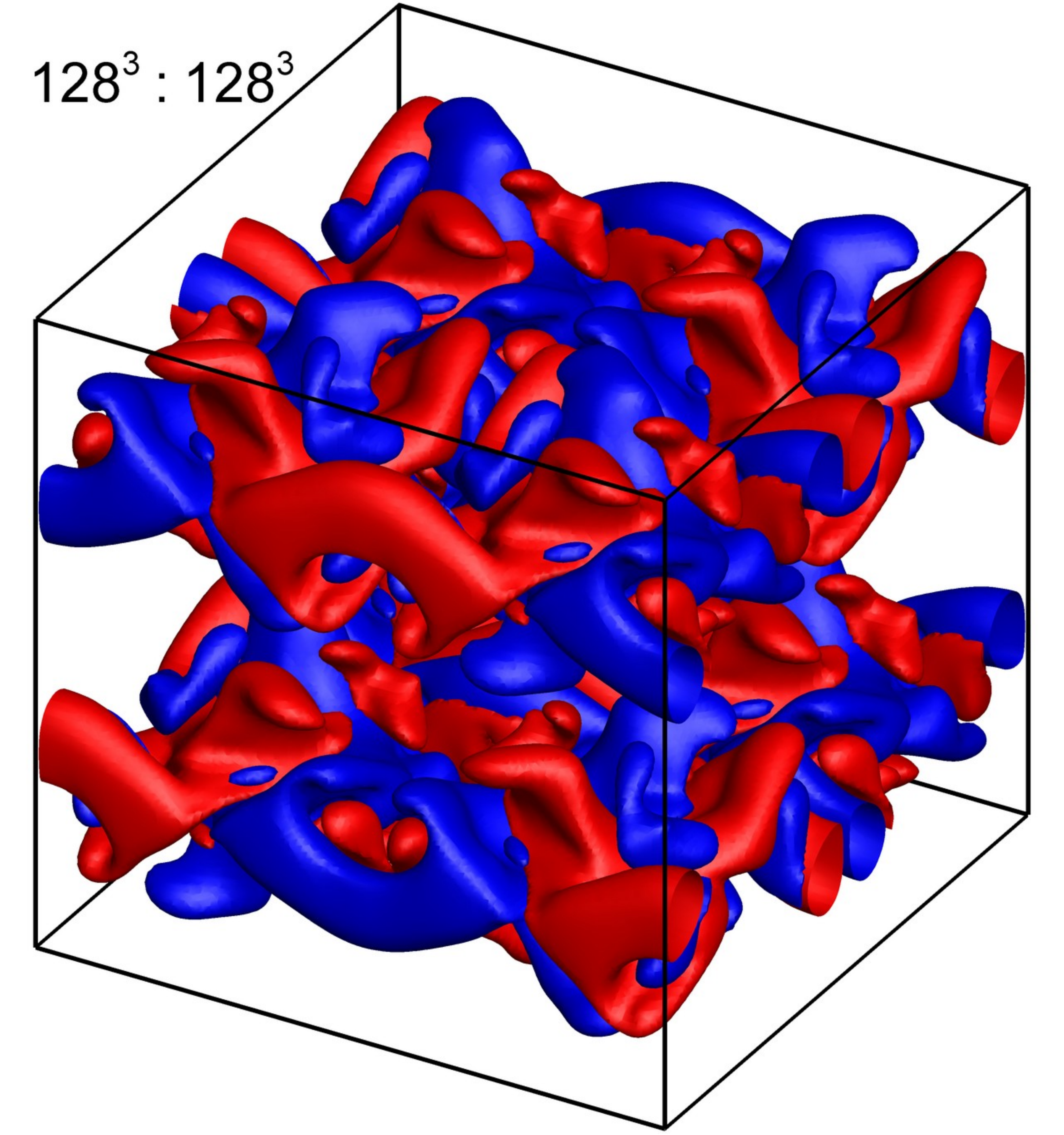}}
\subfigure{\includegraphics[width=0.33\textwidth]{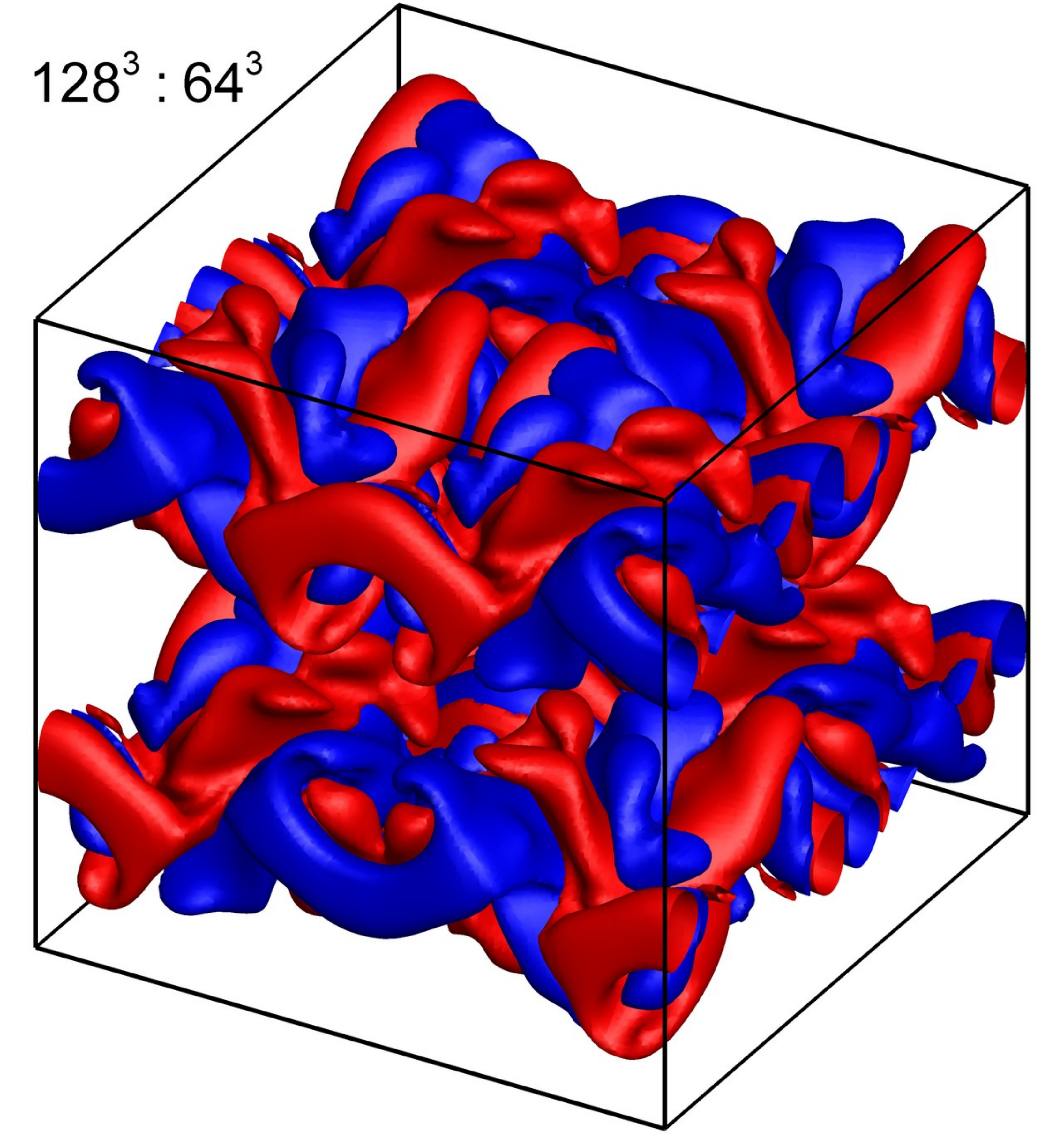}}
\subfigure{\includegraphics[width=0.33\textwidth]{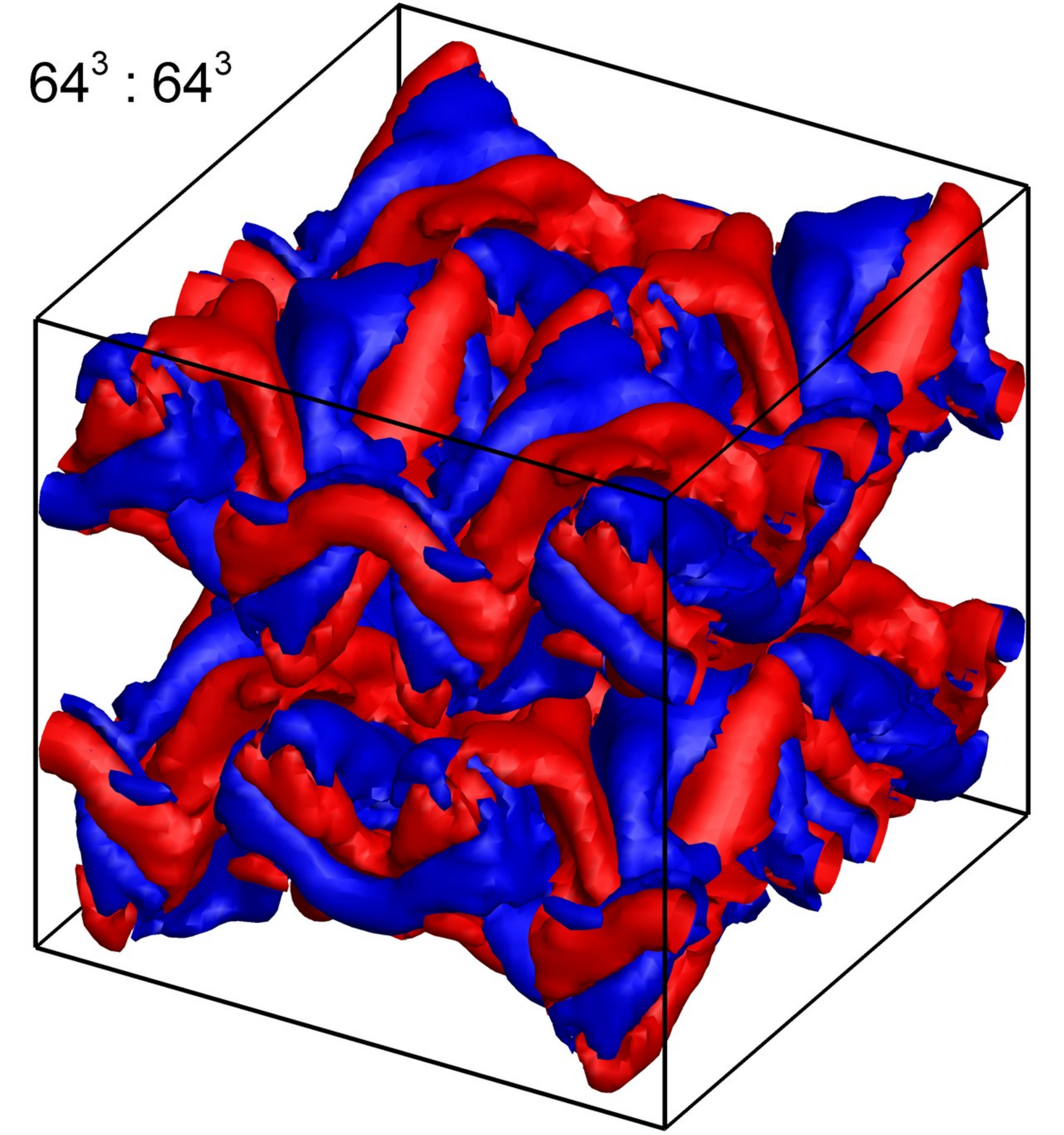}}
}
\caption{Comparison of the $y$-component of the vorticity at $t=10$ for $Re=200$. Iso-surfaces of $\omega_y =-0.5$ (blue) and $\omega_y = 0.5$ (red) are shown. Labels include the resolutions for both parts of the solver in the form $N^3 : M^3$, where $N^3$ is the resolution for the vorticity-transport equations, and $M^3$ is the resolution for the elliptic sub-problems.}
\label{fig:wy-05}
\end{figure*}

\begin{figure*}
\centering
\mbox{
\subfigure{\includegraphics[width=0.33\textwidth]{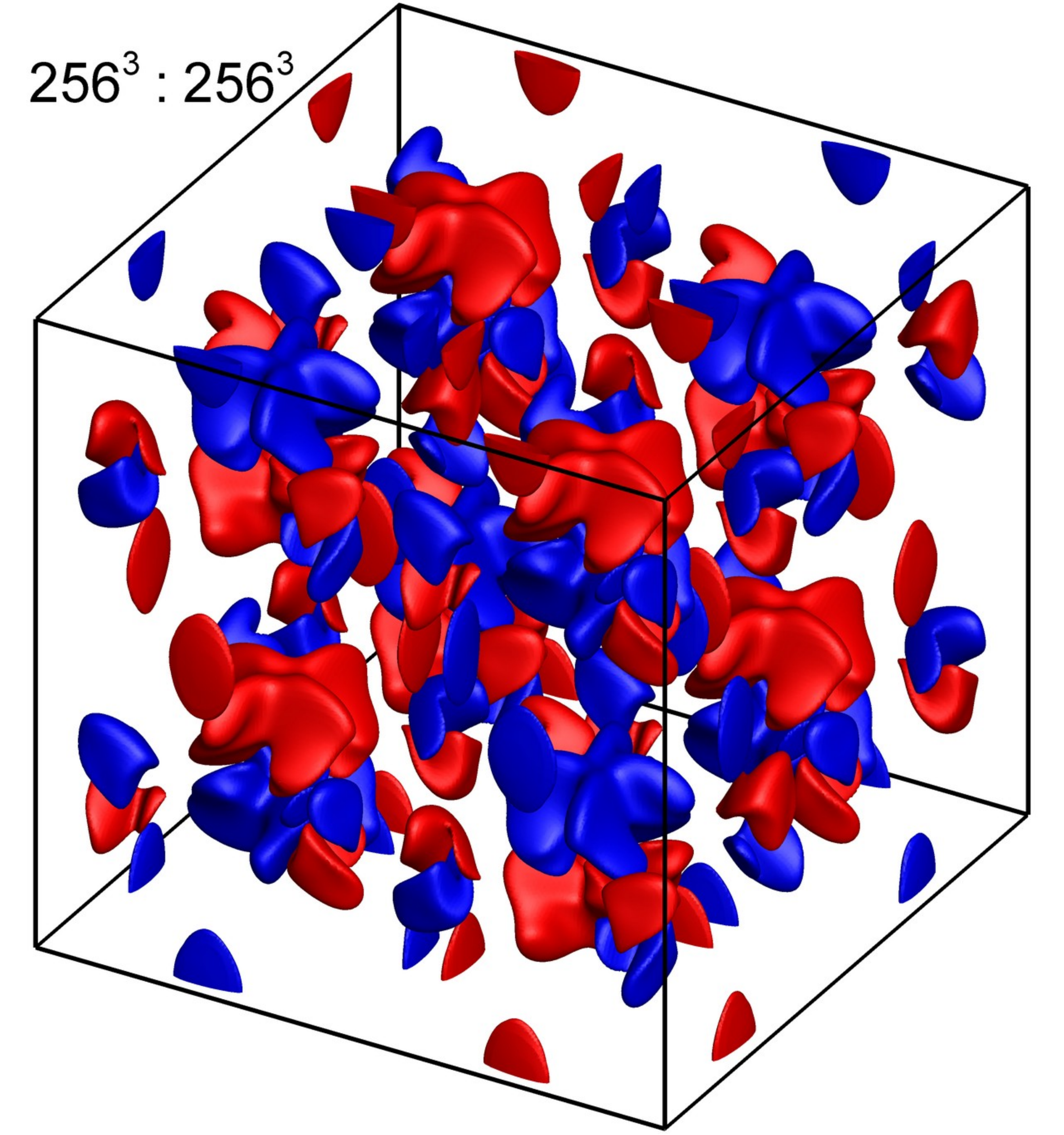}}
\subfigure{\includegraphics[width=0.33\textwidth]{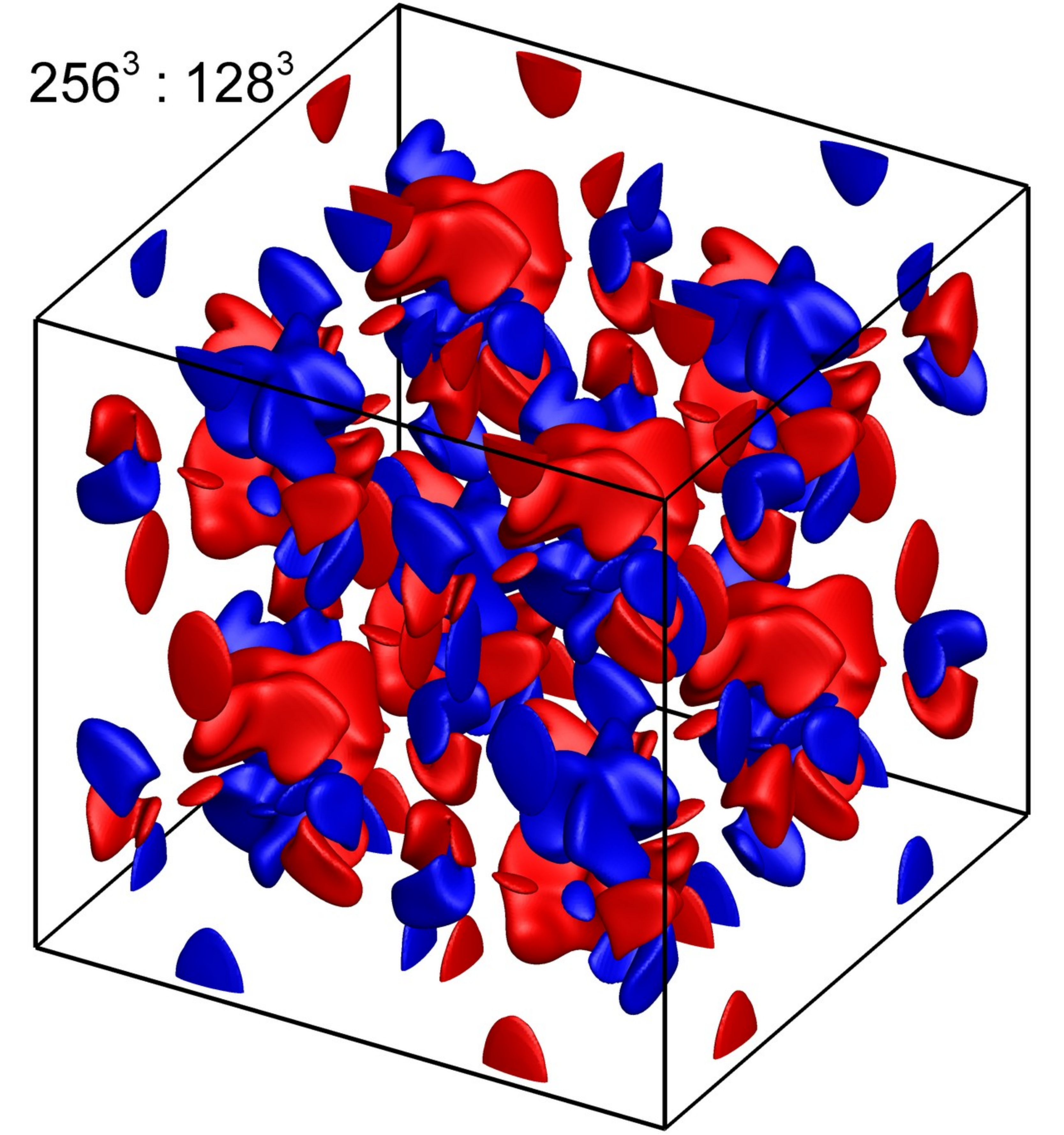}}
\subfigure{\includegraphics[width=0.33\textwidth]{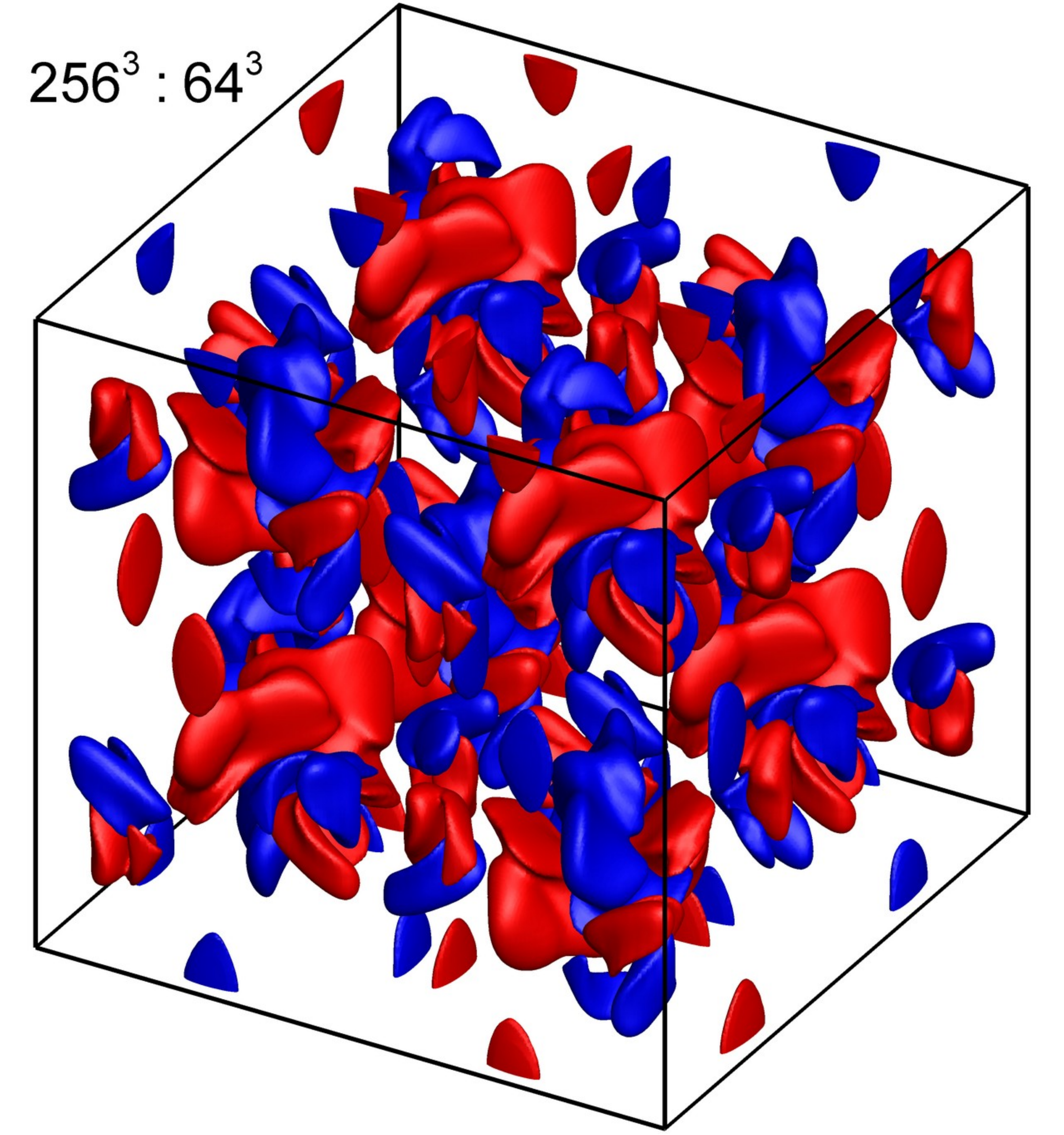}}
}
\\
\mbox{
\subfigure{\includegraphics[width=0.33\textwidth]{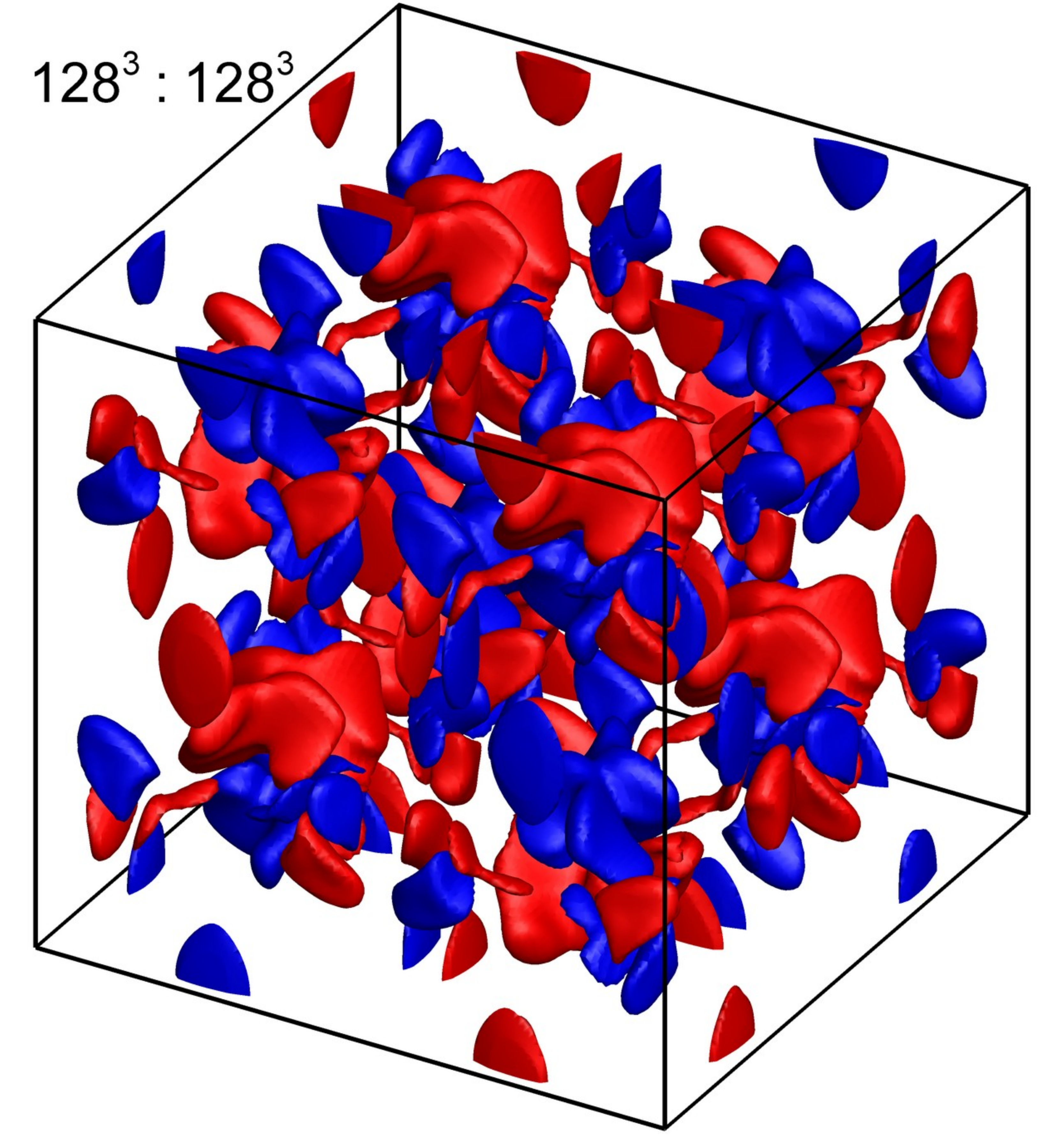}}
\subfigure{\includegraphics[width=0.33\textwidth]{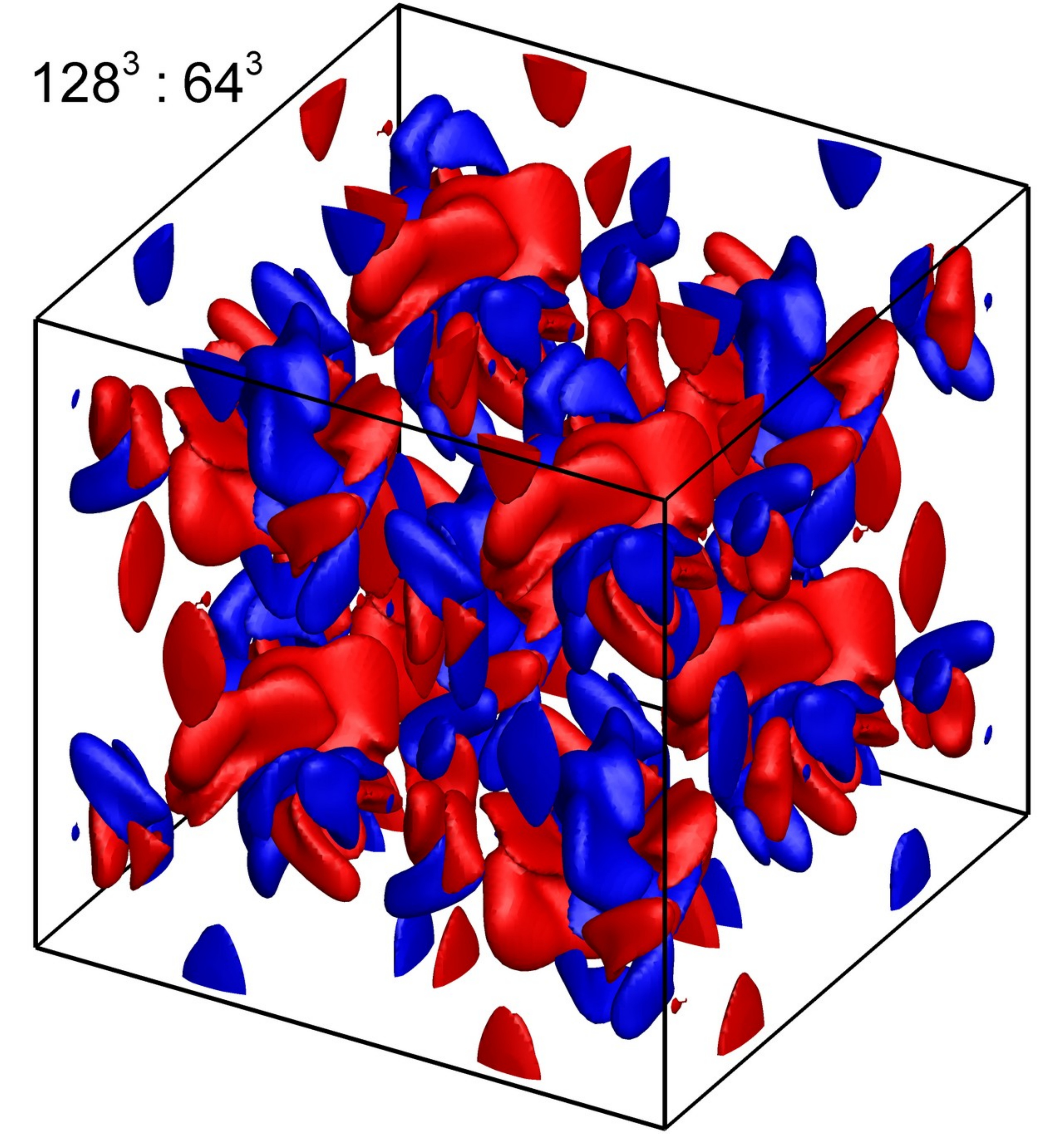}}
\subfigure{\includegraphics[width=0.33\textwidth]{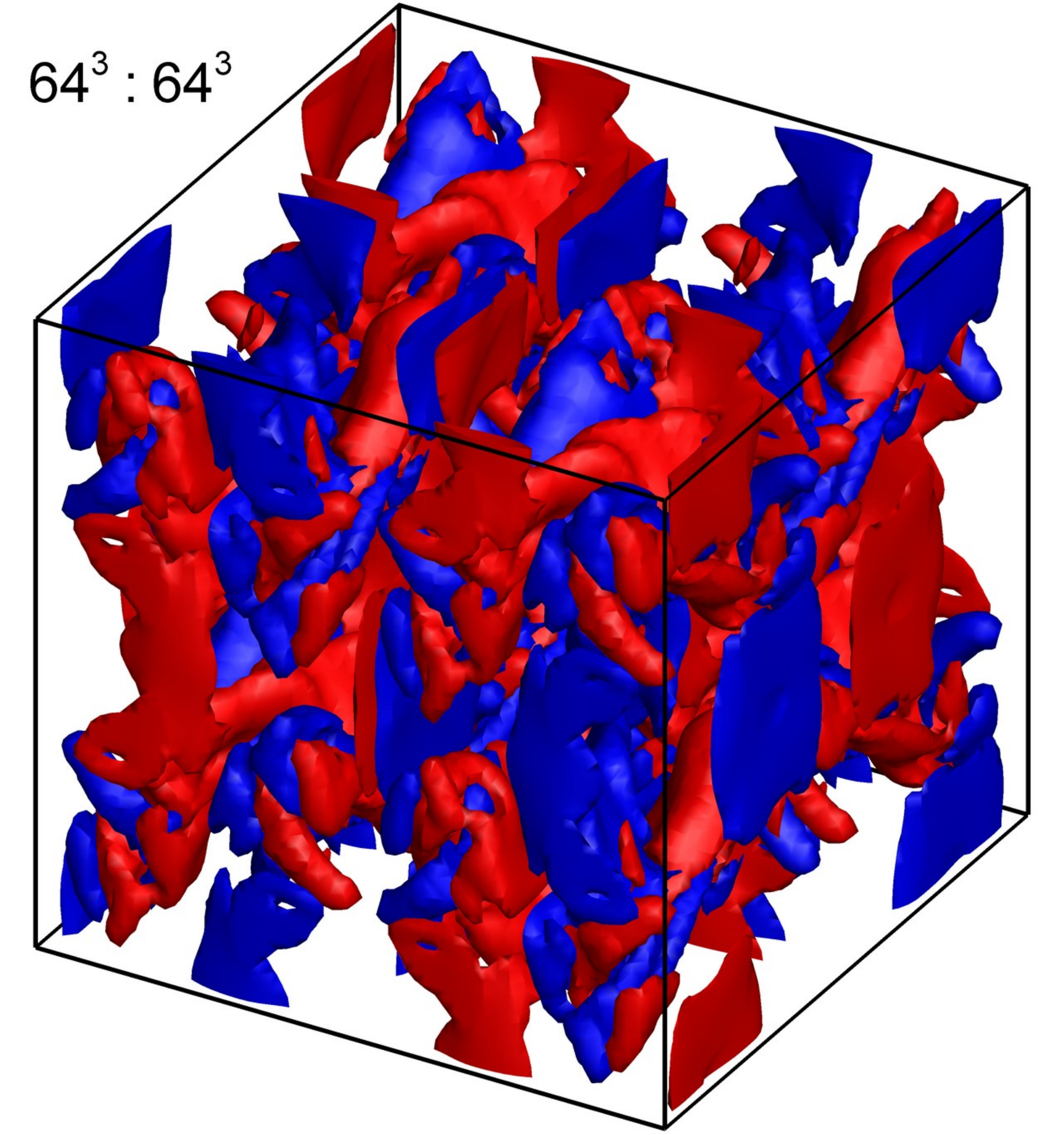}}
}
\caption{Comparison of the $z$-component of the vorticity at $t=10$ for $Re=200$. Iso-surfaces of $\omega_z =-1.0$ (blue) and $\omega_z = 1.0$ (red) are shown. Labels include the resolutions for both parts of the solver in the form $N^3 : M^3$, where $N^3$ is the resolution for the vorticity-transport equations, and $M^3$ is the resolution for the elliptic sub-problems.}
\label{fig:wz-05}
\end{figure*}

\begin{figure*}
\centering
\mbox{
\subfigure{\includegraphics[width=0.33\textwidth]{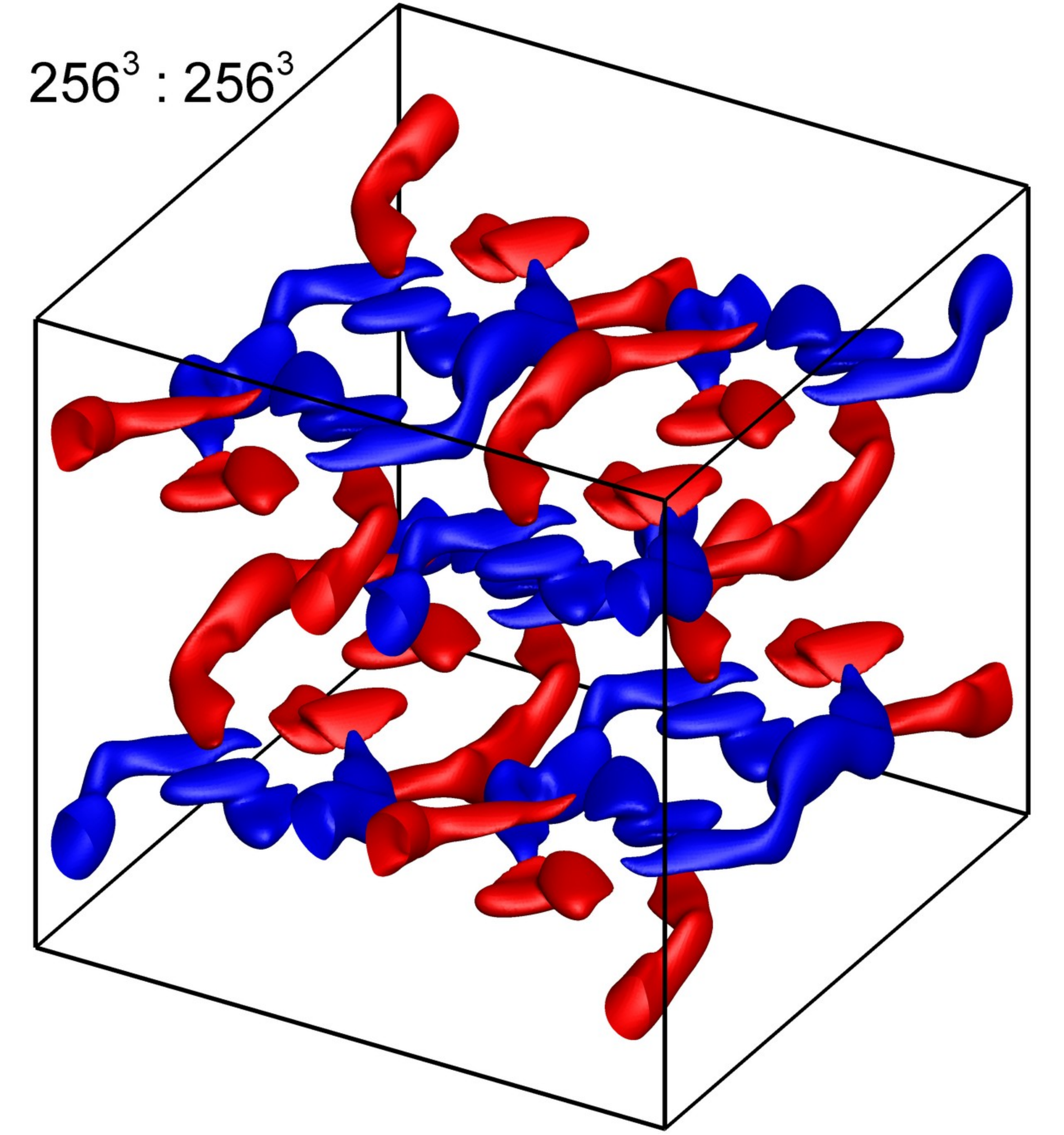}}
\subfigure{\includegraphics[width=0.33\textwidth]{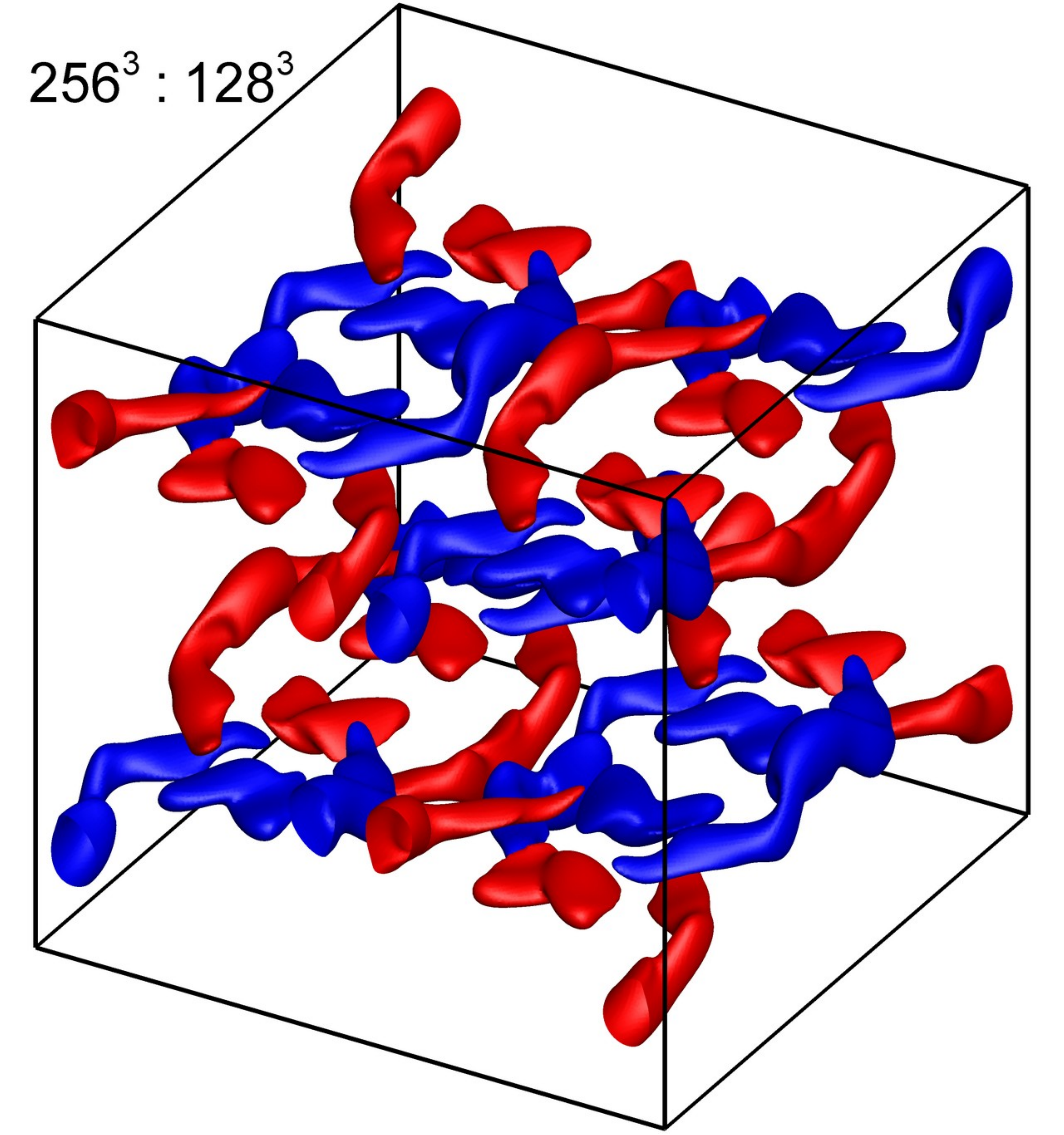}}
\subfigure{\includegraphics[width=0.33\textwidth]{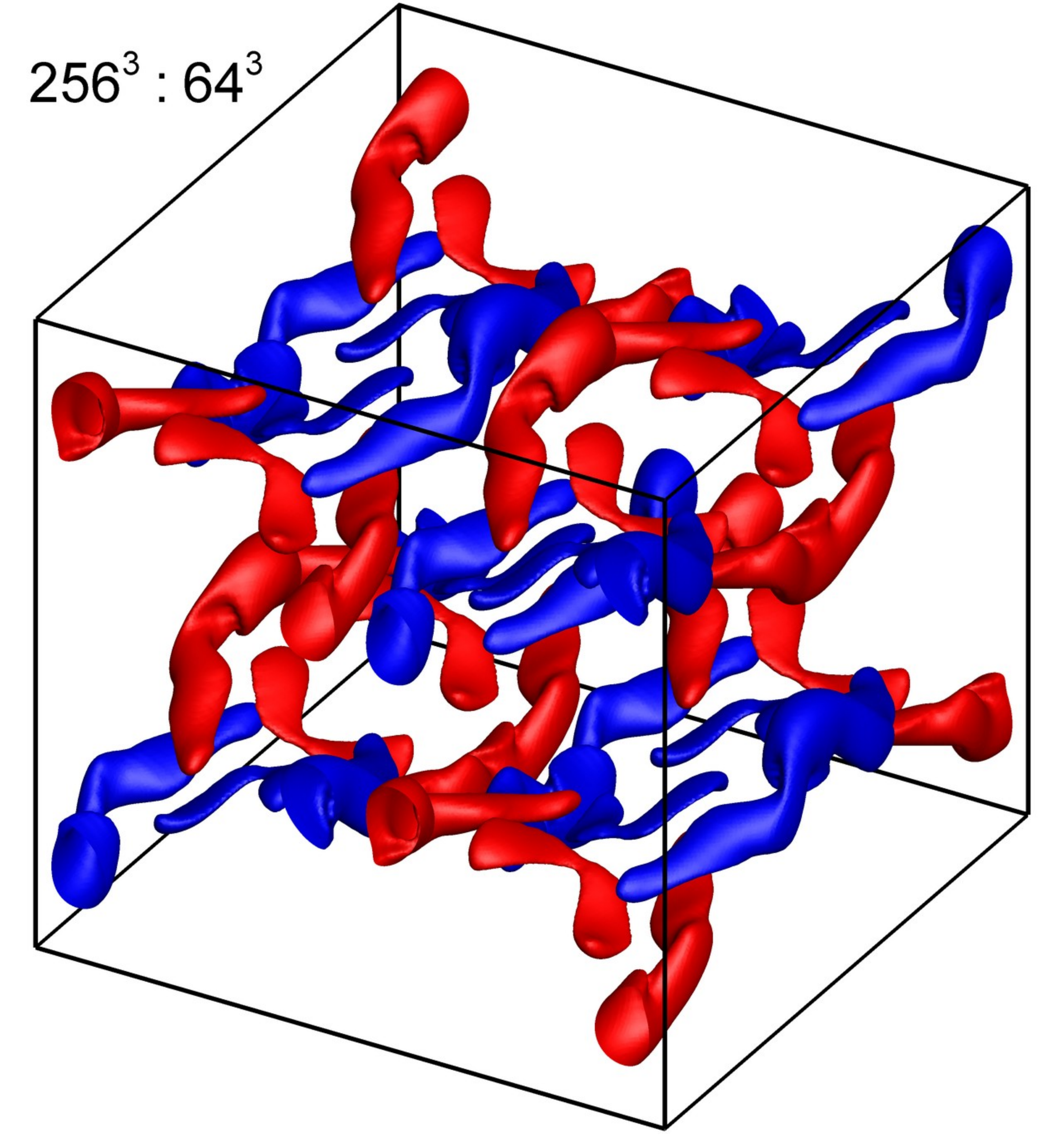}}
}
\\
\mbox{
\subfigure{\includegraphics[width=0.33\textwidth]{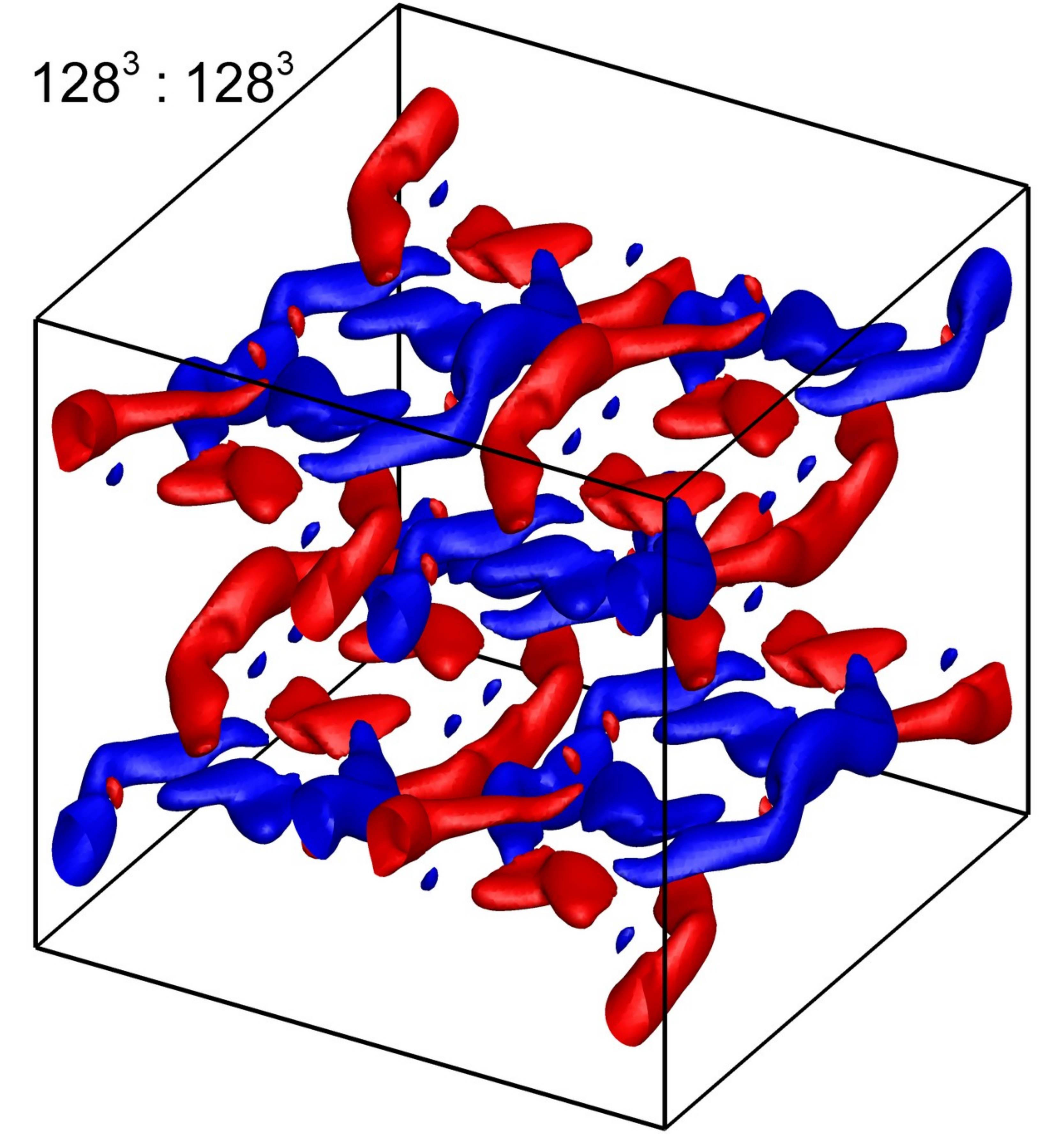}}
\subfigure{\includegraphics[width=0.33\textwidth]{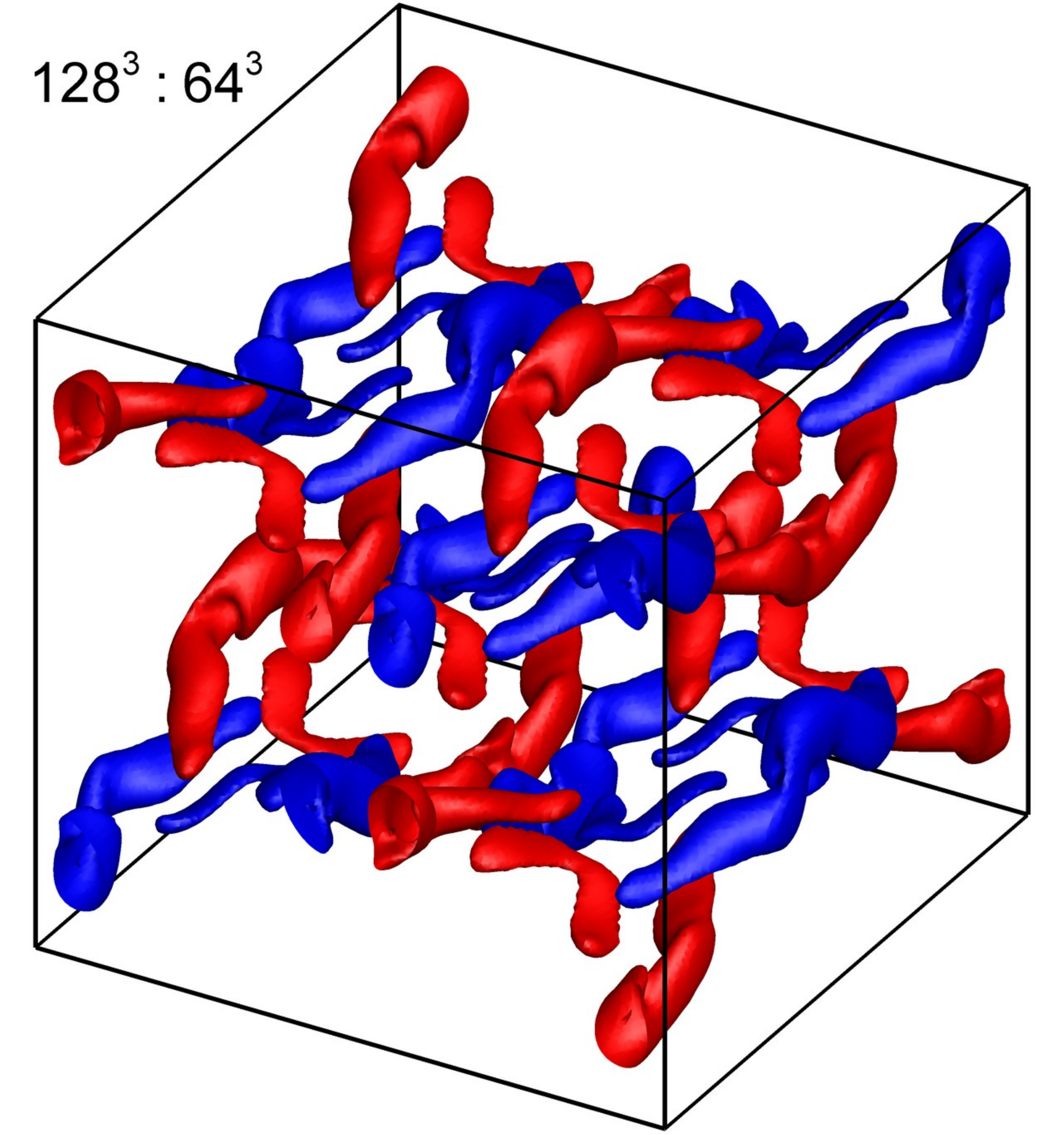}}
\subfigure{\includegraphics[width=0.33\textwidth]{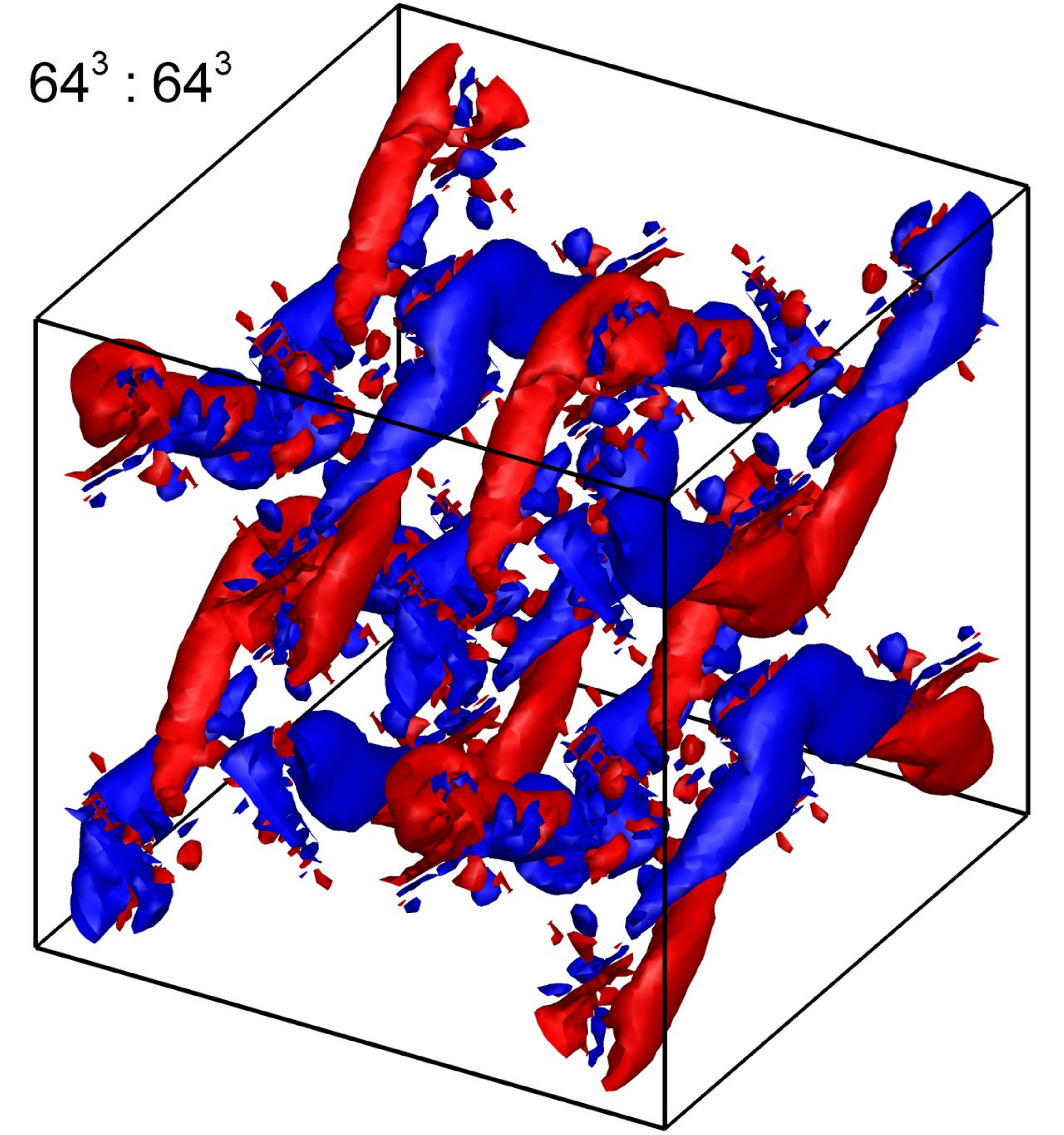}}
}
\caption{Comparison of the $x$-component of the vorticity at $t=10$ for $Re=200$. Iso-surfaces of $\omega_x =-2.0$ (blue) and $\omega_x = 2.0$ (red) are shown. Labels include the resolutions for both parts of the solver in the form $N^3 : M^3$, where $N^3$ is the resolution for the vorticity-transport equations, and $M^3$ is the resolution for the elliptic sub-problems.}
\label{fig:wx-20}
\end{figure*}

The evolution of flow field is shown in Fig.~\ref{fig:time-tgv3} in terms of the $x$-component of vorticity vector. Instantaneous vorticity isosurfaces colored blue for $\omega_x=-0.5$ and red for $\omega_x= 0.5$ are illustrated in this figure showing the generation of small scale structures by vortex stretching. Comparisons of the CGP method with the standard method are shown in Figs.~\ref{fig:wx-05}-\ref{fig:wz-05} for components of the vorticity field at time $t=10$. In order to visualize fine details of the flow field better, we also plot the vorticity isosurfaces for higher values of vorticity in Fig.~\ref{fig:wx-20}. In these figures, labels include the resolutions for both parts of the solver in the form $N^3:M^3$, where $N^3$ is the resolution for the vorticity-transport equations, and $M^3$ is the resolution for the elliptic sub-problems. It is clear from these figures that the results using the CGP method agree well with the results of the fine scale computations using the standard method with a considerable reduction in computational cost as reported in Table~\ref{tab:tgv3}. In this table, we also report the maximum values of $\omega_x$ in the simulation box and the corresponding percent errors computed according to the finest resolution. In these computations we use a linear-cost fast Poisson solver, and the CGP method yields a 5-8 fold reduction in computational cost. As discussed earlier, the speed ups would be greater if we used a quadratic-cost sub-optimal Poisson solver.

\section{Conclusions}
\label{sec:concl}
In the coarse-grid projection (CGP) methodology the cost of incompressible flow computations is reduced by coarsening the number of grid points used for the solution of the Poisson equation in the governing equations of incompressible fluid flows. The CGP framework is applicable to incompressible flow solvers that require solving a Poisson equation at every time step to enforce the divergence-free constraint on the velocity or vorticity field, which requires the bulk of the computational effort. The CGP approach is general and in fact constitutes a family of methods, since in addition to choosing the coarsening and prolongation operators and the time integration scheme, the Poisson solver used in the approach can vary.

In this work, we investigated the performance of a particular CGP method that uses an optimal Poisson solver, the full weighting operation for the coarsening operator, and bilinear interpolation for the prolongation operator in the structured grid framework. We tested the proposed approach for the vorticity-stream function, primitive variable fractional step, and vorticity-velocity formulations of the governing equations using finite differences for spatial derivatives and the third-order Runge-Kutta scheme for temporal integration. Seven different benchmark flow problems were solved to investigate the performance of the method. We first tested the CGP approach for four problems with periodic boundary conditions: the Taylor-Green decaying vortex problem, for which error norms were calculated via exact solutions, a double shear layer, the merging of a pair of co-rotating vortices, and two-dimensional decaying turbulence. We then used CGP for two problems on non-Cartesian grids, the Taylor-Green vortex problem on a distorted grid and the laminar flow over a circular cylinder, to investigate the performance of the CGP method in generalized curvilinear coordinates. Finally, we performed computations for the three-dimensional Taylor-Green vortex problem, which is perhaps the simplest test case in which to study the generation of small scale motions and turbulence, to test the applicability of the CGP method in the presence of vortex stretching and tilting.

Using the CGP method, very similar fine resolution field data were obtained at reduced computational costs for one level of coarsening. If 2$N$:$N$ (advection:Poisson) is the CGP resolution, 2$N$:2$N$ is the fine grid resolution, and $N$:$N$ is the coarse grid resolution, there is a significant benefit to using 2$N$:$N$ grid-based computations in terms of decreasing computational cost and obtaining accuracies close to 2$N$:2$N$ computations and significantly better than $N$:$N$ computations. Strong benefits to using 2$N$:$N$ simulations are demonstrated throughout the manuscript, in the selected problems. In some cases, which are problem and resolution dependent, there is very little loss of accuracy for the 2$N$:$N$ case compared to the 2$N$:2$N$ case with a significant reduction in computational time. This is particulary true for well resolved cases in which the coarsening in the Poisson equation behaves as a low-pass filter and does not reduce the accuracy of the advection-diffusion part as much as pure coarsening does due to the corresponding prolongation operator. For two further levels of grid coarsening (reducing the number of grid points in each direction by factors of 4 and 8, respectively), increased reductions in computational time were found, but the accuracy of the fine resolution field data was reduced at each level of coarsening. Speed up factors between 2 and 42 were found using the CGP method, and the speed-up obtained was found to increase with increasing resolution of the fine level field data, as is to be expected. We also found that the acceleration rate of the computations due to CGP increases with the grid distortion ratio in curvilinear grid settings.

The proposed CGP approach works independently of the choice of Poisson solver and the choice of algorithm for the advection-diffusion part. It uses them as black-box solvers and uses simple interpolations between them. In this work we used optimal Poisson solvers with linear computational cost in all cases, either a FFT-based fast Poisson solver or a V-cycle multigrid Poisson solver, as our black-box solver. We expect the speed-up of the CGP method to increase dramatically for versions of the method that use other, suboptimal, Poisson solvers (which are generally $O(N^2)$). The results of these computations demonstrate that this version of the CGP approach is well suited for accelerating incompressible flow solvers that require the solution of a Poisson equation as a constraint to produce a divergence-free field. This suggests that the method may provide a useful tool for computing incompressible flows both in Cartesian and non-Cartesian domains. The current study yields promising results for structured Cartesian and curvilinear grids. The three-dimensional computations also suggest that the CGP method can be used to accelerate the DNS of turbulence by coarsening the number of grid points for the Poisson solver without affecting the well resolved DNS data, a topic we intend to investigate further in a future study.

\section*{Acknowledgements}
The authors are grateful to the anonymous reviewers for their constructive comments on an earlier version of this manuscript. This research was partially supported by the Institute for Critical Technology and Applied Science (ICTAS) at Virginia Tech via grant number 118709.

\bibliographystyle{elsarticle-num}
\bibliography{ref}

\end{document}